\documentclass[acmtog]{acmart}
\usepackage{booktabs} 
\usepackage[colorinlistoftodos]{todonotes}
\usepackage{overpic}
\usepackage{wrapfig}
\usepackage{subcaption}
\usepackage{verbatim}
\usepackage{amsmath}
\usepackage{stackrel}
\usepackage{makecell}
\usepackage{afterpage}
\usepackage{multirow, makecell}
\usepackage{array}
\usepackage[ruled,vlined, linesnumbered]{algorithm2e}
\usepackage{algpseudocode}
\usepackage{graphicx}
\usepackage{grffile}
\usepackage{hyperref}

\SetKwIF{If}{ElseIf}{Else}{if}{}{else if}{else}{end}%
\SetKwFor{While}{while}{}{end while}%
\SetKwRepeat{Do}{do}{while}
\SetKw{KwGoTo}{go to}

\acmSubmissionID{416}

\usepackage[ruled]{algorithm2e} 

\SetAlFnt{\small}
\SetAlCapFnt{\small}
\SetAlCapNameFnt{\small}
\SetAlCapHSkip{0pt}

\usepackage[normalem]{ulem}

\newcommand\hide[1]{}

\definecolor{cbcolor}{RGB}{210,10,210}
\definecolor{vcacolor}{RGB}{123,50,210}
\definecolor{jccolor}{RGB}{60,145,110}

\usepackage[export]{adjustbox}

\setcopyright{acmlicensed}
\acmJournal{TOG}
\acmYear{2022}
\acmVolume{41}
\acmNumber{6}
\acmArticle{243}
\acmMonth{12} 
\acmDOI{10.1145/3550454.3555498}

\begin{document}
\title{Curl-Flow: Boundary-Respecting Pointwise Incompressible Velocity Interpolation for Grid-Based Fluids}
\author{Jumyung Chang}
\affiliation{%
\institution{NVIDIA}
\city{Toronto}
}
\email{jumyungc@nvidia.com}
\author{Ruben Partono}
\affiliation{%
\institution{University of Waterloo}
\city{Waterloo}
}
\email{rsparton@uwaterloo.ca}
\author{Vinicius C. Azevedo}
\affiliation{%
\institution{ETH Zurich}
\city{Zurich}
}
\email{vinicius.azevedo@inf.ethz.ch}

\author{Christopher Batty}
\affiliation{%
\institution{University of Waterloo}
\city{Waterloo}
}
\email{christopher.batty@uwaterloo.ca}

\begin{abstract}
We propose to augment standard grid-based fluid solvers with \emph{pointwise} divergence-free velocity interpolation, thereby ensuring exact incompressibility down to the sub-cell level. Our method takes as input a discretely divergence-free velocity field generated by a staggered grid pressure projection, and first recovers a corresponding discrete vector potential. Instead of solving a costly \emph{vector} Poisson problem for the potential, we develop a fast parallel sweeping strategy to find a candidate potential and apply a gauge transformation to enforce the Coulomb gauge condition and thereby make it numerically smooth. Interpolating this discrete potential generates a pointwise vector potential whose analytical curl is a pointwise incompressible velocity field. 
Our method further supports irregular solid geometry through the use of level set-based cut-cells and a novel Curl-Noise-inspired potential ramping procedure that simultaneously offers strictly non-penetrating velocities and incompressibility. Experimental comparisons demonstrate that the vector potential reconstruction procedure at the heart of our approach is consistently faster than prior such reconstruction schemes, especially those that solve vector Poisson problems. Moreover, in exchange for its modest extra cost, our overall \emph{Curl-Flow} framework produces significantly improved particle trajectories that closely respect irregular obstacles, do not suffer from spurious sources or sinks, and yield superior particle distributions over time.
\end{abstract}

%
%

\ccsdesc[500]{Computing methodologies~Physical simulation}

%
%

\keywords{divergence-free, stream function, vector potential, velocity interpolation, advection }

\begin{teaserfigure}
	\centering	
	\begin{subfigure}[t]{0.33\textwidth}
		\begin{overpic}[width=\textwidth]{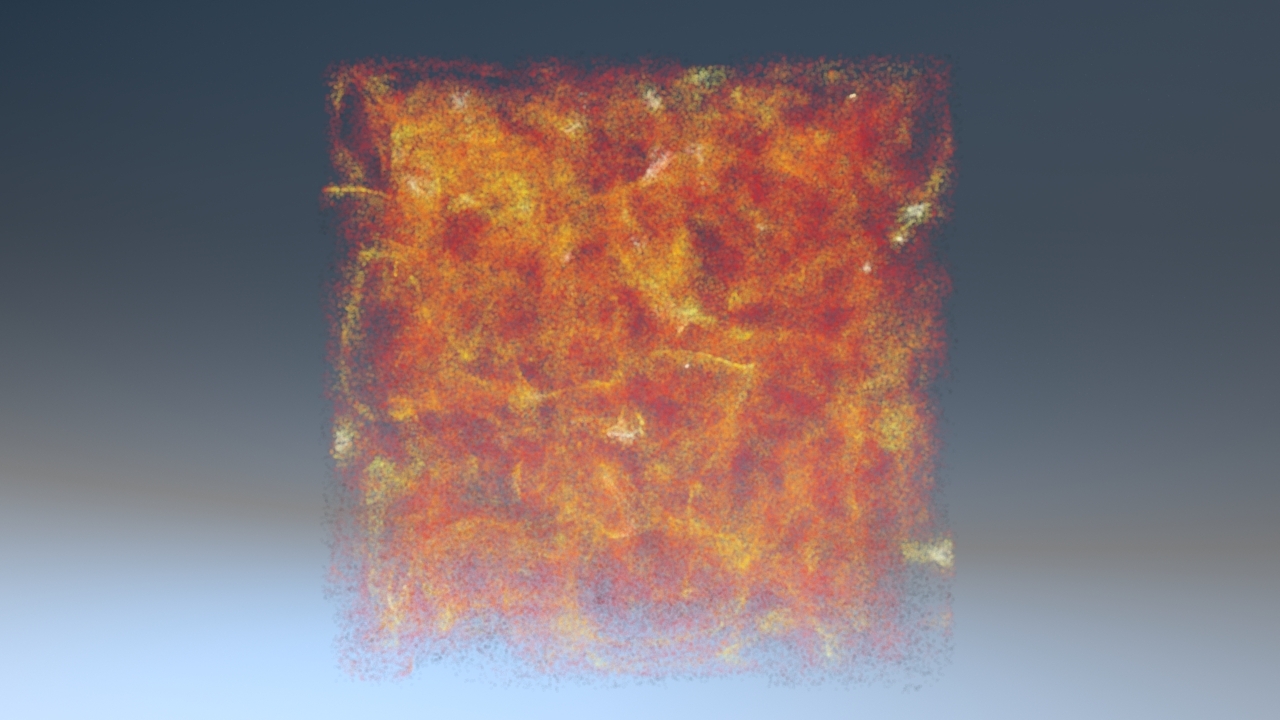}
		\end{overpic}    
	\end{subfigure} 
	\hfill
	\begin{subfigure}[t]{0.33\textwidth}
		\begin{overpic}[width=\textwidth]{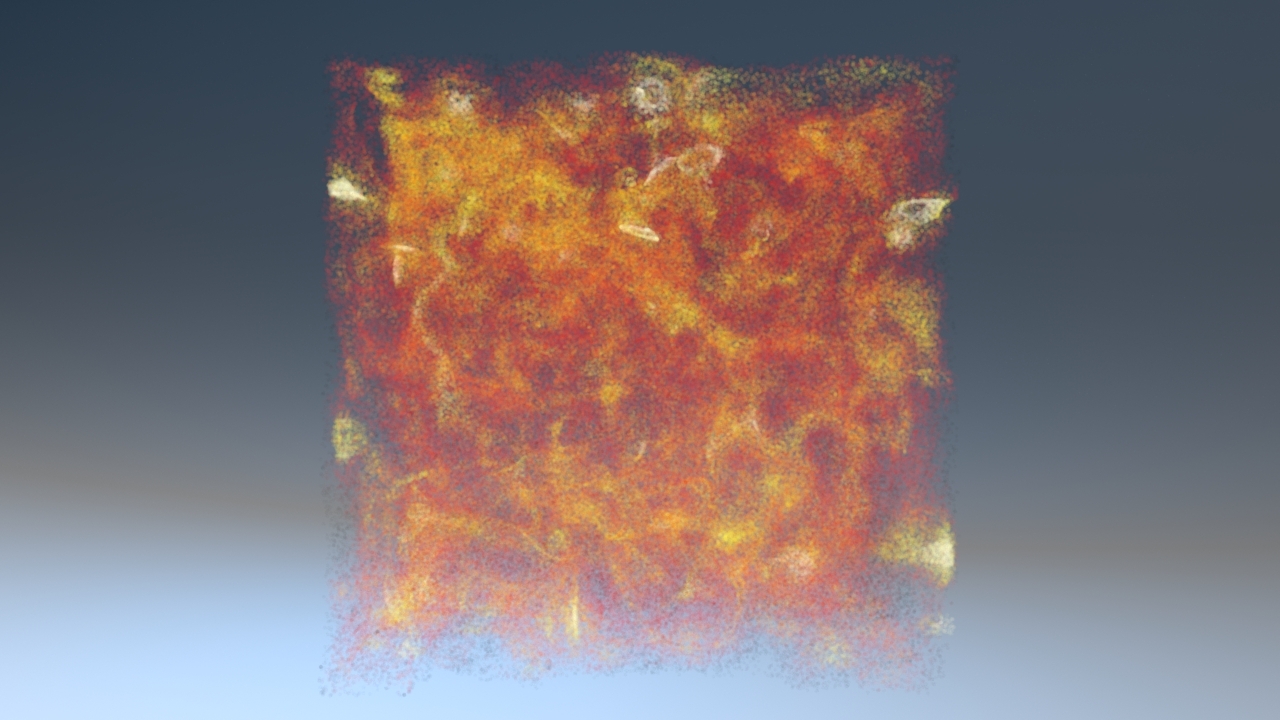}
		\end{overpic}   
	\end{subfigure}
	\hfill
	\begin{subfigure}[t]{0.33\textwidth}
		\begin{overpic}[width=\textwidth]{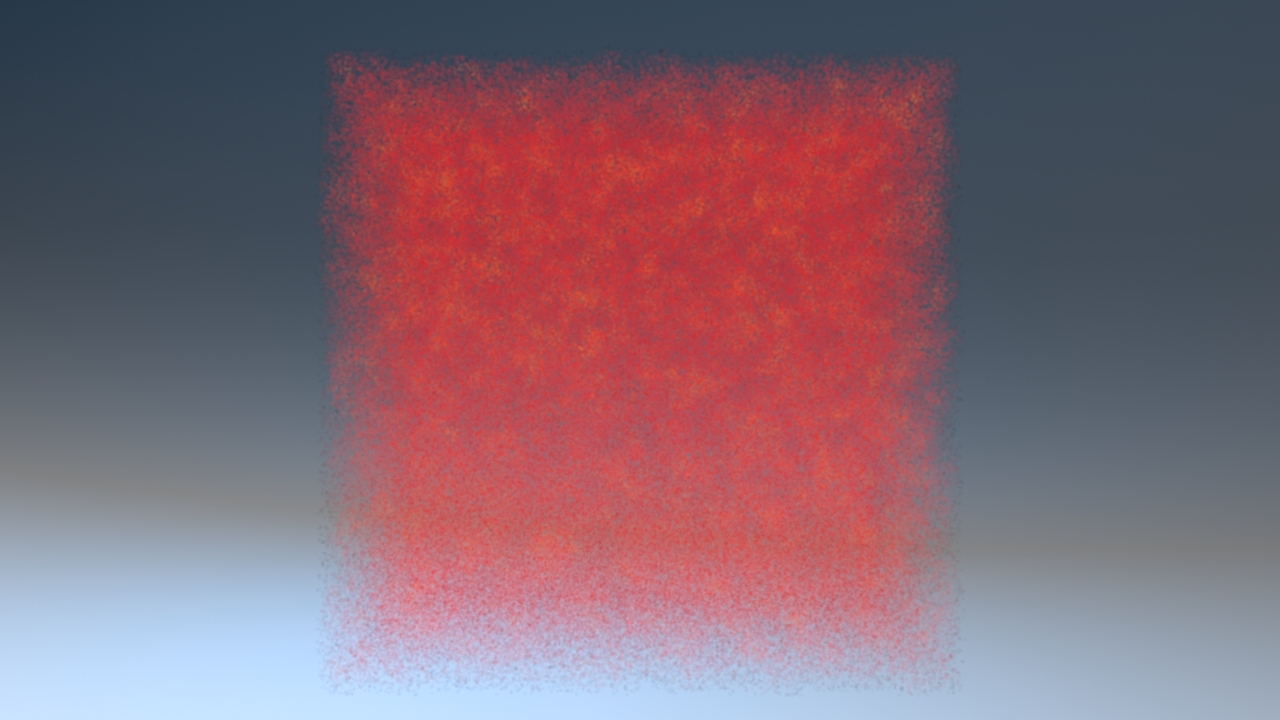}
		\end{overpic}   
	\end{subfigure}
    \caption{ \textbf{Effect of Interpolants on Particle Distribution in 3D:}
		The results of identical, initially uniform dense particle distributions advected through the same $10\times 10\times 10$ \emph{discretely} incompressible field for 160 frames using different velocity interpolants. Left is standard trilinear velocity interpolation, center is monotonic cubic velocity interpolation \cite{Fritsch1980}, and right is our Curl-Flow method. The coloring indicates the per-particle local density estimate, with yellow/white being highest density. Basic polynomial interpolants lead to undesirable severe clustering and spreading due to grid-scale divergence, while our pointwise incompressible Curl-Flow interpolation preserves close-to-uniform particle distributions over long time periods.}
	\label{fig:particleDistribution3D}
\end{teaserfigure}

\maketitle
\section{Introduction}
The assumption of incompressibility is pervasive in computer animation of fluids. Since compressive effects are imperceptible in many (but of course not all) visually relevant liquid and gas scenarios, neglecting fast-moving compression waves is often justifiable in practice and yields significant efficiency gains. Mathematically, incompressibility implies that the fluid velocity field $\mathbf{u}$ should be divergence-free: $\nabla \cdot \mathbf{u} = 0$.
Popular staggered grid-based schemes rely on this assumption,  using finite difference or finite volume ideas, combined with Lagrangian or semi-Lagrangian advection methods, to transform the continuous incompressible flow equations into discrete, computable algorithms \cite{Bridson2015}. This fertile mathematical soil has sprouted diverse numerical tools for visual simulation of drifting cigarette smoke, coiling honey, crashing ocean waves, and more \cite{Larionov2017,Fedkiw2001,Enright2002}, which are widely integrated into industrial software like \emph{Houdini}, \emph{Maya}, and \emph{Blender}. Yet despite the widespread use of grid-based incompressible flow animation techniques, the pointwise velocity vector fields they produce are in fact \emph{not incompressible}.

To understand this potentially surprising statement, one must distinguish \emph{discrete} incompressibility from its \emph{continuous} counterpart. Under a finite volume approach, the discrete velocity components, stored at cell face midpoints, will indeed satisfy discrete incompressibility: the sum of discrete fluxes across each cell's boundary is zero. However, interpolation is often required to provide \emph{pointwise} velocity values everywhere in the simulation domain, to support popular Lagrangian or semi-Lagrangian discretizations of advection for passive tracer particles, density, velocity, temperature, and so on \cite{Stam1999,Zhu2005,Jiang2015}. Applying basic polynomial interpolants on discretely incompressible grid data affords no guarantee that the interpolated velocity fields will be \emph{pointwise analytically incompressible}, and in practice they are not.

Spurious compressibility has non-negligible implications. Advecting particles through vector fields with artificial sources and sinks damages volume conservation and causes uniformly distributed particles to clump and spread. Alternatively, reducing discretization error by significantly increasing grid resolution fails to address the root cause and is too costly regardless: simulation time typically scales at least quartically with grid resolution. Irregular solid boundaries further exacerbate the issue because naive grid-based interpolants are essentially oblivious to obstacles; obstacle-aware cut-cell interpolants \cite{Azevedo2016} improve the no-normal-flow enforcement at the cost of worsening compression artifacts.

To guarantee an analytically divergence-free interpolated velocity field by construction, we instead form the velocity field ${\mathbf{u}}$ from the curl (${\nabla \times}$) of a second vector field, ${\boldsymbol{\psi}}$, known as a \emph{vector potential}:
\begin{equation}
\label{eq:vectorPotential}
\mathbf{u} = \nabla \times \boldsymbol{ \psi}.
\end{equation}
In 2D,  ${\boldsymbol{\psi}}$ reduces to a scalar \emph{stream function}, $\psi$.
Incompressibility is thus enforced by a basic vector calculus identity,
\begin{equation}
\label{eq:divCurl}
\nabla \cdot \mathbf{u} = \nabla \cdot \nabla \times \boldsymbol{ \psi} = 0.
\end{equation}
We exploit this relationship in the discrete and continuous settings to ensure pointwise incompressibility through three steps:
\begin{enumerate}
\item Given discretely incompressible fluid velocity values, construct corresponding discrete vector potential values.
\item Interpolate the discrete vector potential values to yield a pointwise vector potential field.
\item Take the analytical curl of the interpolated vector potential field to yield a \emph{pointwise incompressible} velocity field.
\end{enumerate}
\citet{Bao2017} first proposed this three-step recipe for deriving discrete delta functions in an immersed boundary method for weakly coupled fluid-structure interaction in computational fluid dynamics. However, their method is limited to perfectly uniform rectangular domains (i.e., no irregular solids) with periodic boundaries. Even so, their potential reconstruction approach requires the solution of a costly vector Poisson problem, making it several times more expensive than pressure projection in a standard fluid solver.

We extend this conceptual framework to be practical for animation applications in two and three dimensions by overcoming several challenges, including enabling much more efficient recovery of discrete vector potentials, enforcing exact no-flow or prescribed flux boundary conditions for exterior boundaries, and supporting irregular cut-cell obstacles. Our proposed \emph{Curl-Flow} framework offers the following technical contributions:
\begin{itemize}
\item An efficient parallel sweeping algorithm to recover velocity-consistent discrete vector potentials on uniform and level set-based cut-cell grids;
\item A gauge correction strategy to make the candidate vector potential well-suited to interpolation (i.e., numerically smooth) using a single fast, scalar Poisson solve;
\item The design of an efficient, low-order vector potential interpolant that efficiently provides continuous and pointwise incompressible velocities in free space;
\item A new additive ramping strategy for improved free-slip boundaries in the "Curl-Noise" method of \citet{Bridson2007};
\item The application of this ramping method during velocity interpolation to enforce divergence-free, \emph{exact} no-normal-flow conditions on axis-aligned exterior domain boundaries and interior irregular static obstacles in both 2D and 3D.
\end{itemize}

\section{Related Work}
\label{sec:relatedWork}

\subsection{Within Computer Graphics}
\subsubsection{Recovering Discrete Vector Potentials} A few vector potential-based Eulerian/hybrid solvers have been proposed in the fluid animation literature, but most operate solely in the discrete realm and require solving a costly \emph{vector} Poisson equation. \citet{Elcott2007} solved a vector Poisson problem to recover the vector potential from vorticity within a simplicial discretization of the vorticity equation. \citet{Ando2015} developed a vector Poisson-based alternative to the standard grid-based pressure projection. Notably, \citet{Ando2015} were the first to suggest employing vector potential-based velocity interpolation, in their discussion of future work. For an ideal Cartesian grid with only axis-aligned boundaries and certain choices of boundary conditions, the vector Poisson system can be decoupled into three scalar Poisson problems for efficiency; Sato et al.\ (likely) used this approach to recover the vector potential from a velocity field in fluid control applications \cite{Sato2015,Sato2021}. Our proposed discrete vector potential reconstruction strategy offers a significantly more efficient solution, finding the desired potential at essentially the cost of a single scalar Poisson solve if the input velocity field is divergence-free, or two otherwise.

For 2D flow visualization, \citet{Biswas2016} recovered the discrete (scalar) stream function from velocity data and used marching squares to construct streamlines. They proposed an axis-based sweeping approach for 2D stream function recovery; we generalize this idea to irregular cut-cell boundaries and 3D vector potentials.

Relatedly, discrete vector field (Hodge) decomposition techniques have long been of interest in graphics \cite{Tong2003}. Recently, more elaborate five-component discrete decompositions were considered by \citet{Poelke2016} and \citet{Zhao2019} for vector fields over triangulated surfaces and tetrahedral volumes, respectively, including various gauges and boundary conditions. These approaches determine the vector potential directly from the vector field via a vector Poisson solve, and in the five-component case, further require solving eigenproblems. 

\subsubsection{Pointwise Divergence-Free Fields} Various procedural techniques have used pointwise divergence-free velocity fields. Stam and Fiume \shortcite{Stam1993Turbulent} first explored this idea for gaseous phenomena, treating incompressibility by restricting cross-spectral density functions in the Fourier domain. Curl-Noise \cite{Bridson2007} uses a continuous vector potential to design animated divergence-free vector fields and the related scheme of \citet{DeWolf2006} uses the cross-product of two gradient fields. Divergence-free sub-grid turbulence models subsequently built on these ideas \cite{Kim2008,Schechter2008}. Incidentally, \citet{Schechter2008} also used (but did not describe) a divergence-free hybrid constant-linear velocity interpolant (see their supplemental video at time 2:18-2:27, left), which we show can be derived from bilinear stream function interpolation. This is the \emph{only} prior instance of divergence-free grid-based interpolation in computer animation we know of.
Pointwise divergence-free vector fields have also been used in geometry processing for volume-preserving modeling \cite{VonFunck2006} and shape interpolation \cite{Eisenberger2018}.

\subsubsection{Divergence-Free Simulators} Lagrangian vorticity-based simulation methods also employ a secondary vector variable to construct analytically divergence-free velocity fields; specifically, velocity is computed from vorticity via the Biot-Savart law. These methods come in many forms, including particles \cite{Park2005}, filaments \cite{Angelidis2005}, and sheets \cite{Brochu2012,Pfaff2012}; closely related boundary integral/element (surface-only) methods also offer divergence-free fields \cite{Da2016}. Model-reduction methods based on Laplacian eigenfunctions offer pointwise divergence-free fields via analytical basis functions \cite{DeWitt2012,cui2021spiral}, albeit in simple domains.

\subsubsection{Subdivision Schemes}
Subdivision schemes based on discrete exterior calculus consider how to preserve properties of discrete differential operators on triangulated surface meshes undergoing refinement \cite{Wang2006,DeGoes2016}. These papers focus on surfaces and do not consider volumetric uniform grids or polyhedral cut-cells. More fundamentally, they achieve pointwise incompressibility only in the limit of infinite refinement. It would be interesting to adapt these ideas to polyhedra.

\subsubsection{Enhanced Advection} A wide array of fundamental improvements to advection have been proposed to go beyond basic (trilinear) interpolation and (semi-Lagrangian) advection schemes \cite{Stam1999}, ranging from  monotonic cubic interpolation \cite{Fedkiw2001} to MacCormack advection \cite{Selle2008} to variants of particle-in-cell schemes \cite{Zhu2005,Jiang2015}. Such methods tend to focus on vorticity-preservation, higher order accuracy, or diffusion-reduction. Because they typically use the underlying interpolant as a black box, they are orthogonal, and often complementary, to our scheme, as we demonstrate for FLIP\cite{Zhu2005}.

\subsection{Outside Computer Graphics}
\subsubsection{Divergence-Free Finite Element Methods}
\label{sec:DG}
In applied math, a wide range of discontinuous Galerkin (DG) and other non-conforming finite element methods (FEM) have been developed to offer pointwise divergence-free fields, often by adopting carefully designed incompressible per-element basis functions  (e.g., \cite{Cockburn2004,Lehrenfeld2016,Rhebergen2018}). In particular, \citet{Maljaars2018} combined an exactly divergence-free Hybridized Discontinuous Galerkin scheme with a particle-in-cell method (in 2D, without obstacles) and demonstrated better particle distributions. However, this family of FEM methods tends to have a more complex implementation and do not naturally integrate with standard approaches in animation. Moreover, they possess inter-element field discontinuities by their very nature, which can be problematic for visual applications; \citet{Maljaars2018} observed that reducing the magnitude of these inter-element discontinuities helps to improve particle distributions. \citet{Guzman2014} tackled the more challenging task of designing a \emph{fully} conforming/continuous Stokes finite element method yielding divergence-free solutions on 3D tetrahedra. Doing so required a specialized finite element basis consisting of a combination of cubic polynomials and divergence-free rational functions, and due to the construction's complexity the authors note that "practical significance of the proposed elements may be questionable". In isogeometric analysis, \citet{Evans2013} developed an exactly divergence-free Navier-Stokes simulation framework based on geometrically mapped rectangular B-spline grids; approaches in this vein require complex mesh construction for non-trivial domains, in contrast to the simpler and more efficient cut-cell techniques often preferred in animation. In fact, all of the approaches above would necessitate replacing the entire Navier-Stokes simulator, while ours provides a convenient plug-in upgrade to the advection phase of industry-standard visual effects methods. Thus, somewhat closer to our approach are methods that post-process flow solver solutions to exactly recover a divergence-free velocity field. For example, \citet{Linke2012} converts the solution of a staggered triangulated discretization into a strictly divergence-free field in terms of Raviart-Thomas elements. Similarly, \citet{Lederer2017} proposed a velocity reconstruction operator that maps discretely divergence-free fields of Taylor-Hood and mini elements to exactly divergence-free ones, and suggested using their method as a postprocessing of the discrete solution in the conclusion. However, like the DG methods discussed above, the resulting finite element spaces (i.e., velocity fields) are not continuous between elements.

\subsubsection{Direct Interpolation of Finite Volume Solutions}
\label{sec:directincompressible}
In a computational fluid dynamics context, Jenny and colleagues \cite{Jenny2001,Meyer2004} derived 2D divergence-free node-based velocity interpolants for uniform grids and demonstrated improved particle distributions in particle-in-cell schemes. In geodynamics, \citet{Wang2015} developed a 3D extension and \citet{Pusok2017} adapted it for face-based data by first averaging to nodes. These approaches augment multilinear interpolants with corrective terms to satisfy pointwise incompressibility per cell. 
In astrophysics (magnetohydrodynamics), \citet{Balsara2001,Balsara2004} proposed very similar divergence-free vector field reconstruction strategies based on local piecewise fitting on each cell or tetrahedron; one presupposes a polynomial basis and uses the divergence-free condition to determine constraints. \citet{Balsara2009} presented an extension of this approach to Cartesian grid WENO schemes. These approaches do not consider cut-cell geometries, and exhibit kinks between cells.

Conveniently, the approaches above do not require recovering a vector potential at all. However, the recovered fields exhibit pervasive discontinuities between cells, and to achieve a given order of accuracy or support a particular geometry, they must be developed each time on a case-by-case basis. Our choice to instead recover and exploit an explicit discrete vector potential makes it comparatively straightforward to derive generalizations of our framework to higher order accuracy, global continuity/smoothness, or different element shapes, using relatively \emph{standard} interpolation techniques, including spline interpolants or mesh-free methods. It could also be fruitfully combined with fluid control tools that already rely on vector potentials \cite{Sato2015,Sato2021}.

Concurrent work by \citet{schroeder2022} proposes a direct spline-based interpolation method for staggered regular grid velocity fields that yields pointwise divergence-free fields of different orders. Some variants of their method can also exactly interpolate the cell normal components. This method avoids recovering the vector potential, but is limited to uniform grids without obstacles.

An alternative meshless divergence-free interpolation strategy  \cite{Lowitzsch2005,McNally2011} exploits matrix-valued radial basis functions and properties of the vector Laplacian. While this method offers divergence-free fields, as a meshless point-based method it struggles to precisely enforce boundaries and, more critically, preliminary tests showed it to be massively expensive, since it uses a fully dense global solve to simultaneously determine coefficients for all of the radial basis functions. 
Even attempting to substitute local kernels for the method's global kernels remains expensive: a large support radius (i.e., particle neighborhood) is needed to avoid local oscillations and thus the system matrix is still unattractively dense. 

\subsubsection{Vector Potentials and Gauge Conditions}
The methods most similar to ours are those that use explicit vector potentials. Since the vector potential for a given velocity field is not unique, various gauge conditions are used to select a particular potential from this null space. The gauge choice provides one dimension along which to categorize the schemes discussed below. Another is whether they apply a local approach to find the vector potential, based on a cell-by-cell (possibly branching) traversal, or find a simultaneous global solution by solving a linear system. We straddle these categories: our parallel sweeping approach is a fast cell-by-cell scheme that yields an initial potential, which we correct with a relatively inexpensive scalar Poisson solve (via Discrete Sine Transform or DST) to ensure smoothness by enforcing the Coulomb gauge, $\nabla \cdot \mathbf{\psi}=0$. Furthermore, beyond our innovations in the uniform grid setting, our work is unique in considering solid geometry via cut-cells.

The Coulomb gauge condition picks out a unique smooth vector potential by enforcing that it be divergence-free; it is the most common choice and has been used in graphics (e.g., \cite{Ando2015} among others.) The initial implicit gauge condition used in our 3D parallel sweeping method is conceptually similar to that of \citet{Ravu2016}, who sought to directly construct spline-based vector potentials. Like us, they set the $\psi_z$ component to zero. However, their overall approach differs in that (a) their velocity and vector potential degrees of freedom are colocated at nodes, (b) they directly seek the coefficients of continuous cubic spline functions rather than edge-based discrete vector potential values (hence four times as many unknowns), and (c) compared to our fast parallel sweeping, they solve an expensive global linear system to find the many coefficients.

Another implicit gauge condition falls out from the tree-cotree approach, explored for edge-based finite element methods by \citet{Albanese1990} on grids and \citet{Manges1995} on meshes. This global approach determines a spanning tree of an edge-based grid or mesh that one can safely set to zero, in order to explicitly eliminate the null space of the matrix representing the discrete curl operator. This idea has natural connections to the traversal patterns of cell-by-cell approaches.

\citet{Bao2017} proposed a divergence-free interpolation strategy on uniform staggered grids that reconstructs a discrete vector potential field from velocities, to improve coupling and reduce volume conservation errors in the immersed boundary (IB) method \cite{Peskin2002}. (IB uses smeared delta functions for approximate coupling with solids, in contrast to our "sharp" cut-cells.) Similar to methods in graphics \cite{Tong2003,Ando2015}, they recover the vector potential under the Coulomb gauge using a vector Poisson system. They also assume periodic boundary conditions, which are undesirable for most graphics applications, and only suggest non-periodic domains as future work. Their periodic, obstacle-free setting decouples the vector problem into three scalar Poisson solves, which they treat via the fast Fourier transform (FFT). Our method recovers the same potential with an efficient sweep process followed by just one scalar Poisson solve (which can likewise be done via FFT on simple domains). We further support interior obstacles and closed domains with a fast DST-based solve. For interpolation, they use tensor-product B-splines, similar to ours. \citet{Casquero2018} also considered the immersed boundary method on uniform grids, but like \citet{Evans2013} and in contrast to \citet{Bao2017}, they perform the entire simulation using divergence-free spline basis functions rather, than post-processing staggered data.

\citet{Silberman2019} adopt a staggered uniform grid configuration and seek a discrete edge-based vector potential, similar to our setting, but for applications in electromagnetics. They propose a sequential, cell-by-cell flooding strategy which is more costly than our approach. Specifically, our axis-based sweeping safely assumes zero values in the $z$-component, which reduces the computation per cell, the number of unknowns (by a third), and the number of special cases, while enabling a fast parallel solution via 1D sweeps. Like us, they use a Poisson-based correction to apply the Coulomb gauge, but like \citet{Bao2017} they support only uniform grids, and they use infinite boundary conditions to allow solution by FFT. \citet{Silberman2019} additionally suggest an alternative "global linear algebra" approach that is loosely similar to the vector Poisson method of \citet{Ando2015} in that it both removes divergence and recovers a potential satisfying the Coulomb gauge; they note that it scales worse than their flood-then-correct approach.

In summary, our vector potential recovery approach is most closely related to the uniform staggered grid methods of \citet{Bao2017} and \citet{Silberman2019}, but is more efficient and generalizes these ideas to the more flexible domain boundary conditions and irregular cut-cell solids favored in graphics applications.

\section{Problem Setting}
\label{problemSetting}
We build on a standard staggered grid-based fluid solver for the incompressible Euler equations \cite{Fedkiw2001,Bridson2015}.
We represent solids using node-based level sets, 
leading to simple marching cubes cut-cells, and assume that all solids are static. For cut-cell pressure projection, we use the finite volume Poisson stencil of \citet{Ng2009}. For velocity advection, we use semi-Lagrangian  backtracing with multilinear interpolation and third-order Runge-Kutta, except where noted.

We develop a new velocity interpolant for use during advection,
\begin{equation}
\label{eq:advectionEquation} 
\frac{ \partial q }{ \partial t} + \mathbf{u} \cdot \nabla q = 0,
\end{equation}
where ${\mathbf{u}}$ is an incompressible velocity field and $q$ is any scalar (e.g., density) or vector (e.g., velocity) quantity to be advected.
Common advection discretizations use semi-Lagrangian \cite{Stam1999} or Lagrangian \cite{Zhu2005} methods
that trace trajectories through the flow and require interpolation to query the velocity.

The input discrete velocity field $\mathbf{u}$ to be used for advection is assumed to be discretely divergence-free due to the underlying simulator's incompressibility enforcement (i.e., pressure projection). Discrete velocity components are placed at grid face centers, so the discrete divergence-free condition is
\begin{align}
\sum_{\text{faces}\, f} A_f u_f  = 0,
\end{align}
where $A_f$ is face area and $u_f$ is the face's outward oriented normal velocity. Given such discrete data $\mathbf{u}$, our task is to generate a pointwise vector field that is analytically divergence-free, while ideally being continuous, smooth, and boundary-respecting.

\setlength{\fboxsep}{50pt}%
\begin{figure}
    \centering
	\begin{subfigure}[b]{0.45\textwidth}
	\centering
		\adjustbox{fbox, width=0.33\textwidth}{\includegraphics{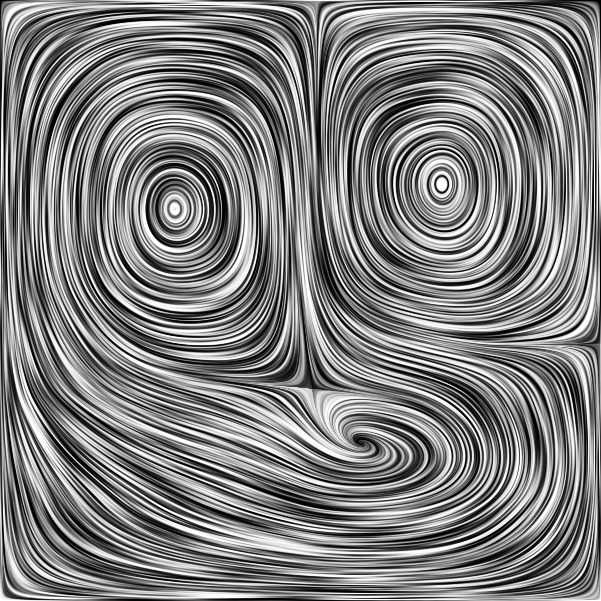}}
		~ \hfill
		\adjustbox{fbox, width=0.33\textwidth}{\includegraphics{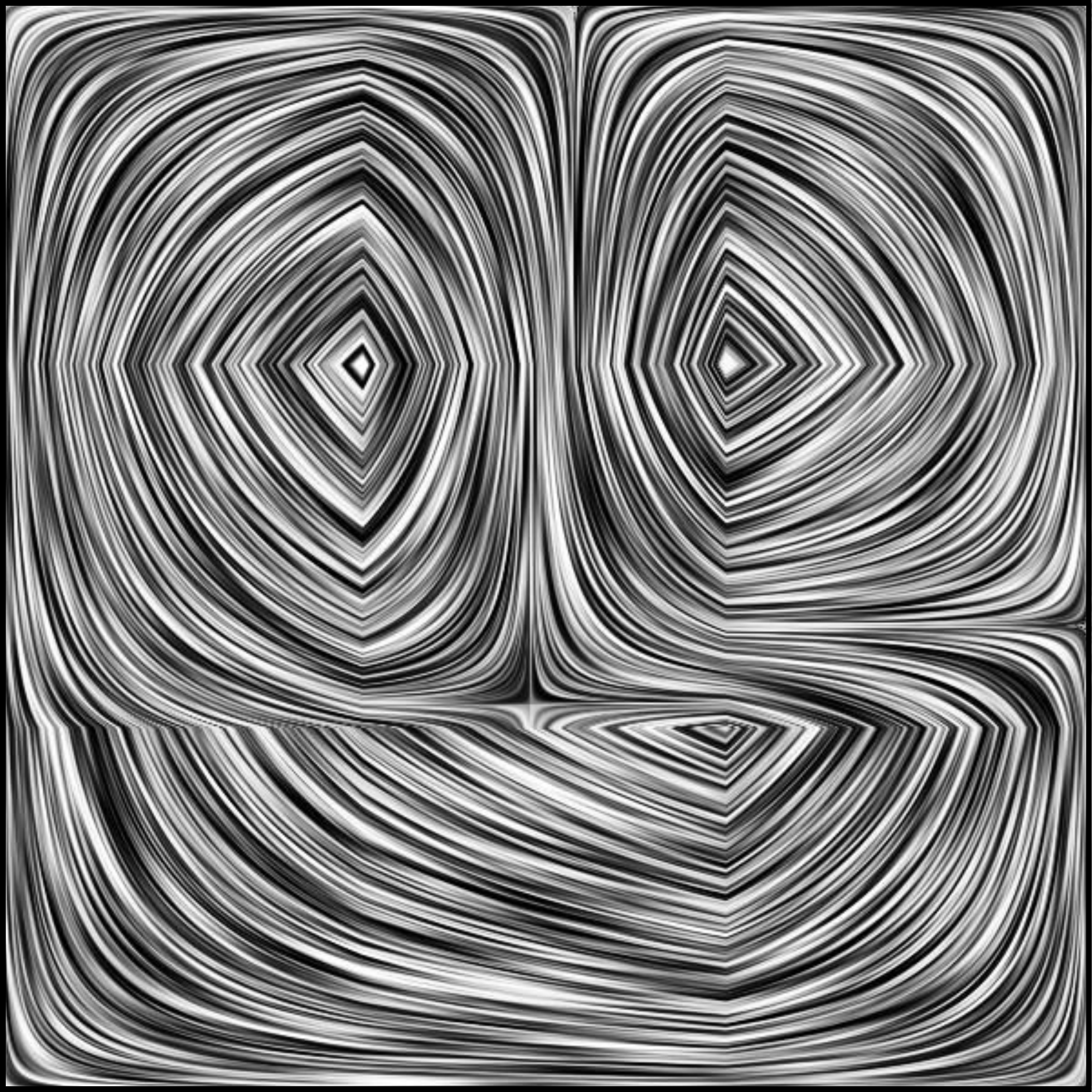}}
		~ \hfill   
		\adjustbox{fbox, width=0.33\textwidth}{\includegraphics{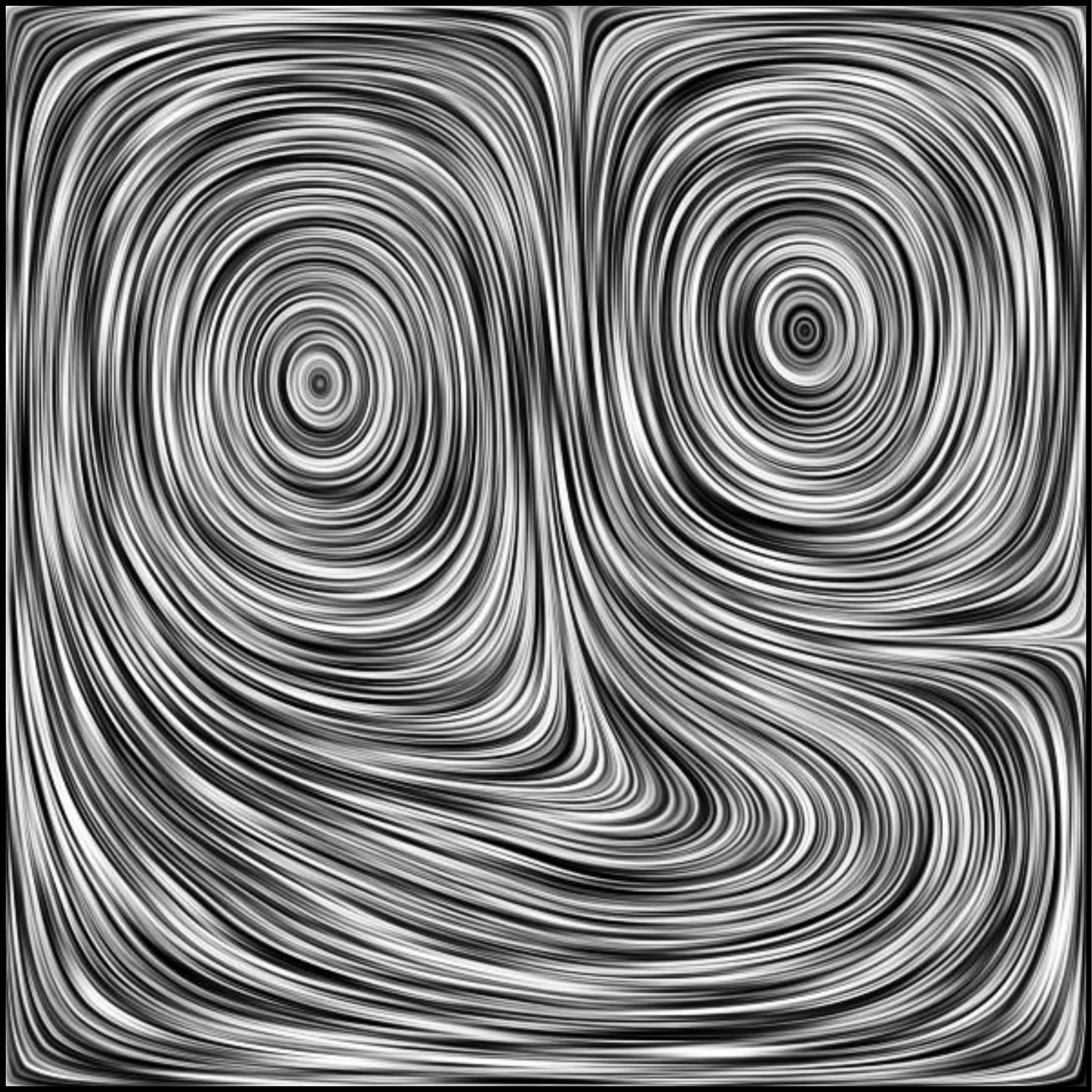}}
	\caption{Interpolated velocity fields using standard bilinear (left), linear Curl-Flow (middle), and quadratic Curl-Flow (right).}
	\end{subfigure}
	\par\medskip
    \begin{subfigure}[b]{0.45\textwidth}
    \centering
		\adjustbox{fbox, width=0.33\textwidth}{\includegraphics{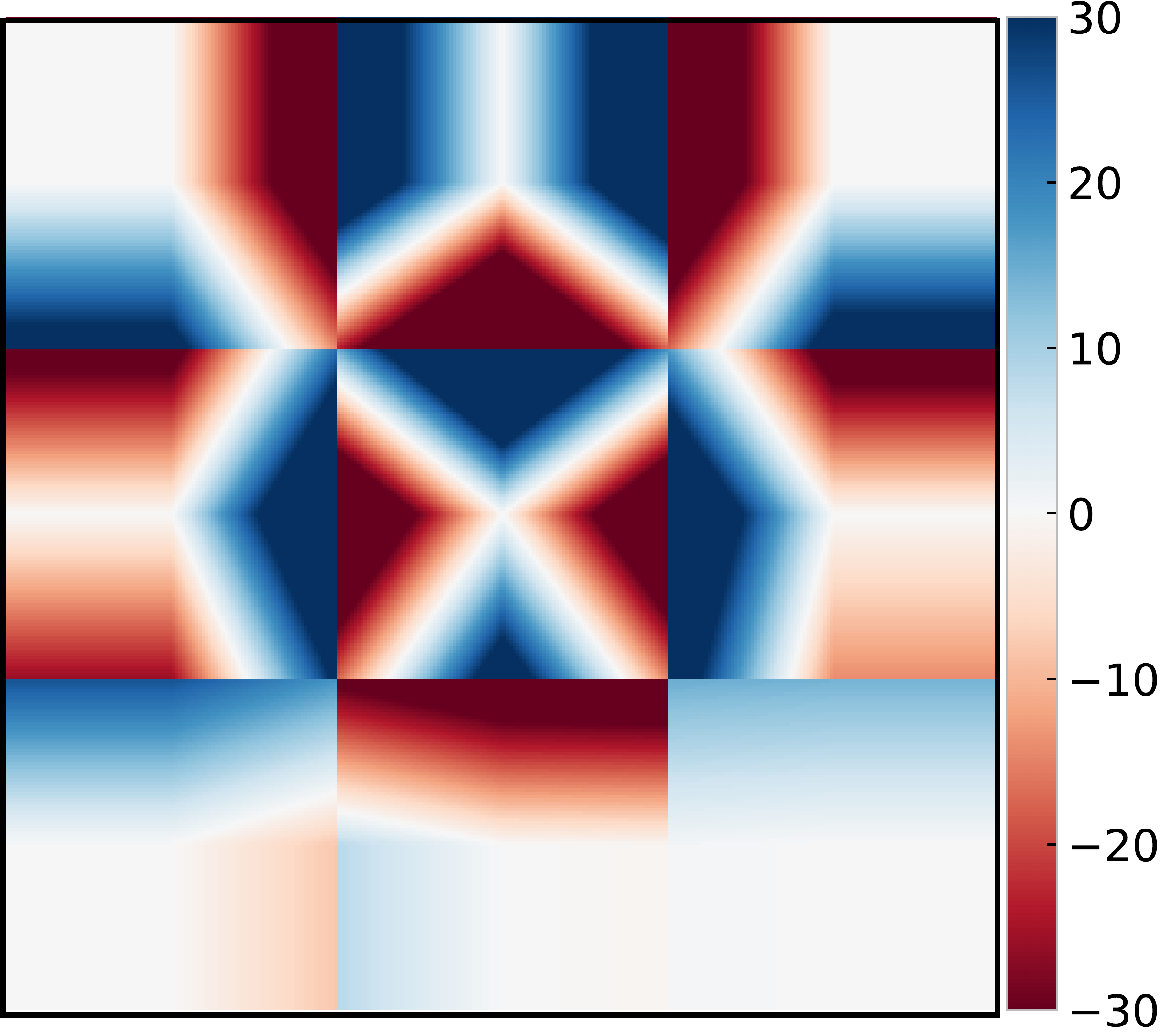}}
		~ \hfill
		\adjustbox{fbox, width=0.33\textwidth}{\includegraphics{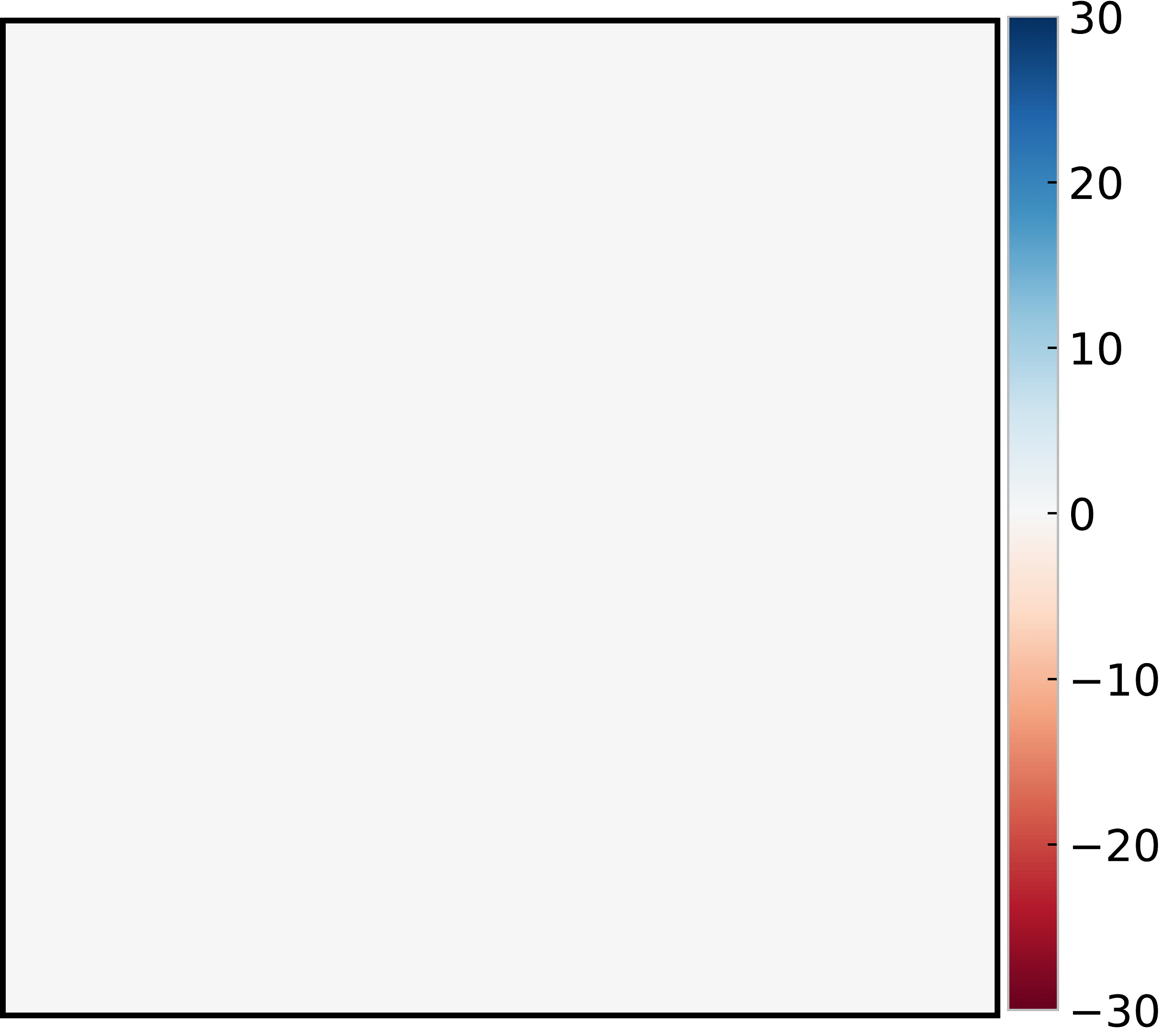}}
		~ \hfill   
		\adjustbox{fbox, width=0.33\textwidth}{\includegraphics{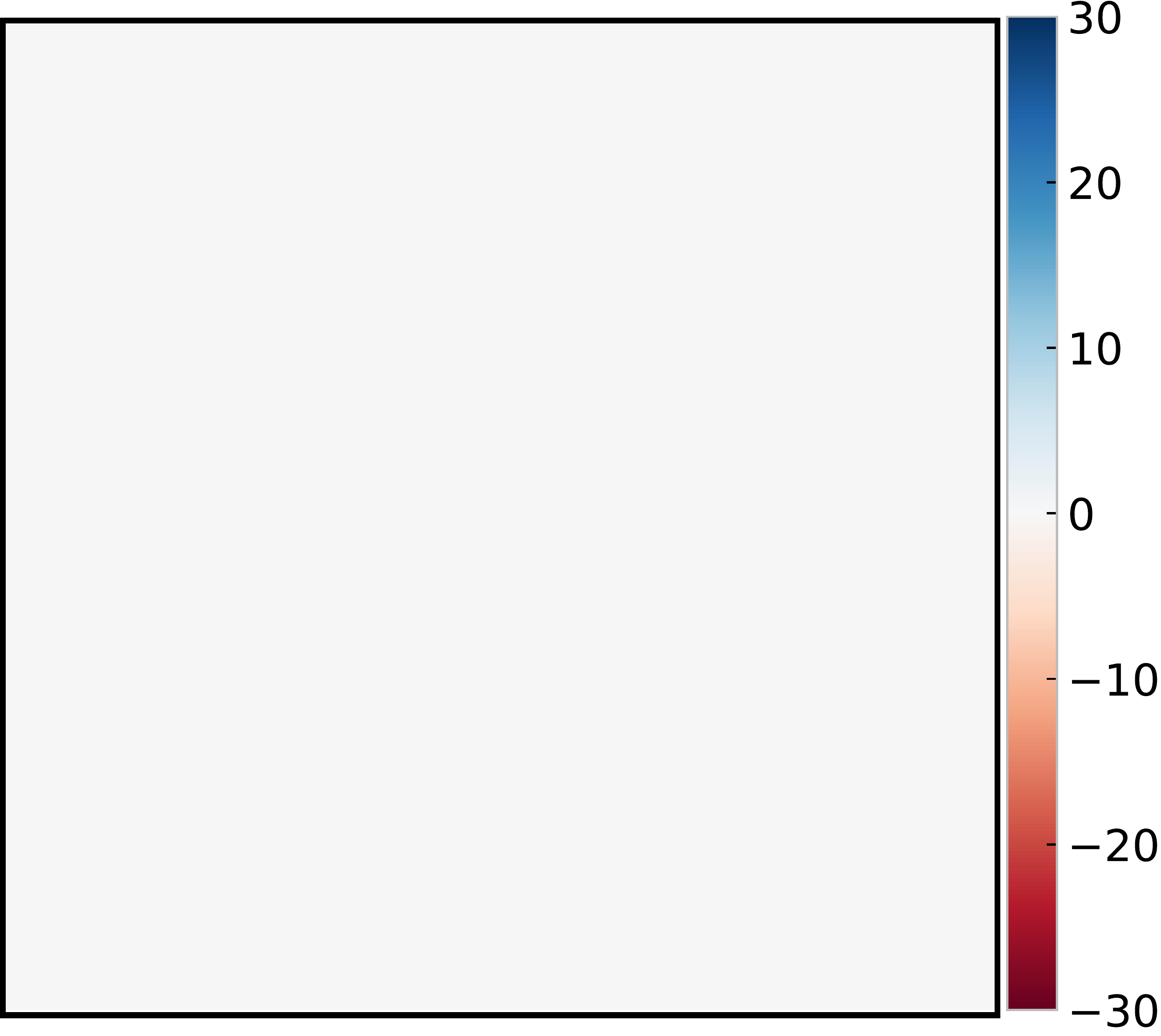}}
	\caption{Corresponding pointwise divergence plots (units: 1/s).}
	\end{subfigure}
	\caption{\textbf{Pointwise Divergence Comparison:} 
	Left: Bilinear interpolation of discretely divergence free velocity data on a $3\times3$ grid. The red box in (a) contains a spurious sink. 
	Middle: Linear Curl-Flow interpolation is divergence-free, but has piecewise constant components that induce kinks.
	Right: Quadratic Curl-Flow interpolation is divergence-free and smooth.
	}
	\label{fig:divComparison}
\end{figure}

\begin{figure}
	\centering	
    \begin{subfigure}[t]{0.45\textwidth}	
		\adjustbox{fbox, width=0.24\textwidth}{\includegraphics{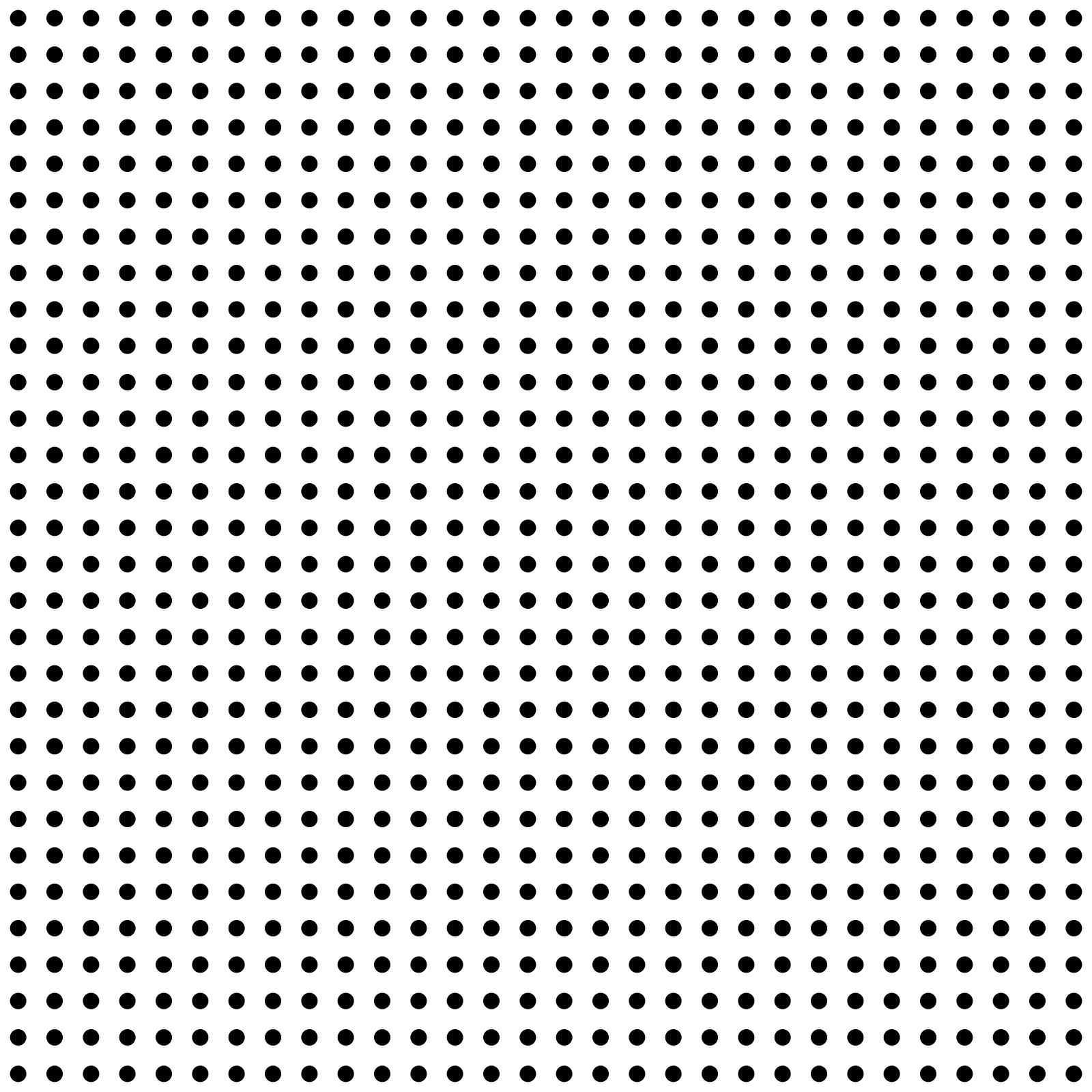}}
		\adjustbox{fbox, width=0.24\textwidth}{\includegraphics{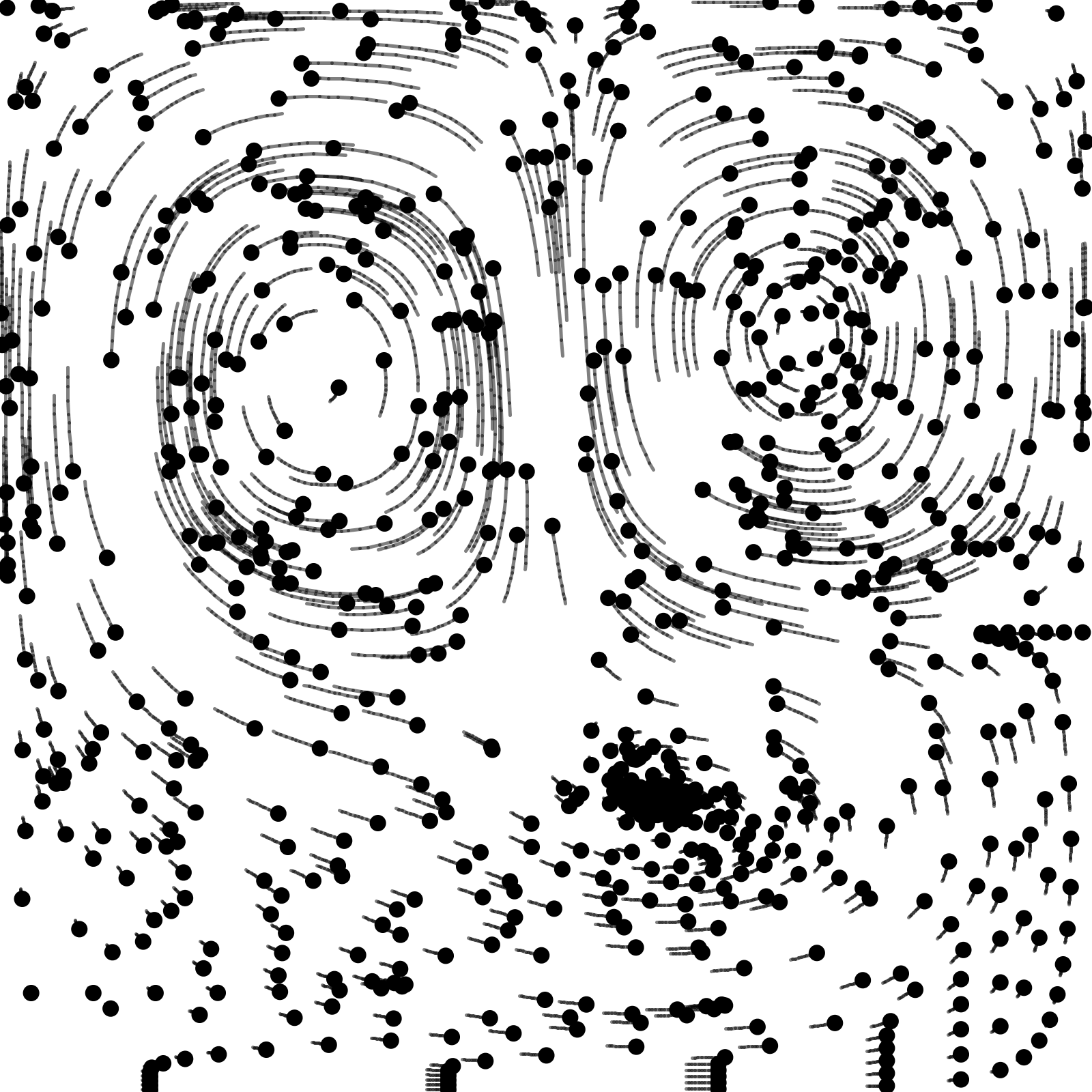}}
		\adjustbox{fbox, width=0.24\textwidth}{\includegraphics{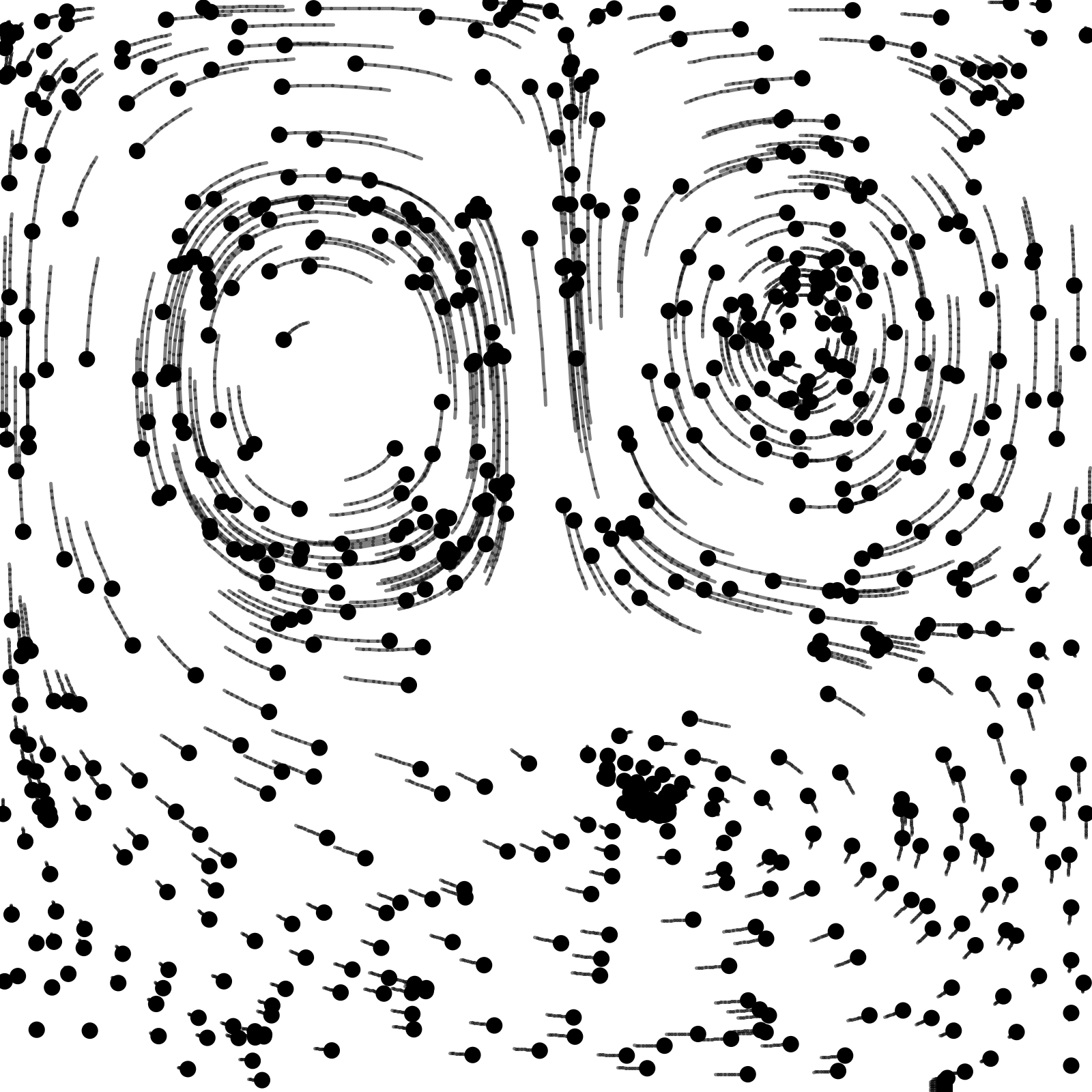}}
		\adjustbox{fbox, width=0.24\textwidth}{\includegraphics{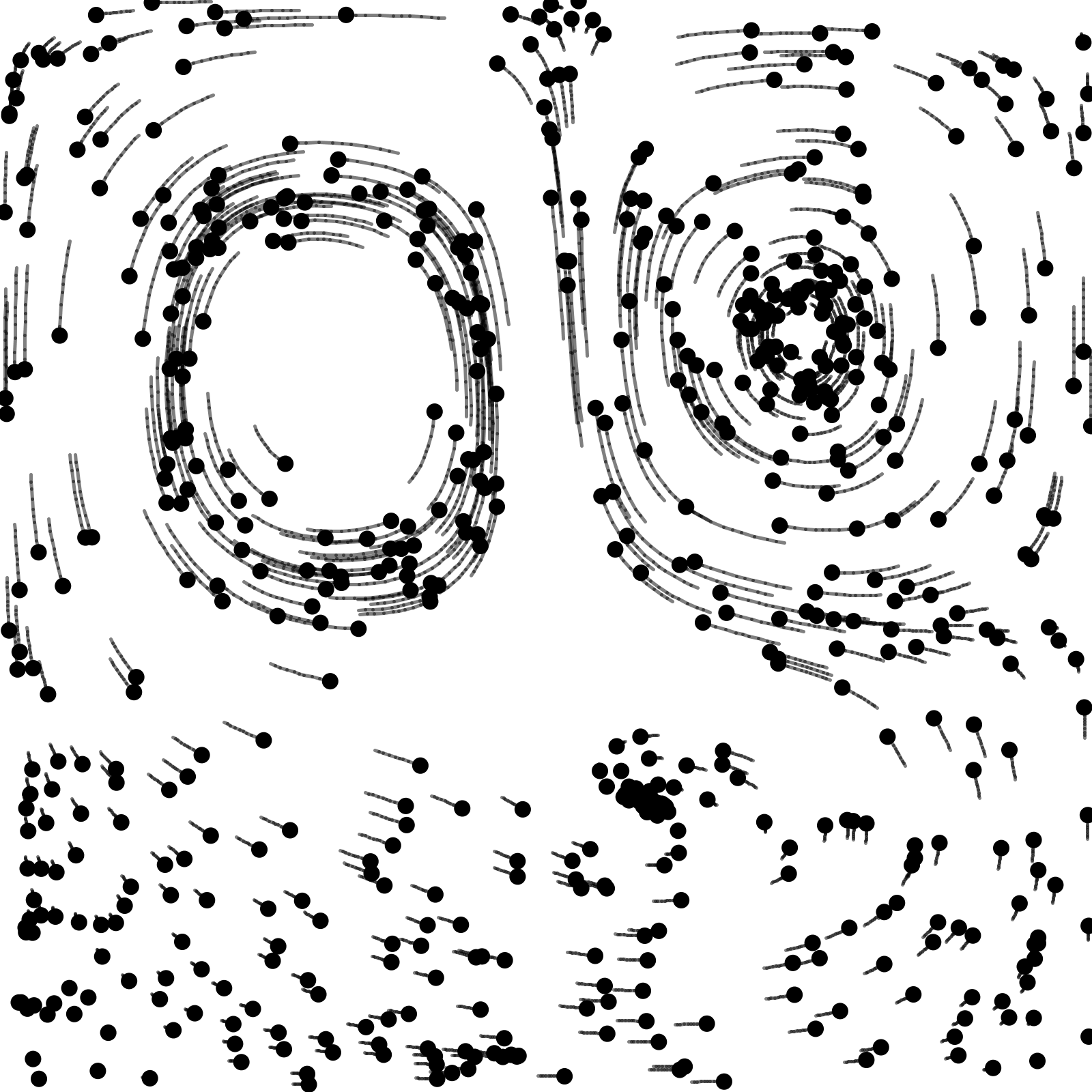}}
    	\subcaption{Bilinear velocity interpolation}
	\end{subfigure}
    \par\medskip
    \begin{subfigure}[t]{0.45\textwidth}	
		\adjustbox{fbox, width=0.24\textwidth}{\includegraphics{img/vel00000}}
		\adjustbox{fbox, width=0.24\textwidth}{\includegraphics{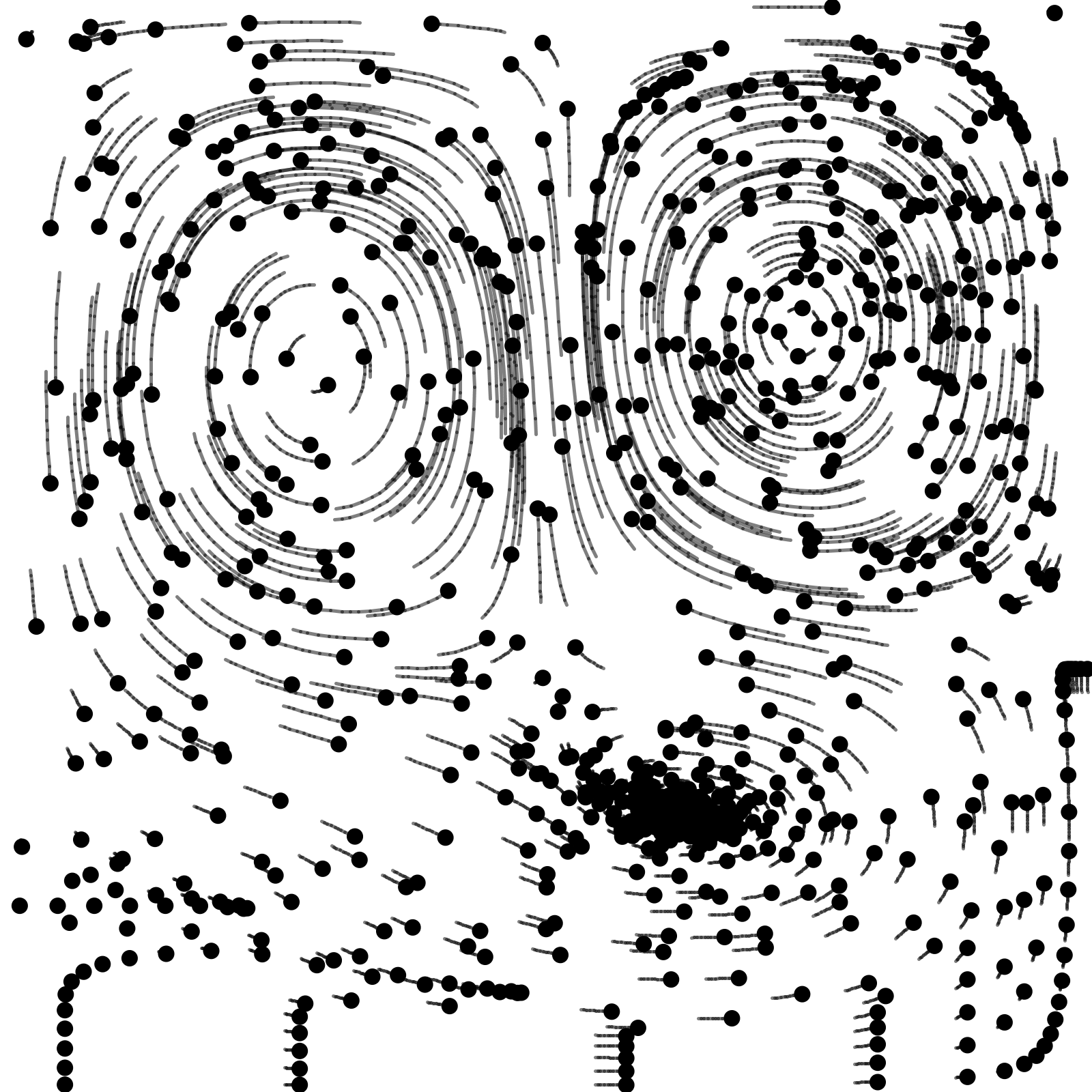}}
		\adjustbox{fbox, width=0.24\textwidth}{\includegraphics{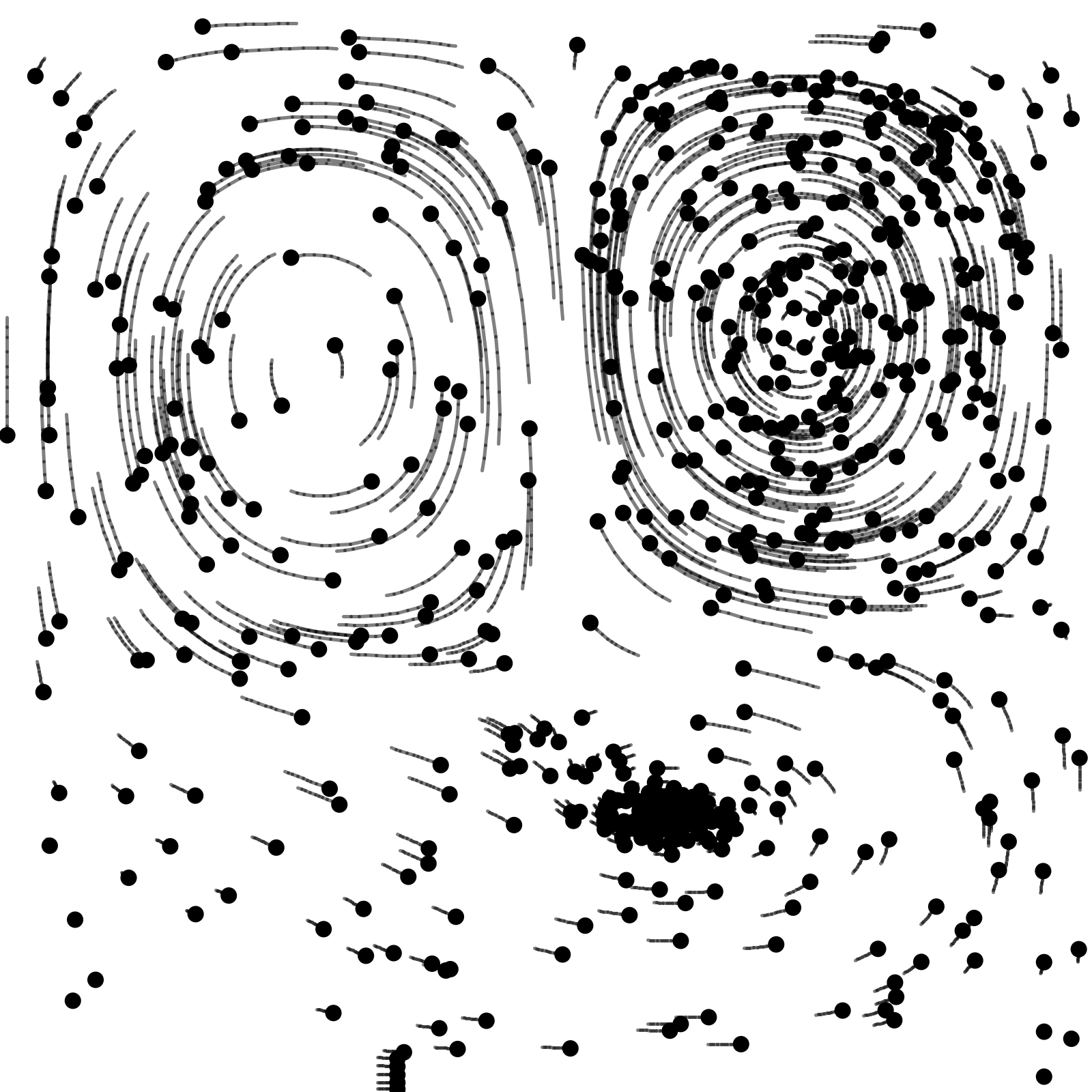}}
		\adjustbox{fbox, width=0.24\textwidth}{\includegraphics{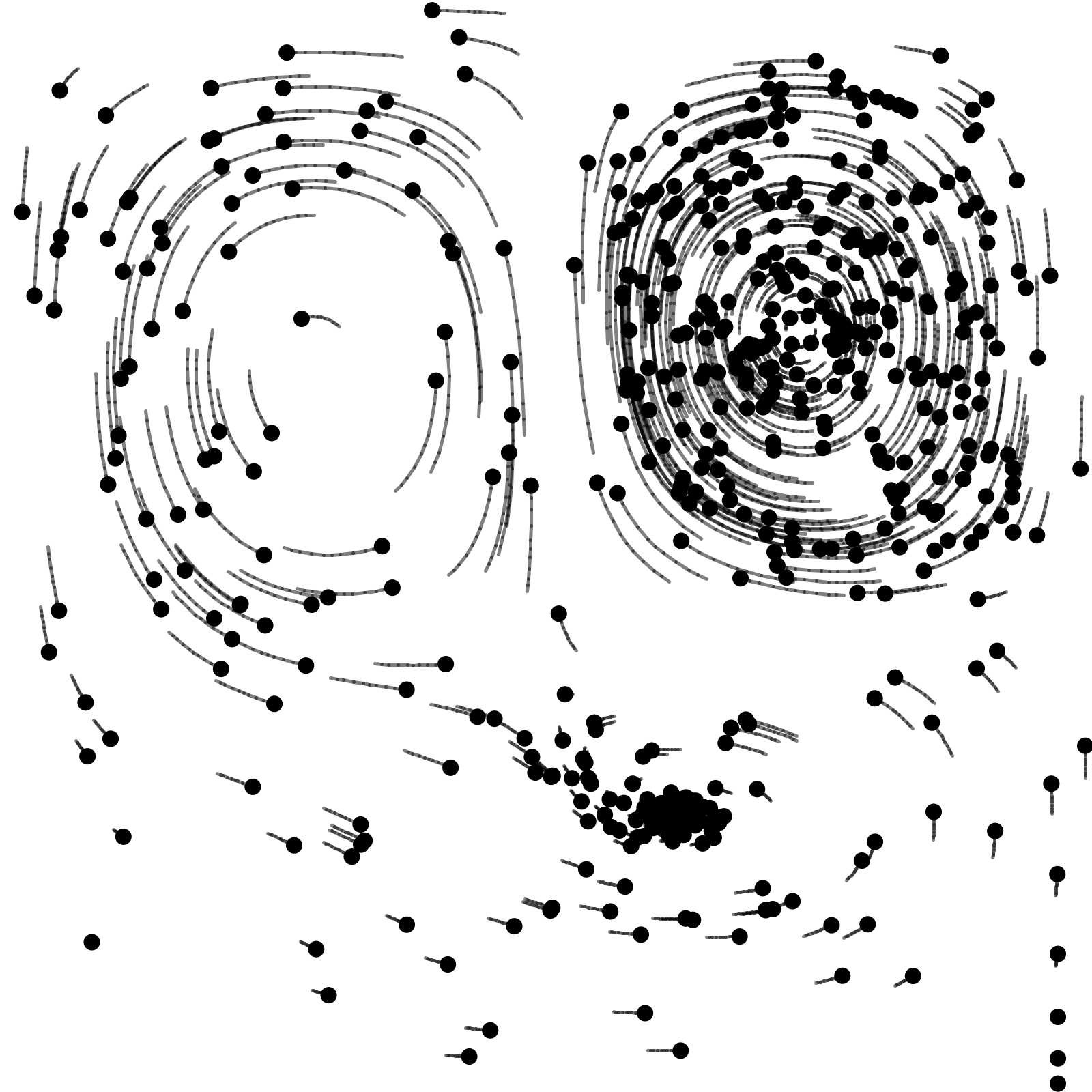}}
    	\subcaption{Bicubic velocity interpolation}
	\end{subfigure}
    \par\medskip
    \begin{subfigure}[t]{0.45\textwidth}	
		\adjustbox{fbox, width=0.24\textwidth}{\includegraphics{img/vel00000}}
		\adjustbox{fbox, width=0.24\textwidth}{\includegraphics{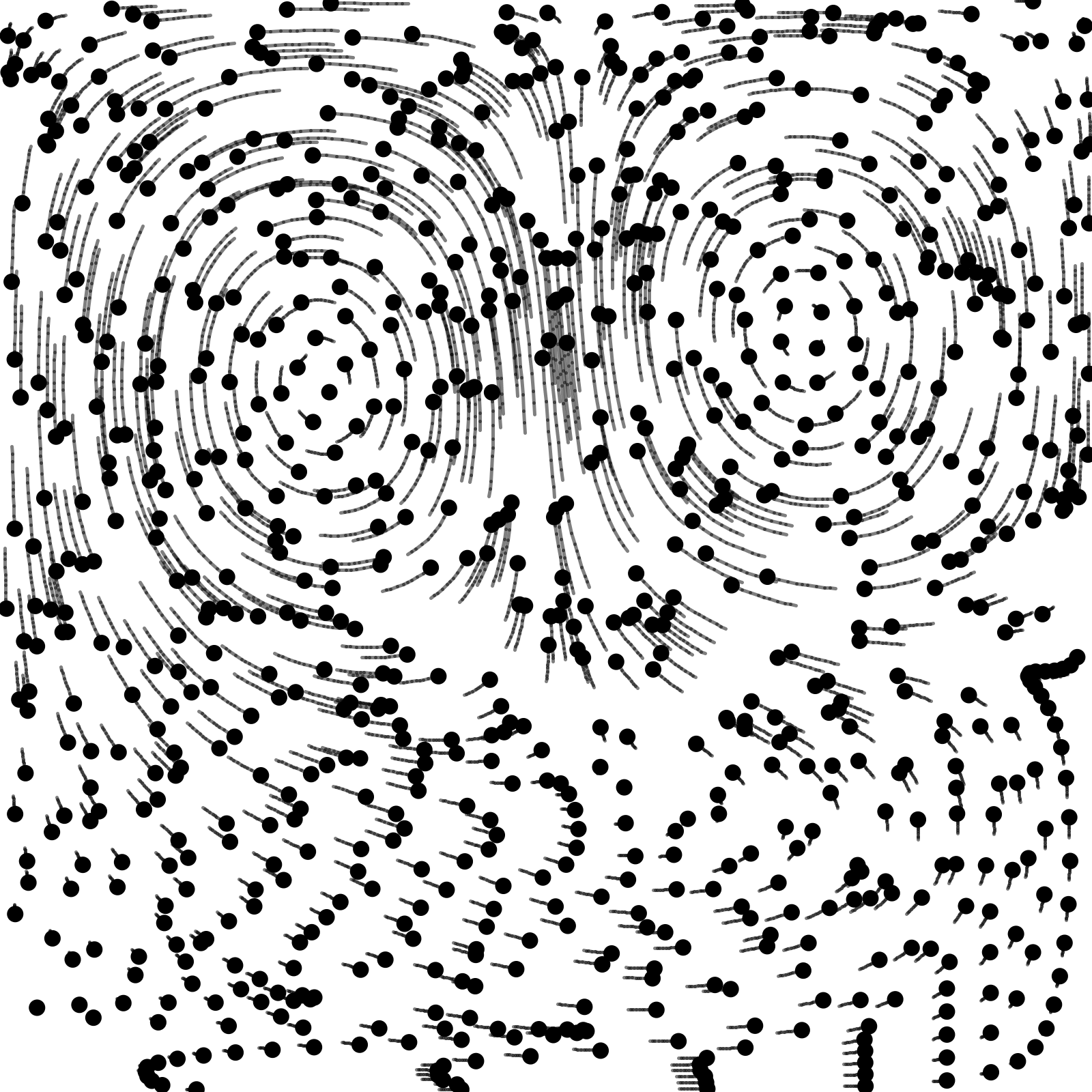}}
		\adjustbox{fbox, width=0.24\textwidth}{\includegraphics{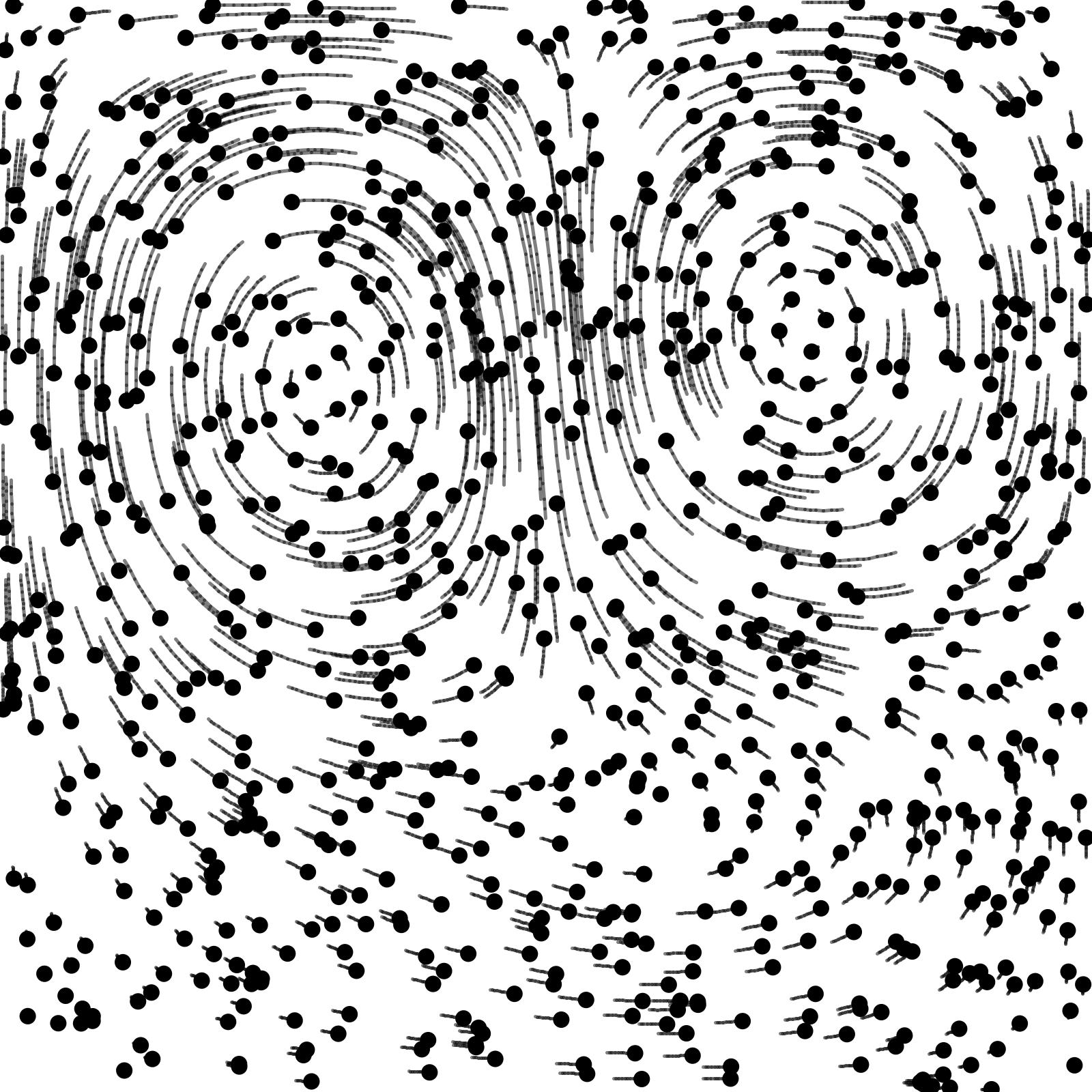}}
		\adjustbox{fbox, width=0.24\textwidth}{\includegraphics{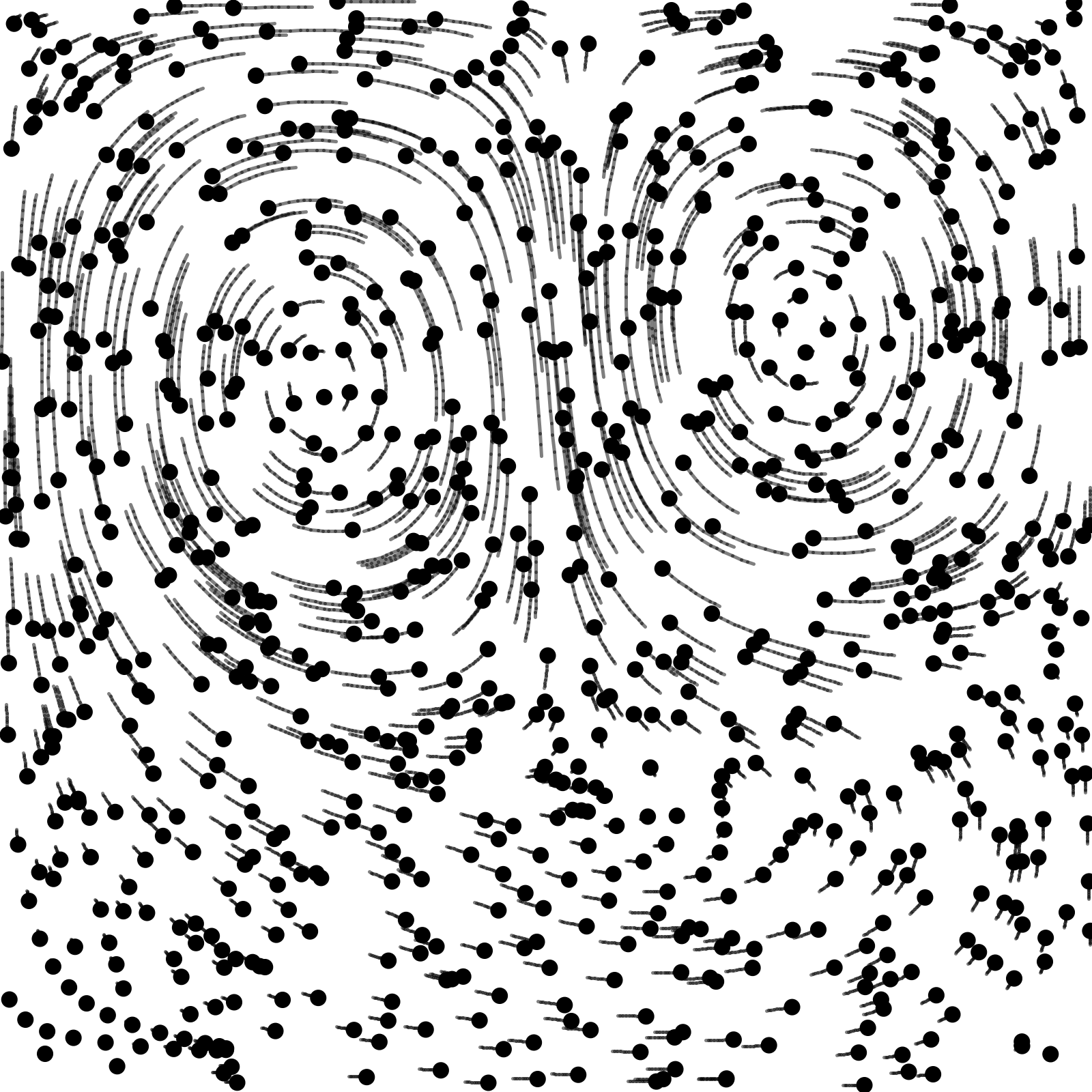}}
    	\subcaption{Curl-Flow interpolation}
	\end{subfigure}
	\caption{ \textbf{Effect on Particle Distribution:} Initially uniform particles advected through a static 2D vector field on a $3\times3$ grid. Frames 1, 100, 200, and 300 are shown left to right.
		(a): With direct bilinear velocity interpolation, particles become clustered, leaving large empty voids.
		(b): With higher order direct velocity interpolation (monotonic cubic~\cite{Fritsch1980,Fedkiw2001}), clustering and spreading remain significant.
		(c): With quadratic Curl-Flow interpolation, particles remain better distributed.
	}
	\label{fig:particleDistribution}
\end{figure}

Bilinear interpolation applied to each staggered velocity component independently is the simplest common interpolant, but the resulting vector field's analytical divergence is nonzero in general (see Section 1 in supplemental material).
Figure~\ref{fig:divComparison}, left column, visualizes such a flow and its divergence; 
if one advects uniformly sampled particles through this field (Figure~\ref{fig:particleDistribution}, top row),
a sink region absorbs a large number of particles 
while many more cluster into large and small rings. Unfortunately, neither higher order interpolation (Figure~\ref{fig:particleDistribution}, middle row) nor using node-based / colocated velocity data will resolve these issues, since those interpolants remain divergence-oblivious. We use staggered multilinear interpolation as our baseline \emph{direct velocity interpolant} throughout the paper, unless stated otherwise.

The middle and right columns in Figure~\ref{fig:divComparison} visualize the linear and quadratic variants of our proposed approach, respectively, developed in the next section. Since the linear case exhibits velocity kinks we ultimately prefer quadratic, but neither flow contains sinks or sources and their pointwise divergence fields are exactly zero.

\setlength{\fboxsep}{0pt}%
\setlength{\fboxrule}{1pt}%

\begin{figure}
	\centering	
	\begin{subfigure}[t]{0.15\textwidth}
		\fbox{\begin{overpic}[width=0.95\textwidth]{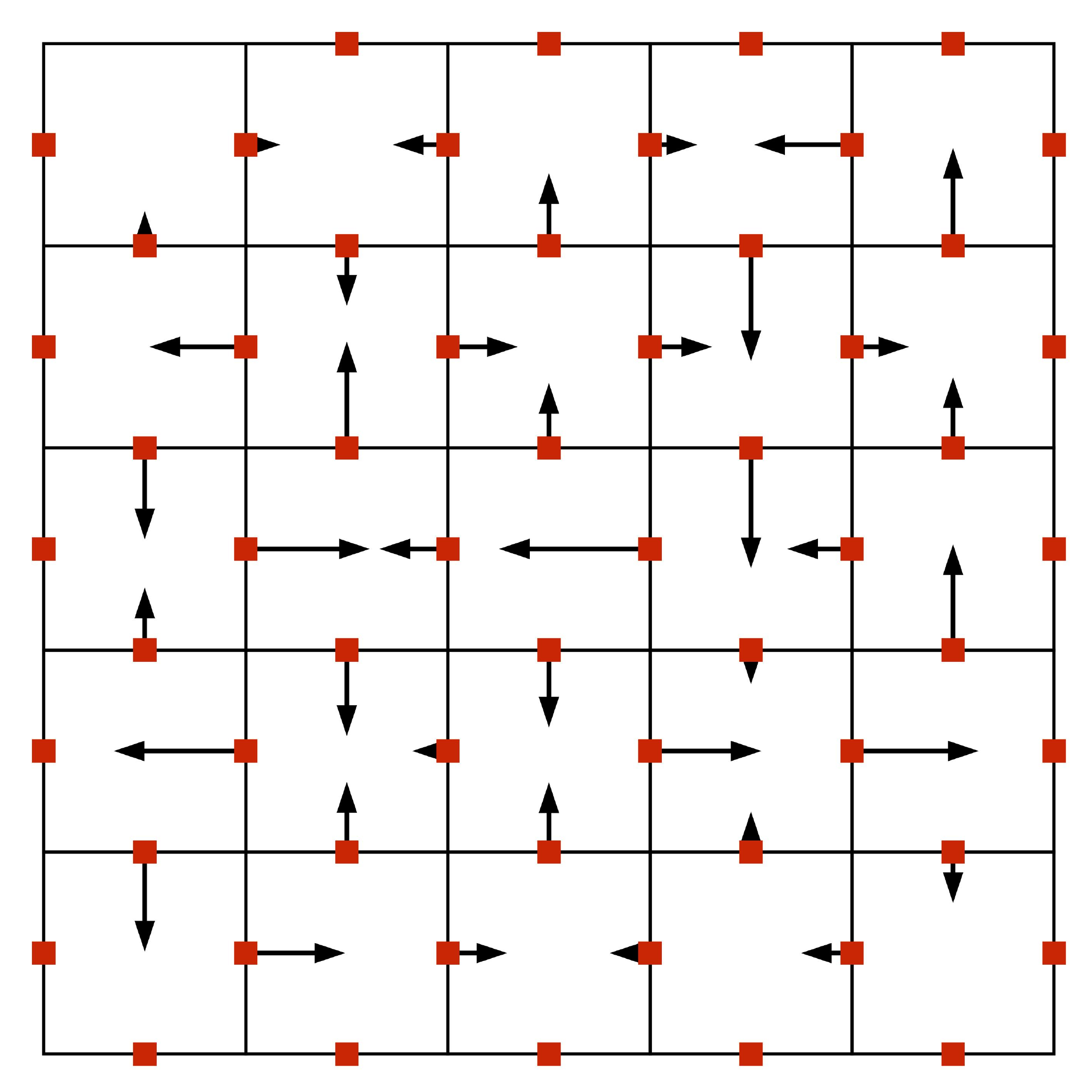}
		\end{overpic}}
		\subcaption{}
	\end{subfigure}
	\begin{subfigure}[t]{0.15\textwidth}
		\fbox{\begin{overpic}[width=0.95\textwidth]{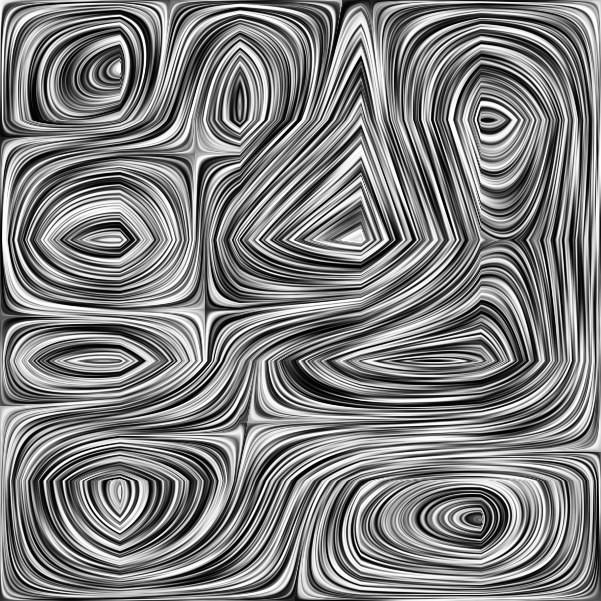}
		\end{overpic}}
		\subcaption{}
	\end{subfigure}
	\begin{subfigure}[t]{0.15\textwidth}
		\fbox{\begin{overpic}[width=0.95\textwidth]{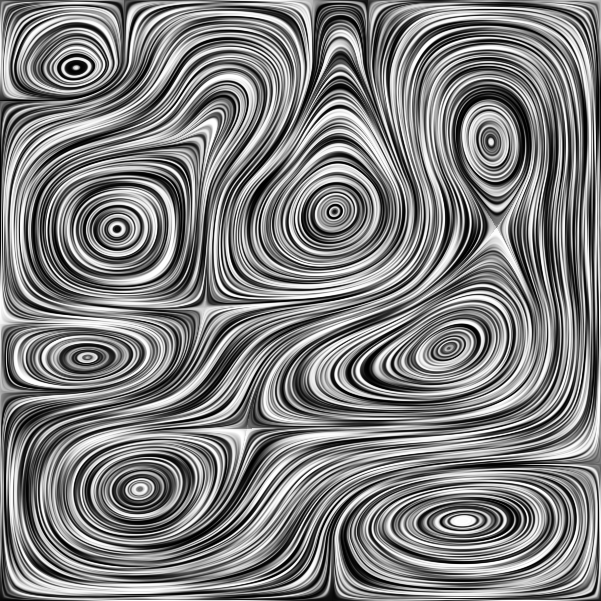}
		\end{overpic}}
		\subcaption{}
	\end{subfigure}
	\caption{\textbf{Direct Incompressible Velocity Interpolation vs. Curl-Flow.} 
	(a) Discrete and discretely divergence-free velocity field to interpolate.
	(b) Direct incompressible velocity interpolation~\cite{Jenny2001, Balsara2001}. The velocity field inside each cell is incompressible, but discontinuities between cells are pronounced.
	(c) Curl-Flow interpolation is pointwise incompressible and smooth both inside and across cells.
	}
    \label{fig:CVI}
\end{figure}

Section \ref{sec:directincompressible} mentioned prior approaches that aim for divergence-free interpolation using velocity data directly, rather than needing intermediate potentials. In particular, \citet{Jenny2001} and \citet{Balsara2001} used the same staggered grid layout as us in 2D (and in practice yield identical fields to one another).
These \emph{direct incompressible velocity interpolants} produce pointwise divergence-free velocities within each cell, but kinks are visible \emph{between} most cells (see Figure~\ref{fig:CVI}) due to their piecewise construction and use of the minmod limiter~\cite{Jenny2001}, unlike our (quadratic) Curl-Flow method.

\setlength{\fboxrule}{0.5pt}%


\section{Curl-Flow Interpolation in 2D}
\label{VP2D}
We first describe our approach in 2D without obstacles. Velocity ${\mathbf{u}}$ has two components, $u$ and $v$, 
and the vector potential has one scalar component, $\psi_z$, called the \emph{stream function} and denoted by non-bolded $\psi$. The relationship \eqref{eq:vectorPotential} simplifies to
\begin{equation}
\label{eq:VP2D}
\begin{aligned}
u = \frac{ \partial \psi }{ \partial y }, && 
v = -\frac{ \partial \psi }{ \partial x }.
\end{aligned}
\end{equation}
That is, $\mathbf{u}$ is a $90^\circ$ rotation of $\nabla \psi$, denoted $\nabla \psi^\perp$.
To discretize, we place $\psi$ samples at cell vertices 
(Figure~\ref{fig:VP2DSingleCell}) and assume $\psi$ varies linearly on edges.
Figure~\ref{fig:VP2DSingleCell}(c) illustrates two nodal $\psi$ values and the normal component of velocity, $v_n$, on a shared edge $e_{ij}$ having unit tangent $\mathbf{e}_{ij}$ and length $l$. The gradient theorem,
\begin{equation}
\label{eq:gradientTheorem}
\begin{aligned}
    \int_{e_{ij}} \nabla \psi (\mathbf{r}) \cdot d\mathbf{r} =
    \psi(\mathbf{x}_j) - \psi(\mathbf{x}_i),
\end{aligned}
\end{equation}
discretized on each edge, gives the relationship
\begin{equation}
\label{eq:VPEdge2D}
\begin{aligned}
    \frac{\psi_j - \psi_i } {l}
    = \nabla \psi \cdot \mathbf{e}_{ij}
    = \nabla \psi^{\perp} \cdot \mathbf{e}_{ij}^{\perp}
    = v_n (\mathbf{n} \cdot \mathbf{e}_{ij}^{\perp}).
\end{aligned}
\end{equation}
Since ${ \mathbf{e}_{i j}^{\perp} }$ is oriented, unit length, and matches $\mathbf{n}$ up to a sign flip, the dot product simply determines the sign.


\subsection{Uniform Grids in 2D}
\label {VPUniformGrid2D}
\subsubsection{Recovering discrete $\psi$}
\begin{figure}
	\centering	
	\begin{subfigure}[t]{0.14\textwidth}
		\begin{overpic}[width=\textwidth]{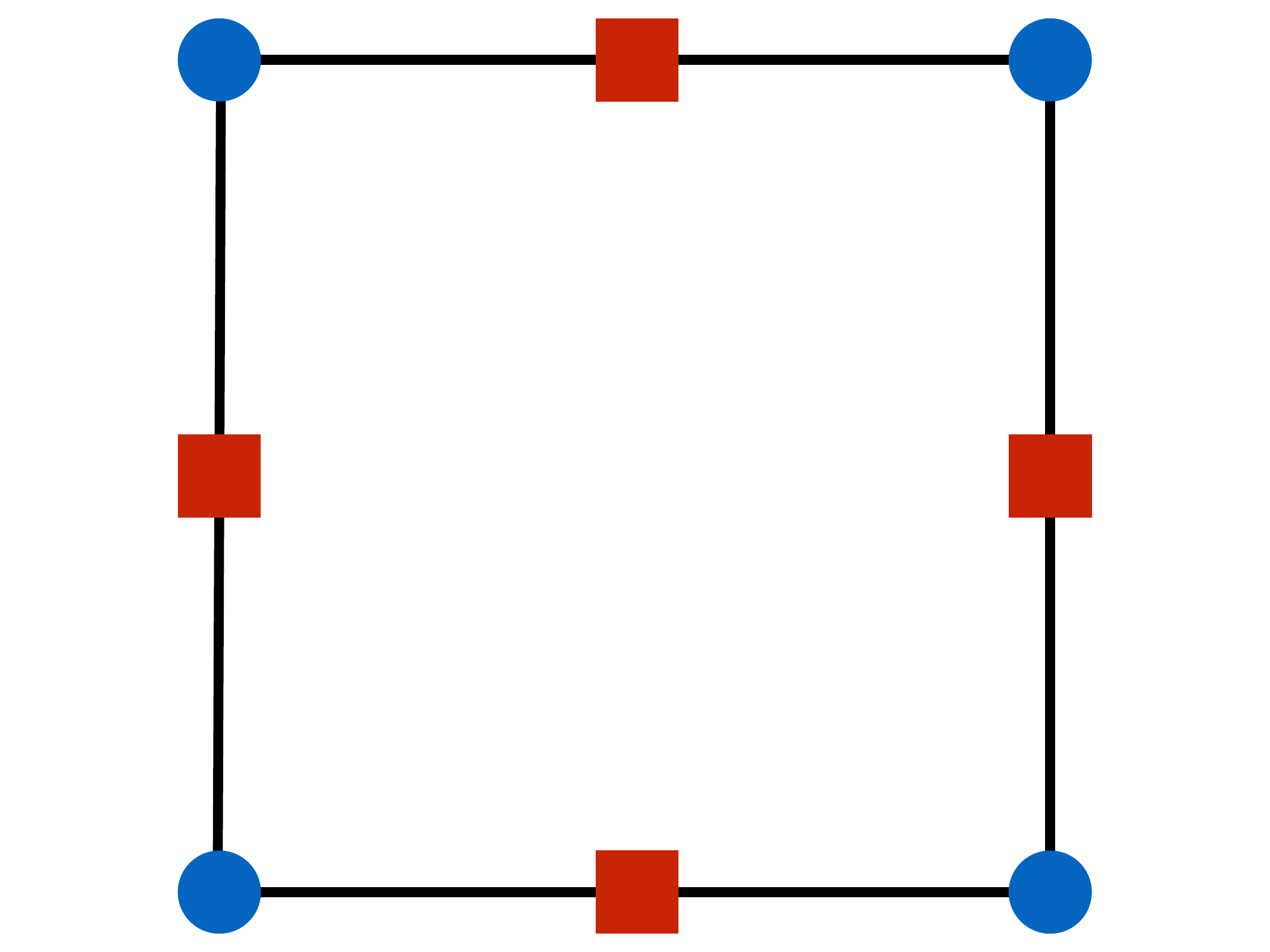}
		\put(0, -2) { ${ \psi_0 }$ }
		\put(85, -2) { ${ \psi_1 }$ }
		\put(0, 75) { ${ \psi_2 }$ }
		\put(85, 75) { ${ \psi_3 }$ }

		\put(0, 35) { ${u_l }$ }
		\put(42, 13) { ${v_b }$ }
		\put(85, 35) { ${u_r }$ }
		\put(42, 78) { ${v_t }$ }		
		\end{overpic}    
	\subcaption {}
	\end{subfigure} 
	\hspace{0.02\textwidth}
	~
	\begin{subfigure}[t]{0.14\textwidth}
		\begin{overpic}[width=\textwidth]{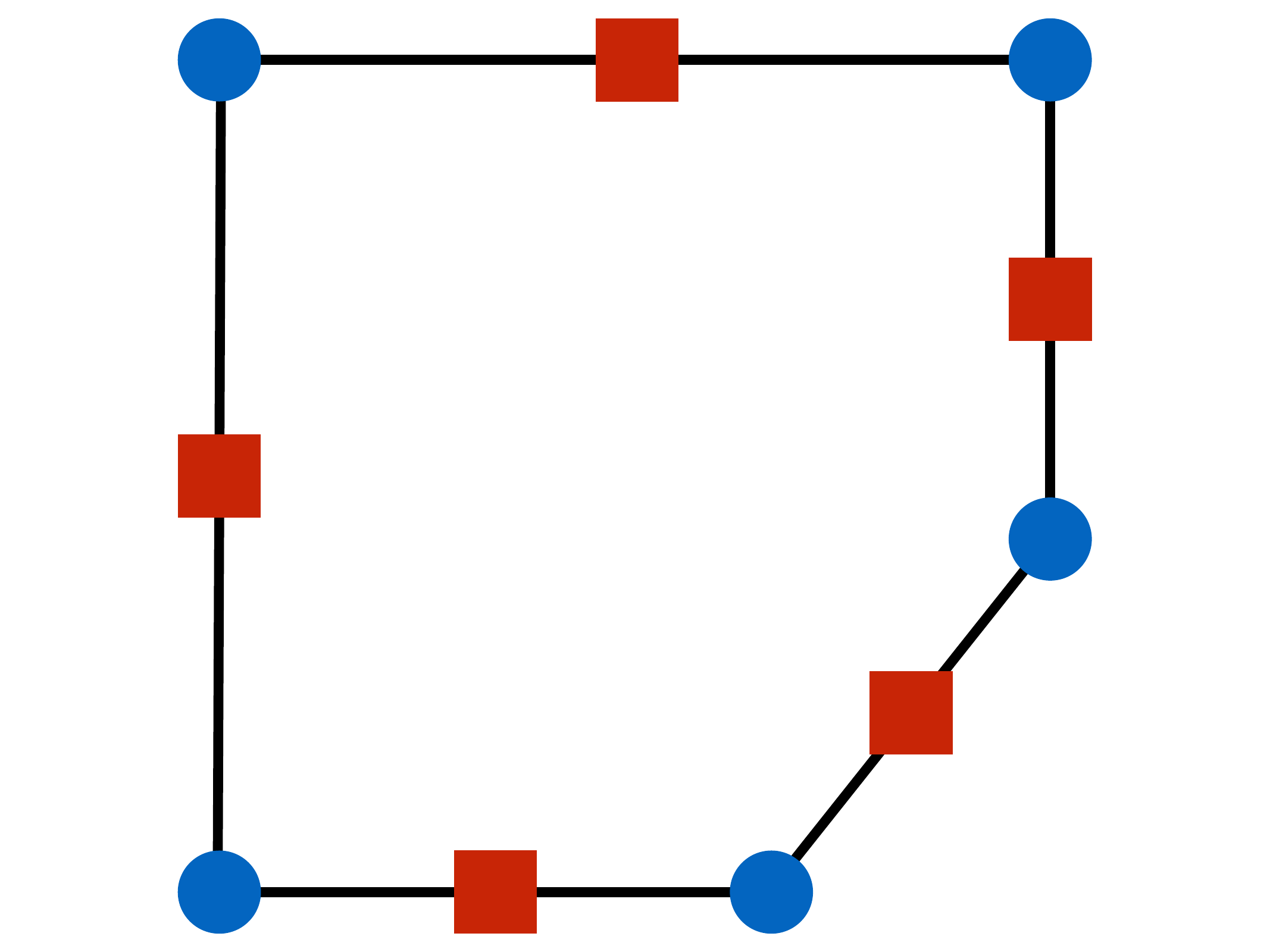}
		\put(0, -2) { ${ \psi_0 }$ }
		\put(62, -2) { ${ \psi_1 }$ }
		\put(85, 31) { ${ \psi_2 }$ }
		\put(0, 75) { ${ \psi_3}$ }
		\put(85, 75) { ${ \psi_4 }$ }

		\put(0, 35) { ${u_l }$ }
		\put(32, 13) { ${v_b }$ }
		\put(85, 50) { ${u_r }$ }
		\put(42, 78) { ${v_t }$ }
		\put(74.5, 14.5) { ${v_n}$ }
		\end{overpic}    
	\subcaption {}
	\end{subfigure}
	\hspace{0.02\textwidth}
	~
	\begin{subfigure}[t]{0.14\textwidth}
		\begin{overpic}[width=\textwidth]{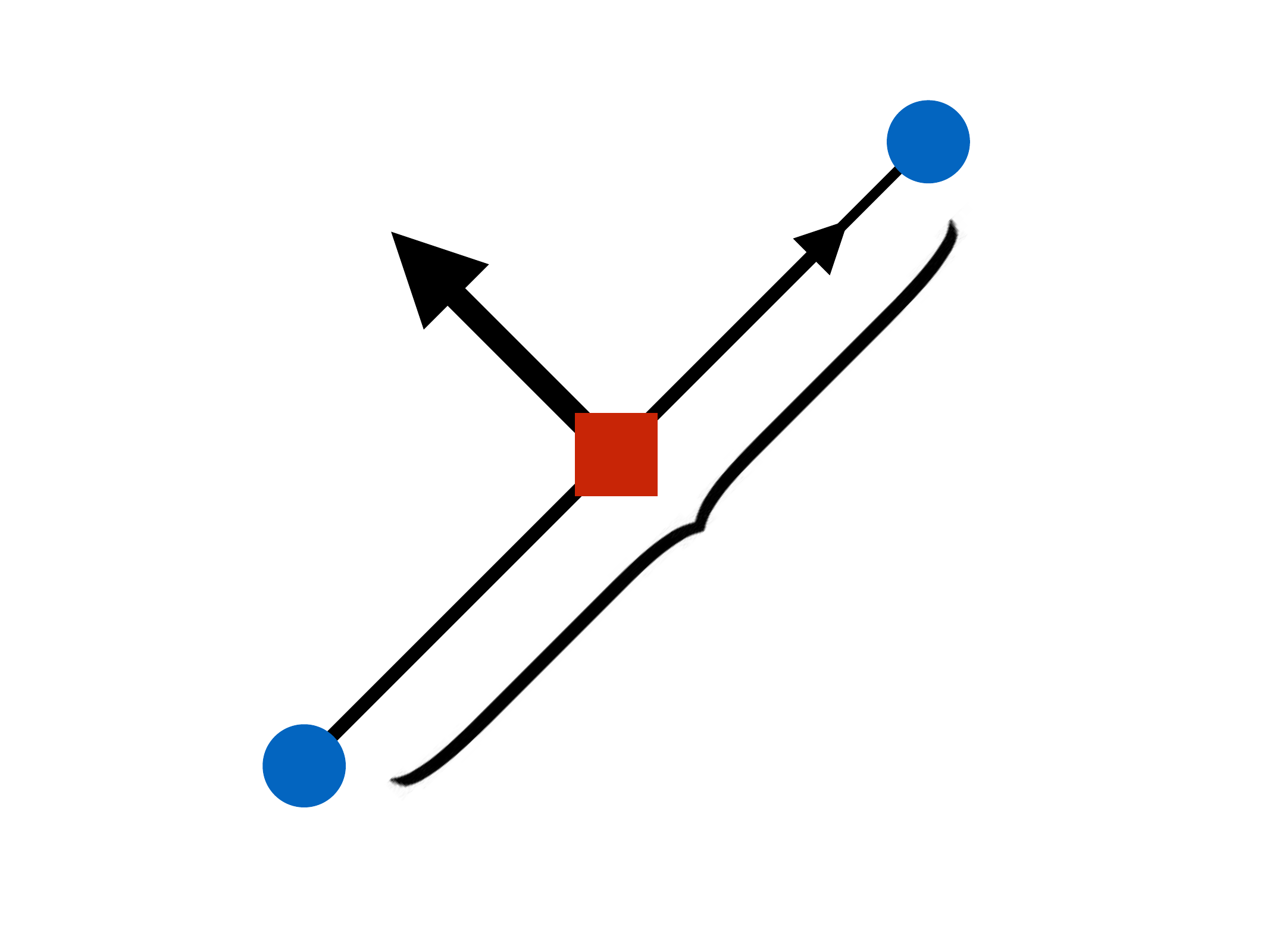}
		\put(-8, -2) { ${ \psi_i (x_i, y_i)}$ }
		\put(19, 59) { ${ \mathbf{n}}$ }
		\put(38, 75) { ${ \psi_j (x_j, y_j) }$ }
		\put(25, 38) { ${ v_n }$ }
		\put(52, 17) { ${ l }$ }
		\end{overpic}    
	\subcaption {}	
	\end{subfigure}
	\caption{ \textbf{Discretization in 2D:} Red squares represent discrete velocity samples and blue circles represent discrete stream function samples.
	(a) A uniform grid cell. (b) A cut-cell induced by clipping with a solid object. (c) The relationship between edge normal velocity and stream function samples. 
	For this case, ${v_n = -(\psi_j - \psi_i) / l}$.
	}
	\label{fig:VP2DSingleCell}
\end{figure}

For a single uniform grid cell as in Figure~\ref{fig:VP2DSingleCell}(a), 
equation~\ref{eq:VPEdge2D} leads to
\begin{equation}
\label{eq:VPUniformGrid2D}
\begin{aligned}
u_l &= \frac{ \psi_2 - \psi_0 }{ h }, && 
u_r = \frac{ \psi_3 - \psi_1 }{ h }, \\
v_b &= -\frac{ \psi_1 - \psi_0 }{ h}, && 
v_t = -\frac{ \psi_3 - \psi_2}{h},
\end{aligned}
\end{equation}
consistent with finite differences on \eqref{eq:VP2D}. We seek $\psi_i$ satisfying the given velocities. These equations have a 1D null space: constant offsets of all $\psi_i$ do not change the velocities. (Physically, discrete incompressibility ensures one flux is the negated sum of the others, implying one equation is linearly dependent.) We select a unique solution by arbitrarily setting ${\psi_0 = 0}$. We determine the other $\psi_i$ by traversing the cell's edge graph, computing each subsequent $\psi$ value from its predecessor on the edge and the edge's velocity component. Discrete incompressibility ensures discrete integrability, i.e., looping back to $\psi_0$ gives a consistent result.

\begin{figure}
	\centering	
	\begin{subfigure}[t]{0.23\textwidth}
		\begin{overpic}[width=\textwidth]{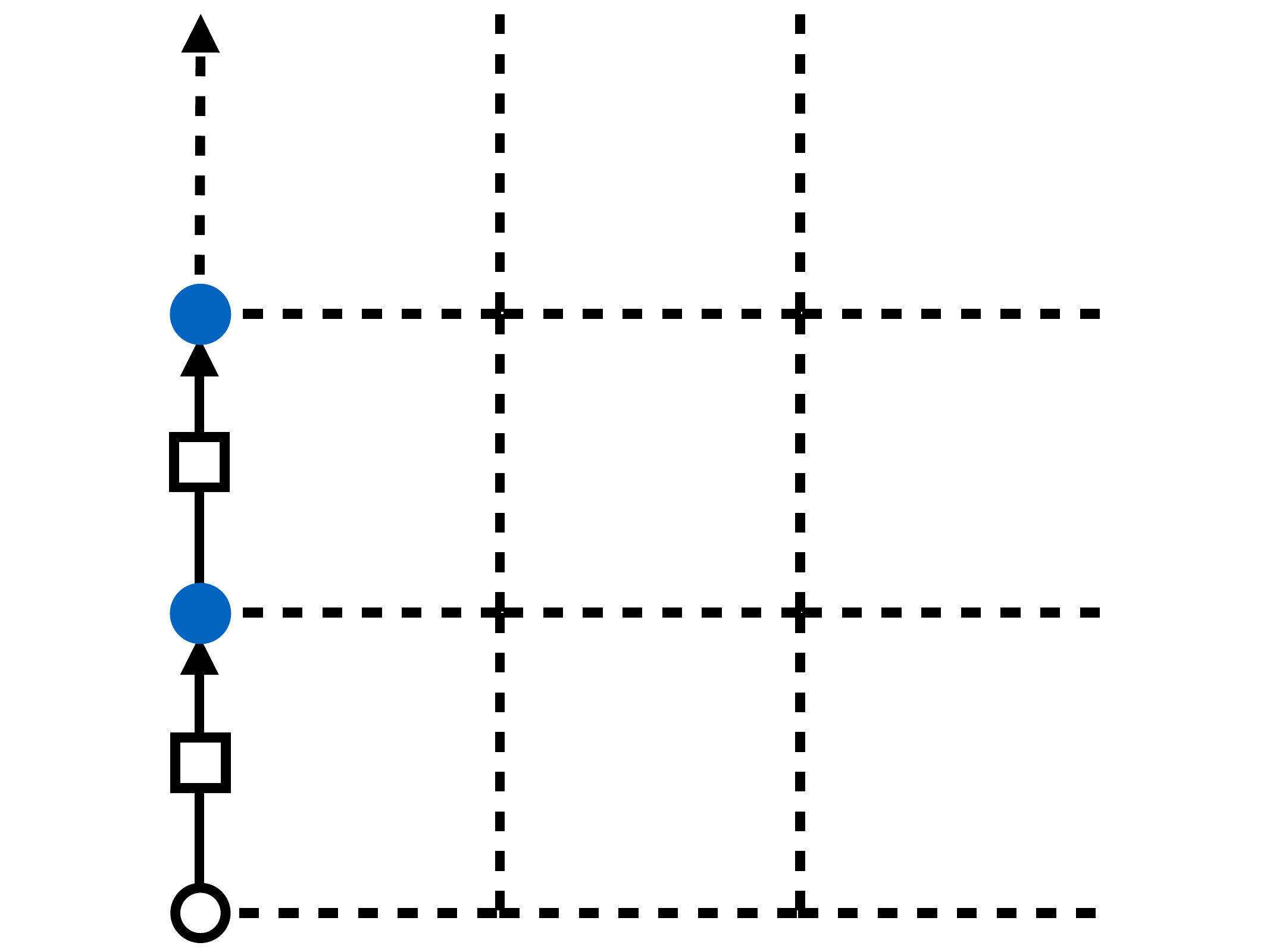}
		\put(2, 1) { ${ \psi_0 }$ }		
		\put(2, 25) { ${ \psi_1 }$ }		
		\put(2, 49) { ${ \psi_2 }$ }		
		\put(18, 37) { ${ u_1 }$ }
		\put(18, 13) { ${ u_0 }$ }		
		\end{overpic}    
	\end{subfigure} 
	~
	\begin{subfigure}[t]{0.23\textwidth}
		\begin{overpic}[width=\textwidth]{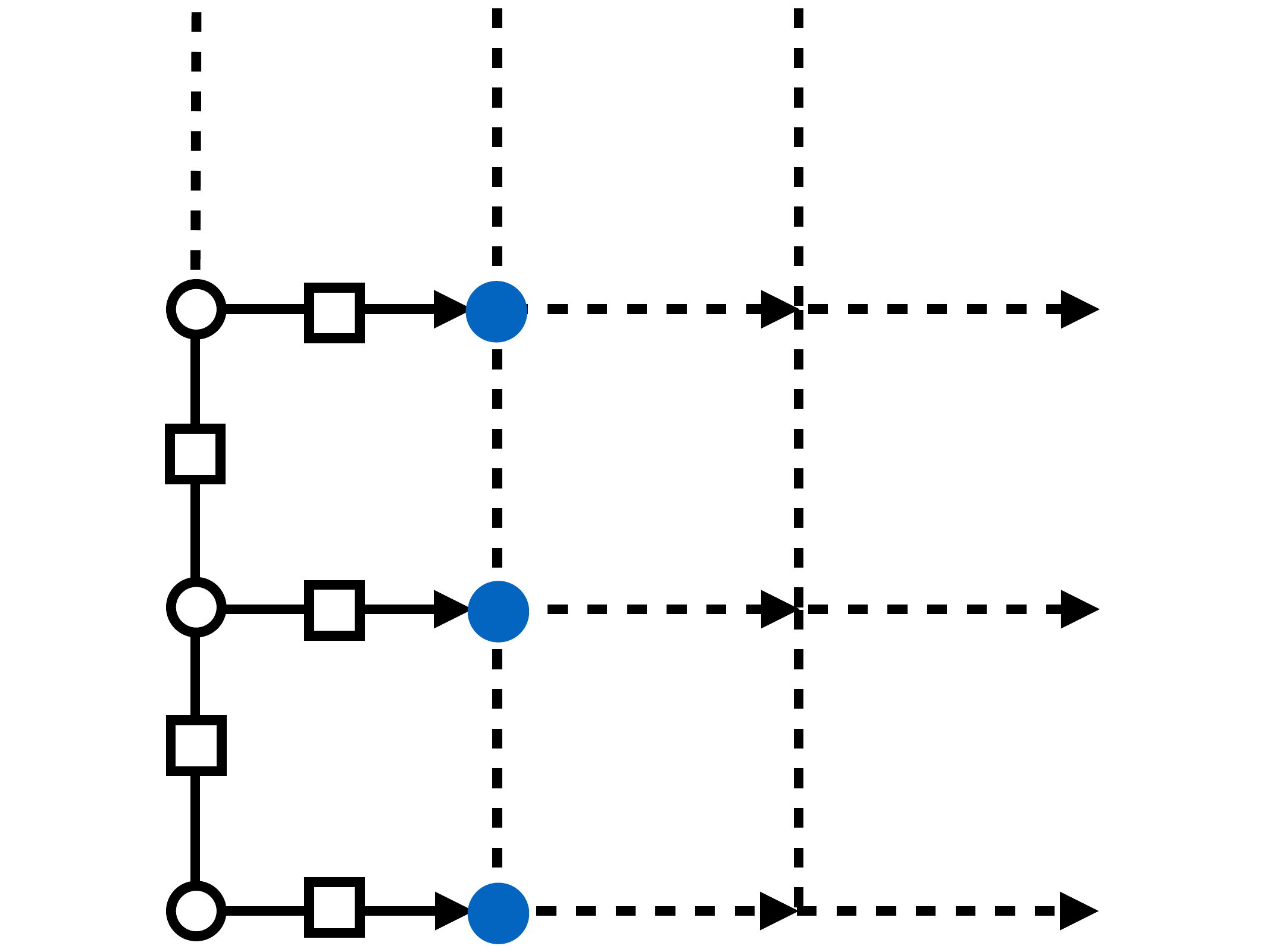}
		\put(2,  0) { ${ \psi_0 }$ }		
		\put(38, 8) { ${ \psi_3 }$ }		
		\put(22, 8) { ${ v_0}$ }		
		\end{overpic}    
	\end{subfigure}
	\caption{\textbf{Parallel Sweeping} efficiently recovers a uniform grid discrete stream function field.
	Left: First, sequentially compute all ${\psi}$ values at ${x = x_0}$, starting from ${\psi_0 = 0}$ and sweeping in the $y$ direction. 
	Right: Next, sweep in the $x$ direction, processing rows in parallel.
	}
	\label{fig:VP2D}
\end{figure}
This graph traversal scheme applies likewise to a whole uniform grid of cells, thereby avoiding a costly global Poisson solve for $\psi$ based on \eqref{eq:divCurl}. Any ordering suffices, so we aim to maximize parallelism.
As proposed by \citet{Biswas2016} and illustrated in Figure~\ref{fig:VP2D}, we first set ${\psi_0 = 0}$, and sweep vertically to obtain all the ${\psi}$ values at ${x = x_0}$ (i.e., ${\psi_1 = \psi_0 + h u_0}$ and so on).
We then sweep horizontally, in parallel, to obtain all remaining ${\psi}$ values. This approach is compatible with any standard exterior domain boundary conditions (inflow/outflow/open/closed), since it only needs the velocity data. 

\begin{wrapfigure}{l}{0.15\textwidth} %
    \centering
    	\begin{overpic}[width=1.0\linewidth]{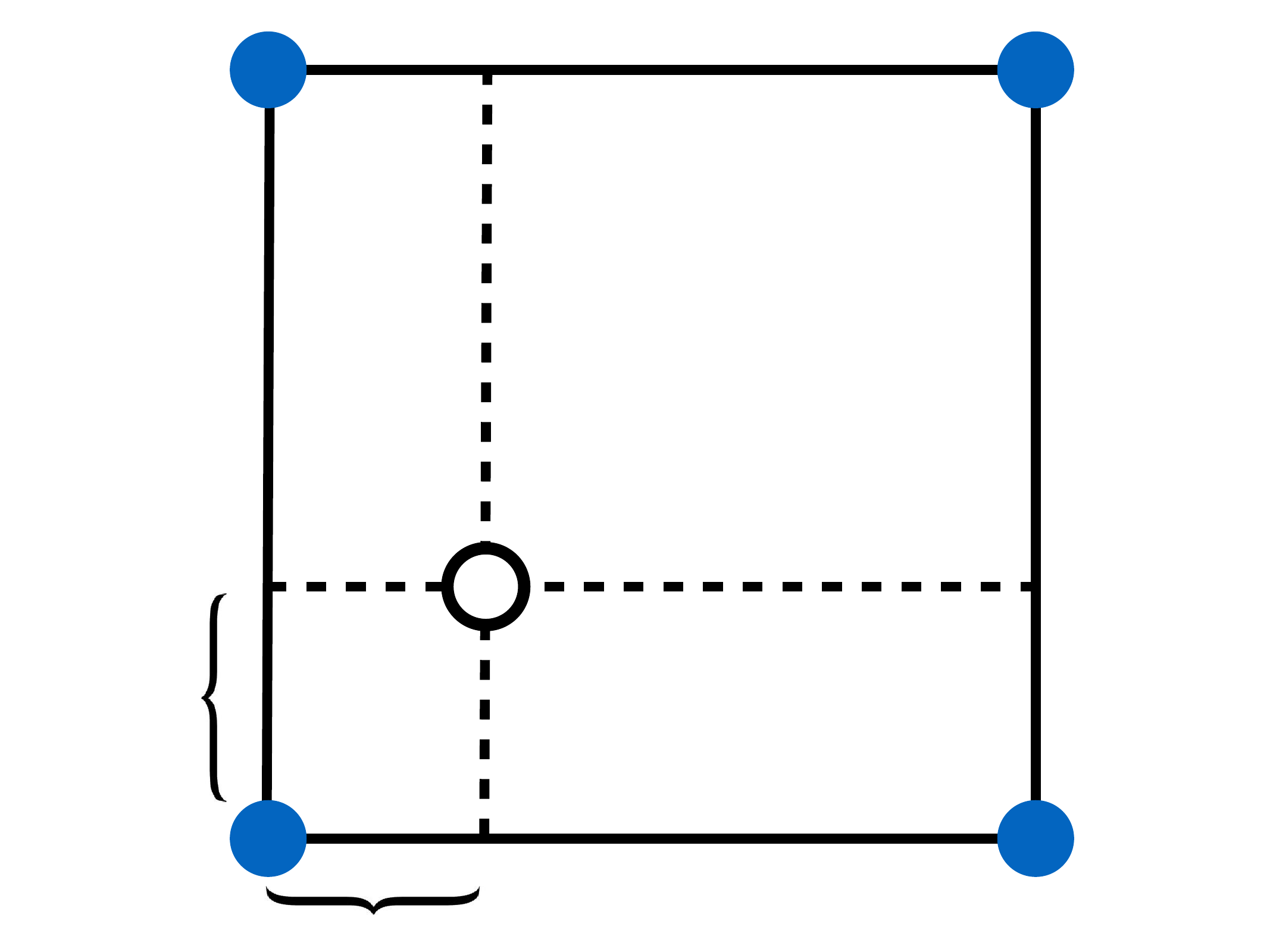}
	        \put(5, -1) { ${ \psi_0}$ }
	        \put(82, -1) { ${ \psi_1}$ }
	        \put(5, 73) { ${ \psi_2}$ }
	        \put(82, 73) { ${ \psi_3}$ }

	        \put(25, -7) { ${ \alpha(x) }$ }
	        \put(-10, 20) { ${ \beta(y) }$ }

	        \put(35,  35) { ${ (x, y)}$ }
       	\end{overpic}    
	\hspace{-12mm}
\end{wrapfigure} 

\subsubsection{Interpolating $\psi$} Interpolation of the grid $\psi$ values provides
an analytical stream function at every point.
Applying the analytical curl, we obtain a velocity field which is \emph{pointwise} divergence-free by construction. 
As a simple example, for bilinear interpolation (referring to the inset figure),
\begin{equation}
\label{eq:bilinearCF}
\begin{aligned}
\psi(x, y) &= lerp(lerp(\psi_0, \psi_2, \beta(y)),lerp(\psi_1,\psi_3, \beta(y)), \alpha(x)), \\
u(x, y) &=  \ \  \frac{lerp(\psi_2-\psi_0, \psi_3-\psi_1, \alpha(x))}{h} = lerp(u_l, u_r, \alpha(x)),\\
v(x, y) &= -\frac{lerp(\psi_1-\psi_0, \psi_3-\psi_2, \beta(y))}{h} = lerp(v_b,v_t, \beta(y)),\\
\end{aligned}
\end{equation}
where $lerp(a,b,t)= (1-t)a+t b$ and the rightmost equalities follow from \eqref{eq:VPUniformGrid2D}.
In this specific case, the analytical divergence is clearly zero because it exactly equals the finite difference divergence: 
\begin{align}
\nabla \cdot \mathbf{u} = \frac{u_r - u_l + v_t - v_b}{h} = 0.
\end{align}
The derivatives in the curl operator inherently induce a different polynomial degree per axis: here, velocity components are piecewise constant in one axis and piecewise linear in the other, causing undesired tangential discontinuities between cells (Figure \ref{fig:divComparison}, middle).

Fortunately, because we recovered the discrete $\psi$, we can achieve velocity continuity simply by upgrading to a higher order interpolant (Figure \ref{fig:divComparison}, right). 
Although many choices are possible, we adopt quadratic dyadic B-spline kernels \cite{Steffen2008,Jiang2016};
they are sufficiently smooth, their analytical curl is straightforward to derive, and they possess small stencils for efficiency.
\citet{Bao2017} similarly used tensor-product B-splines when constructing divergence-aware Dirac delta functions.

For the special case of bilinear $\psi$ interpolation on 2D uniform grids, \eqref{eq:bilinearCF} shows that the velocity interpolant reduces to a simple function of the input discrete $u,v$ velocities, eliminating the need for explicit $\psi$ values. 
This effect occurs because, in 2D, each velocity component is defined by a single partial derivative of $\psi$ and the null space is a constant shift. 
However, deriving simple direct incompressible velocity interpolants in this way becomes unwieldy or impossible in the combined presence of cut-cell geometry, higher order interpolation, and especially 3D,
where each velocity component depends on two vector potential components and the null space is nontrivial. We therefore  prefer to explicitly recover $\boldsymbol{\psi}$.


\subsection{Cut-Cell Obstacles in 2D}
We next extend our scheme to support irregular solid obstacles.
\label {VPCutCell2D}
\subsubsection{Recovering discrete $\psi$}
Considering cut-cell solids during stream function recovery only requires accounting for truncated grid edges and non-axis-aligned "cut edges".
Applying ~\eqref{eq:VPEdge2D} we get an update rule for sweeping through such edges:
\begin{equation}
\label{eq:VPCutCell2D}
\psi_{i + 1} = \psi_i + l_{i, i + 1} v_{i, i + 1} (\mathbf{e}_{i, i + 1}^\perp \cdot \mathbf{n}_{i, i + 1}).
\end{equation}
For the static obstacles we consider, the recovered $\psi$ will be constant on the surface, i.e., the surface is an isocontour that particles slide along in a free-slip manner \cite{Bridson2007}. 
Static obstacles also preserve the remarkable simplicity of our parallel sweeping method: nodal stream function values at the entry and exit point of a grid line passing through an obstacle are identical, so we can sweep \emph{through} obstacles by assuming zero velocities inside.

\subsubsection{Interpolating $\psi$}
\setlength{\columnsep}{7pt} %
\begin{wrapfigure}[9]{l}{0.15\textwidth} %
    \vspace{-3mm}
    	\begin{overpic}[trim = 4cm 0 6cm 0, clip, width=1.0\linewidth]{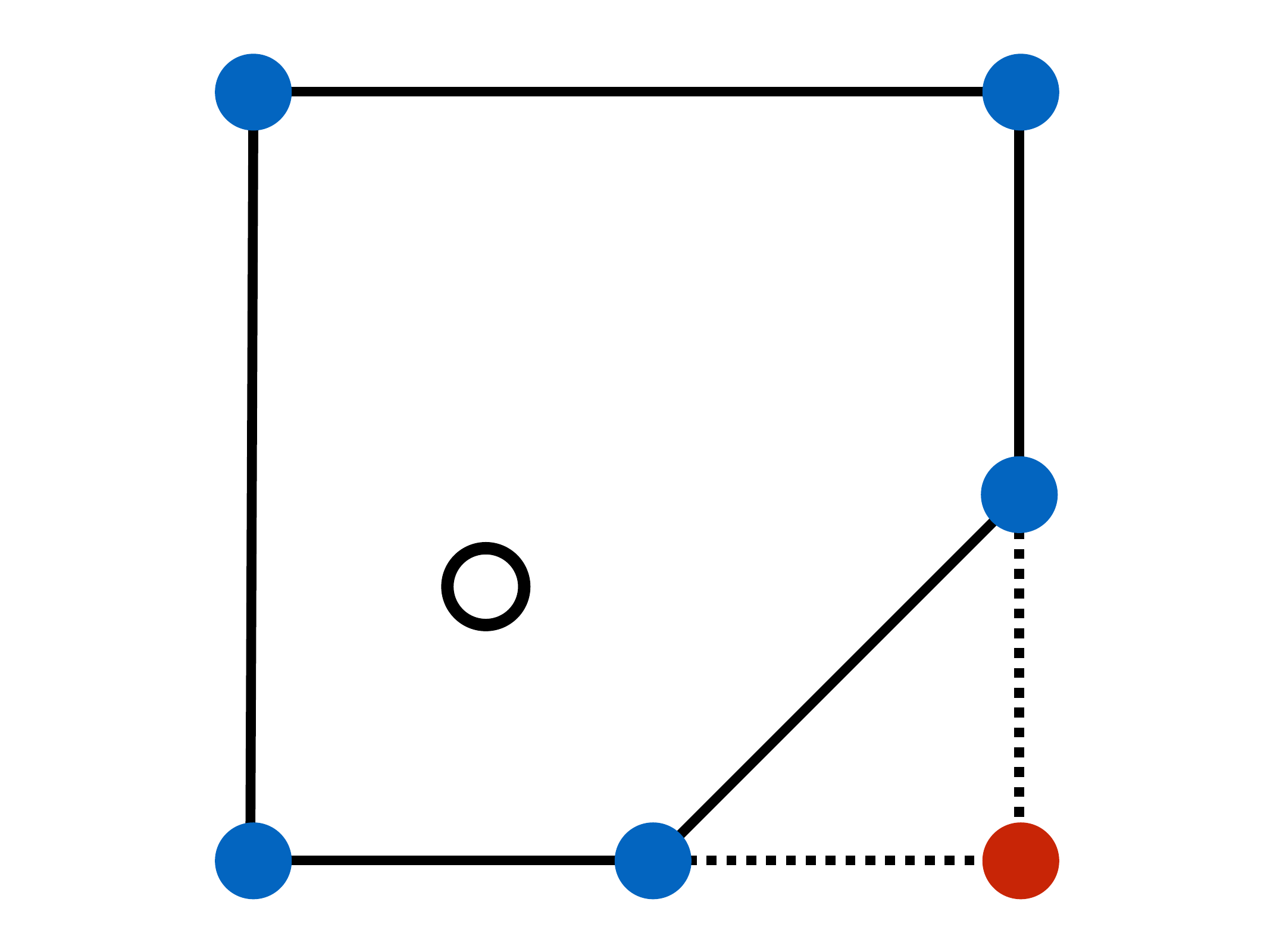}
	        \put(91, -1) { ${ \psi_s}$ }
	        \put(91, 35) { ${ \psi_g}$ }
	        \put(45, -3) { ${ \psi_g}$ }
       	\end{overpic}    
\end{wrapfigure} 
\setlength{\columnsep}{20pt}%
Cut-cell solids cause the nodal $\psi$ values (blue in the inset) to no longer lie in a convenient uniform grid, thereby complicating interpolation. 
One could use moving least squares (MLS) interpolation based on $\psi$ data in some neighborhood, but MLS requires solving small dense linear systems and computing the necessary analytical derivatives is difficult \cite{Huerta2004}. 
Instead we extrapolate $\psi$ values to uniform grid nodes inside the solid (i.e., ${\psi_s}$ in the inset) and use quadratic B-spline kernels as in the uniform grid case.
This extrapolation can be done in various ways, such as using MLS or simply copying (e.g., from ${\psi_g}$ to $\psi_s$), assuming discrete flux is zero on interior solid edges (dashed lines); we adopt the latter for simplicity. 
At the outer axis-aligned domain boundaries, the quadratic stencil is also missing data samples, so we extrapolate an extra uniform layer of $\psi$ samples outside the domain.
However, regardless of these various extrapolation and interpolation choices near boundaries, the desired pointwise no-normal-flow boundary condition is (so far) enforced only approximately; we propose a final correction below.


\subsubsection{A Curl-Noise enhancement for exact 2D boundary enforcement}
\label{ramping2D}
To ensure that the interpolated $\psi$ will be \emph{strictly} constant along the polygonal boundary for static obstacles, we adapt ideas from the Curl-Noise method \cite{Bridson2007}. 
That method modulates the existing $\psi$ value near obstacles, forcing it to a constant zero value on the boundary by multiplying against a smooth ramp function based on boundary proximity. Defining $d_0$ as an influence radius (we use the grid width $h$), $d(\cdot)$ as the distance to the surface, and $ramp(\cdot)$ as a smooth ramp function (we use a classic smoothstep function), 
Bridson's multiplicative correction to the ambient $\psi$ is the product $\psi'(\mathbf{x}) = \alpha \psi(\mathbf{x})$, where $\alpha = ramp(d(\mathbf{x})/ d_0)$.

\begin{figure}
	\centering	
		\fbox{\begin{subfigure}[t]{0.25\textwidth}
		\begin{overpic}[width=\textwidth]{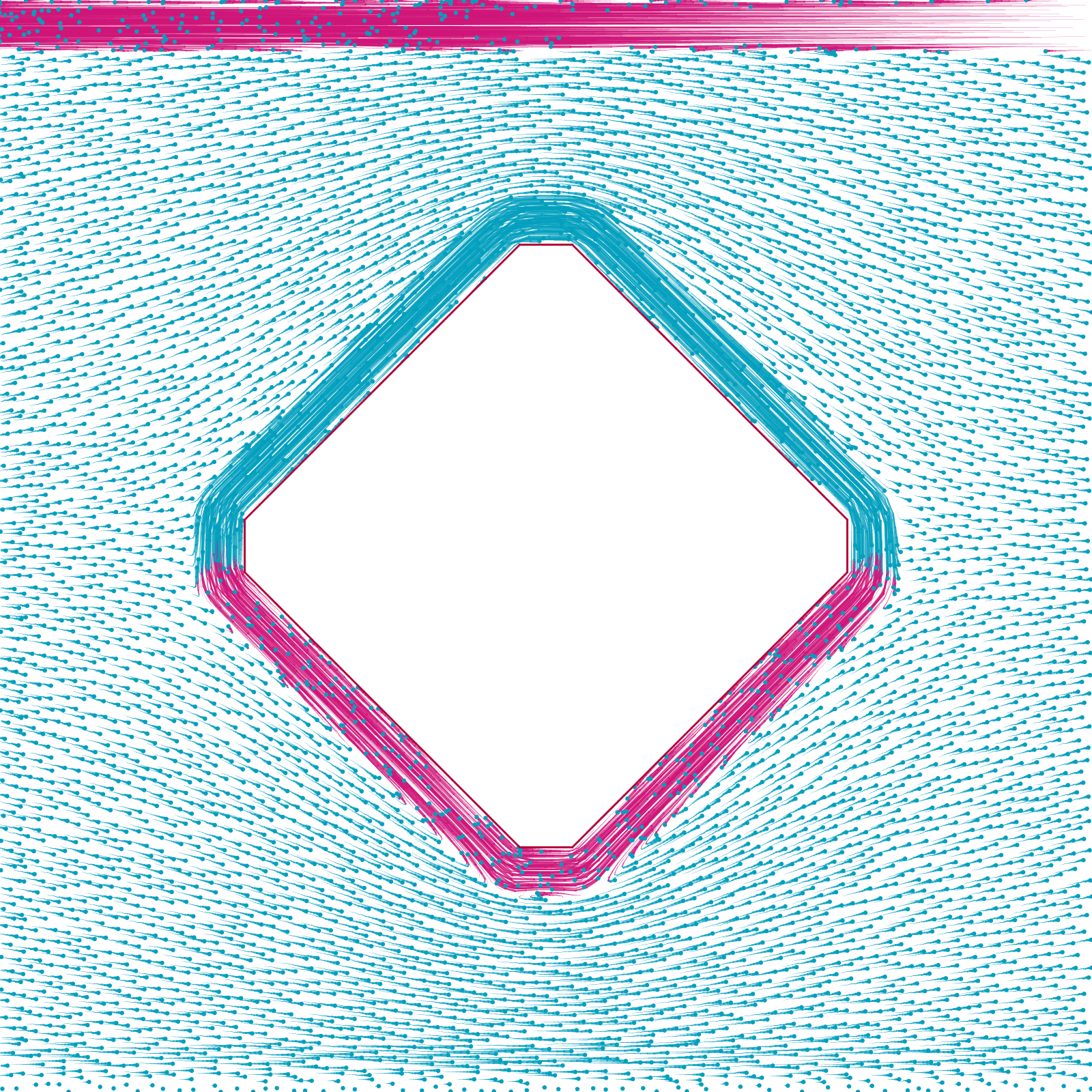}
		\put(-39, 0) { ${ \psi_{bottom} = 0 }$ }
		\put(-34, 96) { ${ \psi_{top} = 0.5 }$ }
		\put(39, 49) { ${ \psi_{t} = 0 }$ }
		\end{overpic}    
	\end{subfigure}}
	\caption{\textbf{Boundary Ramping Issues:} Multiplicative Curl-Noise ramping with a target value of zero (${\psi_g = 0}$) applied to a steady rightward flow, where the constant inflow/outflow rate is described by a vertical linear stream function value from  ${\psi_{top}=0.5}$ to ${\psi_{bottom}=0}$. 
	Blue trails indicate the right direction ($+x$) and red trails indicate (erroneous) trajectories in the reverse ($-x$) direction.
	Because the expected discrete $\psi_g$ value should differ at each surface,
    always ramping to $\psi_g=0$ causes spurious tangential flow near the obstacle, undesired no-slip at the bottom where $\psi$ is already zero, and large reverse flow at the top.}
    \label{fig:ramp_mulzero}
\end{figure}

\begin{figure}
	\begin{subfigure}[t]{0.22\textwidth}
		\fbox{\begin{overpic}[width=\textwidth]{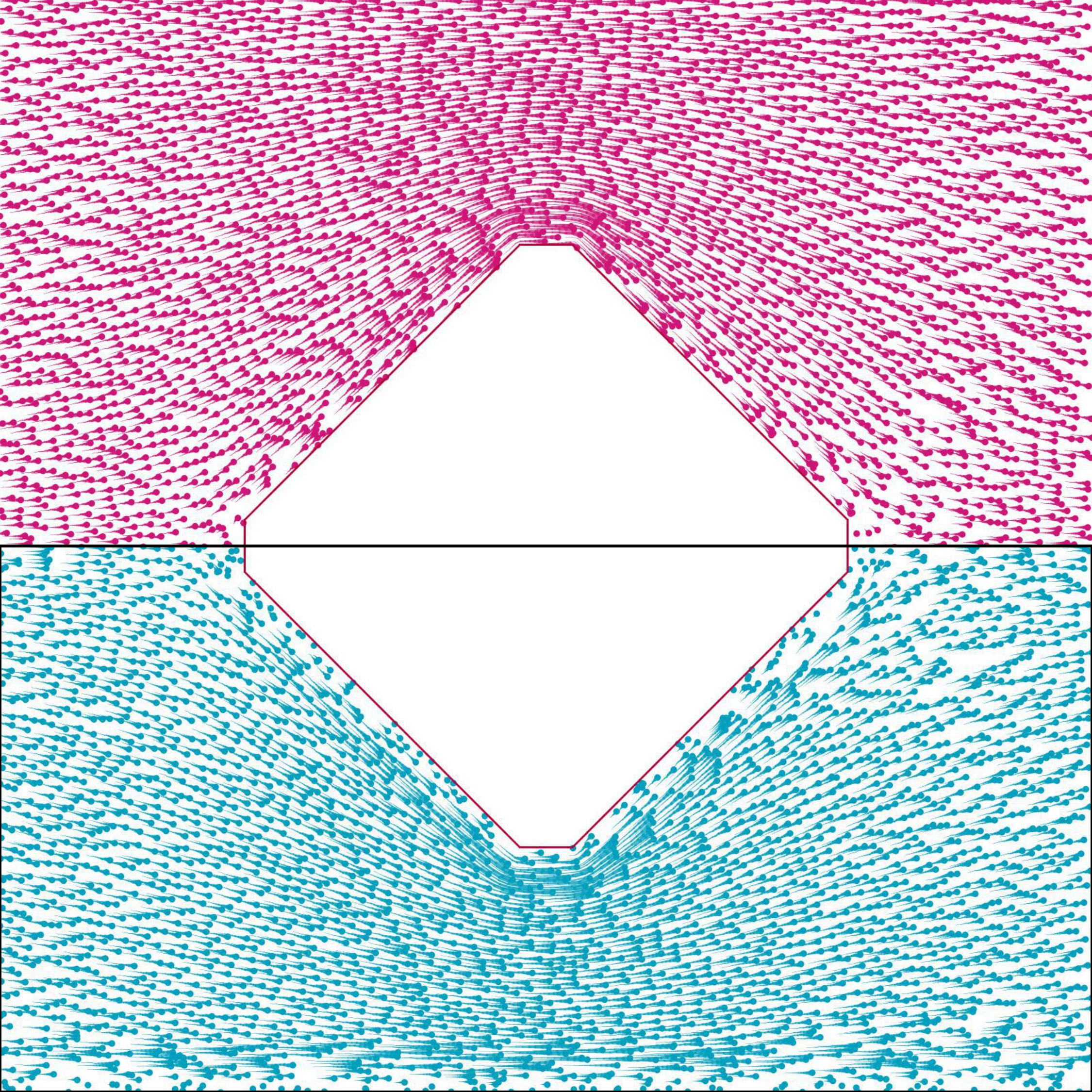}
		\end{overpic}}    
		\subcaption{}
	\end{subfigure} \hspace{3mm}%
	\begin{subfigure}[t]{0.22\textwidth}
		\fbox{\begin{overpic}[width=\textwidth]{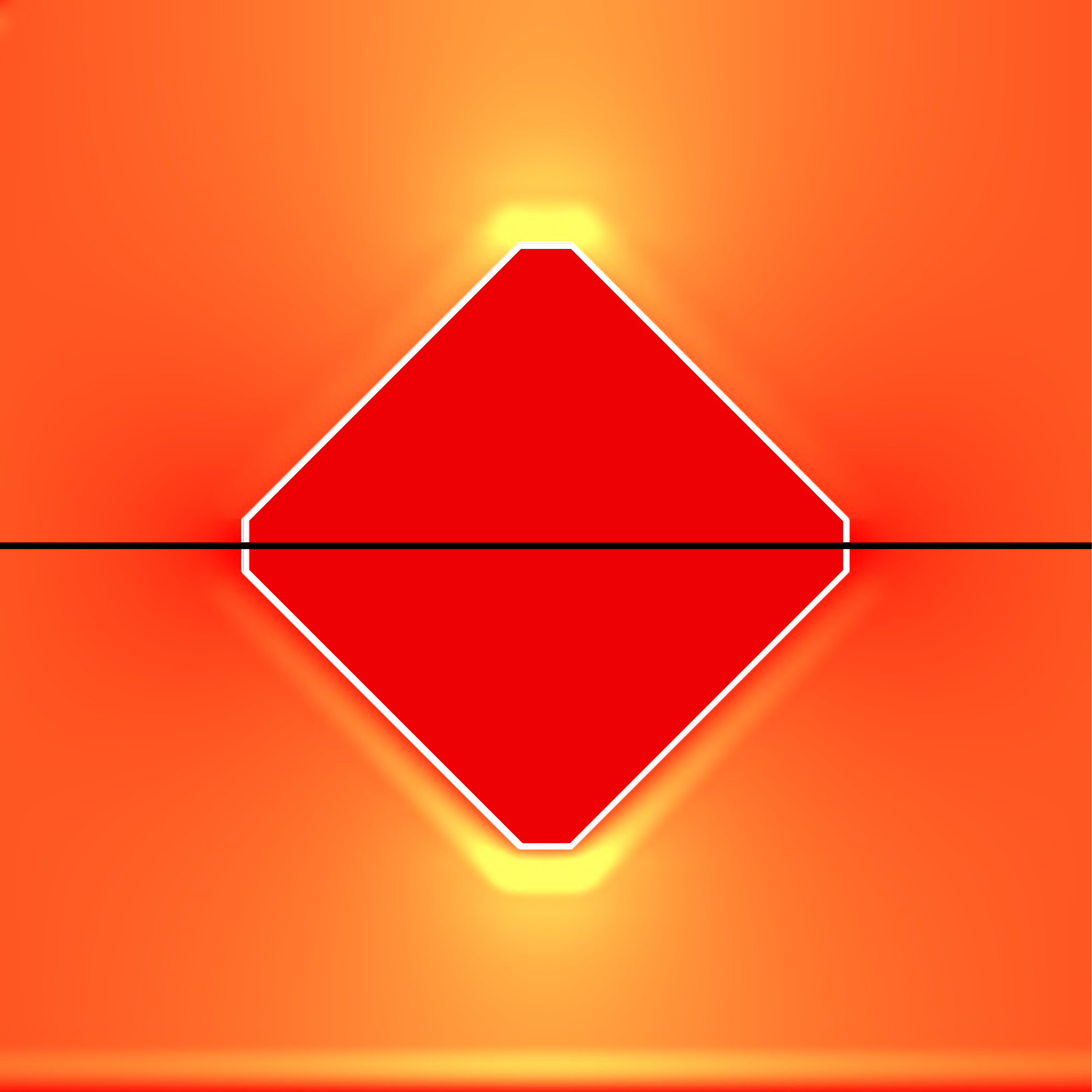}
		\end{overpic}}    
		\subcaption{}
	\end{subfigure}
    \caption{\textbf{Additive vs. Multiplicative Ramping}: A steady rightward flow with boundary $\psi_g$ value adjustment. Left: Particles with speed-dependent trail lengths. Right: Velocity magnitudes.
    Top half: Our additive ramping exactly satisfies the boundary with no apparent damping of free-slip velocities.
    Bottom half: Despite modifying Bridson's multiplicative ramp to target the correct $\psi_g$ value, tangential damping occurs along the bottom wall and near the solid, since $\partial \psi / \partial \mathbf{n}$ is damaged. Note the more pronounced banding in the bottom magnitude plot.}
    \label{fig:rampComparison}
\end{figure}

However, depending on the background $\psi$, the appropriate target boundary value $\psi_g$ is often \emph{not} zero; arbitrarily choosing zero leads to dramatic spurious flow deviations (Figure \ref{fig:ramp_mulzero}). 
One can modify Bridson's method to "ramp in" the nonzero target boundary value $\psi_g$: 
\begin{align} \label{eq:bridsonboundaries}
\psi'(\mathbf{x}) = \alpha \psi(\mathbf{x}) + (1-\alpha)\psi_g.
\end{align}
Unfortunately, the initial multiplication still damages the \emph{normal derivatives} of $\psi$. 
Since $\mathbf{u} = \nabla \psi^\perp$ implies $(\mathbf{u} \cdot \mathbf{t})\mathbf{t} = \frac{\partial \psi}{\partial \mathbf{n}}$, where $\mathbf{t}$ is the boundary tangent vector, \eqref{eq:bridsonboundaries} induces undesired tangential damping, acceleration, or no-slip behavior (Figure \ref{fig:rampComparison}, bottom half).

We propose a novel \emph{additive} ramping procedure that instead enforces $\psi_g$ by adding a compensating offset. 
The required offset is determined by evaluating the existing $\psi$ at the closest boundary point and subtracting it from $\psi_g$. 
Letting $cp(\cdot)$ be a function returning the closest boundary point, we modify $\psi$  as:
\begin{align}
\psi'(\mathbf{x})=\psi(\mathbf{x}) + (\psi_g - \psi(cp(\mathbf{x})))(1-\alpha).
\label{eq:additiveRamp}
\end{align}
This expression ramps the full correction precisely on at the boundary ($d(\mathbf{x}) / d_0 = 0$) and blends it smoothly off at the edge of the influence region ($d(\mathbf{x}) / d_0 = 1$). 
The inset shows an example input $\psi$ curve (black), and the result after using multiplicative (red) and our additive (green) ramping to $\psi_g=0$. 

\begin{wrapfigure}[10]{r}{0.5\linewidth}
  \begin{center}
    \vspace{-0.3cm}
    \hspace{-1.5\columnsep}
    \includegraphics[width=\linewidth,trim=3.8cm 8.6cm 4.7cm 9cm, clip]{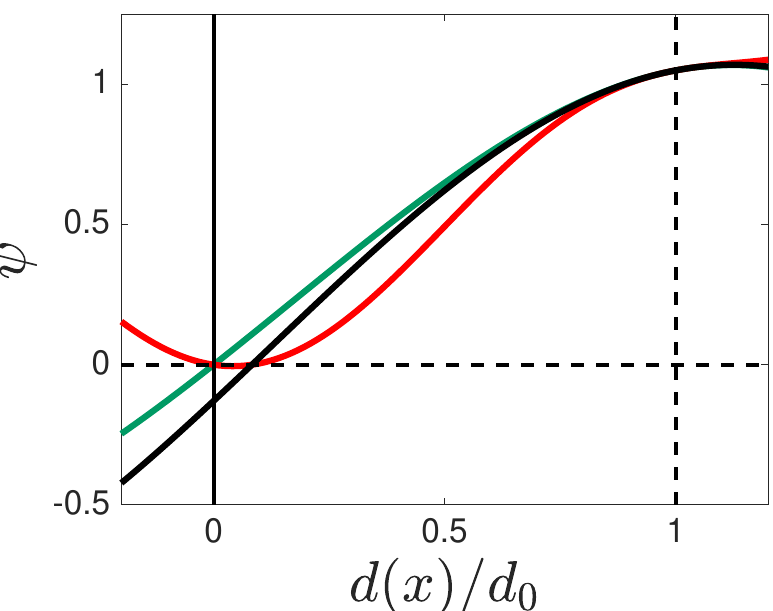}
  \end{center}
\end{wrapfigure}
Both methods correct the value, but additive ramping better preserves the derivative near the boundary and thus yields more faithful free-slip flow (Figure \ref{fig:rampComparison}, top half). We again use the analytical curl to determine this added velocity contribution. 

By its nature, applying ramping on piecewise flat (polygonal or polyhedral) geometry can introduce some discontinuity in velocity near sharp edges or vertices.
This is because the recovered velocity $\mathbf{u'} = \nabla \psi'^\perp$ involves a $\nabla \alpha$ term from \eqref{eq:additiveRamp} (i.e., gradient of the normalized distance field), and $\nabla \alpha$ is not continuous in general. Nevertheless, our Curl-Flow method produces precisely boundary-respecting and divergence-free flows, and the discontinuities are often acceptable in practice, since the applied corrections only occur within one cell-width of obstacles. 

In the context of our overall interpolation scheme, the ambient $\psi$ is the interpolated grid-based $\psi$ field and the target boundary $\psi_g$ values are known at solid surface nodes, since they are recovered during sweeping.
We apply our ramping strategy to closed exterior domain boundaries and static solid surfaces.
In those cases, $\psi_g$ is a constant value, since the boundary's normal flux is zero. 
For exterior domain boundaries that have nonzero inflow/outflow, 
the discrete ${\psi}$ edge endpoint values of a segment differ (e.g., $\psi_{c_0}$, $\psi_{c_1}$).
One can set the ramping target value ${\psi_g}$ to the linearly interpolated value of $\psi_{c_0}$ and $\psi_{c_1}$ at the closest point to achieve the desired constant flow in the normal direction across that edge.
However, for simplicity, we did not apply our ramping-based enhancement to the exterior domain boundaries with inflow/outflow conditions, since exact enforcement there is not visually critical.

\begin{figure}
	\centering	
	\begin{subfigure}[t]{0.14\textwidth}
		\adjustbox{fbox, width=\textwidth, , trim = 15cm 6cm 15cm 6cm, clip}{\includegraphics{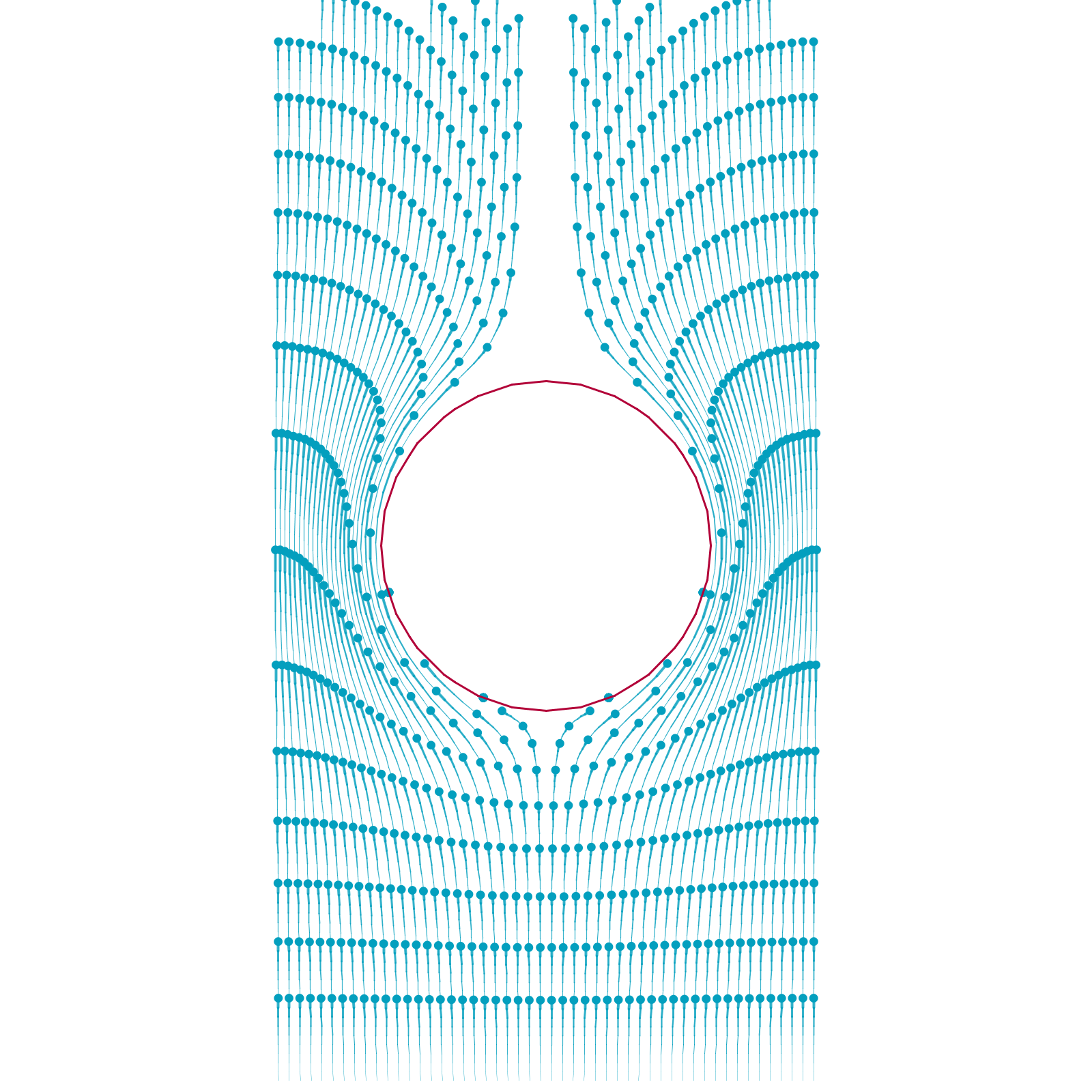}}
		\subcaption{}
	\end{subfigure}
	\begin{subfigure}[t]{0.14\textwidth}
		\adjustbox{fbox, width=\textwidth, , trim = 15cm 6cm 15cm 6cm, clip}{\includegraphics{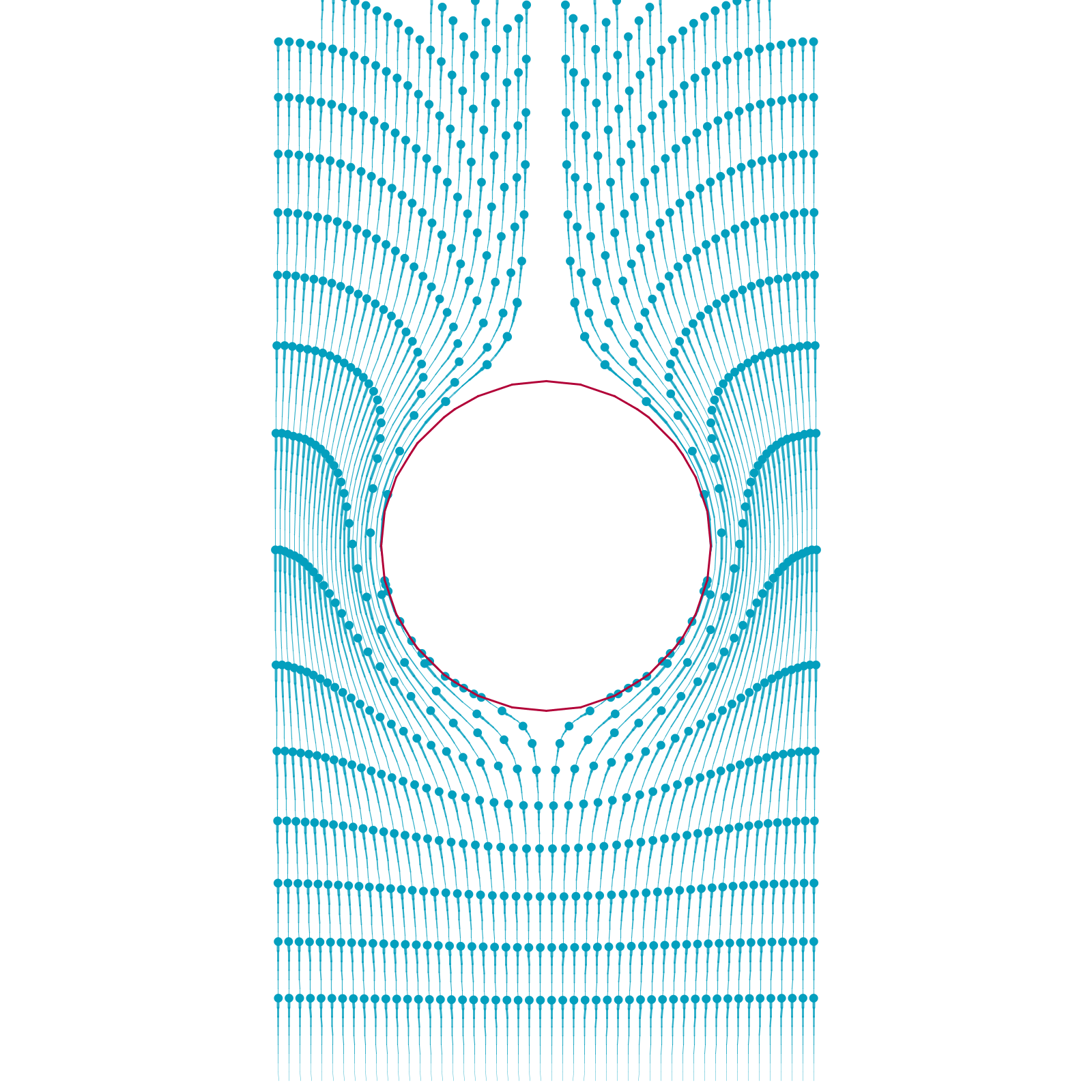}}
		\subcaption{}
	\end{subfigure}
	\begin{subfigure}[t]{0.14\textwidth}
		\adjustbox{fbox, width=\textwidth, trim = 15cm 6cm 15cm 6cm, clip}{\includegraphics{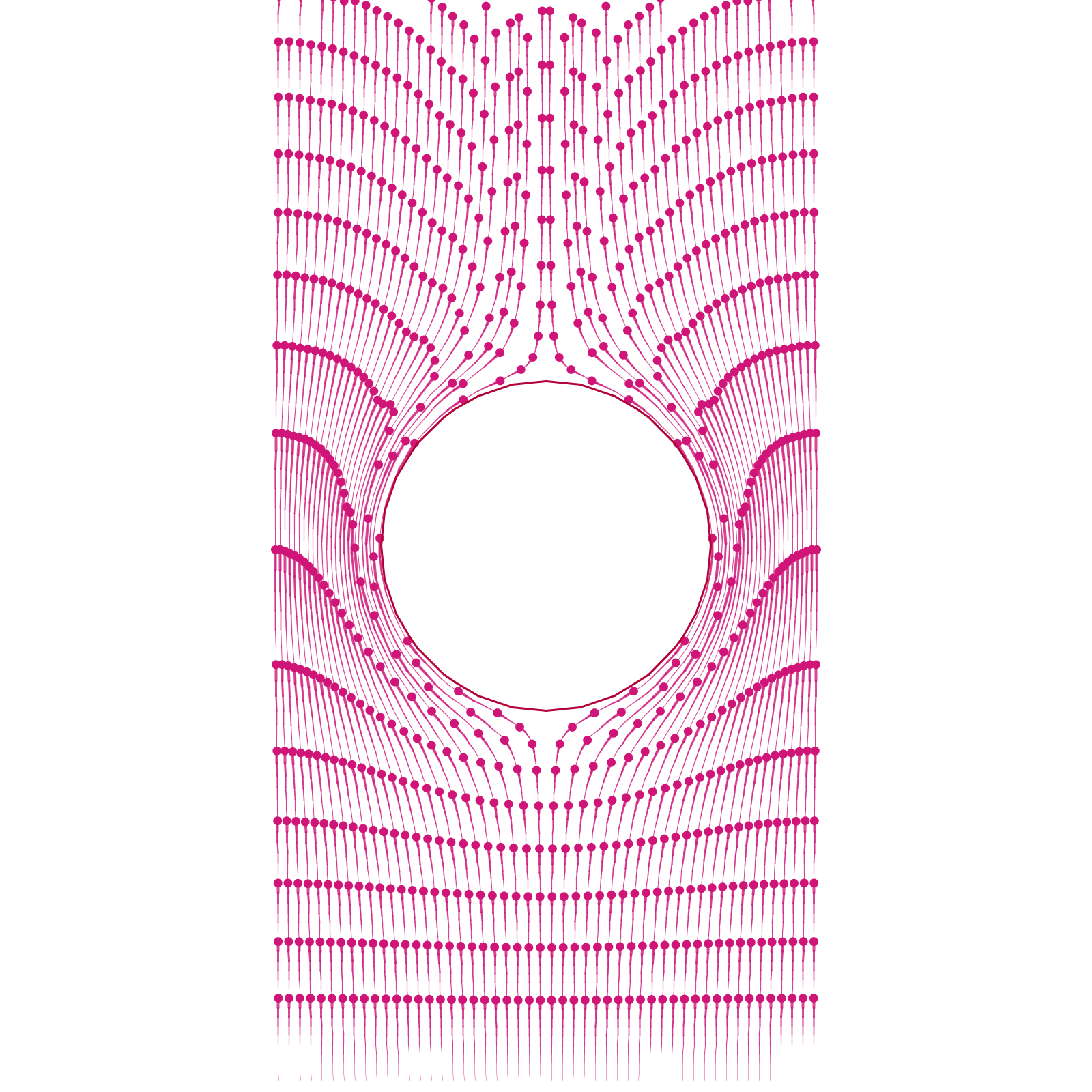}}
		\subcaption{}
	\end{subfigure}
	\caption{\textbf{Flow Near Solids:} Particle trajectories flowing from bottom to top under a fixed (steady state) flow past a solid disk in 2D. Grid resolution: $30 \times 15$. 
	(a) Direct velocity interpolation induces spurious trailing gaps, when trajectories colliding with the solid are terminated.
	(b) With the same velocity interpolant, projecting penetrating particles back to the surface does not significantly help divergent trajectories to "close up" and unwanted particle clumping occurs.
	(c) Our Curl-Flow method better respects solid boundaries and produces less spurious empty space behind the solid. Notice that the trajectories are simply isocontours of the 2D $\psi$ field.
	}
	\label{fig:staticDiskTrajectories}
\end{figure}

Figure \ref{fig:staticDiskTrajectories} compares direct (bilinear) velocity interpolation  vs.\ our full Curl-Flow method (including additive ramping boundary correction) on a simulated steady-state flow past a disk. 
To approximate free-slip flow for the comparison direct method, we extrapolate fluid velocities (with solid-normal components projected out) to samples inside the solid \cite{Houston2003,Rasmussen2004}.

\section{Curl-Flow Interpolation in 3D}
\label{VP3D}
In three dimensions, the scalar stream function $\psi$ is replaced by the vector potential $\boldsymbol{ \psi} = ({\psi_x}, {\psi_y},{\psi_z})$. Each velocity component is now dictated by the \emph{interactions} of two $\psi$ components' derivatives (e.g., $u = \partial \psi_z / \partial y - \partial \psi_y / \partial z$), which complicates boundary enforcement, especially for irregular geometry. The curl operator also possesses a multi-dimensional null space, necessitating enforcement of a \emph{gauge condition} to find an appropriate unique $\boldsymbol{ \psi}$.


\subsection{Uniform Grids in 3D}
\label{gridCells}

\begin{figure}
\begin{overpic}[width=2.5in]{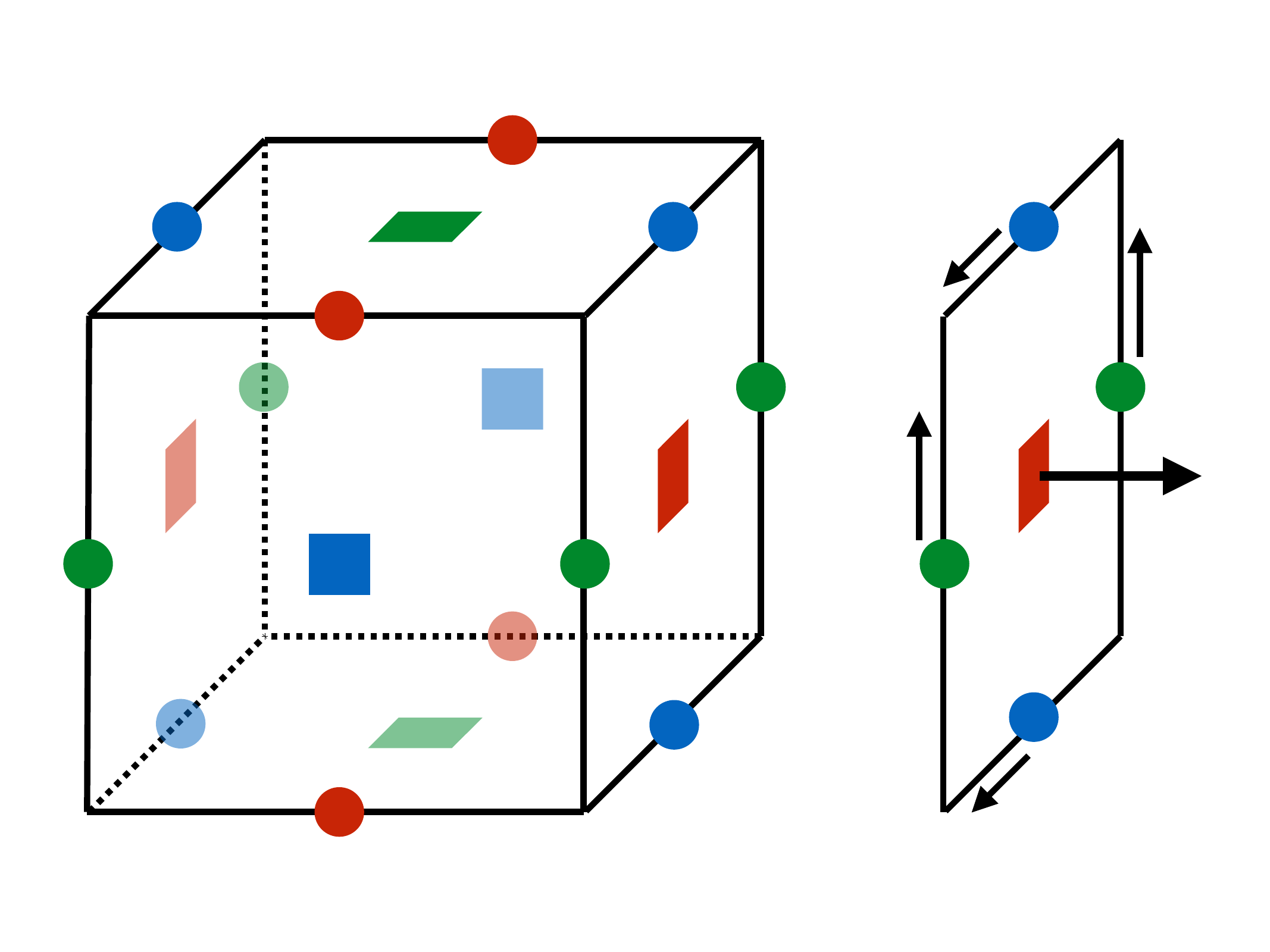}
\put(90, 50) { ${ \psi_{y_0} }$ }
\put(64, 35) { ${ \psi_{y_1} }$ }
\put(80, 12) { ${ \psi_{z_0} }$ }
\put(74, 61) { ${ \psi_{z_1} }$ }
\put(80, 31) { ${ u }$ }
\end{overpic}
\vspace{-0.2in}
\caption{\textbf{Discretization in 3D:} 
Discrete vector potential samples are located on cell edges (circles) and velocity samples are on cell faces (squares). Red, green, and blue represent $x$, $y$, and $z$ components, respectively.}
\label{fig:velToVP}
\end{figure}

We place vector potential components on cell edges and velocity normal components on cell faces \cite{Elcott2007,Ando2015,Bao2017} (Figure~\ref{fig:velToVP}). 
Given discretely divergence-free face velocities, we seek corresponding edge vector potential values.

For simplicity, we assume for now that the input axis-aligned exterior domain boundaries have zero normal velocity. 
We can safely achieve this with a constant value (i.e., 0) for the tangential components of the boundary $\boldsymbol{\psi}$ (denoted $\boldsymbol{\psi}_{tan}$). (Our method extends straightforwardly to prescribed inflow/outflow velocities, with appropriate variations in $\boldsymbol{\psi}_{tan}$.)

With this choice, the continuous problem we are solving is
\begin{equation}
\label{eq:VPCondition}
\begin{aligned}
\nabla \times \boldsymbol{\psi} &=  \mathbf{u} && \text{ in } \ \Omega, \\
\boldsymbol{\psi}_{tan} &= \mathbf{0} && \text{ on } \ \partial \Omega. 
\end{aligned}
\end{equation}
The flux across a given surface, in terms of the vector potential, is
\begin{equation}
\label{eq:continousFaceFlux}
\iint \limits_{S}   \mathbf{u} \cdot d \mathbf{S} 
= \iint \limits_{S} ( \nabla \times \boldsymbol{ \psi} ) \cdot  d \mathbf{S}
= \int \limits_{C} \boldsymbol{ \psi} \cdot d \mathbf{r},
\end{equation}
where ${S}$ is an oriented smooth surface, 
${C}$ is the oriented boundary curve of the surface, and $\mathbf{r}$ is the boundary tangent direction. 
(The second equality holds by Stokes' theorem.)
Discretizing \eqref{eq:continousFaceFlux}  with midpoint quadrature for the single face in Figure~\ref{fig:velToVP}, right, we have
\begin{equation}
\label{eq:discreteFaceFlux}
u_fh^2 = \sum \limits_{e \in E} \boldsymbol{\psi}_e \cdot (h \mathbf{e}_e)
= h(\psi_{y_0} + \psi_{z_1} - \psi_{y_1} - \psi_{z_0}),
\end{equation}
where ${h}$ is a cell width and ${\mathbf{e}_e}$ is the oriented unit vector along an edge.
(Finite differences on \eqref{eq:VPCondition} yields the same.)
Equation \ref{eq:discreteFaceFlux} relates four unknown edge $\psi_e$ values to one face $u_f$ velocity.
Stacking the equations for all grid faces yields a global sparse linear system.

Equation \ref{eq:VPCondition} is an inverse curl problem with infinitely many solutions, e.g., a single cell has a seven-dimensional null space: 12 edge $\psi_e$ and five linearly independent face $u_f$ (the sixth is redundant by incompressibility).
We select a unique solution using the popular \emph{Coulomb} gauge condition (with our boundary condition), which enforces  ${\nabla \cdot \boldsymbol{\psi} = 0}$. 
This gauge offers maximal smoothness of the $\boldsymbol{\psi}$ field, which will be attractive for interpolation.

Directly taking the curl of \eqref{eq:VPCondition} and assuming $\nabla \cdot \boldsymbol{\psi} = 0$ yields
\begin{align}
    \nabla \times (\nabla \times \boldsymbol{\psi}) =
    - \nabla^2 \boldsymbol{\psi} = \nabla \times \mathbf{u},
\end{align}
i.e., a \emph{vector} Poisson problem for $\boldsymbol{\psi}$, around three times larger than the pressure projection (e.g., \cite{Ando2015}). 
Fortunately, unlike Ando et al., our input discrete velocities are \emph{already} incompressible, enabling us to to solve \eqref{eq:VPCondition} more efficiently.
Observe that
\begin{equation}
\label{eq:VPNullSpace}
\mathbf{u} = \nabla \times \boldsymbol{\psi} 
= \nabla \times (\boldsymbol{\psi} + \nabla \phi) 
= \nabla \times \boldsymbol{ \psi}',
\end{equation}
where ${\phi}$ is an arbitrary scalar field. 
Since ${\nabla \times \nabla \phi}$ is always zero (a vector calculus identity), defining 
$\boldsymbol{\psi}' = \boldsymbol{\psi} + \nabla \phi$ gives another valid vector potential field for the same velocity.
Leveraging this characterization of the null space, 
we propose a two step approach. First, use an efficient 3D extension of our 2D parallel sweeping scheme (Section~\ref{gridCellsSweep}) to find a $\boldsymbol{\psi}$ field that is \emph{velocity-consistent but has arbitrary gauge}.
Second, modify this $\boldsymbol{\psi}$ field to satisfy the Coulomb gauge $\nabla \cdot \boldsymbol{\psi}=0$ and our boundary condition ${\boldsymbol{\psi}_{tan} = \mathbf{0}}$ using a carefully constructed scalar potential $\phi$ (Section ~\ref{modifyVP}).
\subsubsection {Recovering a Vector Potential by Parallel Sweeping} 
\label{gridCellsSweep}
\begin{figure}
	\centering	
	\begin{subfigure}[t]{0.23\textwidth}
		\begin{overpic}[width=\textwidth]{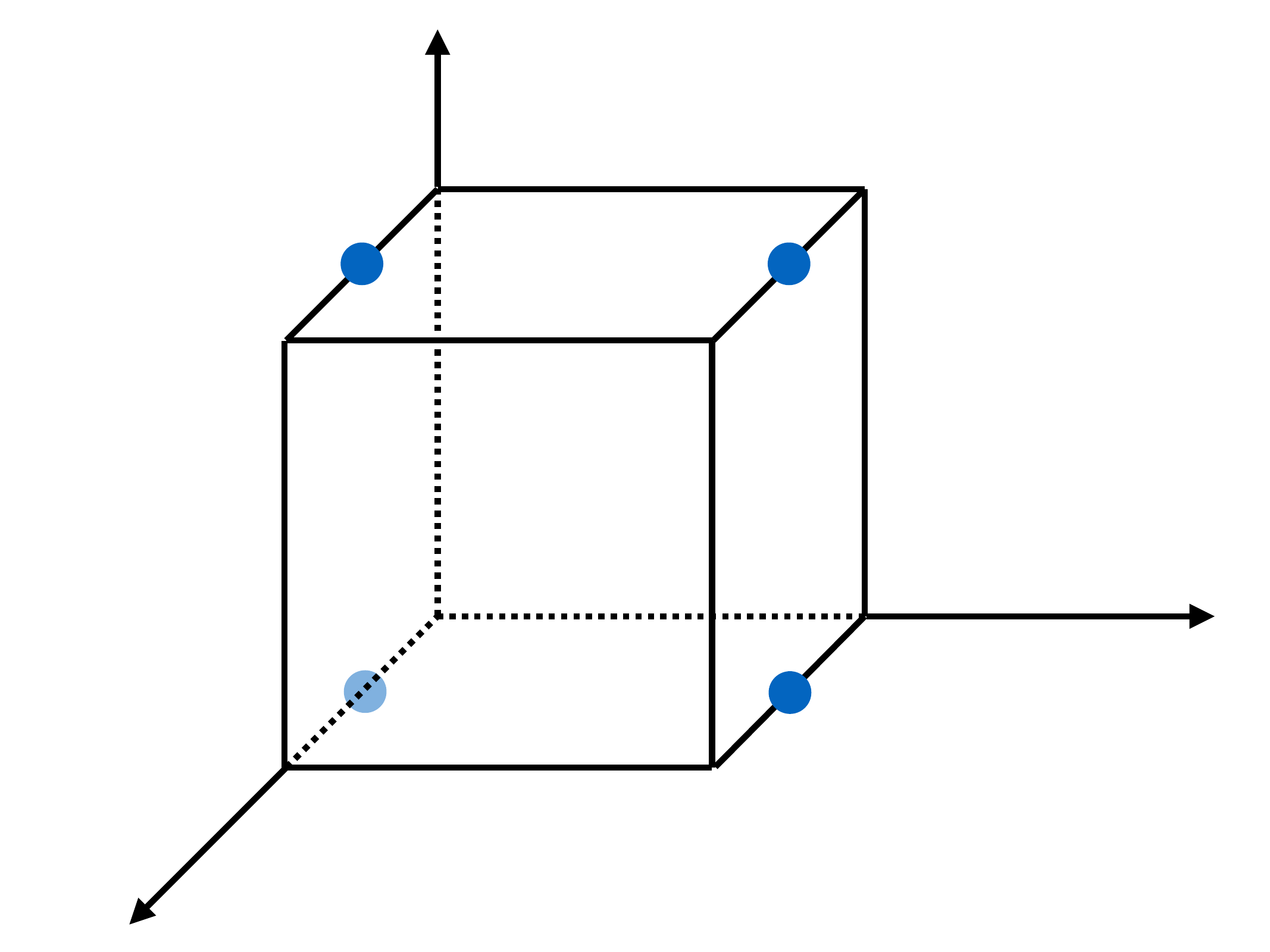}
			\put(15, 17) { 0 }
			\put(15, 52) { 0 }
			\put(67, 17) { 0 }
			\put(67, 52) { 0 }
			\put(88, 20) { x }
			\put(35, 68) { y }
			\put(16, 3) { z }
		\end{overpic}    
	\end{subfigure} 
	~
	\begin{subfigure}[t]{0.23\textwidth}
		\begin{overpic}[width=\textwidth]{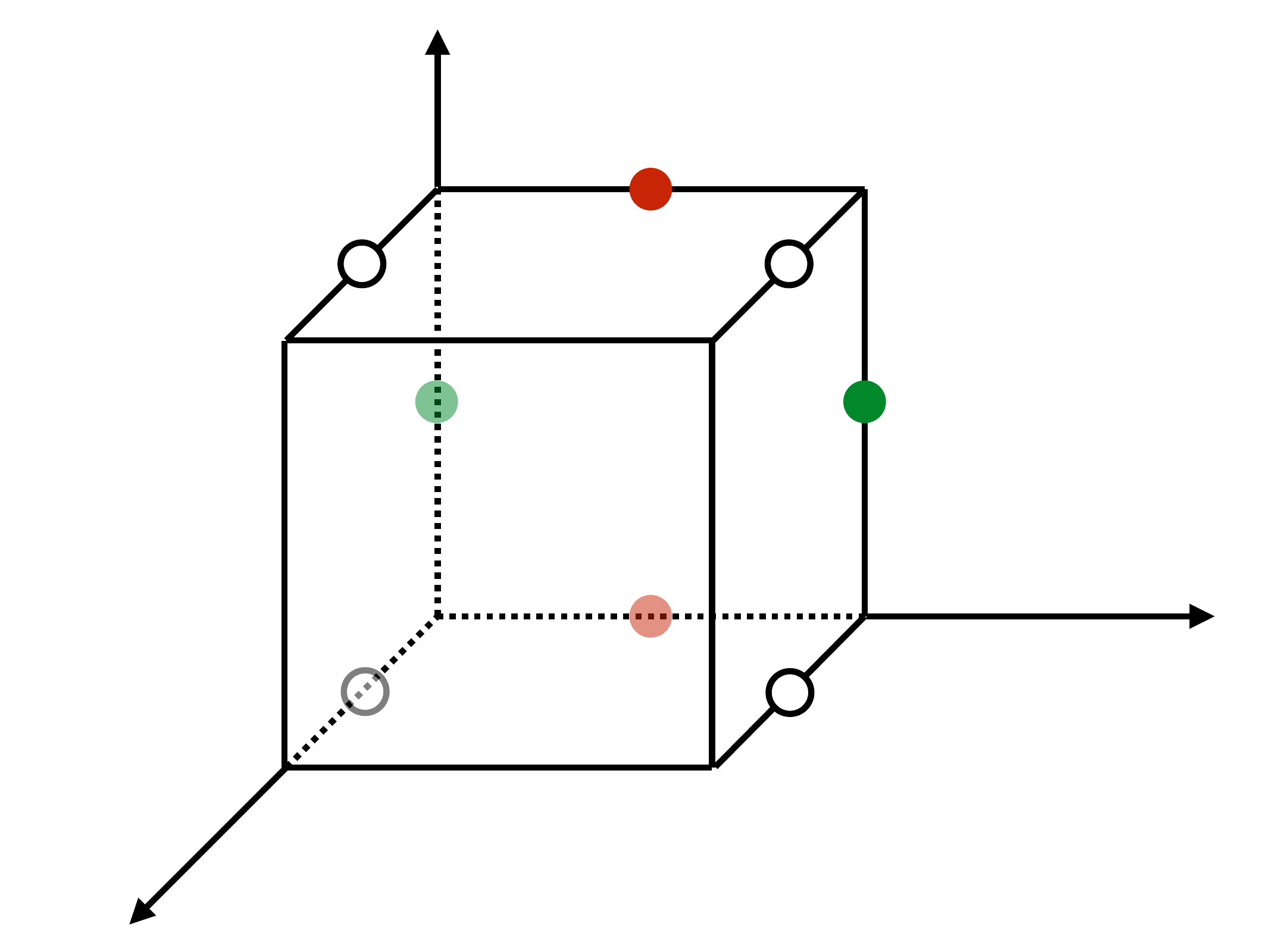}
			\put(45, 30) { 0 }
			\put(45, 65) { 0 }
			\put(26, 40) { 0 }
			\put(70, 40) { 0 }
			\put(88, 20) { x }
			\put(35, 68) { y }
			\put(16, 3) { z }
		\end{overpic}    
	\end{subfigure}

	\begin{subfigure}[t]{0.23\textwidth}
		\begin{overpic}[width=\textwidth]{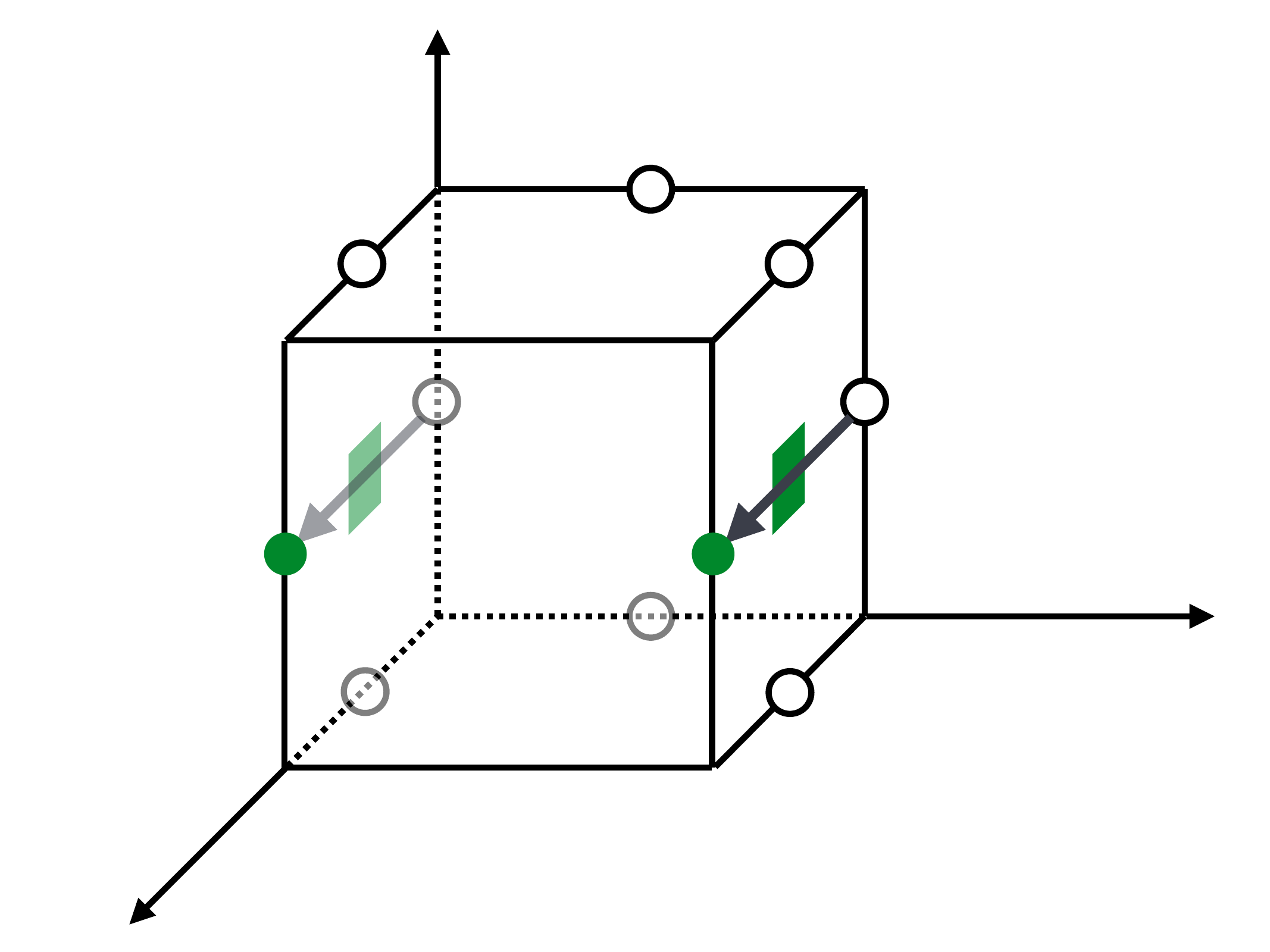}
			\put(70, 45) { ${\psi_{y_0}}$ }		
			\put(43, 35) { ${\psi_{y_1}}$ }		
			\put(56, 43.5) { ${u_r}$ }		
			\put(88, 20) { x }
			\put(35, 68) { y }
			\put(16, 3) { z }
		\end{overpic}    		
	\end{subfigure}
	~
	\begin{subfigure}[t]{0.23\textwidth}
		\begin{overpic}[width=\textwidth]{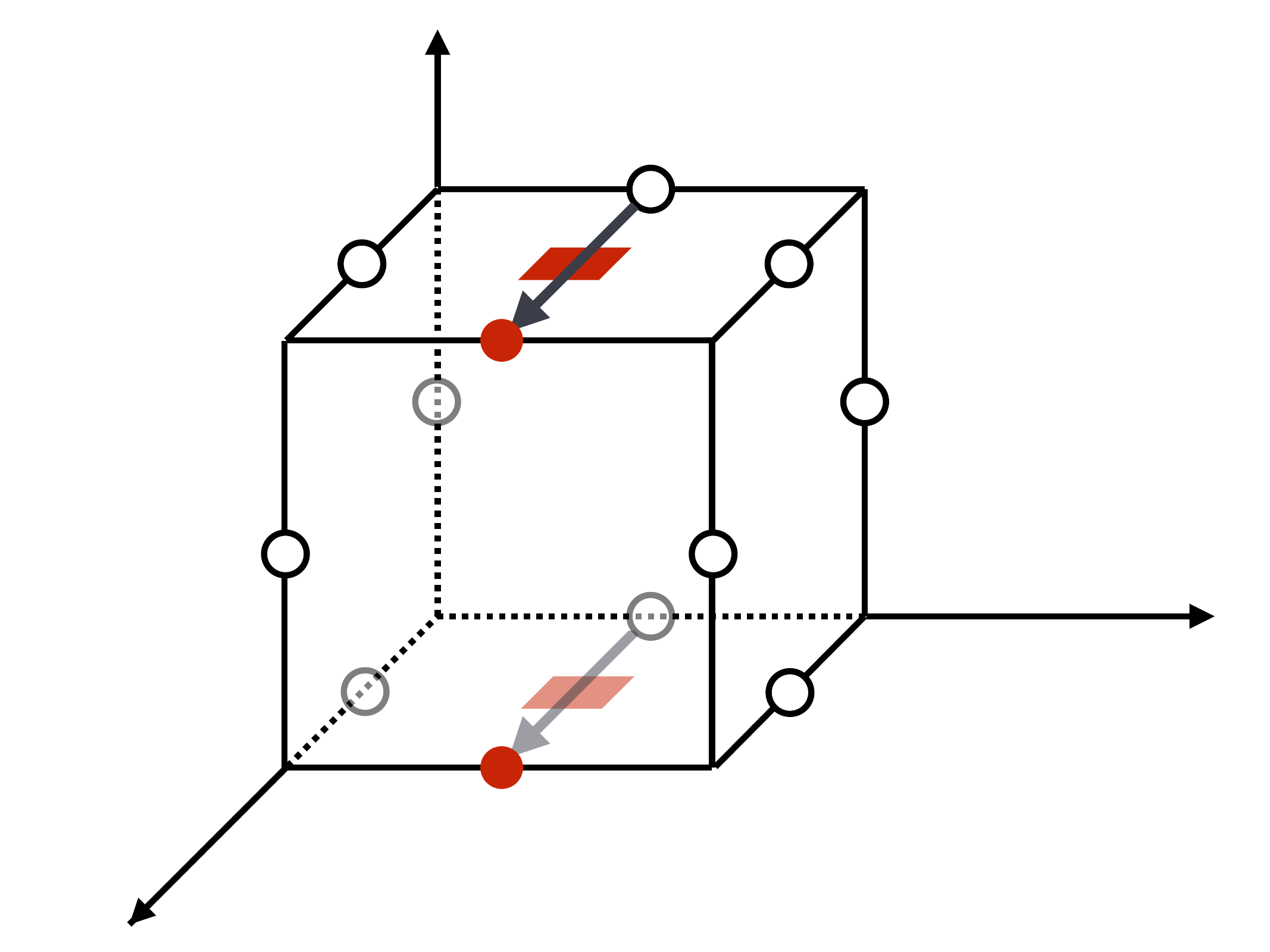}
			\put(51, 65) { ${\psi_{x_0}}$ }
			\put(36, 40) { ${\psi_{x_1}}$ }
			\put(48.5, 53) { ${v_t}$ }
			\put(88, 20) { x }
			\put(35, 68) { y }
			\put(16, 3) { z }
		\end{overpic}    		
	\end{subfigure}

	\caption{\textbf{Parallel Sweeping in 3D: } 
	Top-left: Set all ${\psi_z}$ values to be zero. 
	Top-right: Compute satisfying ${\psi_x}$ and ${\psi_y}$ values at ${z = z_{min}}$. In this example they are all set to zero.
	Bottom-left:
	Starting from ${\psi_y}$ values at ${z = z_{min}}$, we can compute all ${\psi_y}$ values in the entire domain in one sweep using the relation ${\psi_{y_1} = \psi_{y_0} - hu_r }$.	
	Bottom-right: Starting from ${\psi_x}$ values at ${z = z_{min}}$, we can compute all ${\psi_x}$ values in the entire domain in one sweep using the relation ${\psi_{x_1} = \psi_{x_0} + hv_t }$.
	}
	\label{fig:parallelSweepVP}
\end{figure}
Our first task is to efficiently find a velocity-consistent discrete vector potential field on a box-shaped domain, irrespective of boundary conditions or gauge choice.
Our proposed fast 3D parallel sweeping strategy is illustrated in Figure~\ref{fig:parallelSweepVP}.
\begin{enumerate}
\item Set ${\psi_z = 0}$ (or a constant) \emph{everywhere} in the domain (Figure~\ref{fig:parallelSweepVP}, top-left).
This is safe because the remaining two components (${\psi_x}$, ${\psi_y}$) still suffice to represent any set of three velocity components ($u$, $v$, $w$) \cite{Ravu2016}. 
\item Compute velocity-satisfying vector potential values for the ${z = z_{min}}$ boundary plane. 
For zero boundary normal fluxes, we can simply set all ${\psi_x}$ and ${\psi_y}$ to zero (Figure~\ref{fig:parallelSweepVP}, top-right).
\item Compute the remaining vector potential values by parallel sweeping. 
Equation \ref{eq:discreteFaceFlux} and ${\psi_z=0}$ give ${\psi_{y_1} = \psi_{y_0} - hu }$ (Figure~\ref{fig:parallelSweepVP},  bottom-left), and likewise 
${\psi_{x_1} = \psi_{x_0} + hv }$ (Figure~\ref{fig:parallelSweepVP}, bottom-right).
In like manner, values of ${\psi_{x_n}}$ and ${\psi_{y_n}}$, along with the discrete velocities, dictate ${\psi_{x_{n + 1}}}$ and ${\psi_{y_{n + 1}}}$ as we sweep across the entire domain.
\end {enumerate}

When this process ends, the velocity condition \eqref{eq:discreteFaceFlux} is met on all grid faces, ${\psi_z=0}$ everywhere, and we have zero ${\psi_x}$ and ${\psi_y}$ values on all the outer boundaries except for ${z = z_{max}}$. (For nonzero boundary fluxes, more ${\psi_x}$ and ${\psi_y}$ values will be nonzero.) We must next modify this $\boldsymbol{\psi}$ to enforce the desired gauge and boundary conditions.

\subsubsection {Boundary Conditions and Gauge Correction} 
\label{modifyVP}
\begin{figure}
\begin{overpic}[width=2.5in]{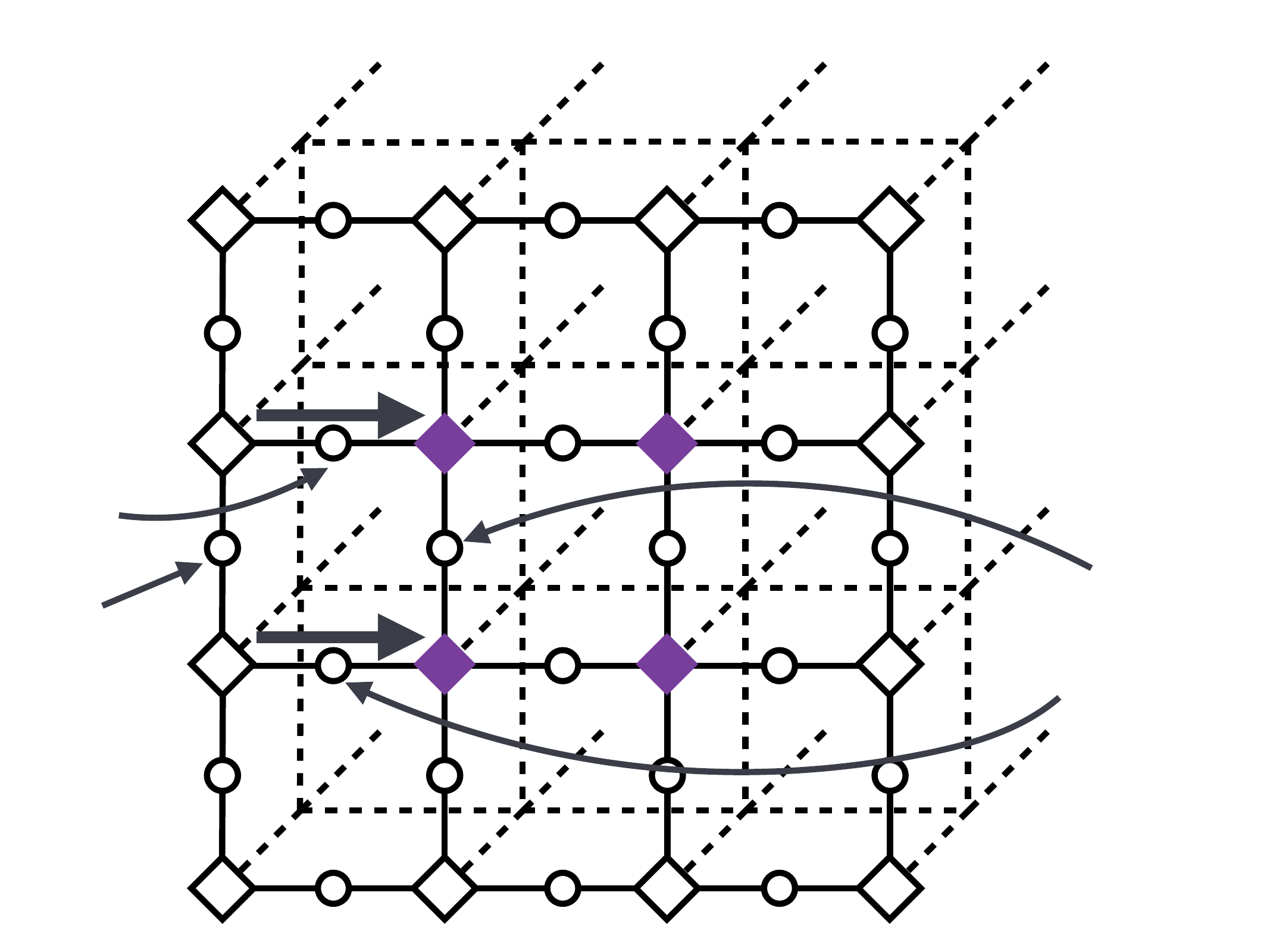}
\put(9, 38) { ${ \phi_2 }$ }
\put(35, 35.5) { ${ \phi_3 }$ }
\put(0, 34.5) { ${ \psi_{x_1} }$ }
\put(9, 18.5) { ${ \phi_0 }$ }
\put(35, 18.5) { ${ \phi_1 }$ }
\put(85, 18.5) { ${ \psi_{x_0} }$ }

\put(0, 25.5) { ${ \psi_{y_0} }$ }
\put(87, 30) { ${ \psi_{y_1} }$ }

\put(75, 3) { ${ z = z_{max} }$ }
\end{overpic}    
\caption{\textbf{Gauge Correction Boundary Conditions:} We construct $\nabla \phi_{BC}$ to update boundary $\boldsymbol{\psi}$ components to satisfy no-normal-flow.
Black/white circles, black/white diamonds, and purple diamonds represent
known vector potential values, pinned ${\phi}$ values, and unknown ${\phi}$ values, respectively.
Unknown ${\phi}$ values are computed from the pinned or previously computed ${\phi}$ values 
(${\phi_1 = \phi_0 - h\psi_{x_0}}$, ${\phi_3 = \phi_2 - h\psi_{x_1}}$).
}
\label{fig:boundaryPhi}
\end{figure}

\begin{figure}
\begin{overpic}[width=2.5in]{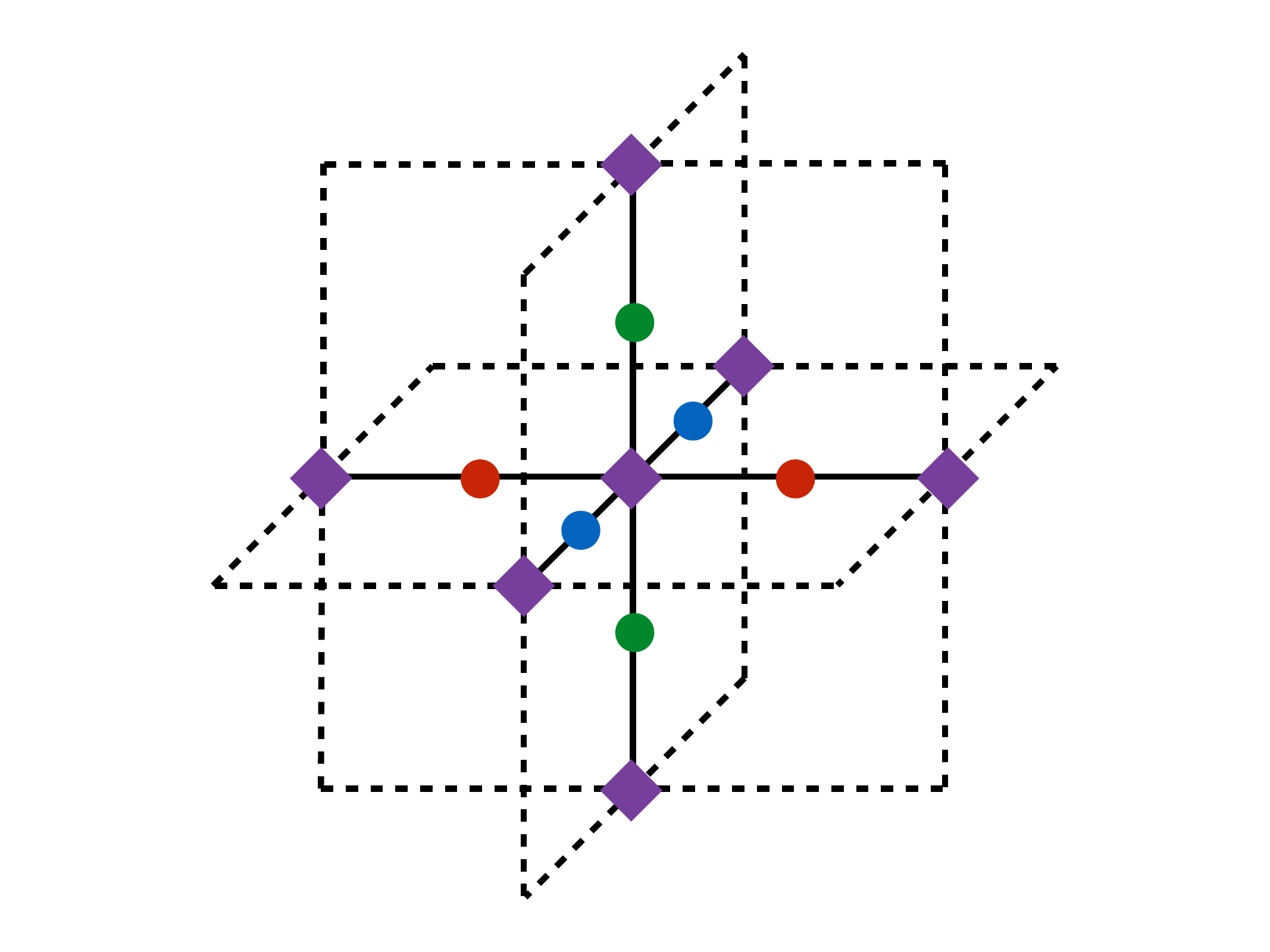}
\end{overpic}    
\caption{\textbf{Gauge Correction:} A uniform grid \emph{scalar} Poisson solve is used to find a nodal scalar field ${\phi}$ (purple diamonds). Adding $\nabla \phi$ to the vector potential values, ${\psi_x}$ (red disks), ${\psi_y}$ (green disks), and ${\psi_z}$ (blue disks),  satisfies the Coulomb gauge condition, ${\nabla \cdot \boldsymbol{\psi}'=\nabla \cdot (\boldsymbol{\psi} + \nabla \phi) = 0}$.
}
\label{fig:interiorPhi}
\end{figure}

To satisfy the boundary condition $\boldsymbol{\psi}_{tan}=\mathbf{0}$,
we first construct a discrete scalar field ${\phi}_{BC}$, which is defined at nodes of cells so components of the discrete ${\nabla \phi_{BC}}$ colocate with edge-based vector potential components. 
The desired boundary condition $\boldsymbol{\psi}_{tan}' = 0$ gives 
\begin{equation}
\label{eq:VPBoundaryCondition}
\begin{aligned}
\boldsymbol{\psi}_{tan}' = \boldsymbol{\psi}_{tan} + \nabla \phi_{BC} = \mathbf{0} \ \text{ on } \partial \Omega.
\end{aligned}
\end{equation}
Given ${\boldsymbol{\psi}_{tan}}$, already known from parallel sweeping, we must find ${\phi}_{BC}$ on boundary nodes. 
Furthermore, since our sweeping process
ensured that the only nonzero boundary ${\boldsymbol{\psi}_{tan}}$ values left to be eliminated are ${\psi_x}$ and ${\psi_y}$ on the ${z = z_{max}}$ plane, finding nonzero ${\phi}_{BC}$ values only on ${z = z_{max}}$ suffices.
We set the outer boundary loop of nodal ${\phi}_{BC}$ values on this plane
(black/white diamonds in Figure~\ref{fig:boundaryPhi}) to zero,
and compute the interior ${\phi}_{BC}$ values in the plane by traversing the interior edges using $\phi_{next} = \phi_{prev} - h \psi_e$.

Adding $\nabla \phi_{BC}$ directly to $\boldsymbol{\psi}$ (including any interior edges touching the nonzero $\phi$ values) would satisfy our exterior boundary condition without breaking consistency between the discrete $\mathbf{u}$ and $\boldsymbol{\psi}$. 
However, the (arbitrary) gauge of the resulting interior field still offers no guarantees on the presence or absence of large, discontinuous jumps \cite{Silberman2019}. 
Thus, while applying interpolation and the analytical curl will yield \emph{some} pointwise divergence-free velocity field, its observed pointwise behavior can be quite irregular depending on interactions between the discrete data jumps and componentwise interpolants, because velocity is determined by differences of vector potential component derivatives. 
Fortunately, we can enforce the Coulomb gauge on the interior to achieve an optimally smooth vector potential field for interpolation.

Defining a new global nodal $\phi$ field, we apply a gauge correction similar to \citet{Silberman2019} (although they consider a Fourier-based solution with exterior boundary conditions at infinity and do not handle interior solids).
The Coulomb condition says that $\nabla \cdot \boldsymbol{\psi}'=\nabla \cdot (\boldsymbol{\psi} + \nabla \phi) = 0$, giving a  node-based \emph{scalar} Poisson problem for ${\phi}$ with Dirichlet boundary conditions:
\begin{equation}
\label{eq:VPInteriorCondition}
\begin{aligned}
\nabla \cdot \nabla \phi &= - \nabla \cdot \boldsymbol{\psi} && \text{ in } \Omega, \\
\phi&=\phi_{BC} && \text{ on }  \partial \Omega.
\end{aligned}
\end{equation}
For the box-shaped domains we consider, this Dirichlet problem is amenable to solution with a classic \emph{fast Poisson solver} \cite{van1992computational} based on the Discrete Sine Transform (DST), even after we incorporate irregular obstacles (Section \ref{cutCells}). After solving for $\phi$ we update the vector potential as ${\boldsymbol{\psi}' = \boldsymbol{\psi}+\nabla \phi}$.

\subsubsection{Inflow/outflow exterior boundaries} The extension to exterior domain boundaries with nonzero fluxes (inflow and outflow) can be carried out similarly to the 2D case.
For example, to match prescribed $u$ velocities on the $x$ boundary,
we can use linear $\psi_z$ values in the $y$ direction on the boundary, while keeping $\psi_y$ values at zero.
The sweeping and gauge correction steps are easily modified to preserve these boundary values throughout.


\subsubsection{Interpolation}
Similar to 2D, we handle 3D uniform grid interpolation using low-order dyadic spline kernels, separately on each staggered component of $\boldsymbol{\psi}$, but 
choose their polynomial degrees to ensure continuous velocities as follows.
The curl operator applied to one component of $\boldsymbol{\psi}$ (e.g., ${\psi_x}$) involves its partial derivatives in the other two axes (e.g., $\partial \psi_x / \partial y$ and $\partial \psi_x / \partial z$). 
To ensure the resulting velocity is at least (piecewise) linear (and continuous) in all directions, we use a mix of linear and quadratic kernels (e.g., for ${\psi_x}$, linear $x$, quadratic $y$ and $z$). 
Uniformly quadratic or higher order interpolants would also suffice, in exchange for higher cost.


\subsection{Cut-Cells in 3D}
\label{cutCells}
The interference of cut-cell solids requires adaptations to our parallel sweeping and gauge correction steps, and a modified 3D ramping strategy to exactly enforce the desired boundary behavior.

\subsubsection {Parallel Sweeping with 3D Cut-Cells} 
\label{gridCellsSweepWithCutCell}
\begin{figure}
	\centering	
	\begin{subfigure}[t]{0.45\textwidth}
		\begin{overpic}[width=\textwidth]{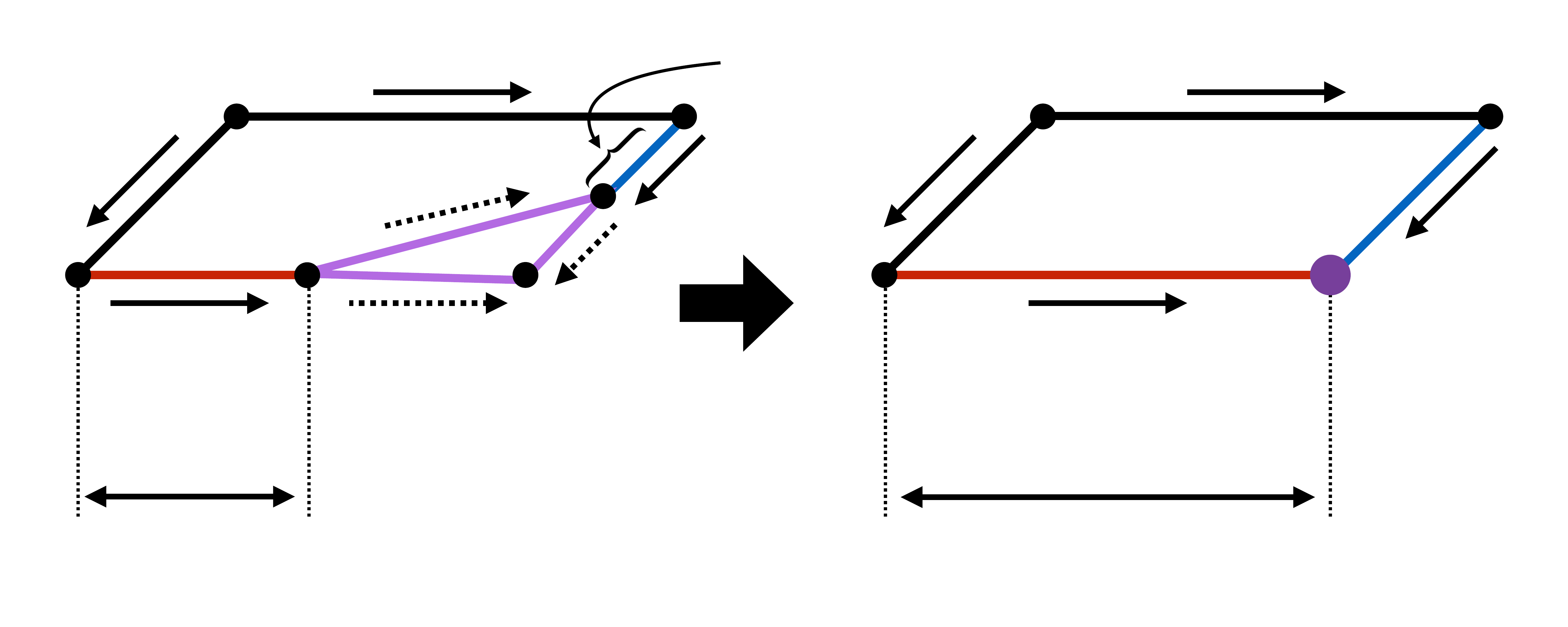}
		\put(69, 4) { ${ h }$ }
		\put(10, 4) { ${ l_0 }$ }
		\put(45, 35) { ${ l_1 }$ }

		\put(0, 30) { ${ \psi_{z_0} }$ }
		\put(43, 28) { ${ \psi_{z_1} }$ }
		\put(25, 37) { ${ \psi_{x_0} }$ }
		\put(9, 16) { ${ \psi_{x_1} }$ }

		\put(52, 30) { ${ \psi_{z_0} }$ }
		\put(93, 25) { ${ \psi_{z_1}' }$ }
		\put(77, 37) { ${ \psi_{x_0} }$ }
		\put(67, 16) { ${ \psi_{x_1}' }$ }
		\end{overpic}    
	\subcaption{Purple triangle (left) is collapsed to the purple point (right) and the new $\psi'$ values in regular cells are length fractions times the original $\psi$ values
	(${\psi_{x_1}' = \frac{l_0}{h} \psi_{x_1}}$, ${ \psi_{z_1}' = \frac{l_1}{h} \psi_{z_1}}$).}
	\end{subfigure} \vfill \vspace{5mm}
	\begin{subfigure}[t]{0.45\textwidth}
		\begin{overpic}[width=\textwidth]{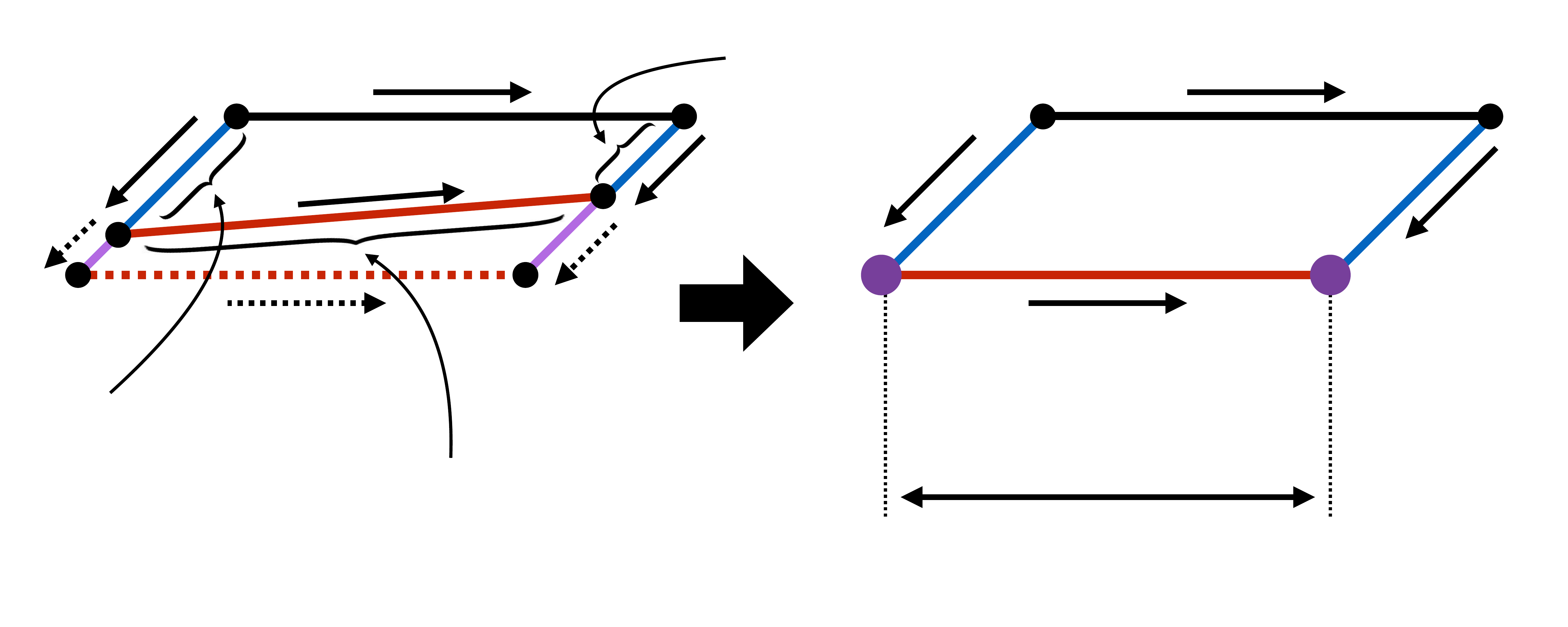}
	    \put(69, 4) { ${ h }$ }
		\put(2, 12) { ${ l_0 }$ }
		\put(46, 35) { ${ l_1 }$ }
		\put(26, 6) { ${ l_2 }$ }

		\put(1, 31) { ${ \psi_{z_0} }$ }
		\put(44, 28) { ${ \psi_{z_1} }$ }
		\put(25, 36) { ${ \psi_{x_0} }$ }
		\put(19, 29.3) { ${ \psi_{x_1} }$ }
		\put(16, 16) { ${ \psi_{x_1'} }$ }

		\put(52, 30) { ${ \psi_{z_0}' }$ }
		\put(93, 25) { ${ \psi_{z_1}' }$ }
		\put(77, 37) { ${ \psi_{x_0} }$ }
		\put(67, 16) { ${ \psi_{x_1}' }$ }
	\end{overpic}    
	\subcaption{Two purple edges (left) are collapsed (right) and the $\psi$ values are rescaled accordingly
	(${\psi_{z_0}' = \frac{l_0}{h} \psi_{z_0}}$, ${ \psi_{z_1}' = \frac{l_1}{h} \psi_{z_1}}$, ${ \psi_{x_1}' = \frac{l_2}{h} \psi_{x_1}}$).}
	\end{subfigure}
	\caption{\textbf{Adapting Parallel Sweeping to 3D Cut-Cells:} Fluid cut-edges are converted to uniform edges by assigning zero $\psi$ to solid purple edges and (conceptually) collapsing them, enabling uniform grid sweeping to proceed.
	}
	\label{fig:voxelizedCutCell}
\end{figure}

By carefully removing redundant DOFs and choosing traversal orders in Sec. \ref{gridCellsSweep}, 
we efficiently obtained velocity-consistent $\boldsymbol{\psi}$ values on uniform grids.
With cut-cells present, the number and orientation of vector potential components on some faces change, and thus our sweeping approach must be adapted. Conceptually, we achieve this by converting the irregular cut-cell geometry to a nearby voxelized state (Figure \ref{fig:voxelizedCutCell}), applying uniform grid parallel sweeping, and transforming back.

To achieve this mathematically, we rely on two assumptions. First, our earlier assumption of static obstacles implies cut-faces interior to the solid have zero normal flux (purple triangle in Figure~\ref{fig:voxelizedCutCell}(a)). Second, we assume that solid portions of partially cut grid-edges 
(e.g., the two axis-aligned purple segments in Figure~\ref{fig:voxelizedCutCell}(a)) have zero vector potential (the fluid parts of the edge are unchanged). Then, applying \eqref{eq:continousFaceFlux} to the solid triangle implies zero vector potential on the sloped purple edge too. Since the solid edges contribute nothing, we collapse them and rescale the fluid edges.

For example, we can now discretize the (pentagonal) top fluid face in Figure~\ref{fig:voxelizedCutCell}(a), left, using \eqref{eq:continousFaceFlux} as
\begin{equation}
\label{eq:faceFluxCutCell1}
v_t A W_t = h \psi_{z_0} + l_0 \psi_{x_1} - l_1 \psi_{z_1} - h \psi_{x_0} + 0
\end{equation}
where ${v_t}$ is the velocity normal component on the top face, ${A}=h^2$ is the area of a regular (non-cut) cell face, the $l_i$ are the lengths of partial fluid edges, ${W_t}$ is the fluid area fraction of the top face, and the 0 comes from the diagonal purple edge.
Through a change of variables, we can finally rewrite this in a form consistent with the uniform case of \eqref{eq:discreteFaceFlux}. We hold the weighted contribution of each edge and face fixed, but rescale potentials by length and velocity by area. Letting ${\psi_{x_1}' = \frac{l_0}{h} \psi_{x_1}}$, ${ \psi_{z_1}' = \frac{l_1}{h} \psi_{z_1}}$ and $v_t' = v_tW_t$, \eqref{eq:faceFluxCutCell1} becomes:
\begin{align}
v_t'A = h (\psi_{z_0} + \psi_{x_1}' - \psi_{z_1}' - \psi_{x_0}).
\end{align}
With these rescalings applied to our input, we can perform uniform grid sweeping across such faces, and then undo the scalings to recover the desired discrete vector potential field.

This face-centric approach extends naturally to cover all geometries of grid faces cut by level set solids (i.e., the marching squares cases). 
As another example, in Figure~\ref{fig:voxelizedCutCell}(b), the original top face consists of a fluid part (nonzero flux) and a solid part (zero flux).
These two (sub-)faces imply two equations,
\begin{equation}
\label{eq:faceFluxCutCell2}
\begin{aligned}
v_t A W_t &= l_0 \psi_{z_0} + l_2 \psi_{x_1} - l_1 \psi_{z_1} - h \psi_{x_0}, \\
0 &= l_2 \psi_{x_1} - h \psi_{x_1}'.
\end{aligned}
\end{equation}
where in the second equation we again assume zero potential on the two axis-aligned purple solid edges. Rescaling and sweeping suffices to recover the potentials, as before.

\subsubsection {Approximate Gauge Correction with 3D Cut-Cells} 
\label{gaugeCorrectionWithCutCell}
Applying gauge correction is more difficult with polyhedral cut-cell solids, since discrete vector potential edge components can correspond to arbitrary, rather than Cartesian, directions. A polyhedral PDE solver could be used, e.g., mimetic finite differences \cite{Lipnikov2006} or the diamond Laplacian \cite{bunge2021diamond}. 
However, such a solver is more costly and interpolating the resulting non-Cartesian edge components would also be unwieldy. Moreover, as in 2D, we still require a ramping correction to precisely respect solid boundaries. Since our motivation for gauge correction is simply to ensure smoothness, minor local deviations from the Coulomb gauge are acceptable.
We therefore enforce the gauge in a simple but approximate manner: we use the rescaled data from the voxelized view of our geometry determined above (Figure~\ref{fig:voxelizedCutCell}), and perform gauge correction on the \emph{entire} uniform grid (including through the interior of solids). This voxelized approach also still lets us retain the use of the fast DST Poisson solver.
Thus, our method efficiently provides smooth axis-aligned $\boldsymbol{\psi}$ values everywhere, which we use for B-spline kernel interpolation as in the uniform case.


\subsection{Precise boundary enforcement in 3D}
\label{ramping3D}

\begin{figure}
	\centering	
	\begin{subfigure}[t]{0.23\textwidth}
		\begin{overpic}[width=0.95\textwidth]{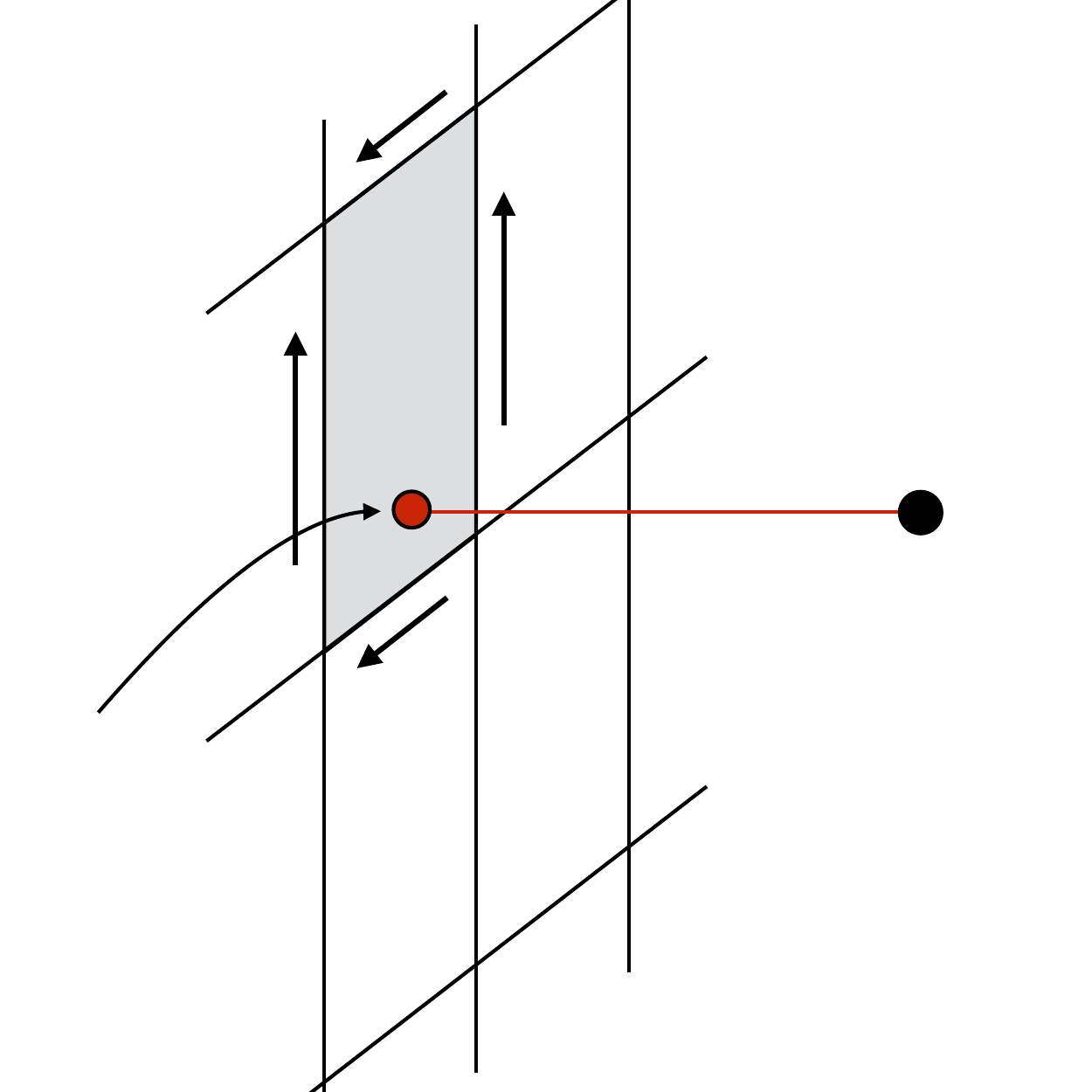}
		\put(45, 74) { ${ \psi_{y_0} }$ }
		\put(14, 54) { ${ \psi_{y_1} }$ }

		\put(30, 32) { ${ \psi_{z_0} }$ }
		\put(30, 95) { ${ \psi_{z_1} }$ }
		
		\put(-8, 25) { ${ cp(\mathbf{x}) }$ }
		\put(81, 44) { ${ \mathbf{x} }$ }
		
		\put(60, 12) { ${ \partial \Omega }$ }
		\end{overpic}
		\subcaption{Exterior boundary}
	\end{subfigure}
	\begin{subfigure}[t]{0.23\textwidth}
		\begin{overpic}[width=0.95\textwidth]{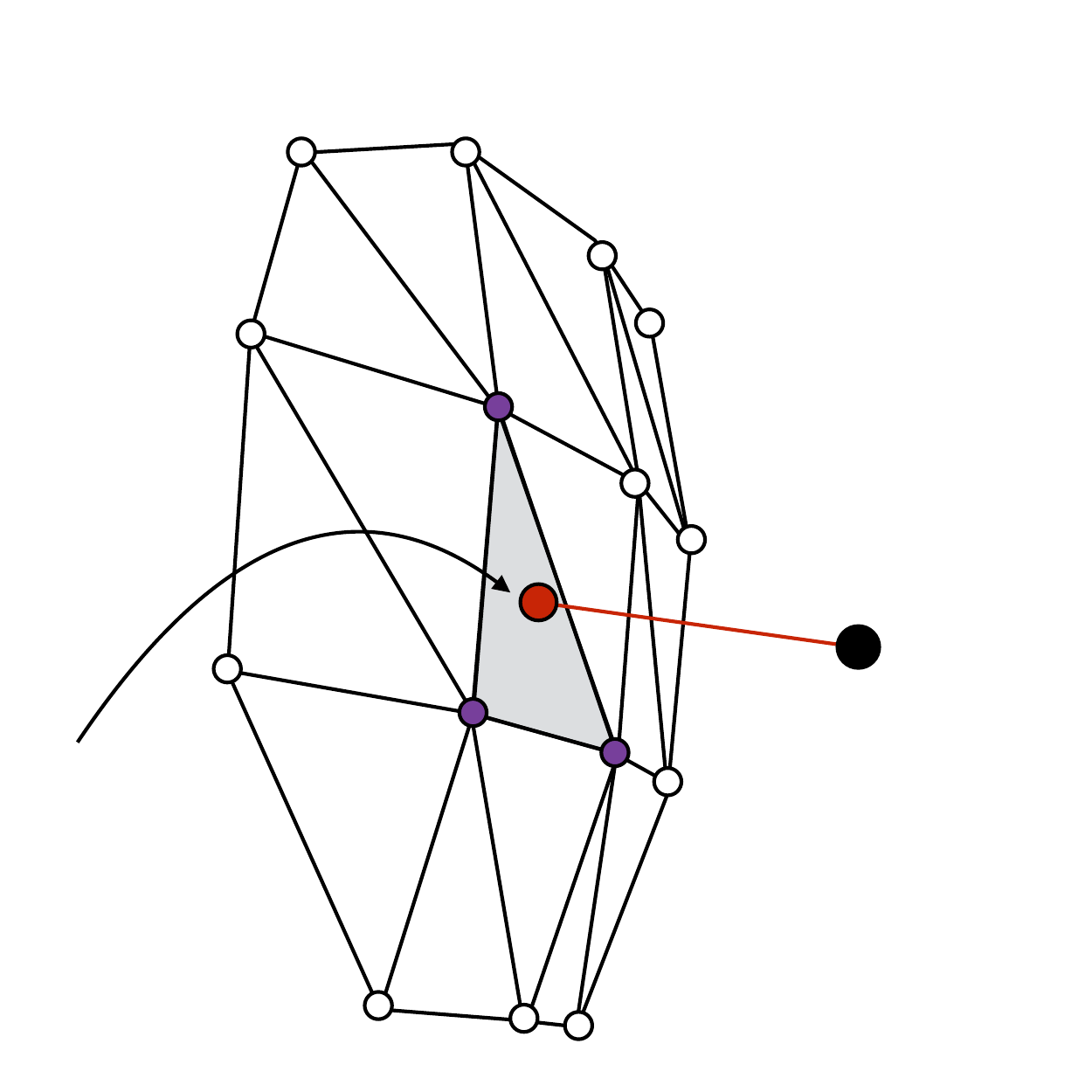}
		\put(28, 40) { ${ \boldsymbol{\psi}_0 }$ }
		\put(44, 25) { ${ \boldsymbol{\psi}_1 }$ }
		\put(32, 58) { ${ \boldsymbol{\psi}_2 }$ }
		
		\put(-8, 25) { ${ cp(\mathbf{x}) }$ }
		\put(80, 33) { ${ \mathbf{x} }$ }
		
		\put(60, 12) { ${ \partial S }$ }
		\end{overpic}
		\subcaption{Solid boundary}
	\end{subfigure}
	\caption{\textbf{Boundary enforcement:} The black and red disks represent a particle near the boundary and its closest point on the boundary, respectively.
	(a) For planar exterior boundaries, the discrete $\psi$ values (e.g., $\psi_{y0}$, $\psi_{y1}$, $\psi_{z0}$, $\psi_{z1}$) are zero by our boundary conditions, and the relevant axis components ($\psi_y (\mathbf{x})$ and $\psi_z (\mathbf{x})$) are ramped to zero for exact enforcement.
	(b) For triangulated solids,
	we first find discrete ${\boldsymbol{\psi}}$ values at the triangle vertices (e.g., ${\boldsymbol{\psi}_0}$, ${\boldsymbol{\psi}_1}$, ${\boldsymbol{\psi}_2}$)
	which imply a perfectly tangential surface velocity, under barycentric interpolation of nodal ${\boldsymbol{\psi}}$. 
	When querying velocities, we ramp the full $\boldsymbol{\psi}$ vector towards this interpolated surface $\boldsymbol{\psi}$ at the closest point.
	}
	\label{fig:boundaryEnforcement3D}
\end{figure}

As in 2D, we propose an additive ramping-based approach for exact boundary enforcement in 3D, applied as a correction atop the approximate $\boldsymbol{\psi}$ values produced by grid interpolation. The 3D adaptation requires caution in treating the interacting $\boldsymbol{\psi}$ components.


\subsubsection {Exterior domain boundaries} 
\label{exteriorBoundary}
The domain exterior consists of axis-aligned planes with two tangential components of $\boldsymbol{\psi}$ lying on each boundary plane (Figure~\ref{fig:boundaryEnforcement3D}(a)). 
Recall that for a $u$-face, the flux is determined by ${\psi_y}$ and ${\psi_z}$ through
$u = \frac{\partial \psi_z }{\partial y} - \frac{\partial \psi_y}{\partial z}.$
To achieve exactly zero flux everywhere on the plane and avoid leakage, these terms must precisely cancel: we set the discrete $\psi_y$ and $\psi_z$ values to zero for simplicity, 
and consequently ramp the pointwise $\psi_y$ and $\psi_z$ values to zero as a particle approaches the border. 
(A careless choice of the ramp values can badly distort the velocity, as we saw in 2D in Figure~\ref{fig:ramp_mulzero}.)
Since we apply ramping to each component of $\boldsymbol\psi$ independently, this ramping strategy is identical to the 2D case (Section~\ref{ramping2D}). As in 2D, we do not enforce exact ramping on inflow/outflow boundaries as its effects are not visually apparent.


\subsubsection {Solid obstacles} 
\label{solidBoundary}
For our explicit representation of the level set-based solid obstacles, we use triangle meshes obtained from marching cubes. A key component of exact boundary enforcement with ramping is the proper choice of target ${\boldsymbol{\psi}}$ values (i.e., ${\boldsymbol{\psi}_g}$) on the solid surface. 
Specifically, ${\boldsymbol{\psi}_g}$ should produce a zero normal velocity,
\begin{equation}
\label{eq:normalFlux}
\begin{aligned}
\mathbf{u} \cdot \mathbf{n} = (\nabla \times \boldsymbol{\psi}_g)\cdot \mathbf{n} = 0,
\end{aligned}
\end{equation}
where $\mathbf{n}$ represents the solid normal vector. 

For exterior boundaries, we were able to simply set constant discrete ${\boldsymbol{\psi}_{tan}}$ values in the solve, and by using those same constant values for the target
${\boldsymbol{\psi}_g}$ during ramping, \eqref{eq:normalFlux} was trivially satisfied.
However, one cannot in general find a single constant vector $\boldsymbol{\psi}_g$ that precisely compensates the existing (spatially varying) ambient interpolated $\boldsymbol{\psi}$ on irregular solids. 
Moreover, recall that an \emph{arbitrary} choice of $\boldsymbol{\psi}_g$ will still severely distort the nearby tangential velocities, even if it satisfies \eqref{eq:normalFlux}.

We instead leverage the characteristics of triangular barycentric coordinates: the gradient of the barycentric interpolant is constant, thus $\nabla \times \boldsymbol{\psi}_g$ yields a constant velocity (per triangle). (We also tried other interpolants satisfying \eqref{eq:normalFlux} continuously on a triangulated surface (piecewise constant, Whitney), but found that barycentric produced the smoothest velocity fields.)
Therefore, an appropriate choice of $\boldsymbol{\psi}$ at the triangle vertices ($\boldsymbol{\psi}_0$, $\boldsymbol{\psi}_1$, $\boldsymbol{\psi}_2$ in Figure~\ref{fig:boundaryEnforcement3D}(b)), can satisfy~\eqref{eq:normalFlux}, exactly and continuously.
At the same time, we wish to minimize the perturbation of the velocity field induced by ramping, which we can do by encouraging the discrete $\boldsymbol{\psi}_i$ values at triangle vertices to be as close to the ambient $\boldsymbol{\psi}$ 
(i.e., the interpolated pointwise $\boldsymbol{\psi}$ before applying ramping) as possible. 
This yields an equality constrained quadratic minimization problem:
\begin{equation*}
\begin{aligned}
    \stackrel[\mathbf{y}]{}{\text{argmin}}\text{ \ \ \ \ \ } &\frac{1}{2}(\mathbf{y} - \boldsymbol{\psi}_{amb})^T
    (\mathbf{y} - \boldsymbol{\psi}_{amb}) \\
    \text{subject to \ \ \ \ \ } &
    \bigg(\nabla \times \sum \limits_{i \in \{0, 1, 2\}} w_i(\mathbf{x}) \mathbf{y}_{t_i} \bigg) \cdot \mathbf{n}_t = 0 \ \ \ \ \ \text{for} \ t \in T,
\end{aligned}
\end{equation*}
where $T$ is a set of (connected) solid triangles, 
$w_i(\mathbf{x})$ is the barycentric coordinate function for the $i^{th}$ vertex for a spatial position $\mathbf{x}$, 
$t_i$ is the global vertex index of the $i^{th}$ vertex in the $t^{th}$ triangle, $\mathbf{n}_t$ is the triangle normal, 
and $\boldsymbol{\psi}_{amb}$ is a stack of the ambient $\boldsymbol{\psi}$ values
at the triangle vertices (before ramping).
If we write the constraints as $A \mathbf{y} = \mathbf{0}$, the optimality conditions yield the linear system
\begin{equation}
    \label{eq:solidKKT}
    \begin{aligned}
    \begin{bmatrix}
        I & A^T \\
        A & 0
    \end{bmatrix}
    \begin{bmatrix}
        \mathbf{y} \\
        \boldsymbol{\lambda}
    \end{bmatrix}
    = 
    \begin{bmatrix}
        \boldsymbol{\psi}_{amb} \\
        \mathbf{0}
    \end{bmatrix}
    \end{aligned}
\end{equation}
or ${A A^T \boldsymbol{\lambda} = A \boldsymbol{\psi}_{amb}}$. 
We can find the desired discrete $\boldsymbol{\psi}_i$ at the solid vertices via
$\mathbf{y} = \boldsymbol{\psi}_{amb} - A^T \boldsymbol{\lambda}$.
Since the normal component of velocity only depends on the tangential components of $\boldsymbol{\psi}$, one need only ramp the tangential components using the target values of ${\boldsymbol{\psi}_{g, tan} = \boldsymbol{\psi}_{g} - (\boldsymbol{\psi}_g\cdot\mathbf{n})\mathbf{n}}$.
For ramping, $\boldsymbol{\psi}_g$ and $\boldsymbol{\psi}_{g, tan}$ can be queried by barycentric interpolation ${\mathbf{y}}$ in the closest triangle.

\begin{figure}
    \centering
	\begin{subfigure}[t]{0.48\textwidth}
		\fbox{\begin{overpic}[trim=600 0 0 0, clip, width=0.33\linewidth]{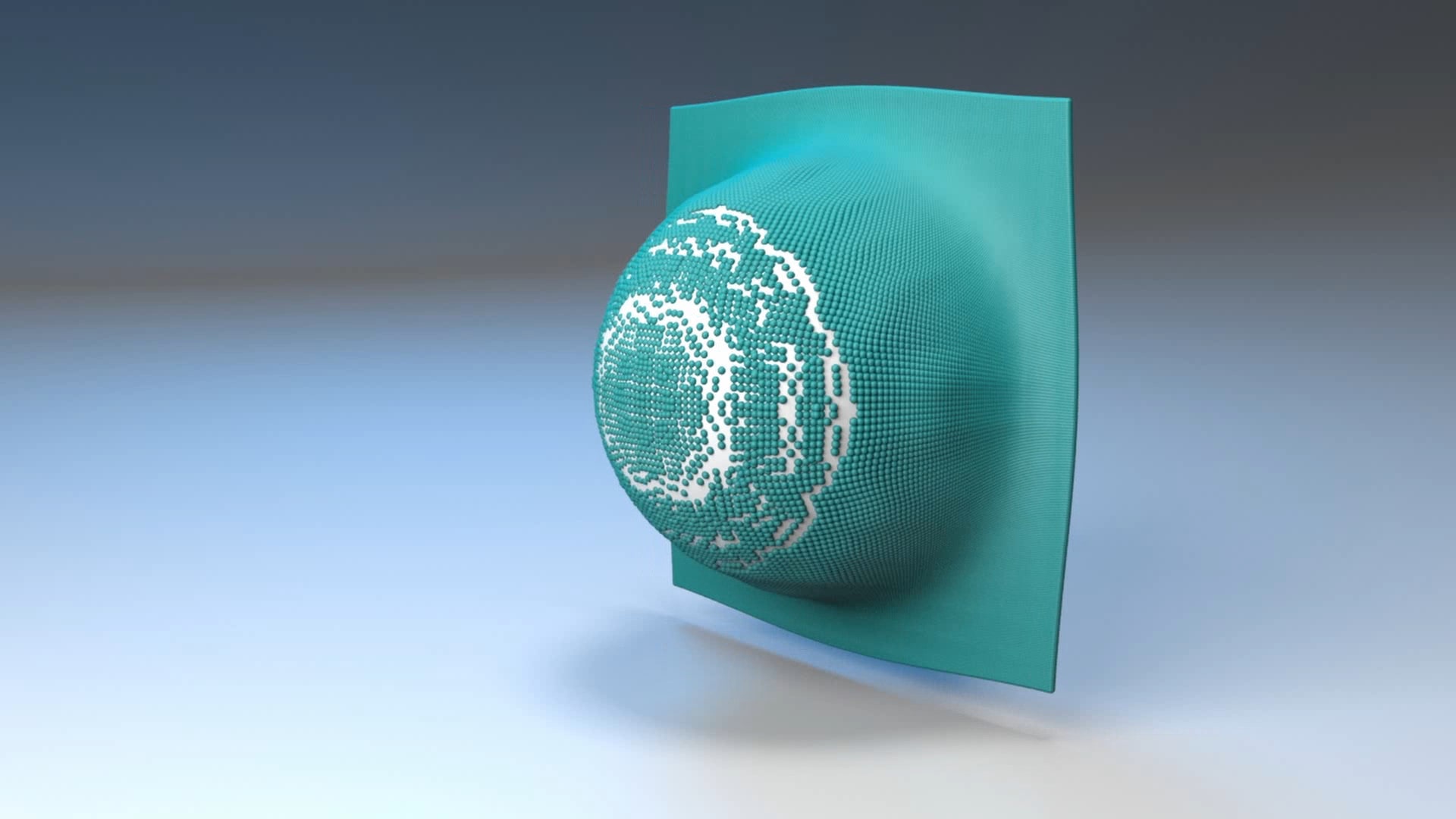}
		\end{overpic}}~    	
		\fbox{\begin{overpic}[trim=600 0 0 0, clip, width=0.33\linewidth]{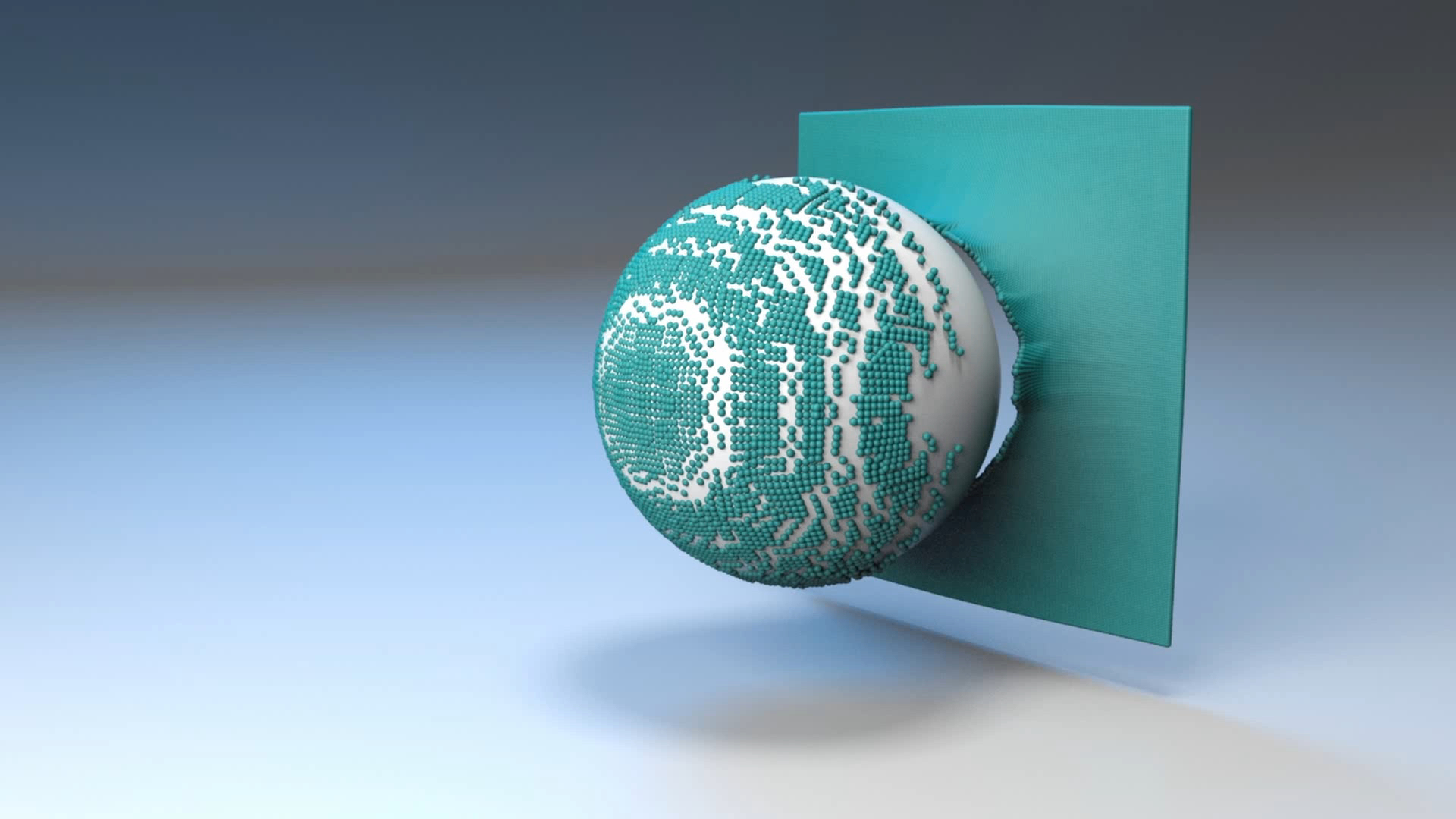}
		\end{overpic}}~    
		\fbox{\begin{overpic}[trim=600 0 0 0, clip, width=0.33\linewidth]{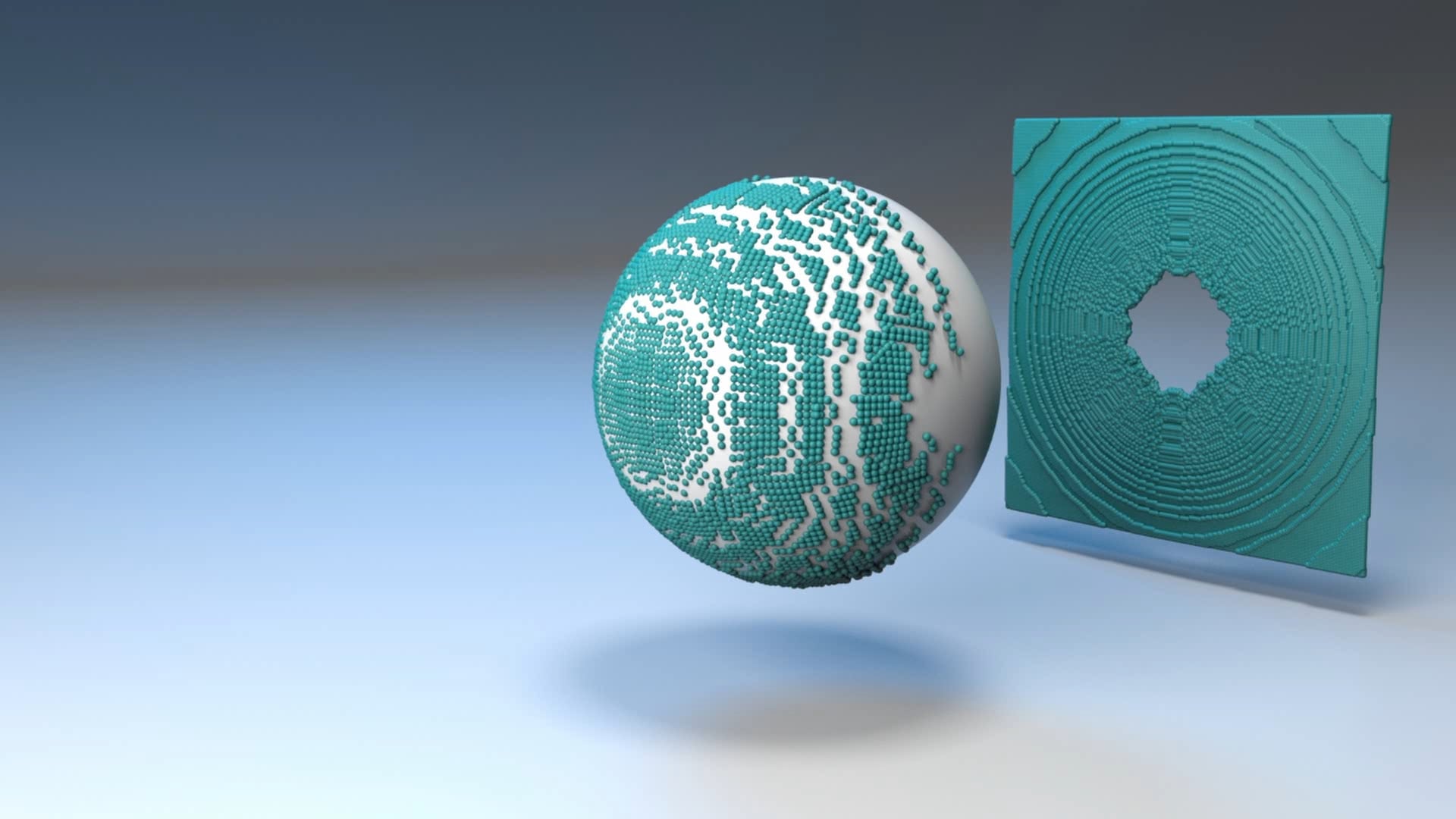}
		\end{overpic}}    
		\caption{Direct Velocity Interpolation (Halt)}
	\end{subfigure}
 \\
	\begin{subfigure}[t]{0.48\textwidth}
		\fbox{\begin{overpic}[trim=600 0 0 0, clip, width=0.33\linewidth]{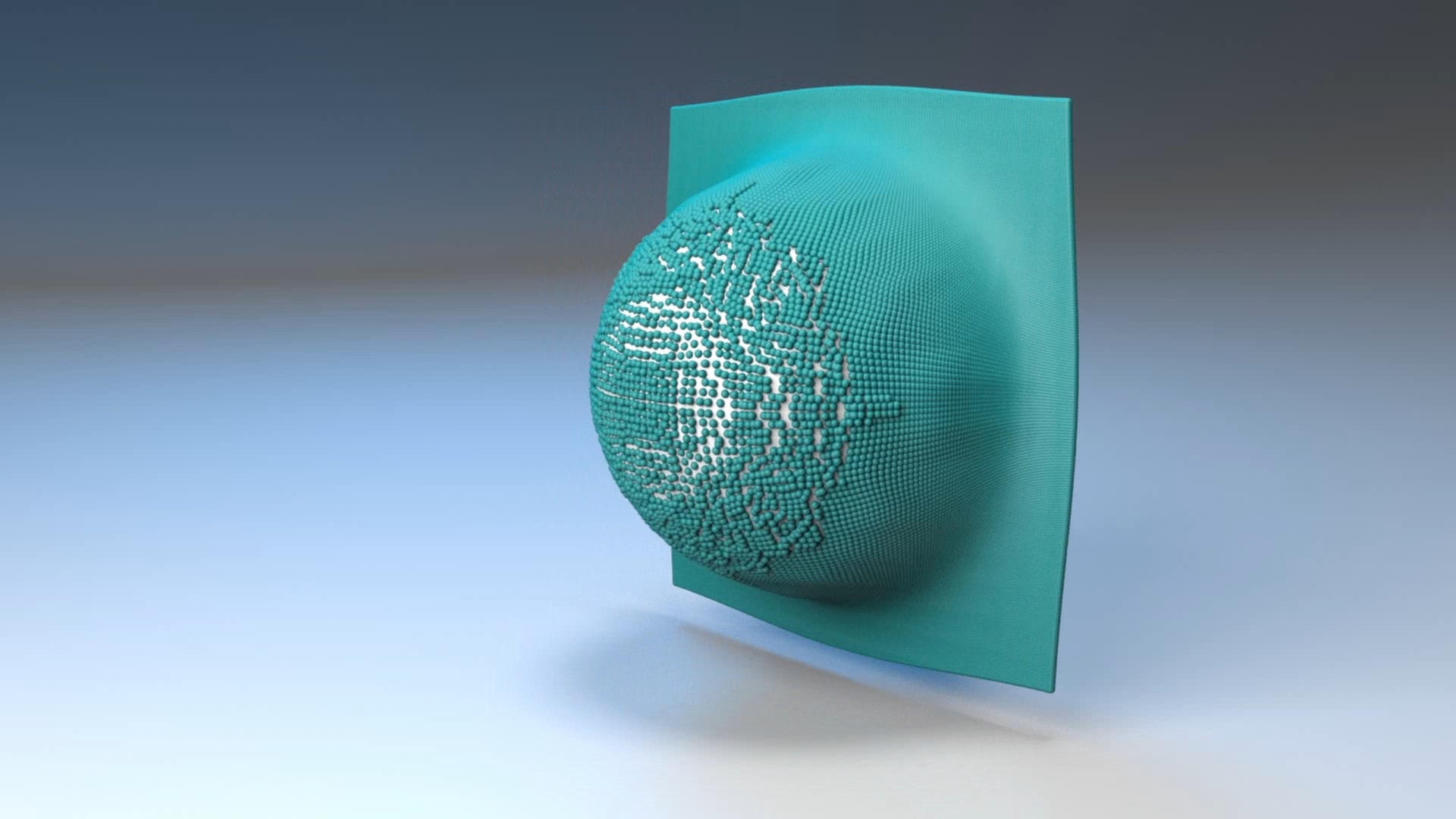}
		\end{overpic}}~    		\fbox{\begin{overpic}[trim=600 0 0 0, clip, width=0.33\linewidth]{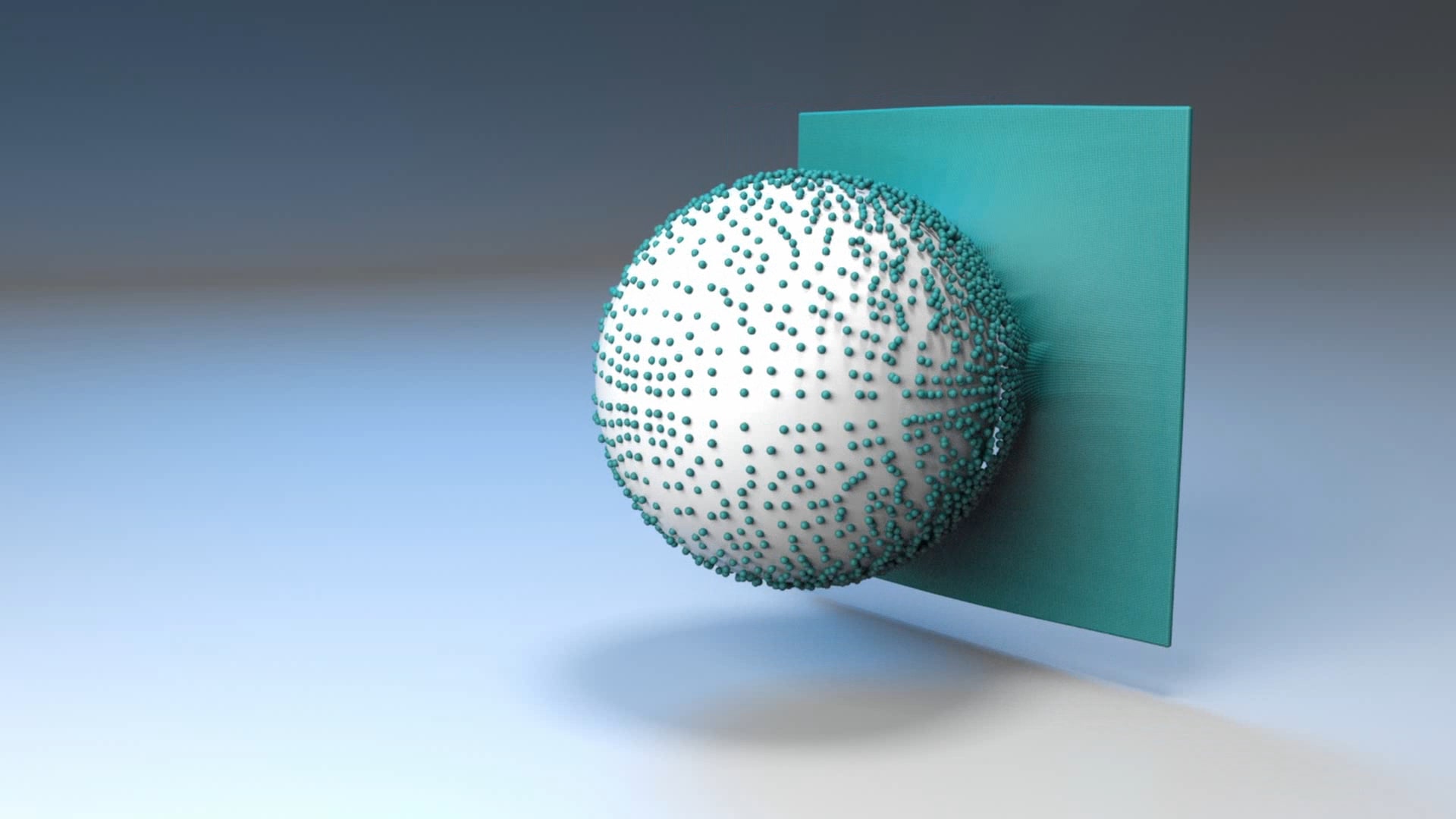}
		\end{overpic}}~    
		\fbox{\begin{overpic}[trim=600 0 0 0, clip, width=0.33\linewidth]{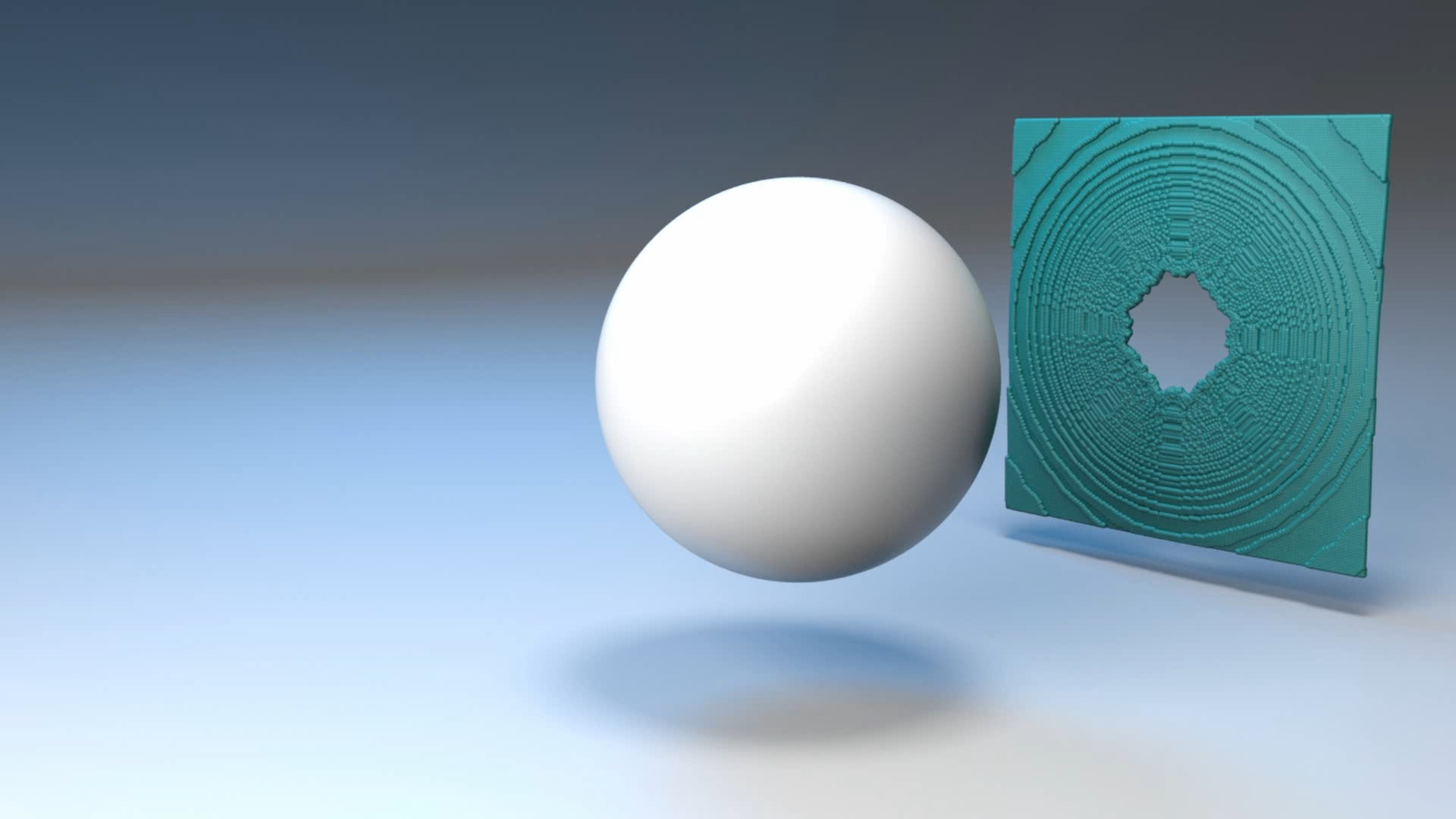}
		\end{overpic}}    
		\caption{Direct Velocity Interpolation (Push)}
	\end{subfigure}
 \\
	\begin{subfigure}[t]{0.48\textwidth}
		\fbox{\begin{overpic}[trim=600 0 0 0, clip, width=0.33\linewidth]{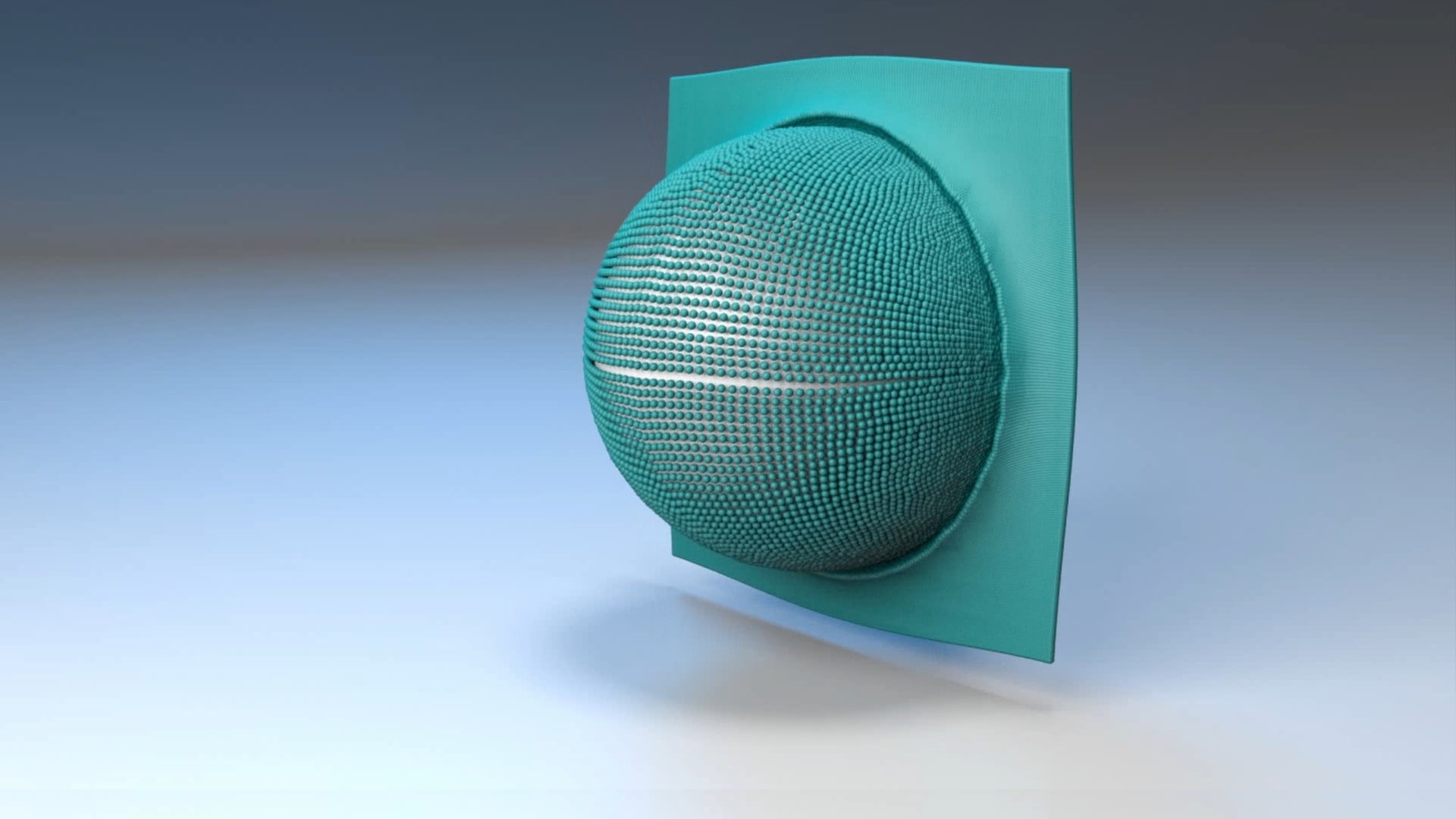}
		\end{overpic}}~	
		\fbox{\begin{overpic}[trim=600 0 0 0, clip, width=0.33\linewidth]{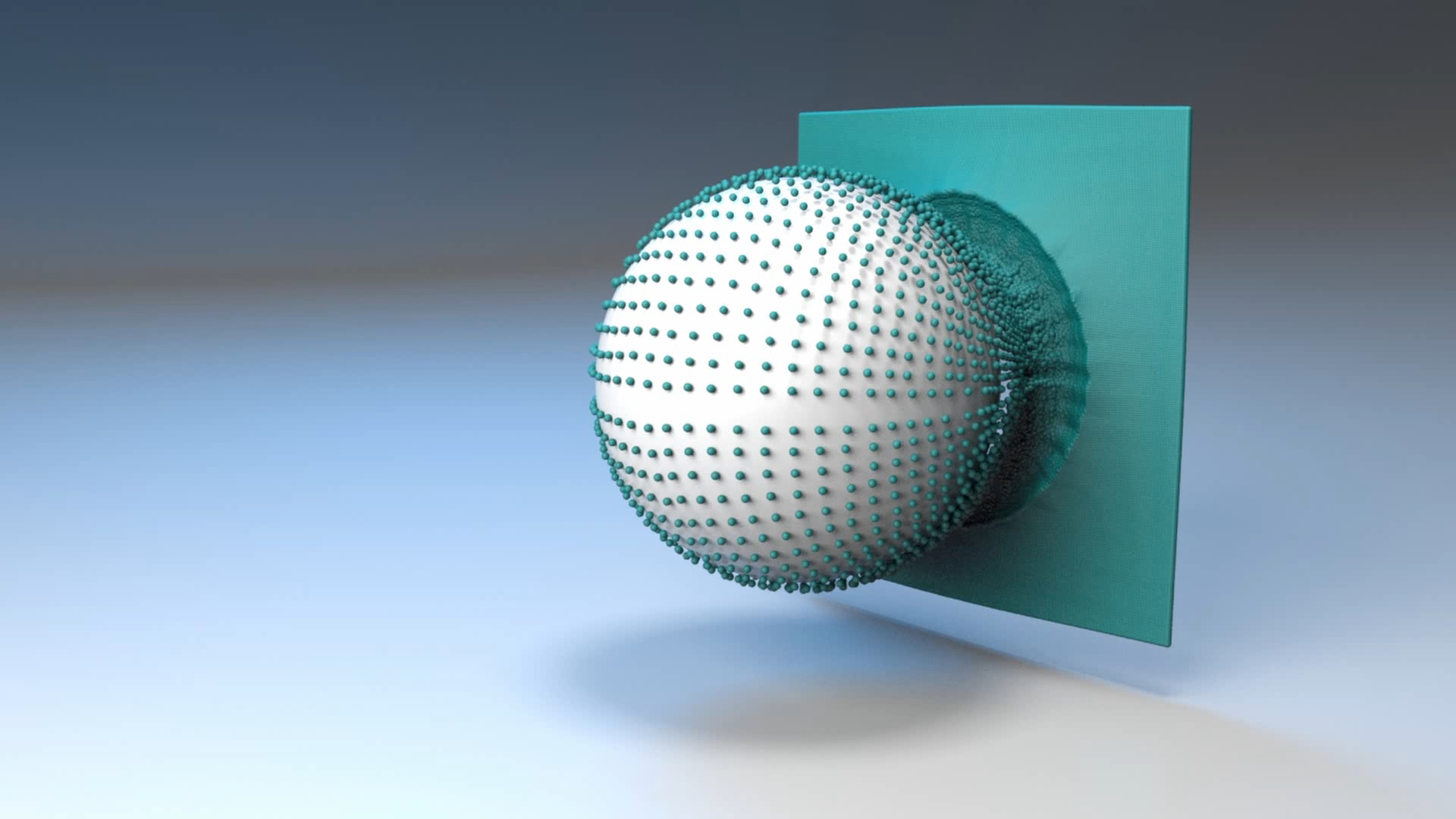}
		\end{overpic}}~    
		\fbox{\begin{overpic}[trim=600 0 0 0, clip, width=0.33\linewidth]{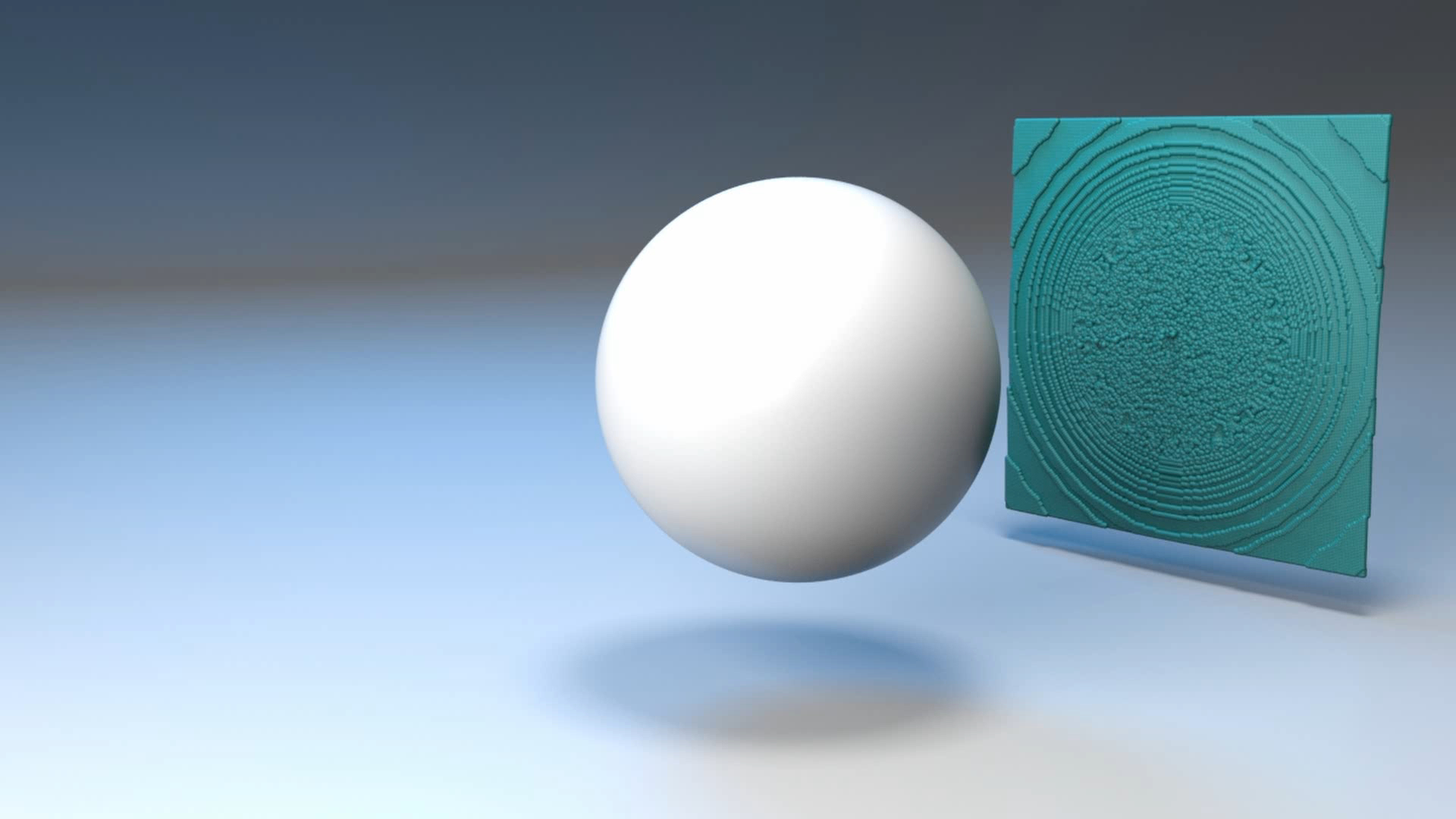}
		\end{overpic}}    
		\caption{Curl-Flow (Halt)}
	\end{subfigure}~
    \caption{
        \textbf{Flow Past A Sphere:} 
            A uniform layer of particles travels left to right under a steady state flow field past a sphere. Particles that exit the domain are frozen in place to emphasize their final distribution (right column).
    	    Grid resolution: $40 \times 20 \times 20$.
            (a) Direct velocity interpolation, with colliding particles halted on contact: Collided particles densely blanket the sphere. The hole in the final particle layer (right) highlights the erroneous trailing gaps in the flow behind the solid.
            (b) Direct velocity interpolation, with colliding particles projected back out: Repeated collision and projection leads to irregular particle patterns and the severe hole in the particle distribution persists.
            (c) Curl-Flow interpolation, with colliding particles halted on contact: Particles smoothly traverse the sphere with very few colliding. The flow also tightly follows behind the solid, so the final particle layer remains much more uniform (right).
    }
    \label{fig:particleSweep3D}
\end{figure}

Letting $\boldsymbol{\psi}_{nor}(\mathbf{x})$ be the normal component of the ambient $\boldsymbol{\psi}$, the final modified pointwise $\boldsymbol{\psi}$ with ramping is
\begin{equation}
\label{eq:additiveRamp3D}
\begin{aligned}
\boldsymbol{\psi}'(\mathbf{x}) &=
\boldsymbol{\psi}_{tan}'(\mathbf{x}) + \boldsymbol{\psi}_{nor}(\mathbf{x}) \\
&= \boldsymbol{\psi}_{tan}(\mathbf{x}) + 
(\alpha(\mathbf{x}, \textbf{cp}(\mathbf{x})) - 1)
(\boldsymbol{\psi}_{tan}(\textbf{cp}(\mathbf{x})) -  \boldsymbol{\psi}_{g, tan}(\textbf{cp}(\mathbf{x}))) \\
&+ \boldsymbol{\psi}_{nor}(\mathbf{x}) 
\end{aligned}
\end{equation}
Again, we use $cp(\mathbf{x})$ to represent the closest point on the solid triangles
and $\alpha(\mathbf{x}) = ramp(d(\mathbf{x} / d_0))$ for a smooth transition toward $\boldsymbol{\psi}_{g, tan}$.
Here, $d_0$ is the influence radius ($h$ in our case), and 
d($\cdot$) is the distance to the closest point $cp(\mathbf{x})$.
We again employ the smoothstep function for $ramp(\cdot)$. 
This approach is much like \eqref{eq:additiveRamp} in 2D, 
except that the target $\boldsymbol{\psi}_{g, tan}$ is not a constant since it is computed from barycentric interpolation. 

The corresponding free-slip velocity is
\begin{equation}
\begin{aligned}
\label{eq:modifiedVelocity3D}
\mathbf{u}'(\mathbf{x}) 
&= \nabla \times (\boldsymbol{\psi}_{tan}'(\mathbf{x}) + \boldsymbol{\psi}_{nor}(\mathbf{x})) \\
&= \nabla \times \boldsymbol{\psi}(\mathbf{x}) + 
\nabla \alpha(\mathbf{x}, \textbf{cp}(\mathbf{x})) \times (\boldsymbol{\psi}_{tan}(\textbf{cp}(\mathbf{x})) -  \boldsymbol{\psi}_{g, tan}(\textbf{cp}(\mathbf{x}))) \\ 
&+ (\alpha(\mathbf{x}, \textbf{cp}(\mathbf{x})) - 1) 
\{
\nabla \times (\boldsymbol{\psi}_{tan}(\textbf{cp}(\mathbf{x})) -  \boldsymbol{\psi}_{g, tan}(\textbf{cp}(\mathbf{x}))) 
\}
\end{aligned}
\end{equation} 
As the position $\mathbf{x}$ approaches $\textbf{cp}(\mathbf{x})$, 
$\nabla \times \boldsymbol{\psi}_{tan}(\mathbf{x})$ and 
$\nabla \times \boldsymbol{\psi}_{tan}(\textbf{cp}(\mathbf{x}))$
cancel each other out, and due to barycentric interpolation,
$\nabla \times \boldsymbol{\psi}_{g, tan}(\textbf{cp}(\mathbf{x}))$
is constant per triangle, 
which we constrained to have zero normal velocity in \eqref{eq:solidKKT}; thus exact zero flux on the solid surface.
(Other components do not influence normal velocity.)

Due to the piecewise constant component from barycentric interpolation in \eqref{eq:modifiedVelocity3D}, it is possible for additional velocity discontinuities to be introduced,
beyond those coming from non-smooth changes in $\textbf{cp}(\mathbf{x})$ (Section \ref{ramping2D}).
However, we found these effects to be a reasonable tradeoff for incompressibility and precise boundary satisfaction.  Moreover, the correction applied by our scheme is often small and any possible discontinuities can only occur within one cell width of solid obstacles or domain boundaries, whereas the discontinuities mentioned in related work (e.g., DG methods (Section \ref{sec:DG}) and most direct incompressible velocity interpolants (Section \ref{sec:directincompressible})) arise between adjacent elements or cells everywhere throughout the domain, even in the absence of obstacles. 

To illustrate the improved behavior of 3D Curl-Flow with cut-cells in a basic scenario, we consider a flow past a solid sphere with a steady state velocity field (Figure~\ref{fig:particleSweep3D}), and release a single planar layer of passive particles. With trilinear advection, many of the particles collide with the obstacle, leaving a large gap in the particle field. With Curl-Flow, the particles instead flow more naturally around the obstacle.

\section{Results and Discussion}
To consistently compare our Curl-Flow method against direct velocity interpolation while isolating interpolation-related effects, most of our comparisons change only the interpolation of velocity used for \emph{passive} particle tracing, keeping the underlying simulated flow data the same. (The exception are tests that apply Curl-Flow to tracing the FLIP particle paths for FLIP advection.)
Whenever a particle penetrates a solid object due to poor quality velocity fields, large timesteps, or insufficiently accurate path integration, a typical choice is to "resolve" the collision by projecting the particle back out of the solid, although this exacerbates clumping. To visually highlight the pervasiveness of such errors, we freeze penetrating particles in place, unless otherwise noted (such as in Figure \ref{fig:particleSweep3D}(b)). Simulation timings were gathered on a 2.8 GHz, 4-Core Intel Core i7 processor.

\subsection{Particle Distribution Comparisons in 2D}
The 2D results of Figures \ref{fig:particleDistribution}, \ref{fig:ramp_mulzero}, \ref{fig:rampComparison}, and \ref{fig:staticDiskTrajectories} used static velocity fields for illustration; below, we consider time-evolving (i.e., dynamic) flow simulations.

\begin{figure}
	\centering	
    \begin{subfigure}[t]{0.45\textwidth}	
		\adjustbox{fbox, width=0.24\textwidth}{\includegraphics{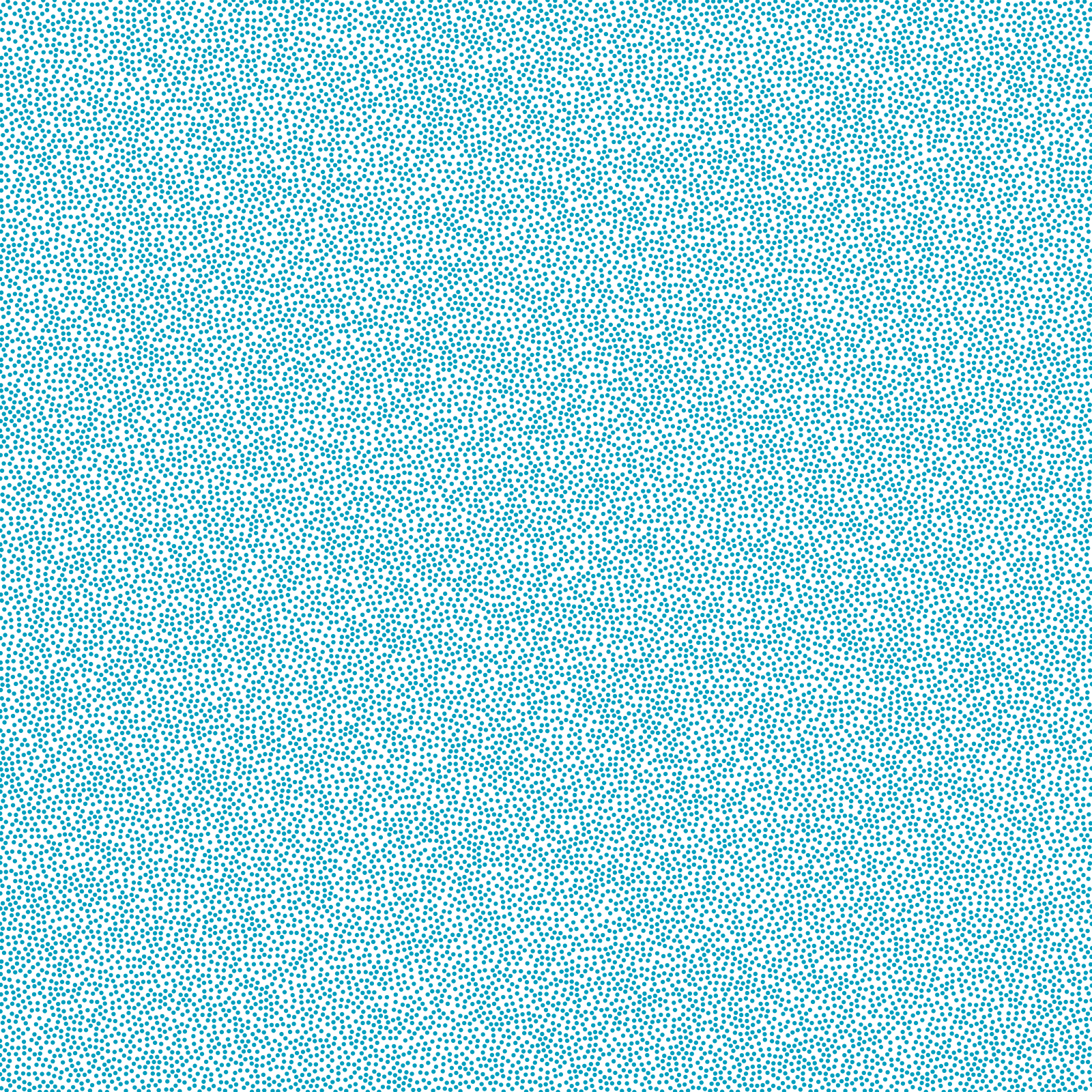}}
		\adjustbox{fbox, width=0.24\textwidth}{\includegraphics{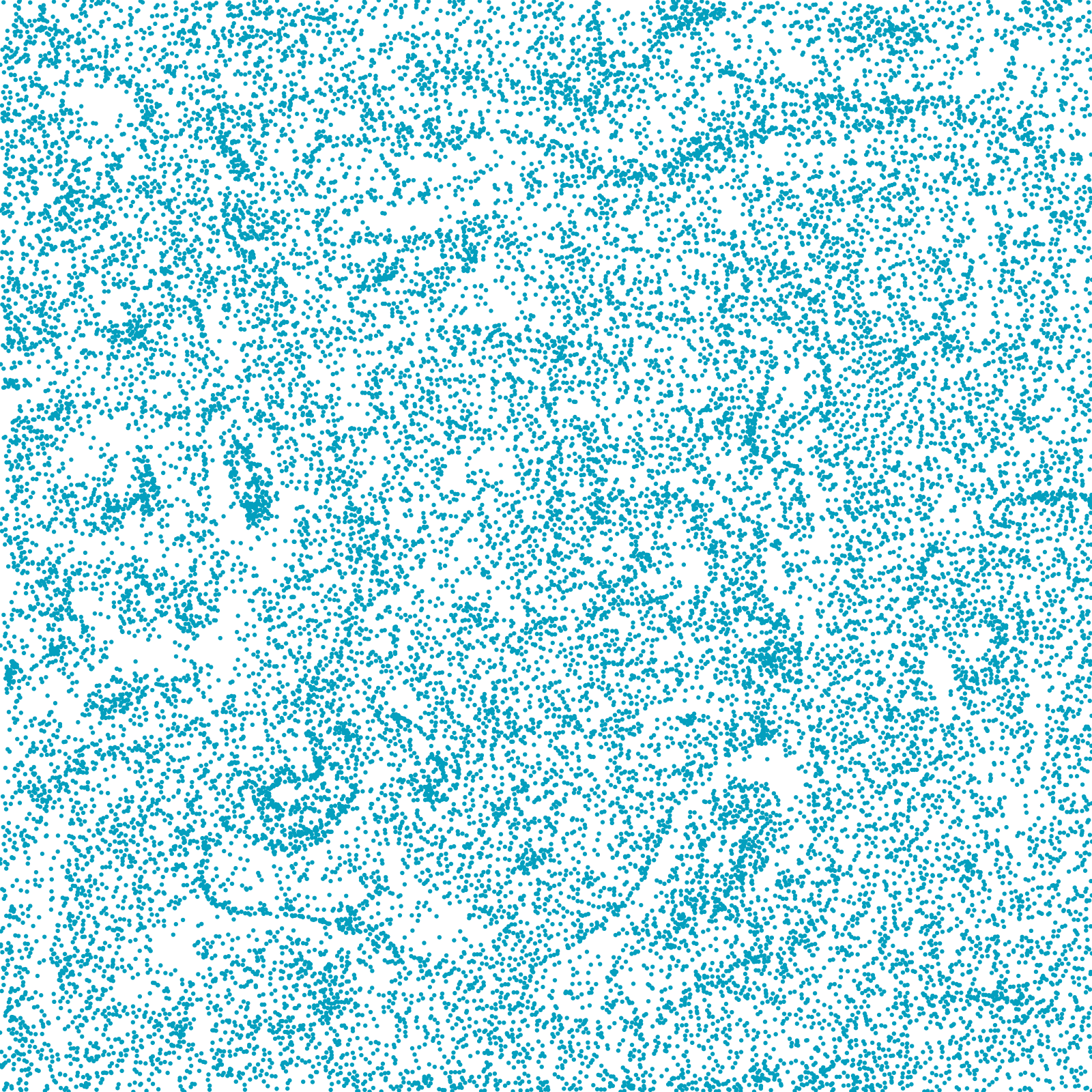}}
		\adjustbox{fbox, width=0.24\textwidth}{\includegraphics{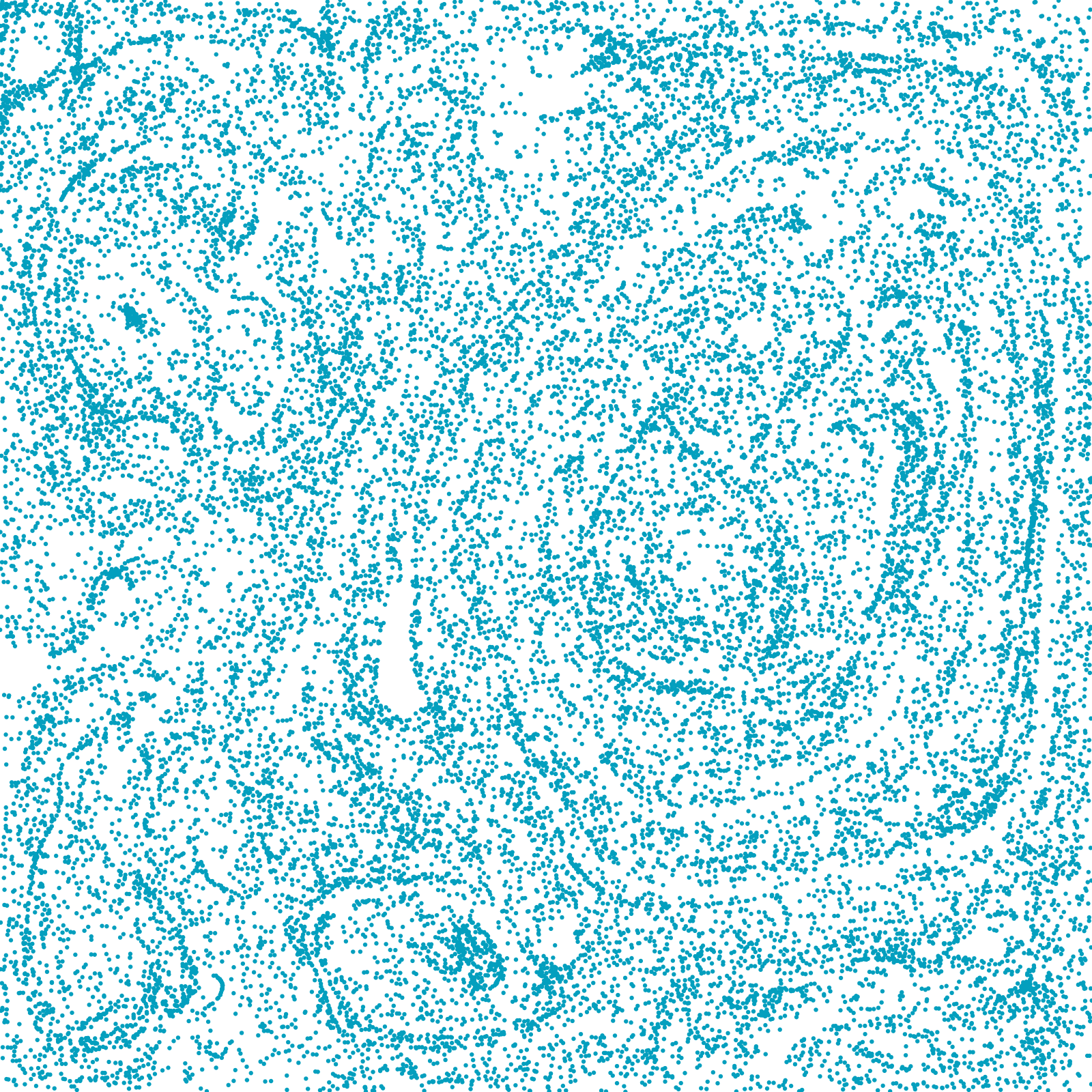}}
		\adjustbox{fbox, width=0.24\textwidth}{\includegraphics{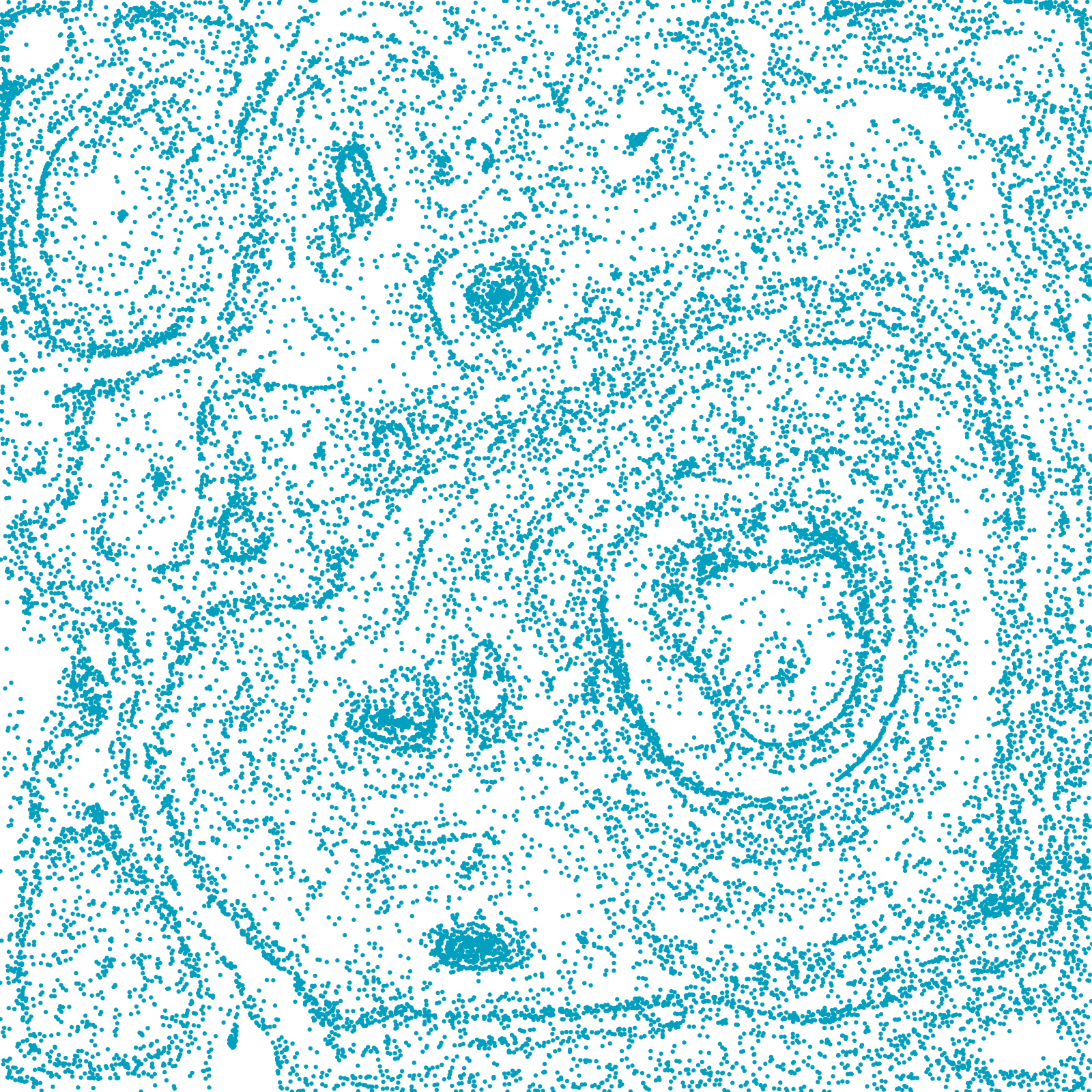}}
    	\subcaption{Bilinear velocity interpolation}
	\end{subfigure}
    \par\medskip
    \begin{subfigure}[t]{0.45\textwidth}	
		\adjustbox{fbox, width=0.24\textwidth}{\includegraphics{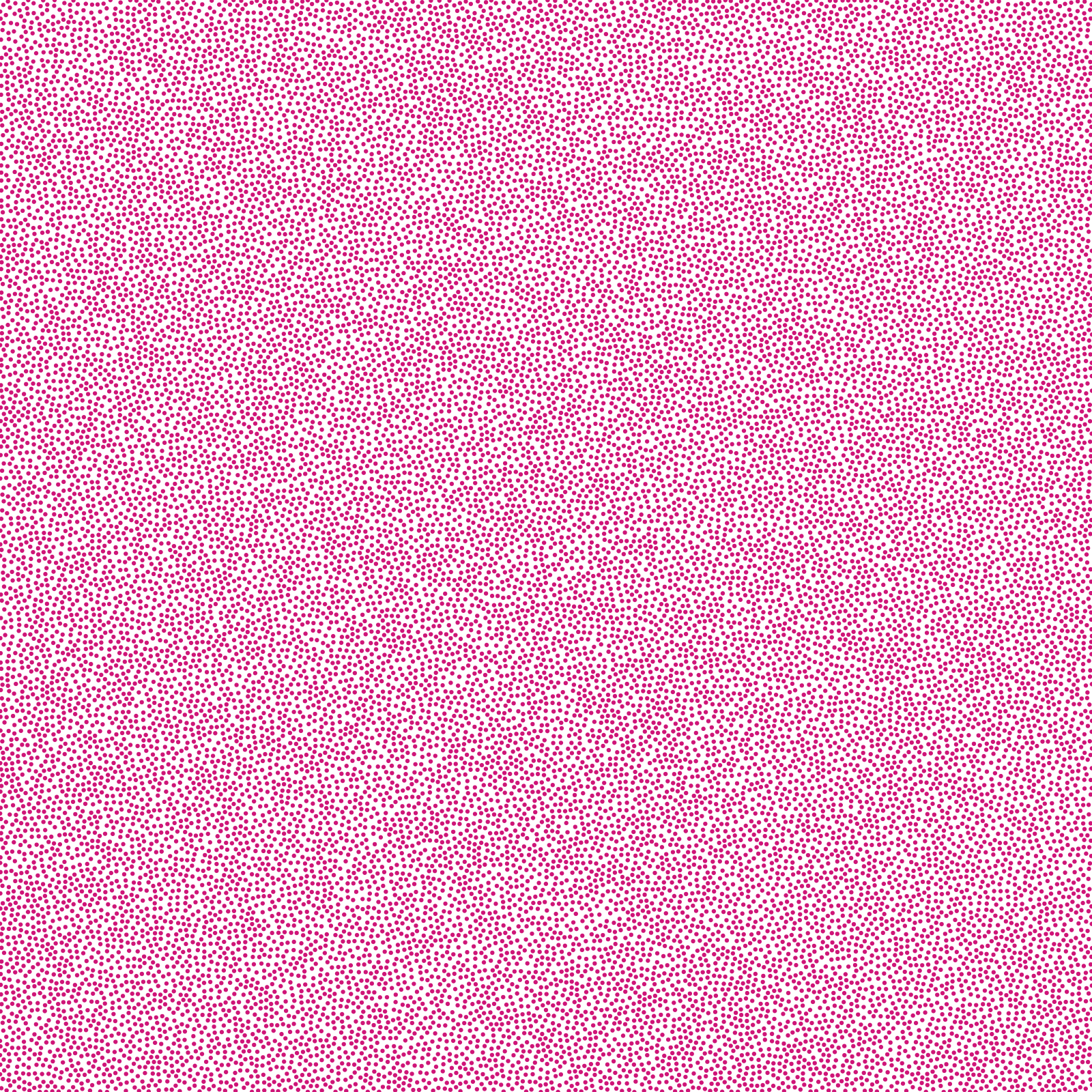}}
		\adjustbox{fbox, width=0.24\textwidth}{\includegraphics{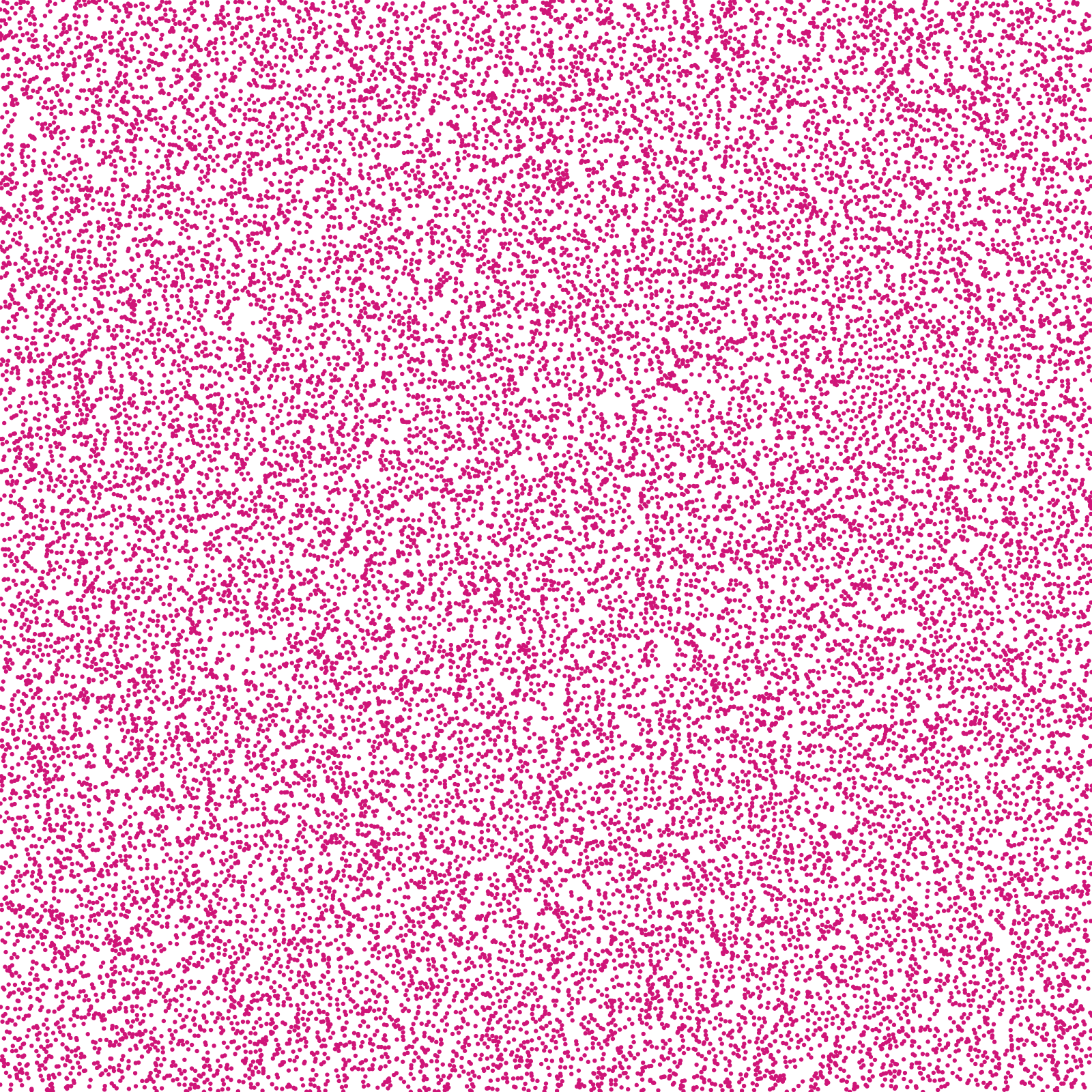}}
		\adjustbox{fbox, width=0.24\textwidth}{\includegraphics{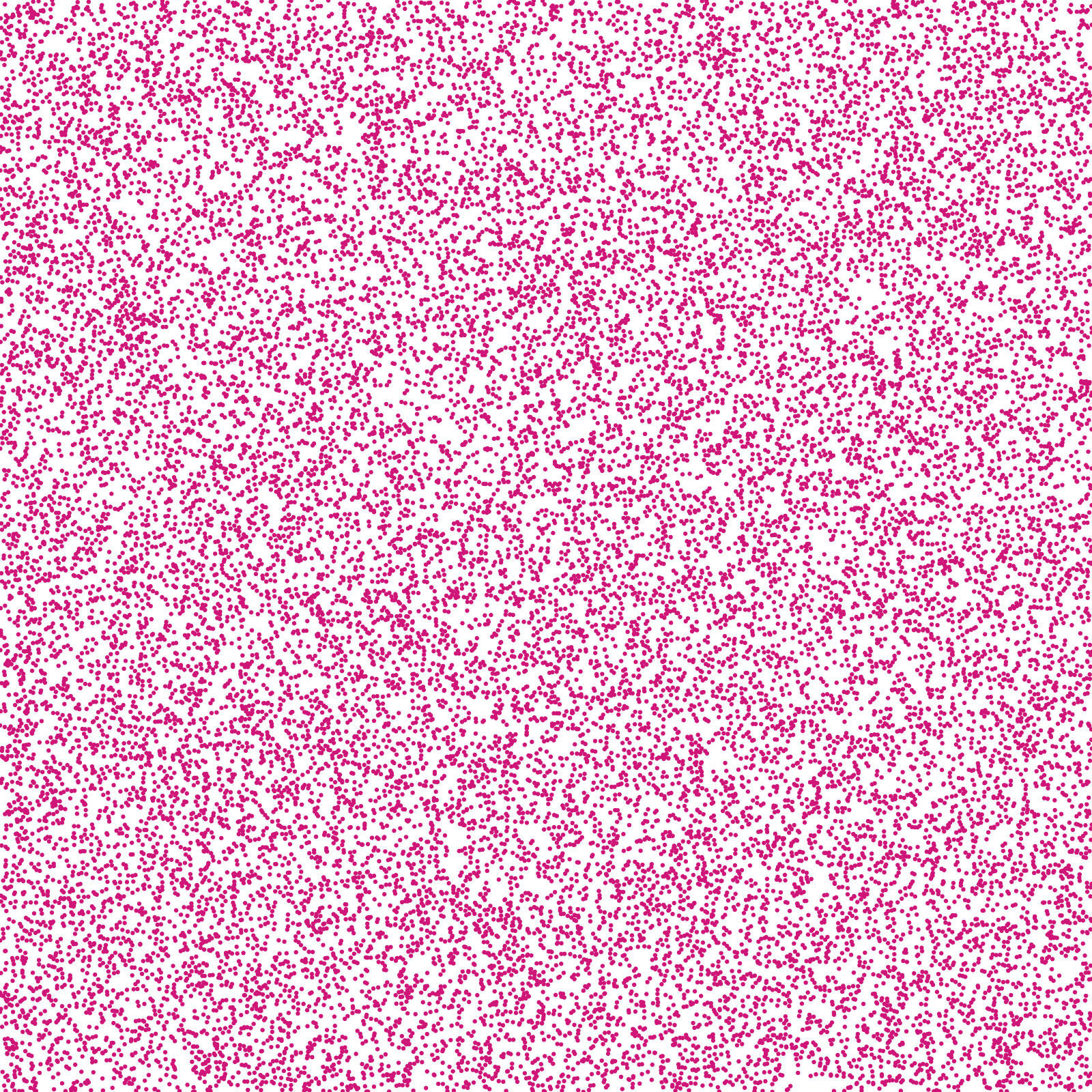}}
		\adjustbox{fbox, width=0.24\textwidth}{\includegraphics{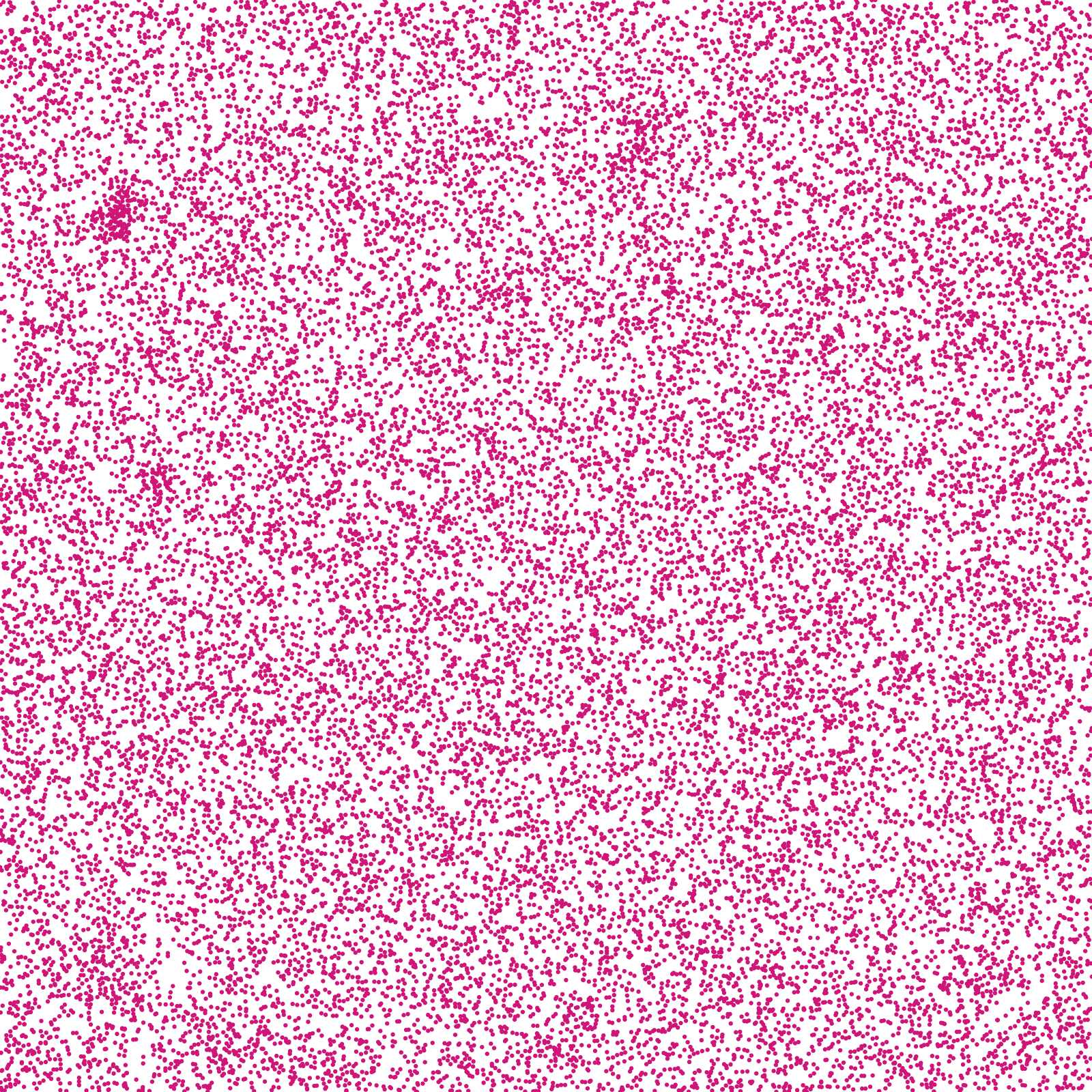}}
    	\subcaption{Curl-Flow interpolation}
	\end{subfigure}
	\caption{ \textbf{Dynamic 2D Flow Comparison:} 
	Uniformly sampled particles advected through a \emph{dynamic} simulated 2D vector field with turbulent forcing on a $20\times20$ grid. Frames 1, 50, 100, and 150 are shown from left to right.
	Particle distribution remains more uniform with Curl-Flow.
	}
	\label{fig:particleDistribution_dynamic}
\end{figure}

\begin{figure}
	\centering	
	\begin{subfigure}[t]{0.15\textwidth}
		\begin{overpic}[trim=0 0 0 0, clip, width=\textwidth]{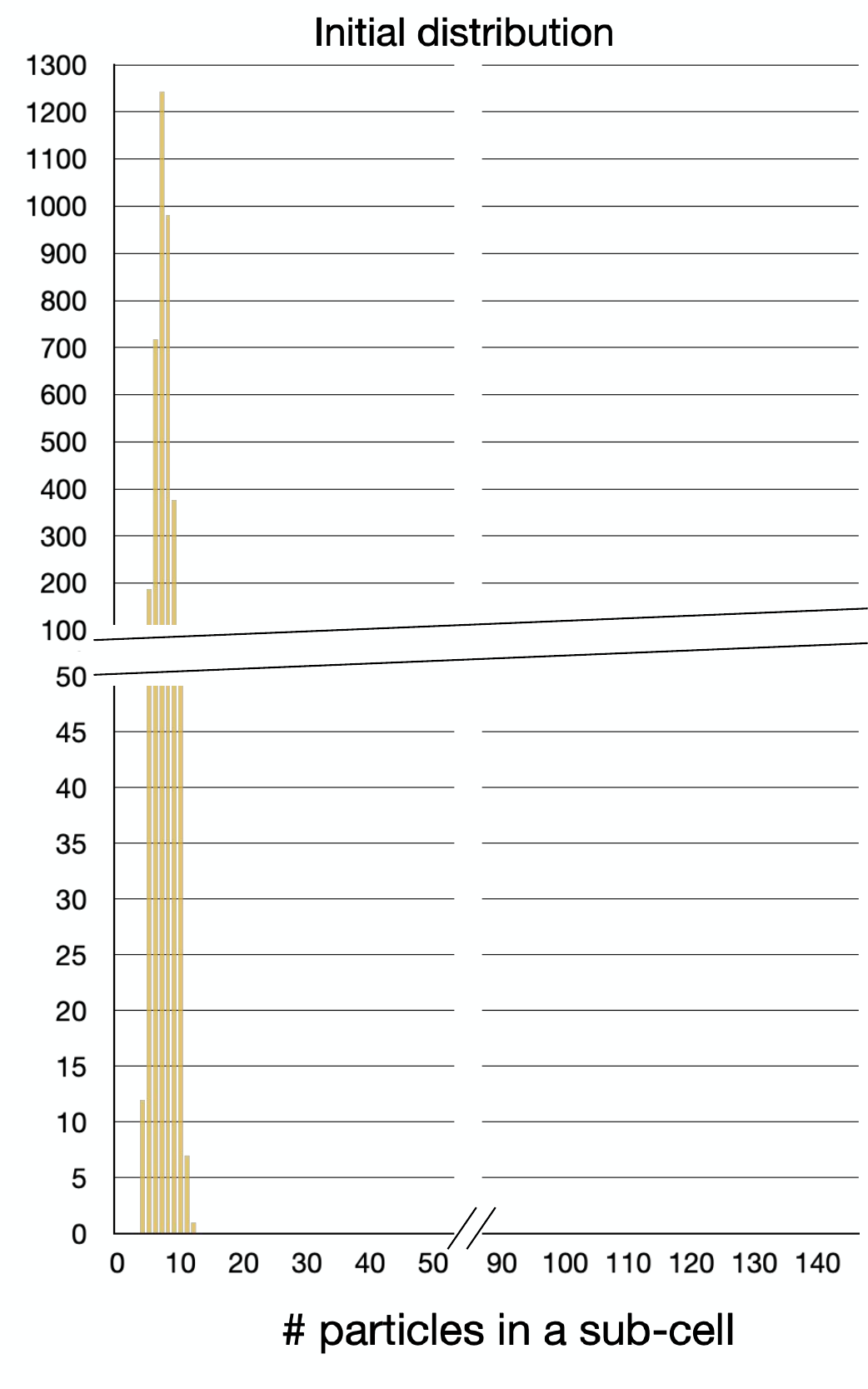}
		\end{overpic}    
	\end{subfigure} 
	\begin{subfigure}[t]{0.15\textwidth} 
		\begin{overpic}[trim=0 0 0 0, clip, width=\textwidth]{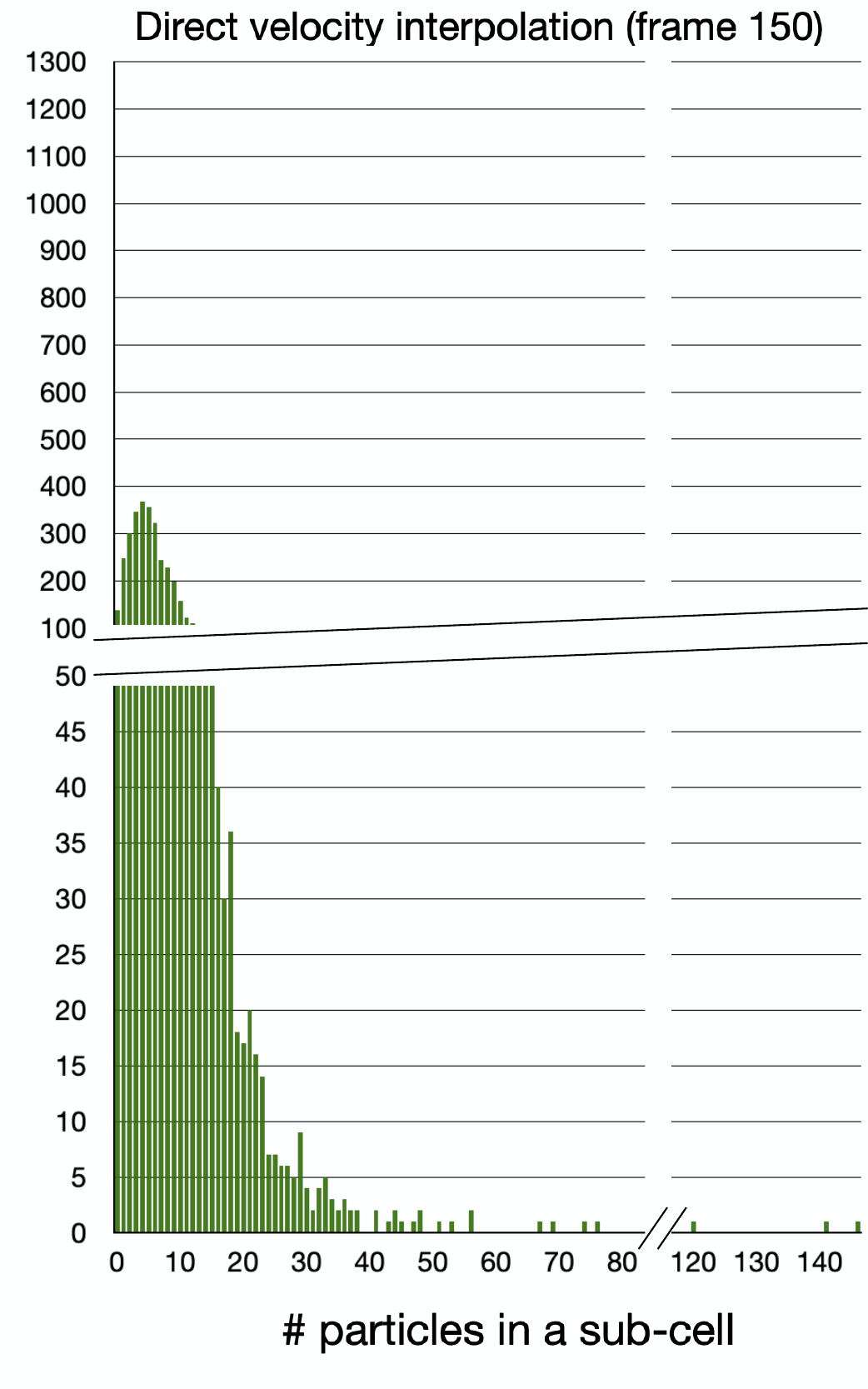}
		\end{overpic}    
 	\end{subfigure} 
	\begin{subfigure}[t]{0.15\textwidth}
		\begin{overpic}[trim=0 0 0 0, clip, width=\textwidth]{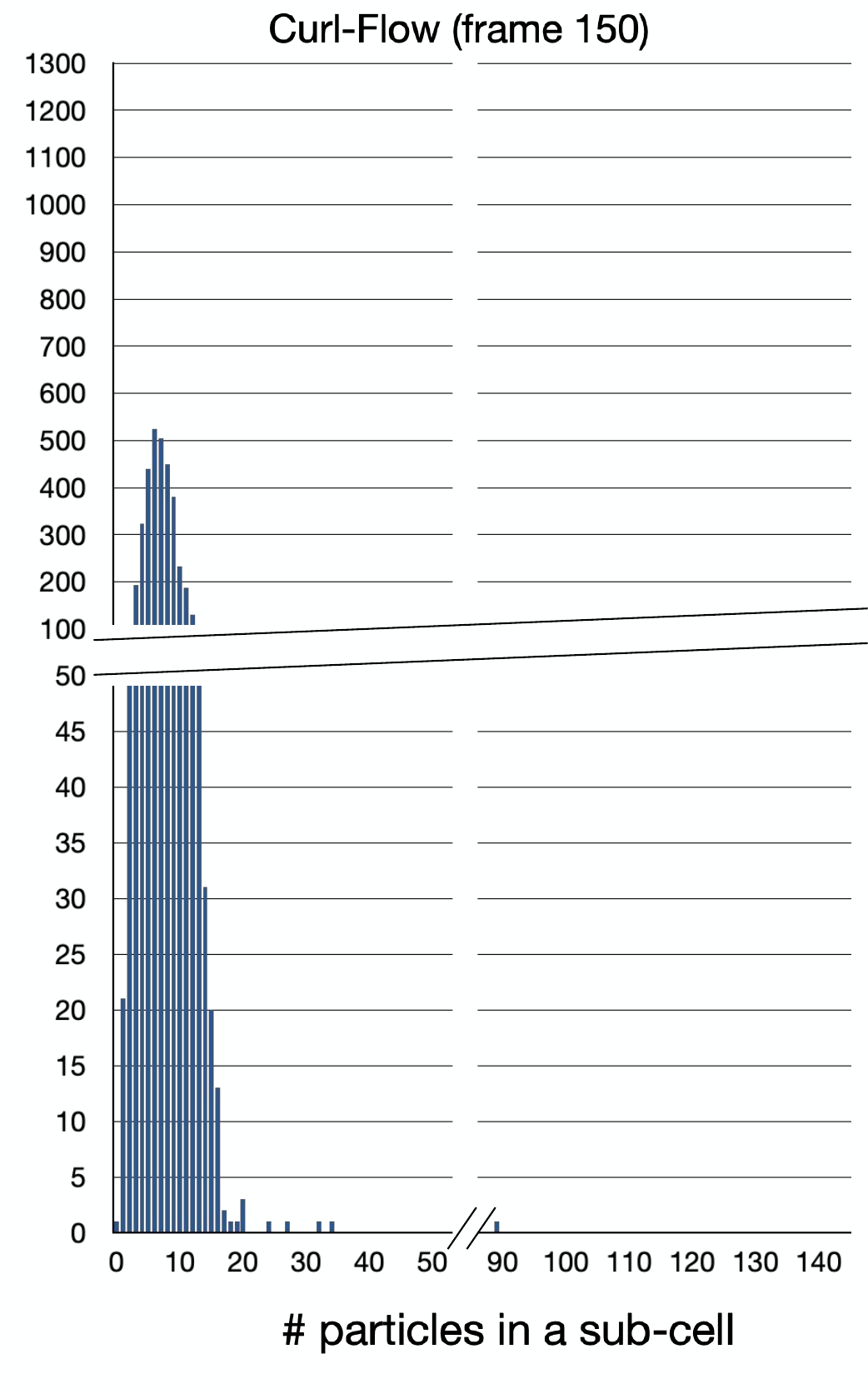}
		\end{overpic}    
	\end{subfigure} 
    \caption{\textbf{Particle Density Histograms:} 
    Histograms of particles per sub-cell for the first and last (150th) frames of the example in Figure \ref{fig:particleDistribution_dynamic}. 
    The sub-cell width for measurement is 1/3 the simulation cell width.
    The x-axis indicates the number of particles in a sub-cell, 
    and the y-axis represents the number of sub-cells having that particle count. 
    The graph for direct velocity interpolation has a longer and thicker tail of high density sub-cells, and dramatically more sub-cells with no particles at all.
    \label{fig:particleDensityChart}
}
\end{figure}

In Figure \ref{fig:particleDistribution_dynamic}, we densely seed passive particles with blue noise sampling and advect them under a dynamic flow on a coarse $20 \times 20$ grid.
We apply an upward force at the center of the domain and a smoothly varying random (noise) force over the entire domain in each frame for a turbulent effect.
Visually, the resulting distribution again remains more uniform under our method. 
To better quantify this effect, we conceptually lay a three times finer grid of sub-cells over the domain and plot histograms of the sub-cells'
particle counts in the first and last (150th) frame of the same example (Figure \ref{fig:particleDensityChart}). 
Initially, most sub-cells contain about 5--10 particles, and the distribution is tightly clustered.
The final Curl-Flow graph remains closely clustered, indicating still fairly uniform particle distribution, while the direct velocity interpolation graph spreads out severely. At the extreme ends, the direct case has 137 empty sub-cells and one sub-cell containing 149 particles.

\begin{figure*}
  \setlength{\tabcolsep}{0em}
  \centering
  \begin{tabular}{| c | c | c | c | c |}
    \hline
     & \thead{ Direct Velocity Interpolation \\ ($60 \times 20$) } 
     & \thead{ Curl-Flow \\ ($60 \times 20$) }
     & \thead{ Direct Velocity Interpolation \\ ($180 \times 60$) } 
     & \thead{ Curl-Flow \\ ($180 \times 60$) } \\ \hline
    \thead{ Passive \\ \ Particles \ \\ (Disk) }
    &
    \begin{minipage}{.23\textwidth}
      \includegraphics[trim=5.5cm 20cm 13cm 18.5cm, clip,width=\textwidth]{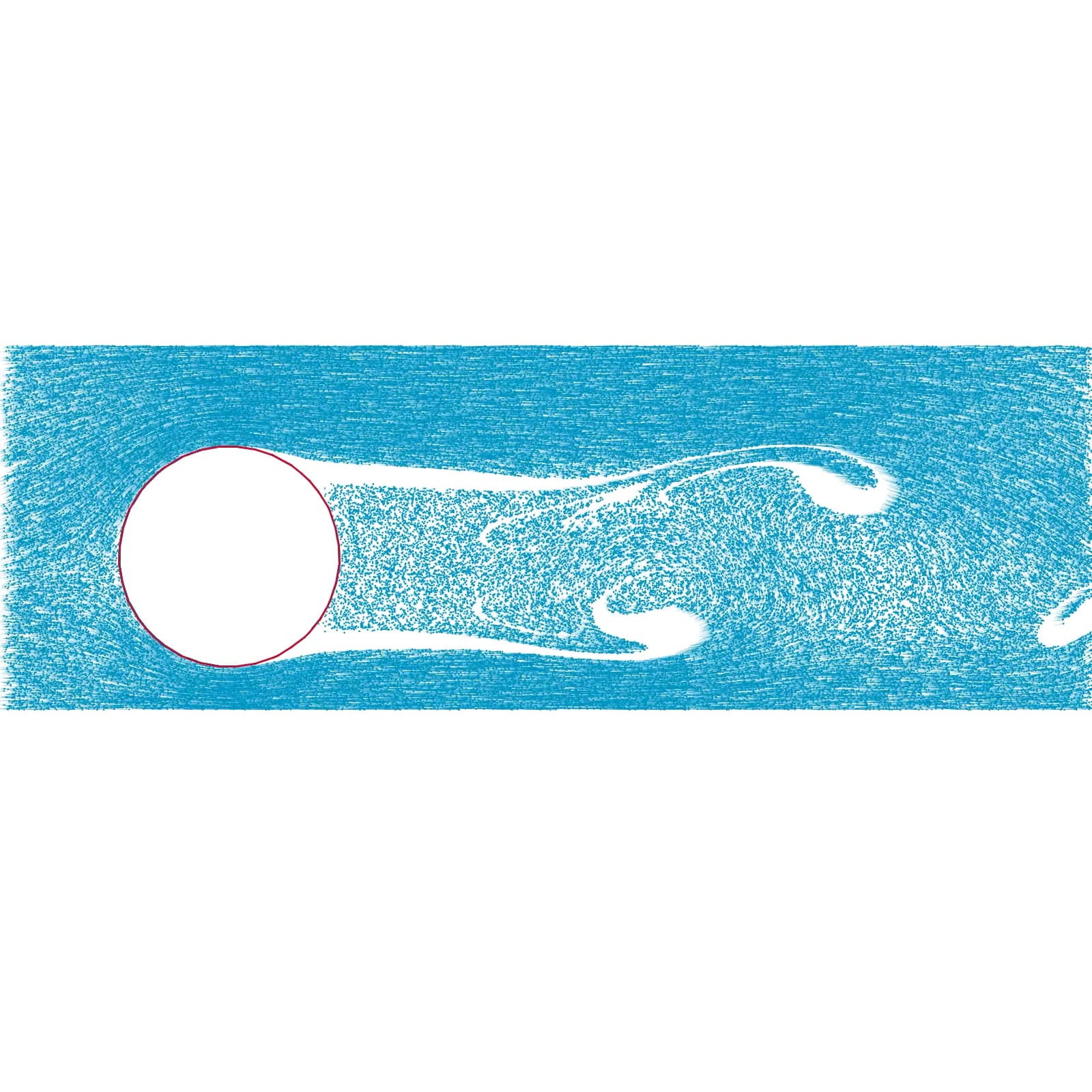}
    \end{minipage}
    &
    \begin{minipage}{.23\textwidth}
      \includegraphics[trim=5.5cm 20cm 13cm 18.5cm, clip,width=\textwidth]{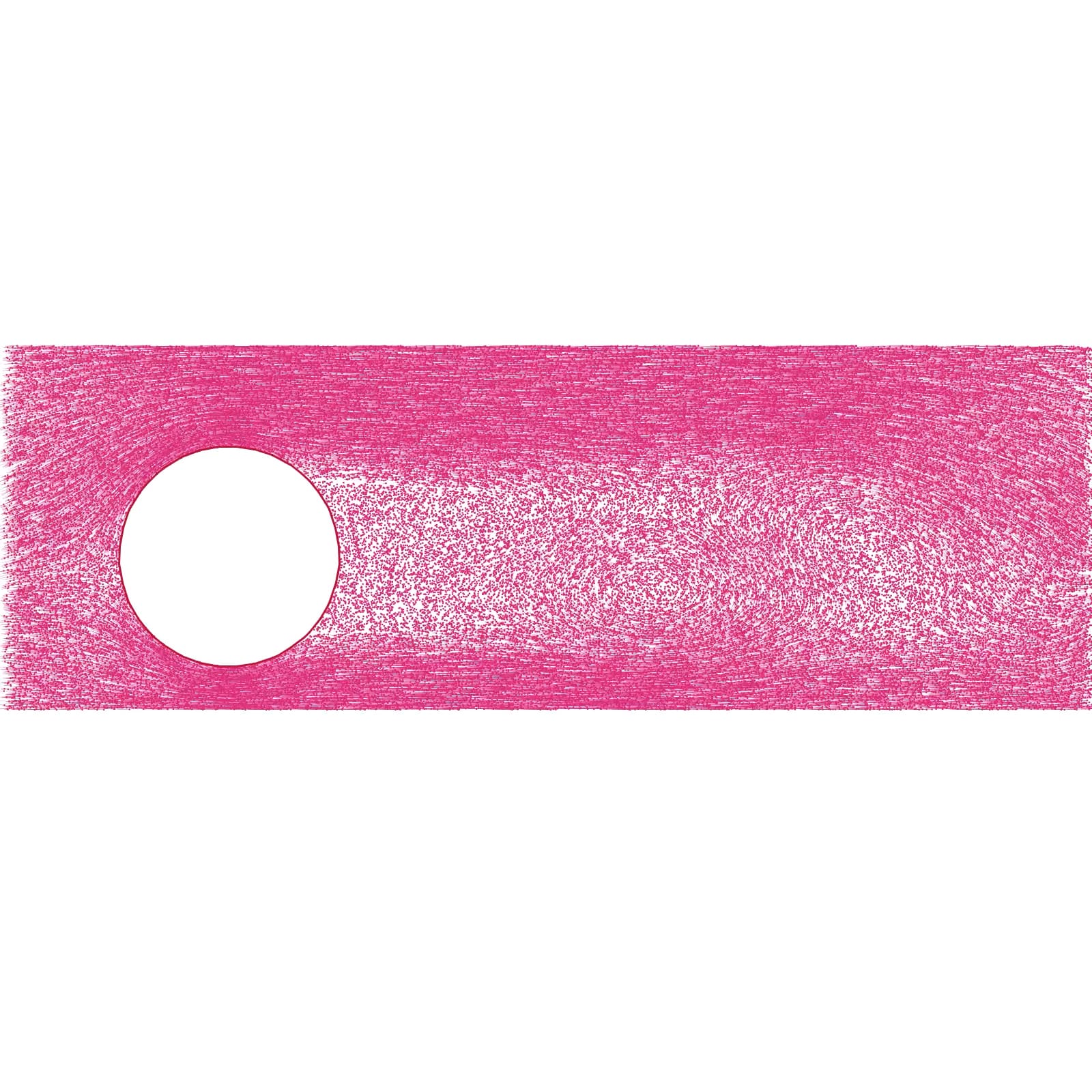}
    \end{minipage}
    & 
    \begin{minipage}{.23\textwidth}
      \includegraphics[trim=5.5cm 20cm 13cm 18.5cm, clip,width=\textwidth]{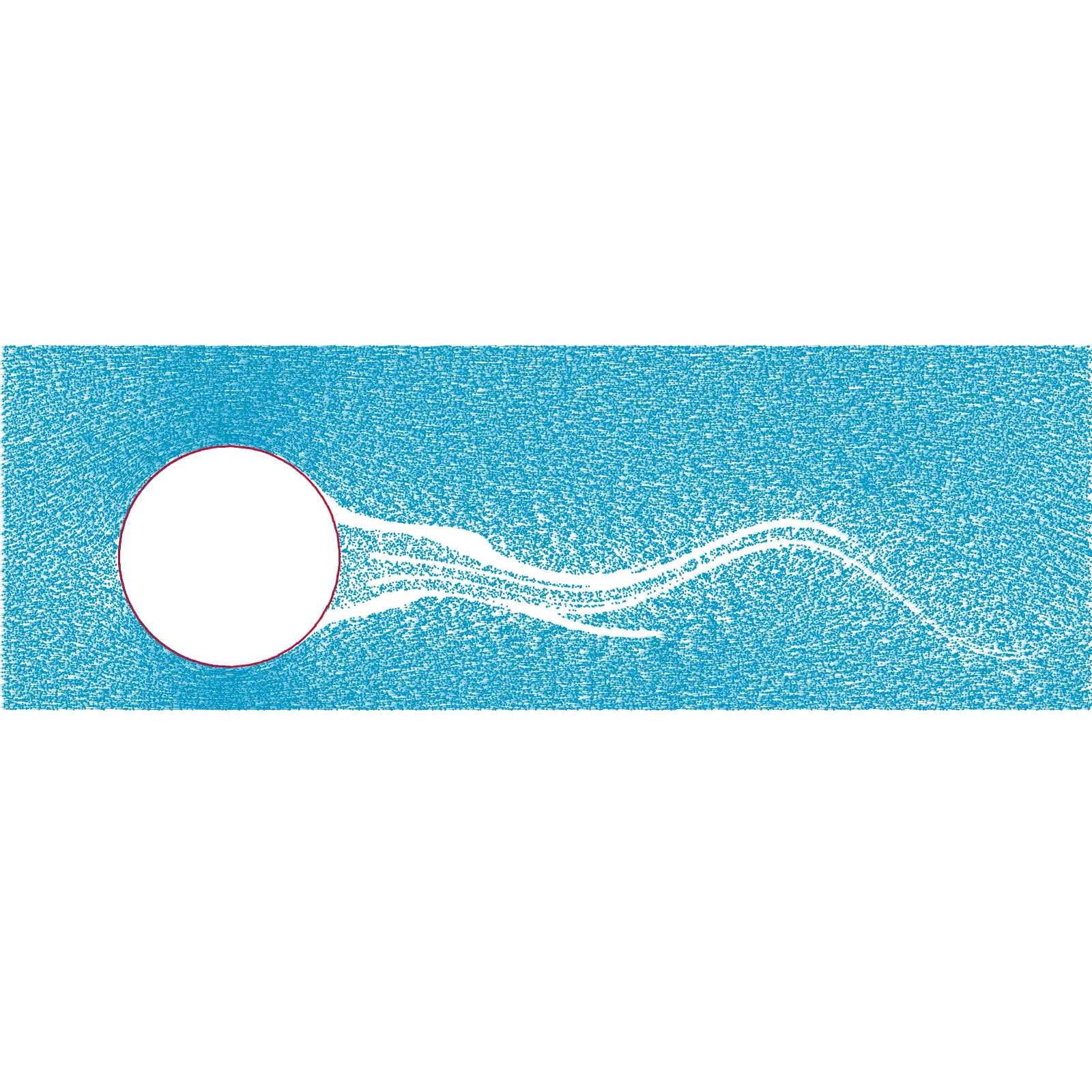}
    \end{minipage}
    & 
    \begin{minipage}{.23\textwidth}
      \includegraphics[trim=5.5cm 20cm 13cm 18.5cm, clip,width=\textwidth]{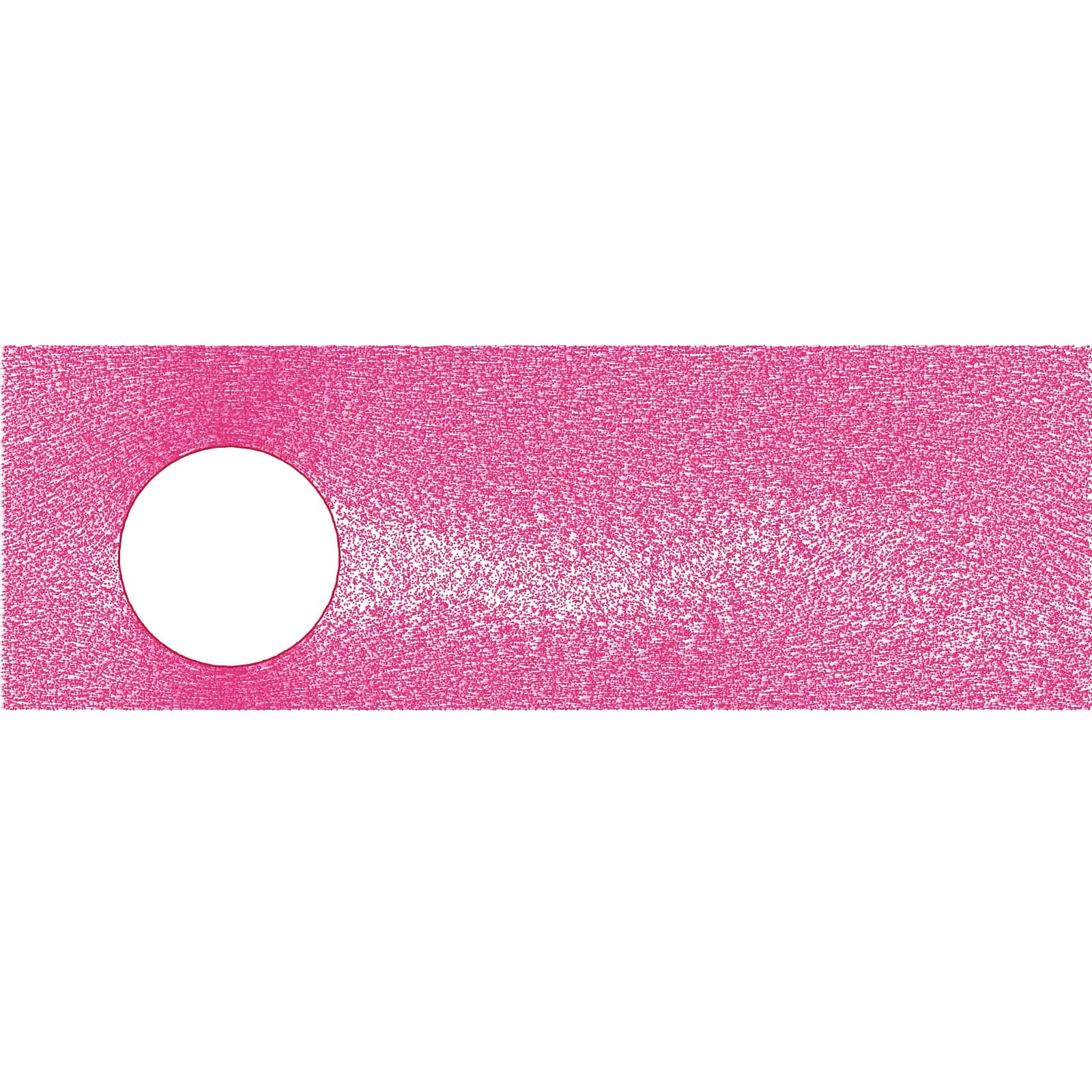}
    \end{minipage}
    \\ \hline
    \thead{ Passive \\ \ Particles \ \\ (Jagged) }
    &
    \begin{minipage}{.23\textwidth}
      \includegraphics[trim=2cm 16cm 2cm 16cm, clip,width=\textwidth]{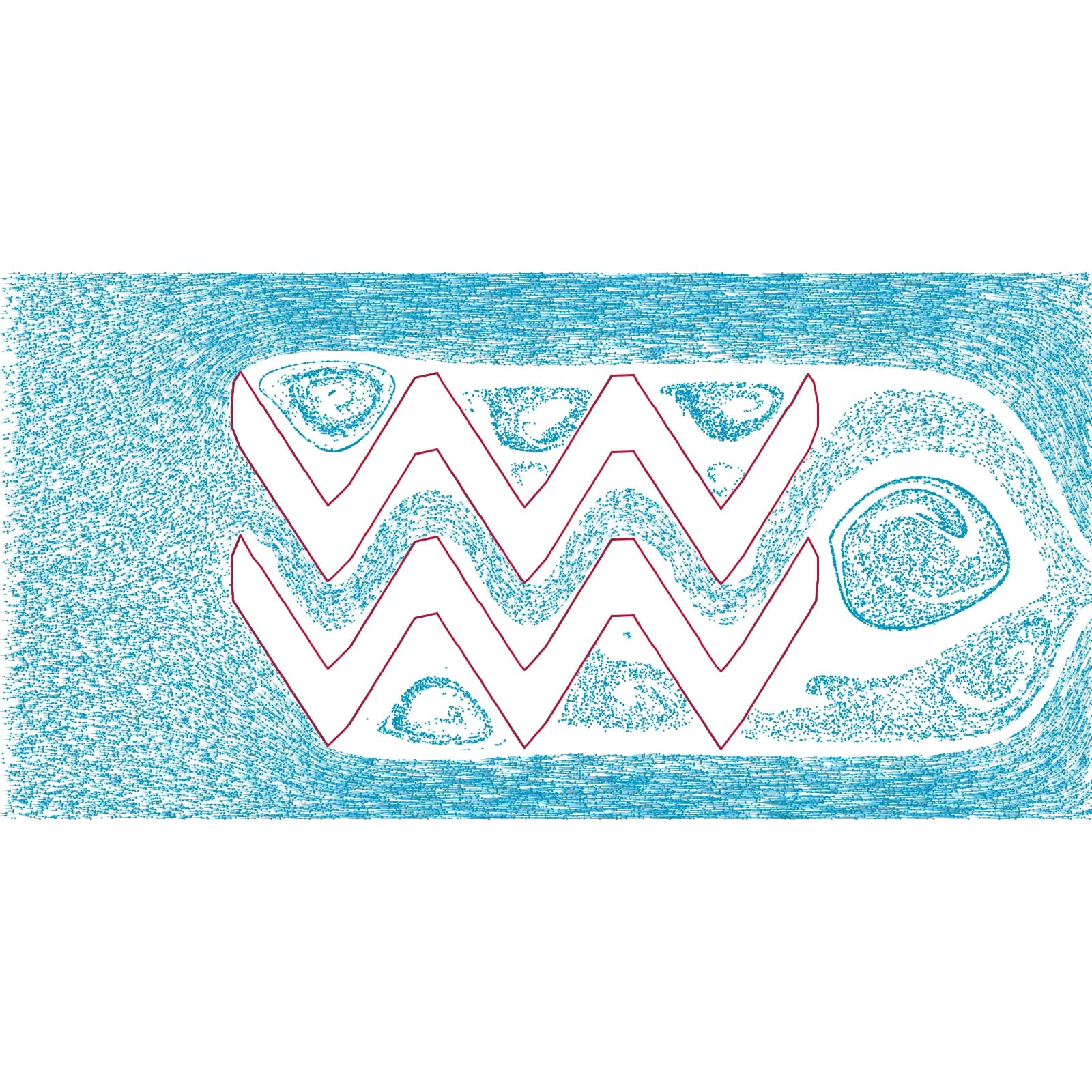}
    \end{minipage}    
    &
    \begin{minipage}{.23\textwidth}
      \includegraphics[trim=2cm 16cm 2cm 16cm, clip,width=\textwidth]{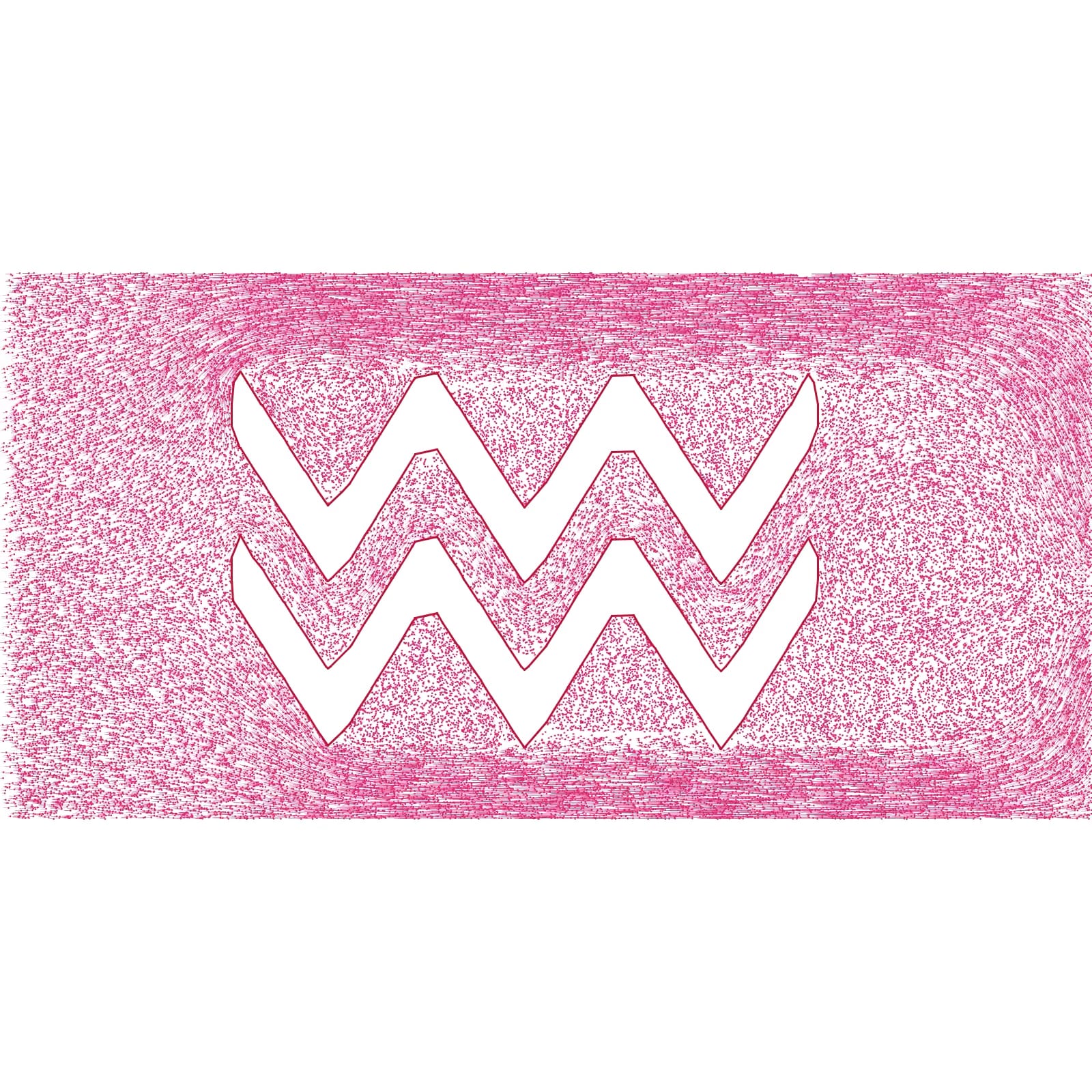}
    \end{minipage}    
    & 
    \begin{minipage}{.23\textwidth}
      \includegraphics[trim=2cm 16cm 2cm 16cm, clip,width=\textwidth]{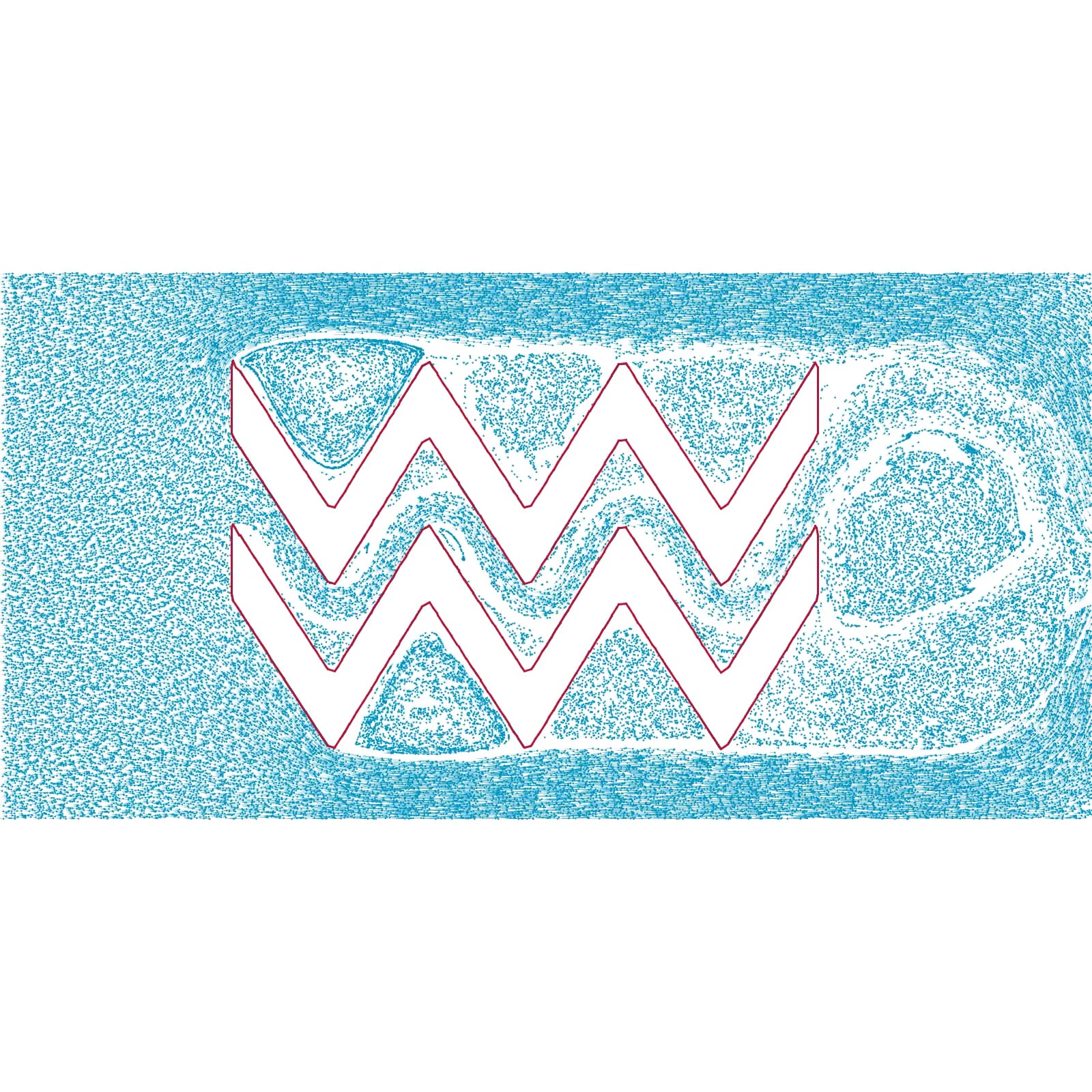}
    \end{minipage}
    & 
    \begin{minipage}{.23\textwidth}
      \includegraphics[trim=2cm 16cm 2cm 16cm, clip,width=\textwidth]{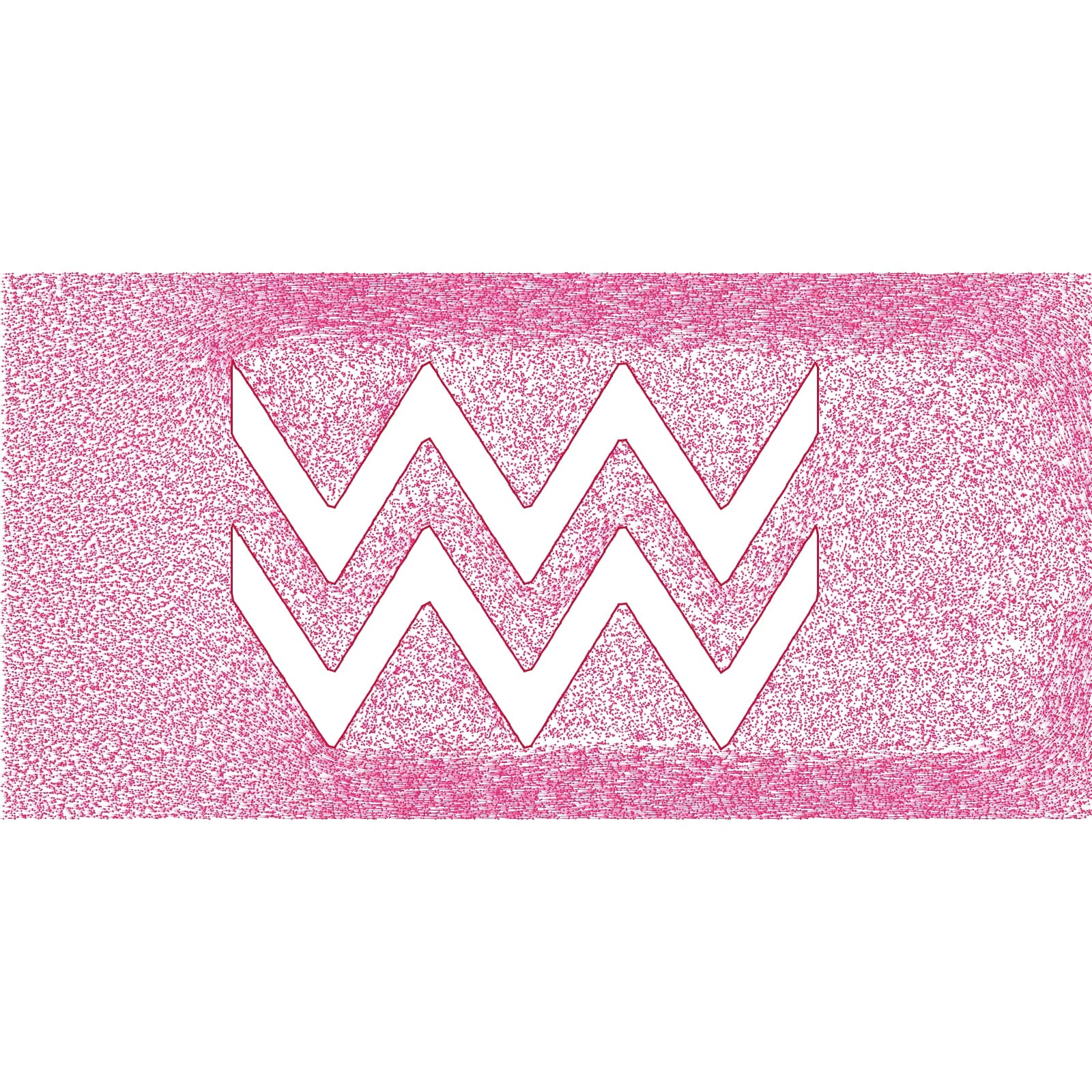}
    \end{minipage}
    \\
    \hline
    \hline
    \thead{ FLIP \\ \ Particles \ \\ (Disk) }
    &
    \begin{minipage}{.23\textwidth}
      \includegraphics[trim=5.5cm 20cm 13cm 18.5cm, clip,width=\textwidth]{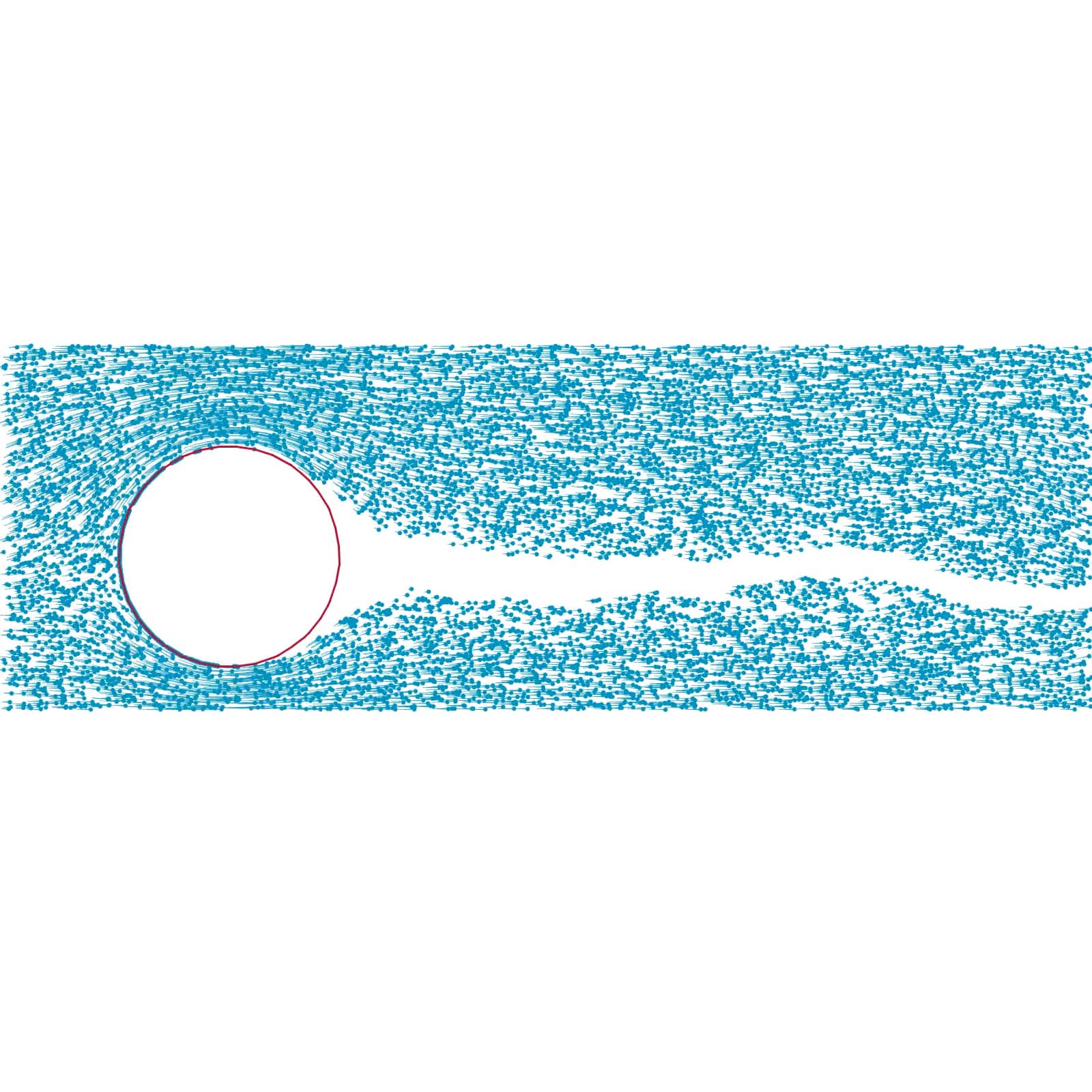}
    \end{minipage}    
    &
    \begin{minipage}{.23\textwidth}
      \includegraphics[trim=5.5cm 20cm 13cm 18.5cm, clip,width=\textwidth]{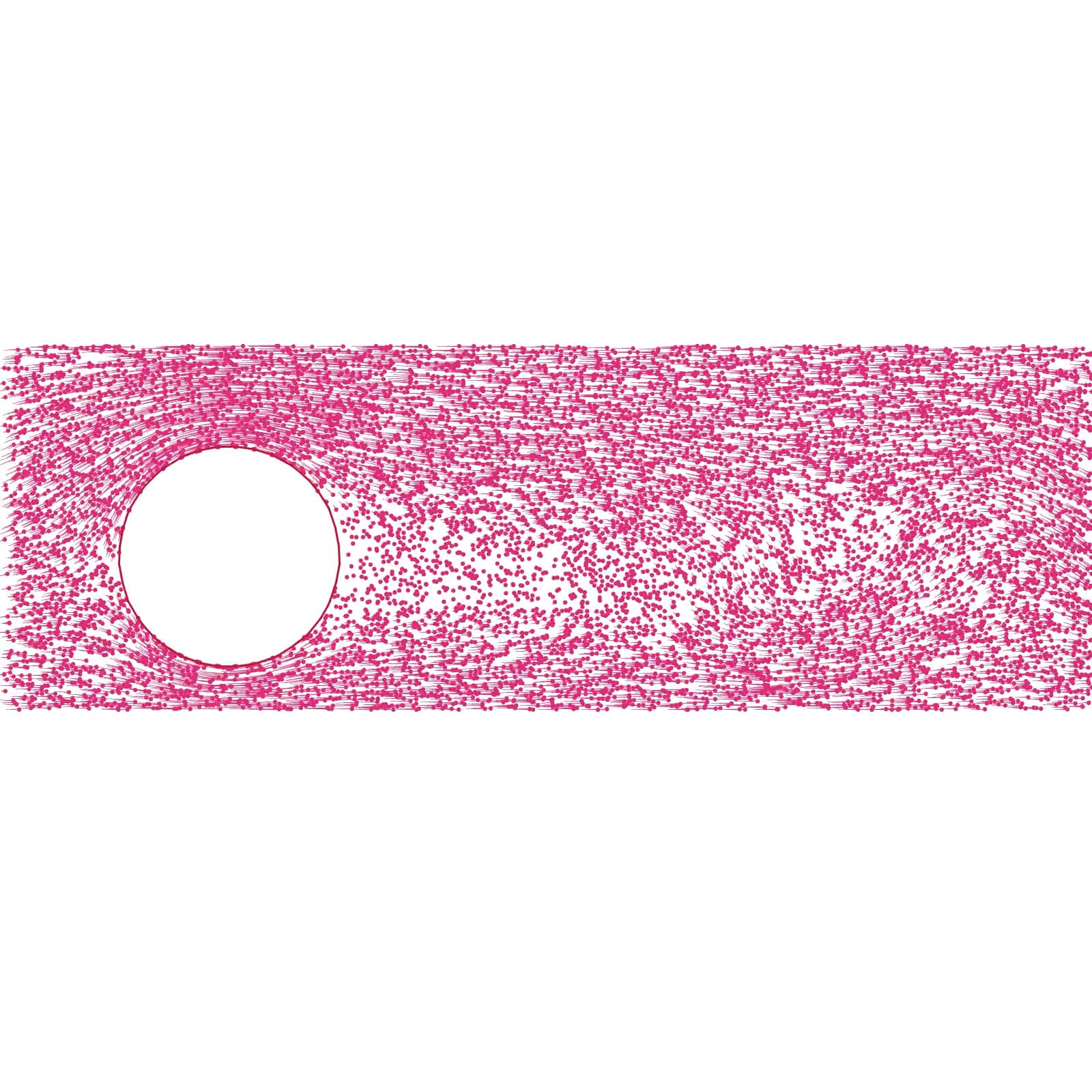}
    \end{minipage}
    & 
    \begin{minipage}{.23\textwidth}
      \includegraphics[trim=5.5cm 20cm 13cm 18.5cm, clip,width=\textwidth]{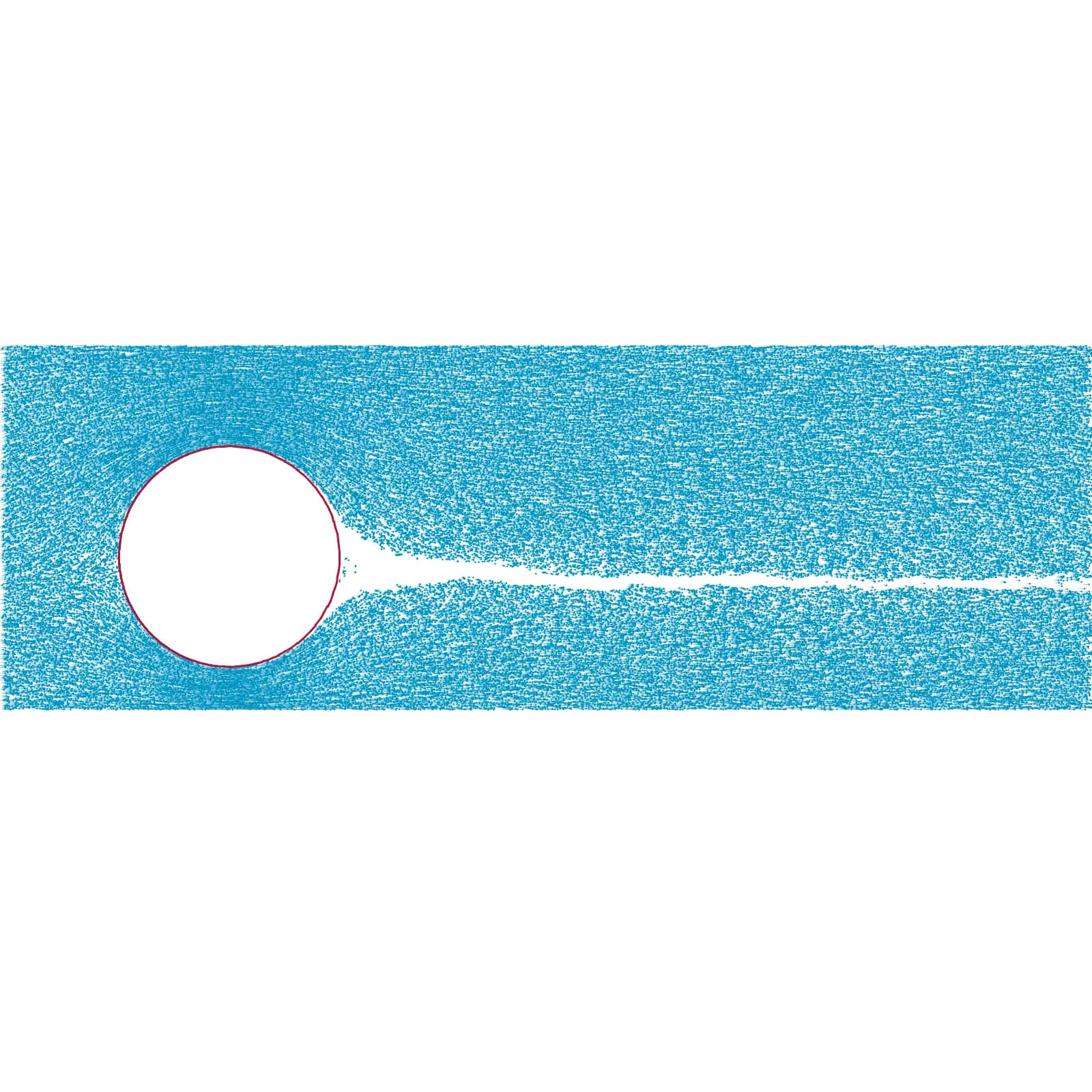}
    \end{minipage}    
    & 
    \begin{minipage}{.23\textwidth}
      \includegraphics[trim=5.5cm 20cm 13cm 18.5cm, clip,width=\textwidth]{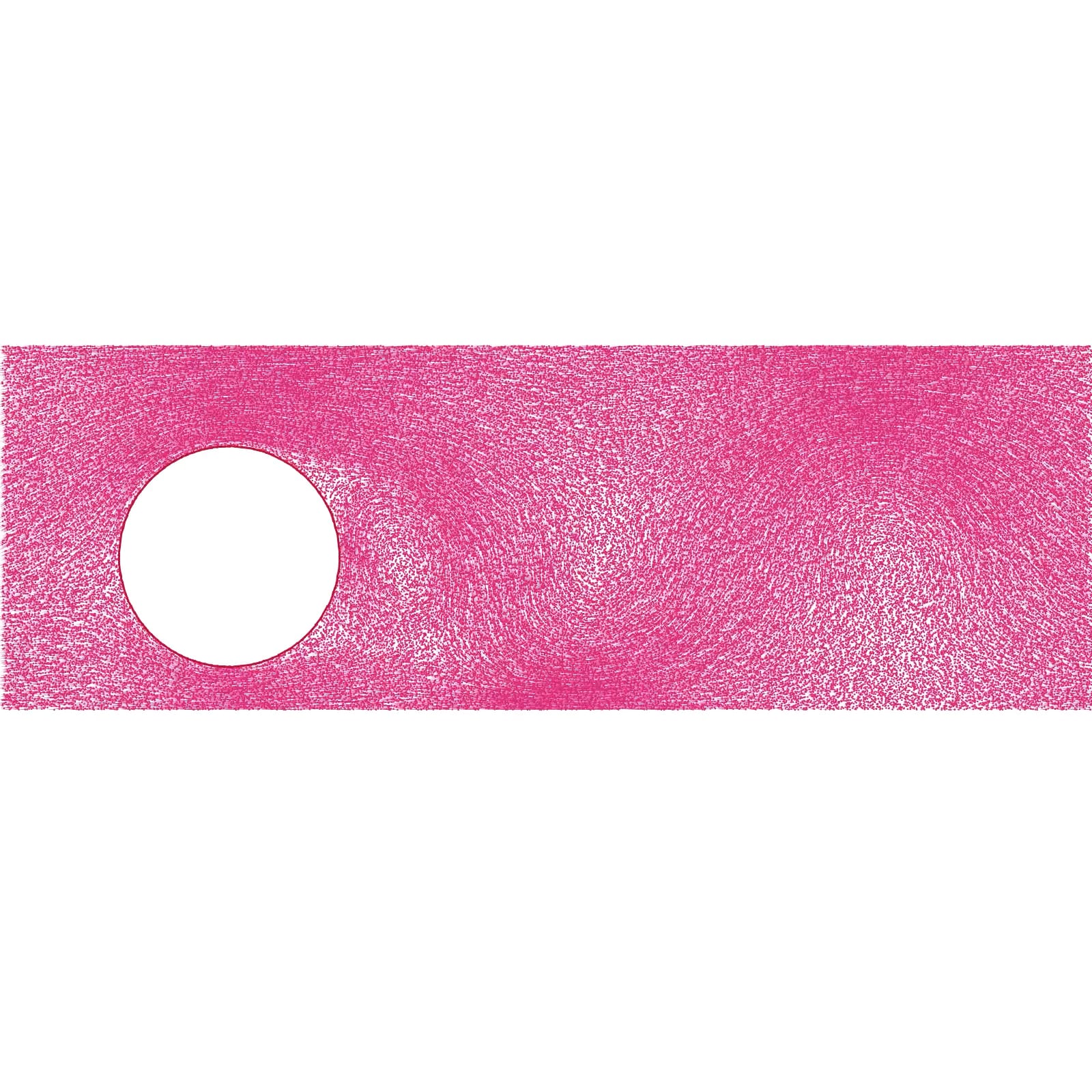}
    \end{minipage}    
    
    \\
    \hline
    \thead{ FLIP \\ \ Particles \ \\ (Jagged) }
    &
    \begin{minipage}{.23\textwidth}
      \includegraphics[trim=2cm 16cm 2cm 16cm, clip,width=\textwidth]{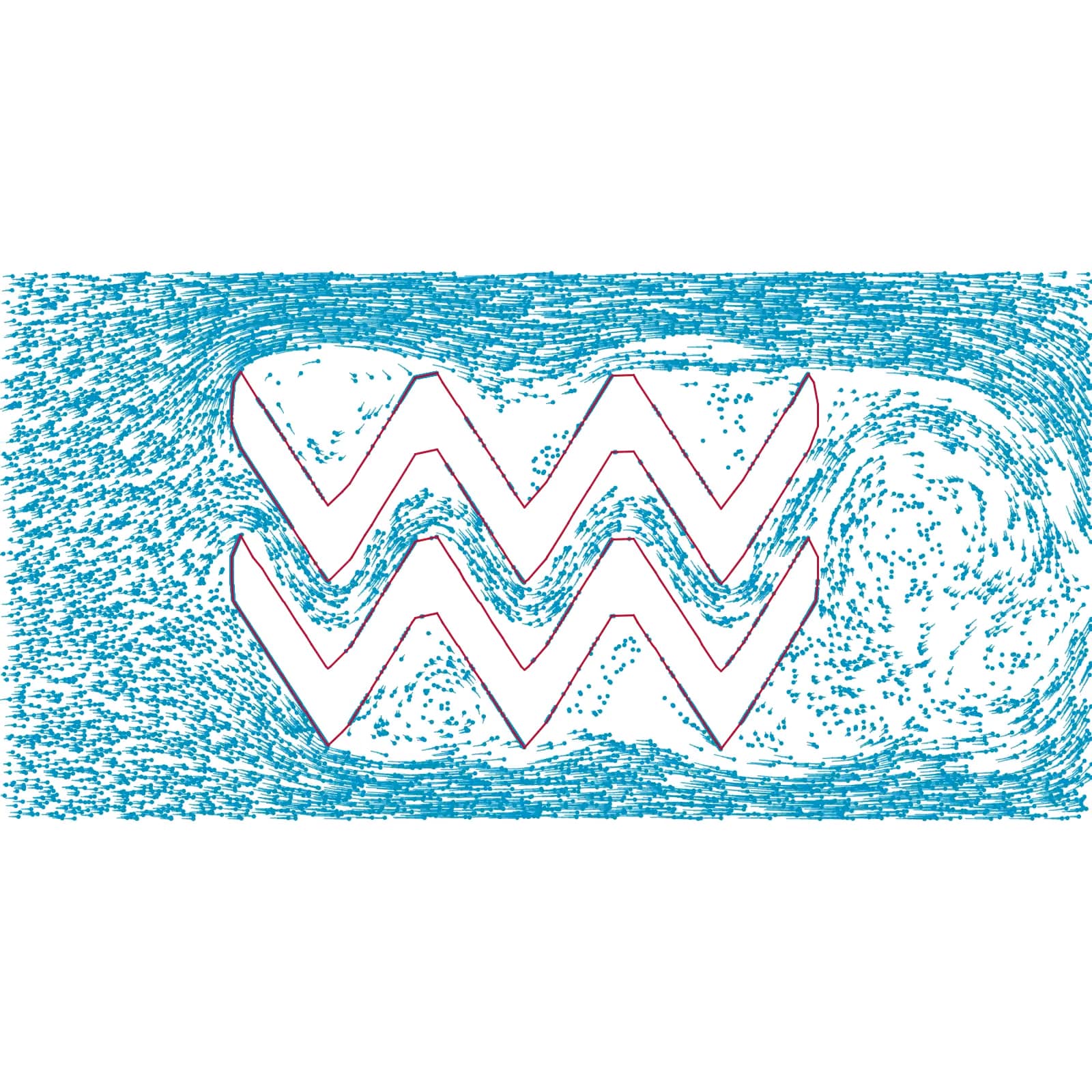}
    \end{minipage}    
    &
    \begin{minipage}{.23\textwidth}
      \includegraphics[trim=2cm 16cm 2cm 16cm, clip,width=\textwidth]{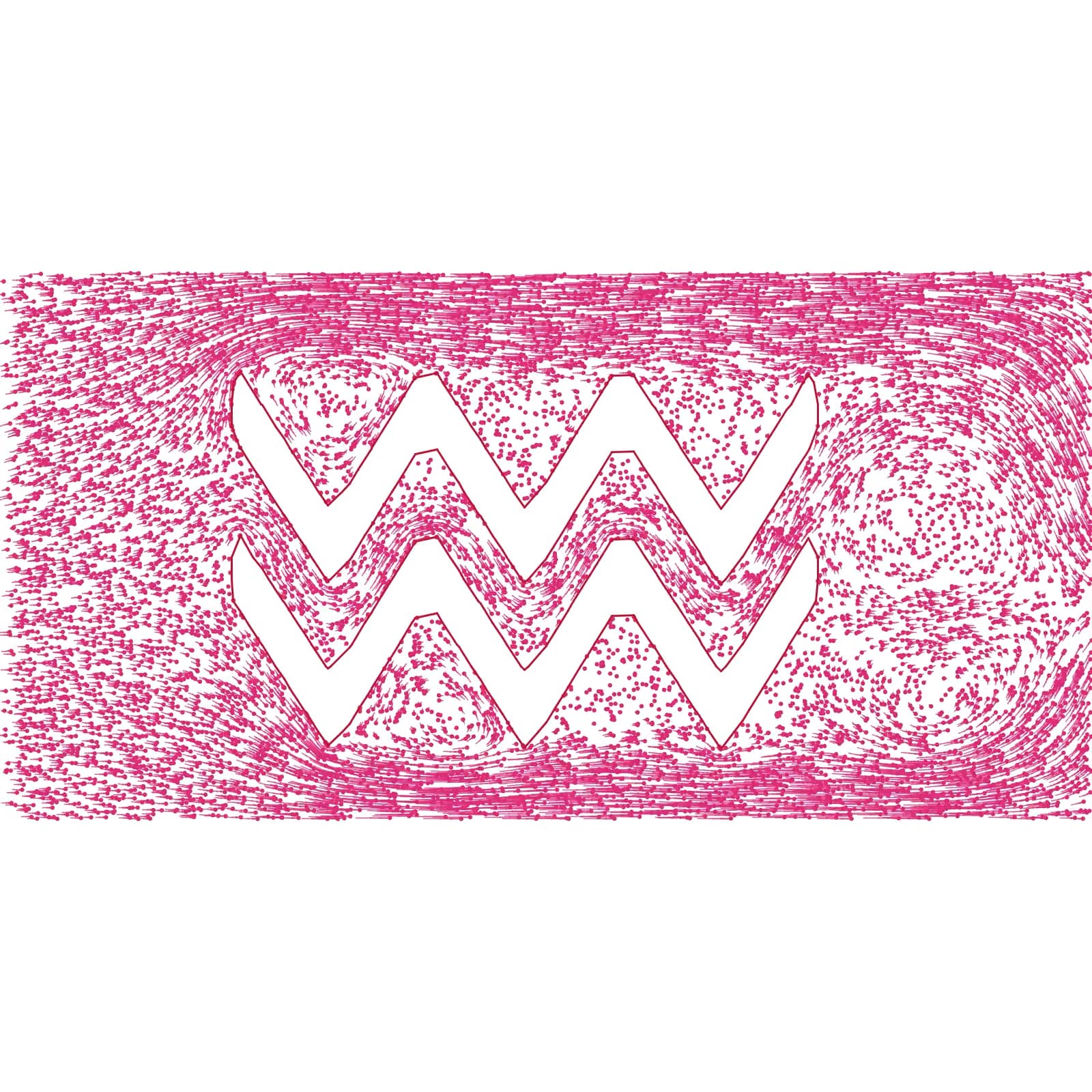}
    \end{minipage}    
    & 
    \begin{minipage}{.23\textwidth}
      \includegraphics[trim=2cm 16cm 2cm 16cm, clip,width=\textwidth]{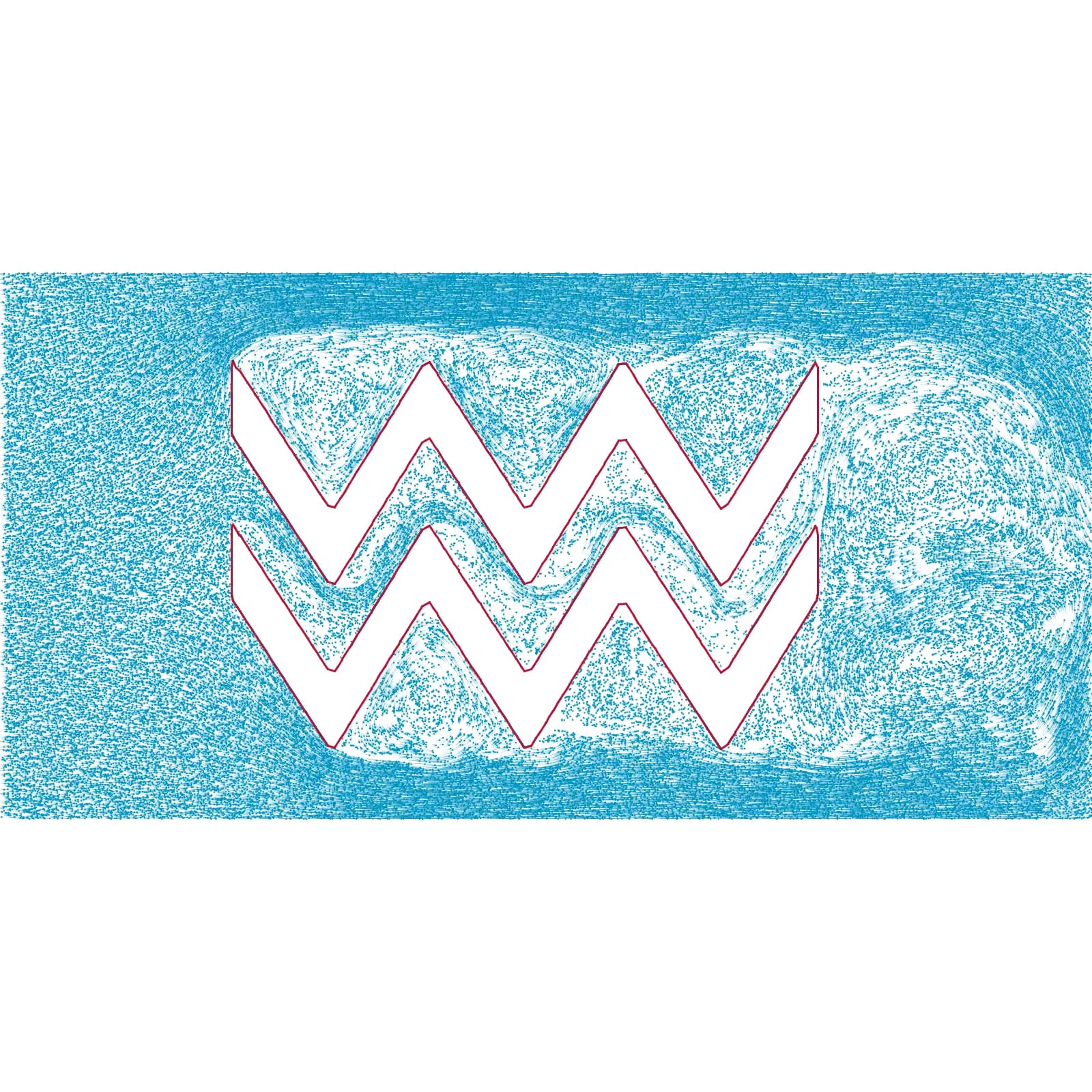}
    \end{minipage}
    & 
    \begin{minipage}{.23\textwidth}
      \includegraphics[trim=2cm 16cm 2cm 16cm, clip,width=\textwidth]{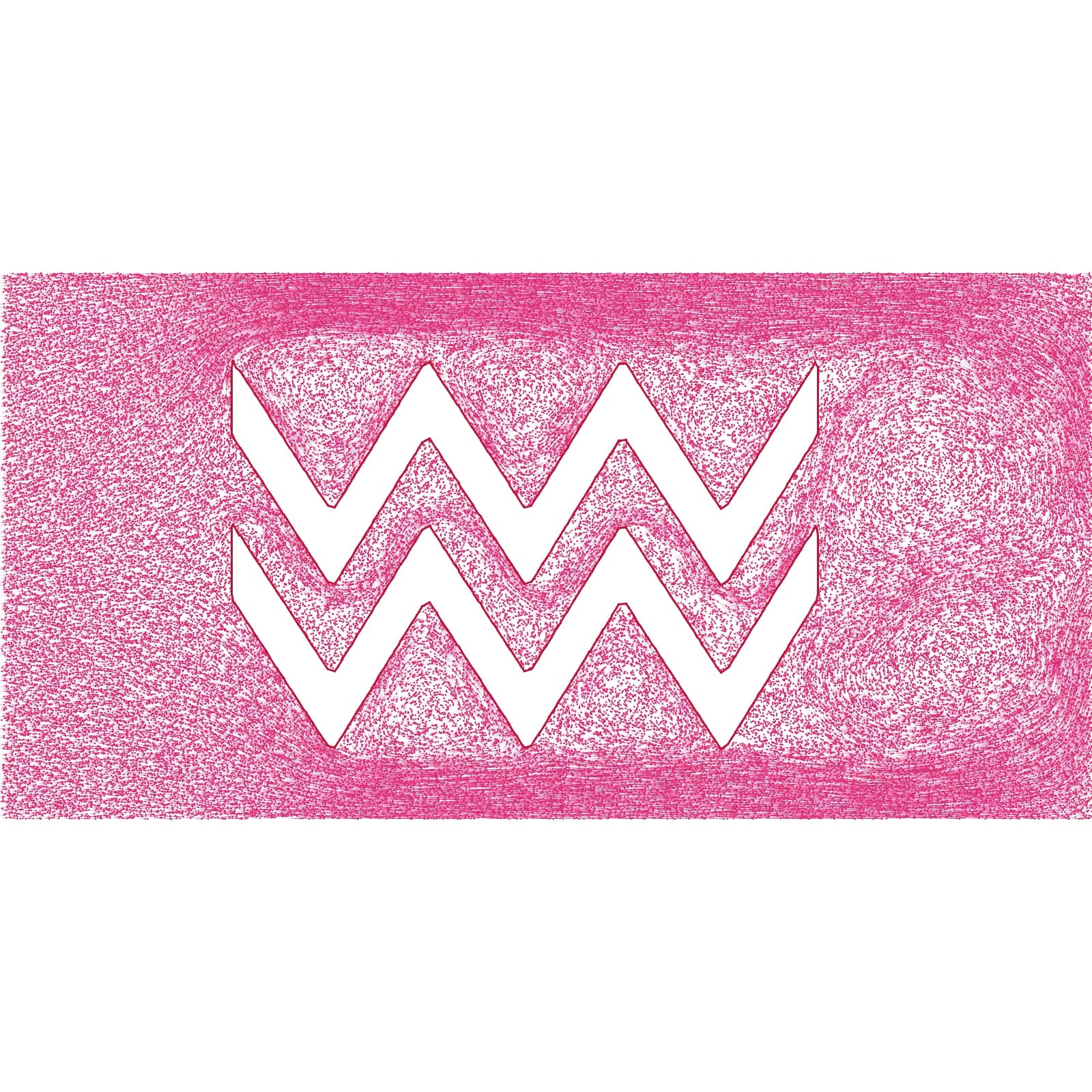}
    \end{minipage}
    \\
    \hline
    \end{tabular}
	\caption{
	\textbf{Dynamic 2D Flow in A Wind Tunnel Past A Solid:} 
	Passive tracer particles (top two rows) and FLIP particles (bottom two rows) undergo a dynamic horizontal flow past static obstacles. 
	Particles colliding into objects are halted in place.
	Direct velocity interpolation creates spurious gaps (blue), even at triple the grid resolution; Curl-Flow interpolation tightly and incompressibly follows the solid objects, significantly reducing the gaps in the flow (red).
	}
	\label{fig:solid2D}
\end{figure*}

In Figure \ref{fig:solid2D} (top half), we exercise our 2D cut-cell treatment by observing passive particles under a dynamic horizontal flow past circular and jagged solids.
Spurious gaps in the initially dense particle sampling quickly arise with direct velocity interpolation.
Curl-Flow closely obeys the boundary while being strictly incompressible, sharply reducing the occurrence of gaps.
Because interpolation errors depend on cell size, tripling the grid resolution (Figure \ref{fig:solid2D}, right pair of columns) proportionally reduces the direct method's gaps, but is costly and such errors persist for any finite resolution.

In FLIP-type simulations \cite{Zhu2005}, particles are not merely passive tracers but are instead used to transport velocity; as a result, particle density errors can be equally or more problematic in that setting. Figure \ref{fig:solid2D} (bottom half) applies FLIP using Curl-Flow for particle tracing (but \emph{not} particle-to-grid (P2G) or grid-to-particle (G2P) transfers). Aside from FLIP particle seeding at the start and left inflow (8 particles per grid cell), no artificial density interventions were applied, e.g., reseeding, particle position adjustment, etc. Since P2G fails for empty cells, such cells fall back to semi-Lagrangian advection. We observe trailing gaps with direct velocity interpolation and, furthermore, its motion is more damped compared to FLIP with Curl-Flow.

\subsection{Particle Distribution Comparisons in 3D}

\paragraph*{Static Flow In A Box} 
To explore incompressibility's effects in 3D at near- and sub-grid scales, Figure \ref{fig:particleDistribution3D} compares our full Curl-Flow method (including exterior boundary ramping) against direct velocity interpolants with (componentwise) trilinear and monotonic tricubic schemes \cite{Fedkiw2001, Fritsch1980}, as these are simple and common choices. The data is a static, extremely coarse $10\times 10\times 10$ discretely incompressible discrete vector field. 
The direct interpolants quickly exhibit particle clustering and spreading, forming visible low density regions and dense ring-like structures.
By contrast, our result remains more uniformly distributed over time, despite no particle resampling, positional perturbations, or other remedies being applied. Clearly our method succeeds at enforcing incompressibility \emph{at even the finest scales}.

\begin{figure*}
  \setlength{\tabcolsep}{0em}
  \centering
  \begin{tabular}{| c | c | c | c | c |}
    \hline
     & \thead{ Direct Velocity Interpolation
     \\ Sphere: $40 \times 20 \times 20$ 
     \\ Jagged: $40 \times 30 \times 30$  } 
     & \thead{ Curl-Flow 
     \\ Sphere: $40 \times 20 \times 20$ 
     \\ Jagged: $40 \times 30 \times 30$  } 
     & \thead{ Direct Velocity Interpolation 
     \\ Sphere: $80 \times 40 \times 40$ 
     \\ Jagged: $80 \times 60 \times 60$  }
     & \thead{ Curl-Flow 
     \\ Sphere: $80 \times 40 \times 40$ 
     \\ Jagged: $80 \times 60 \times 60$  } 
     \\ \hline
    \thead{ Passive \\ \ Particles \ \\ (Sphere) }
    &
    \begin{minipage}{.23\textwidth}
      \includegraphics[trim=0cm 0.5cm 0cm 0.5cm, clip, width=\textwidth]{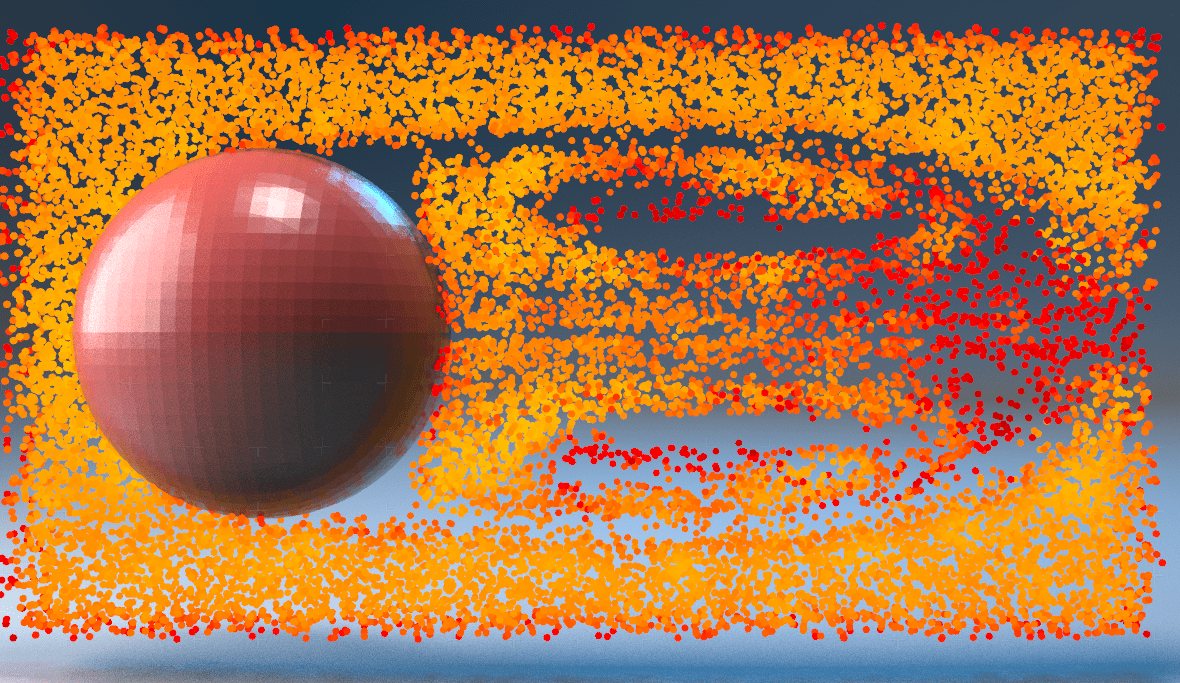}
    \end{minipage}
    &
    \begin{minipage}{.23\textwidth}
      \includegraphics[width=\textwidth]{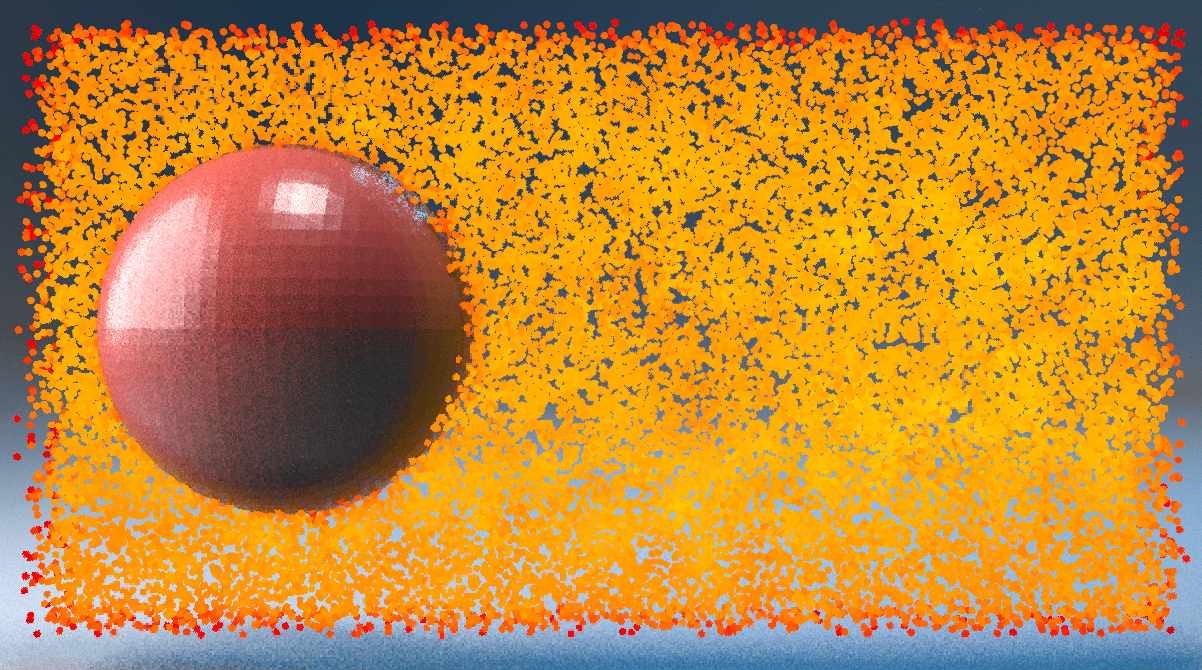}
    \end{minipage}
    & 
    \begin{minipage}{.23\textwidth}
      \includegraphics[trim=0cm 0.5cm 0cm 0.5cm, clip, width=\textwidth]{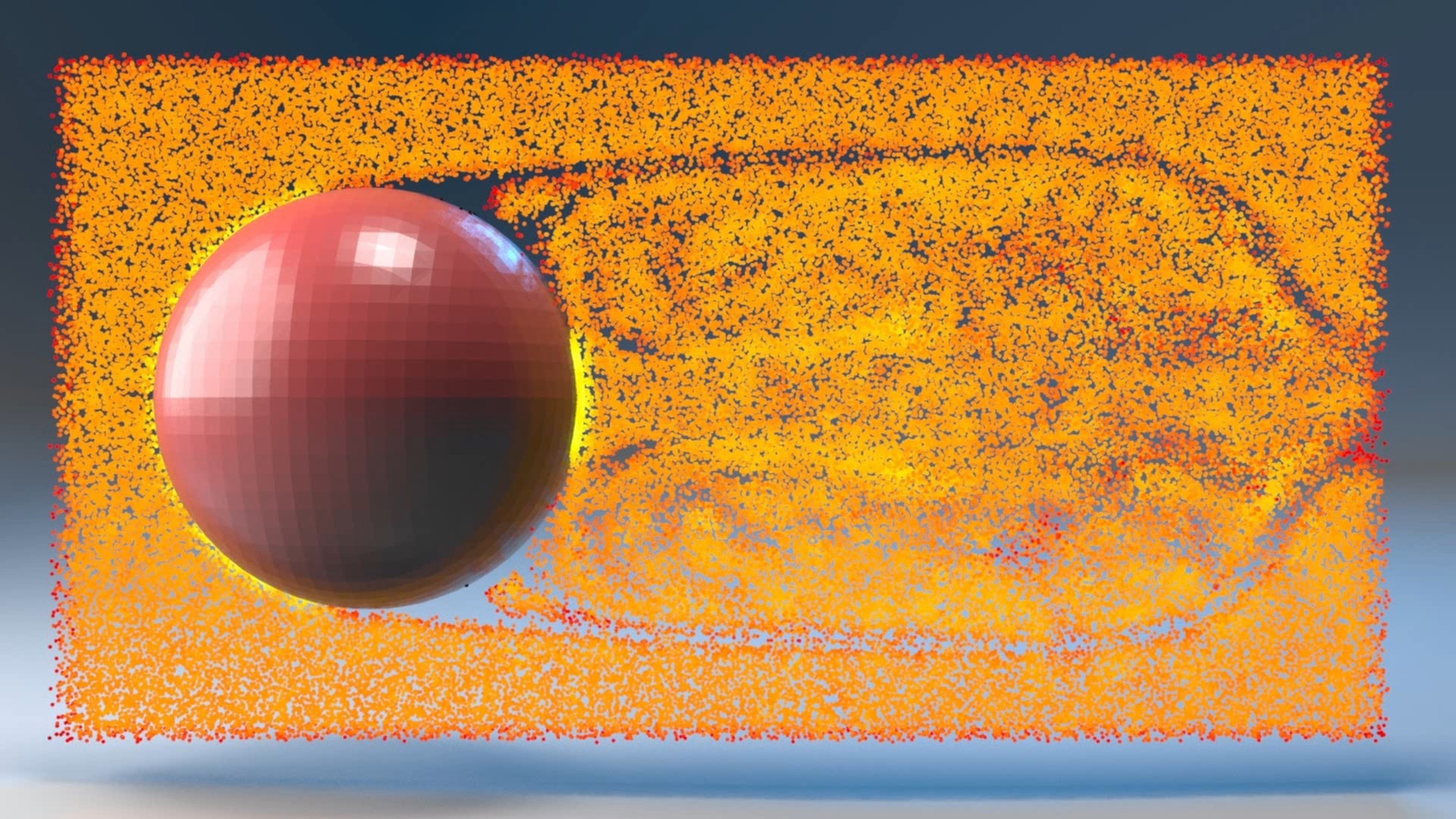}
    \end{minipage}
    & 
    \begin{minipage}{.23\textwidth}
      \includegraphics[trim=0cm 0.5cm 0cm 0.5cm, clip, width=\textwidth]{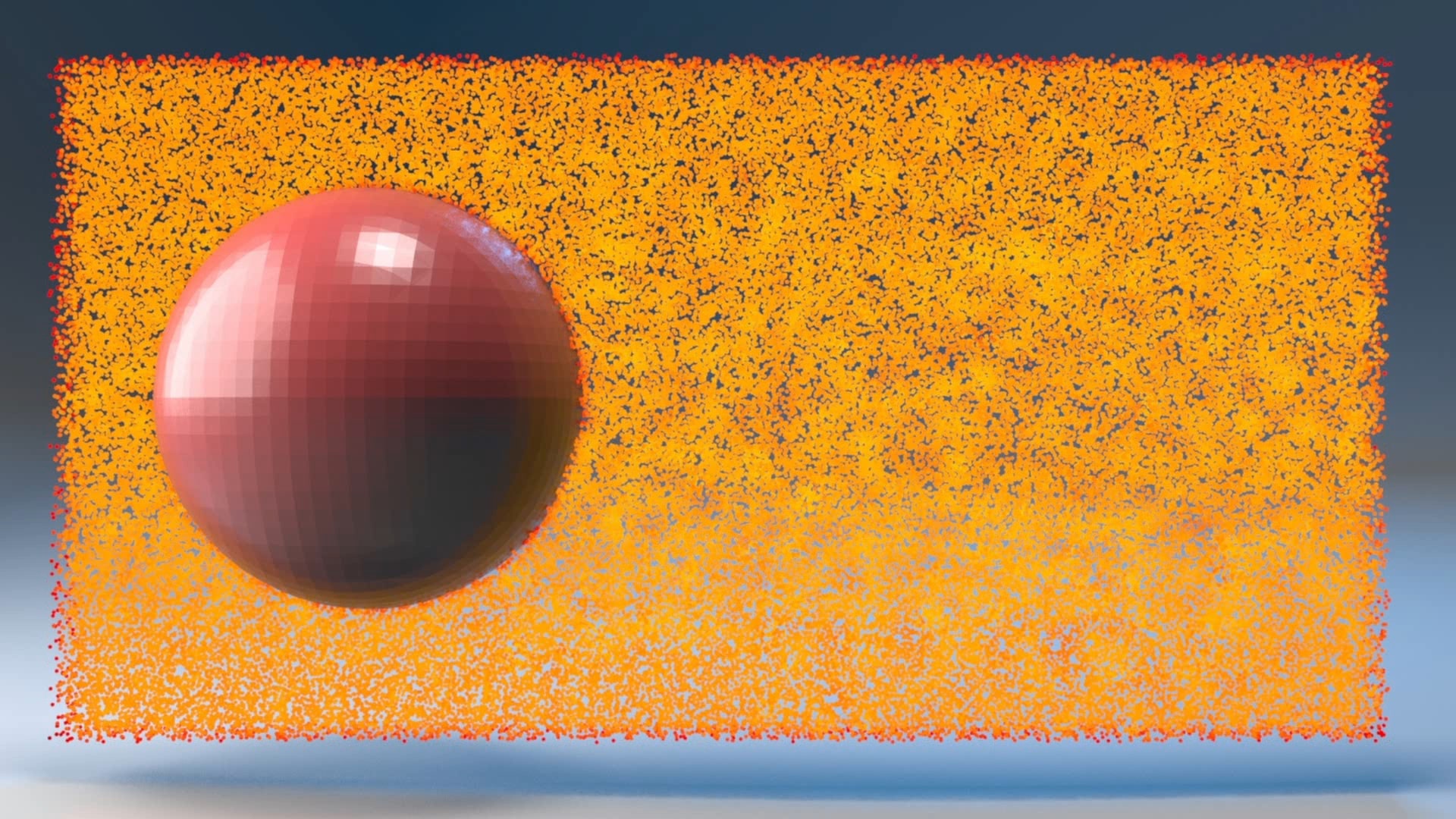}
    \end{minipage}
    \\
    \hline
    \thead{ Passive \\ \ Particles \ \\ (Jagged) }
    &
    \begin{minipage}{.23\textwidth}
      \includegraphics[trim=7cm 1cm 7cm 1cm, clip, width=\textwidth]{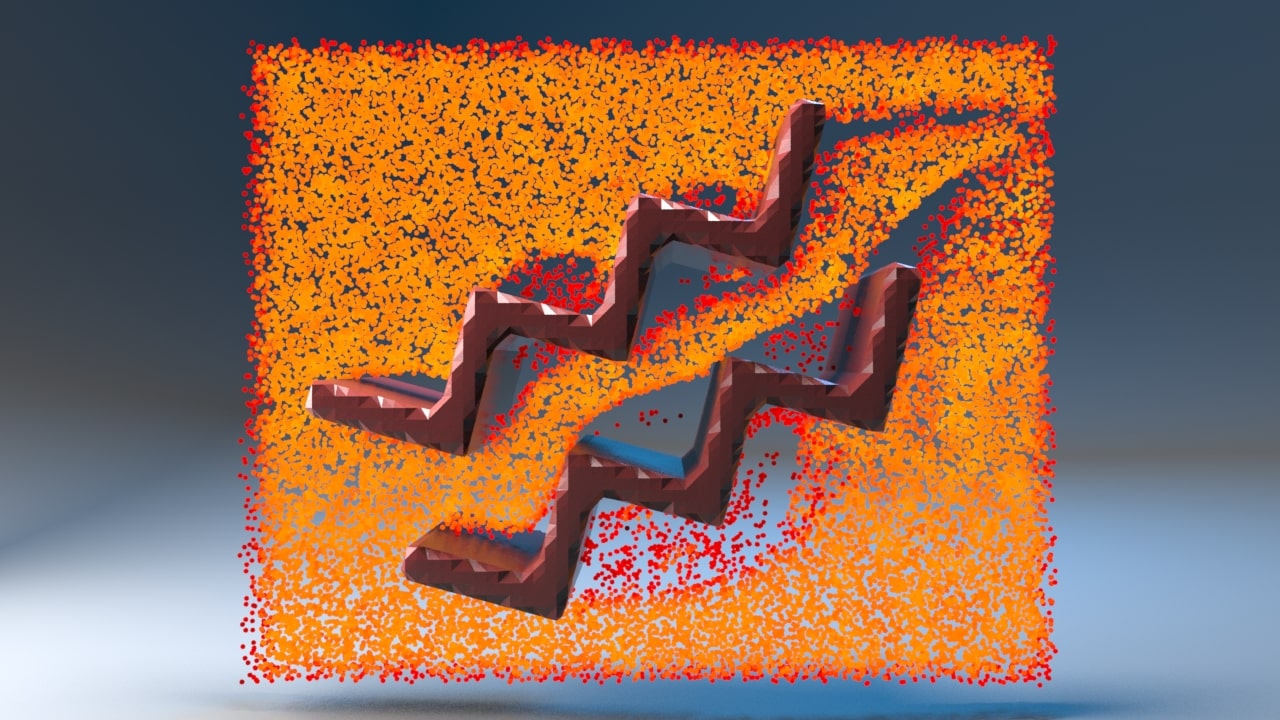}
    \end{minipage}    
    &
    \begin{minipage}{.23\textwidth}
      \includegraphics[trim=7cm 1cm 7cm 1cm, clip, width=\textwidth]{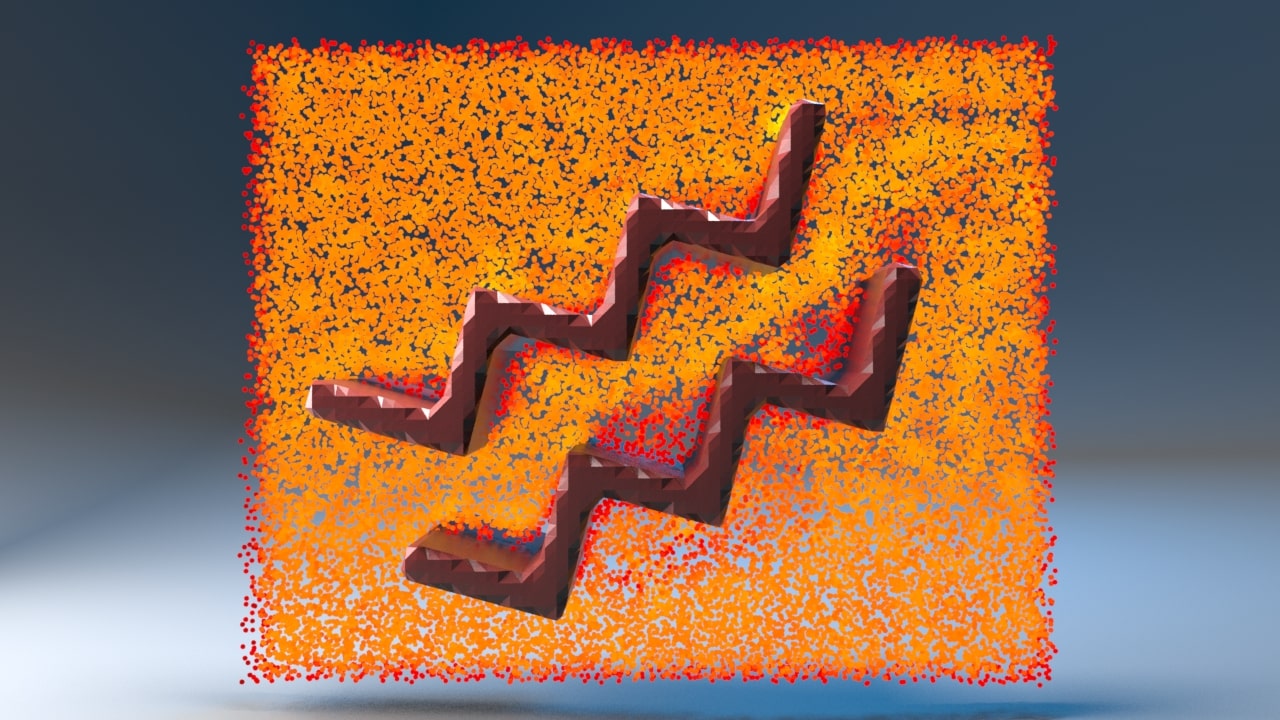}
    \end{minipage}
    & 
    \begin{minipage}{.23\textwidth}
      \includegraphics[trim=7cm 1cm 7cm 1cm, clip, width=\textwidth]{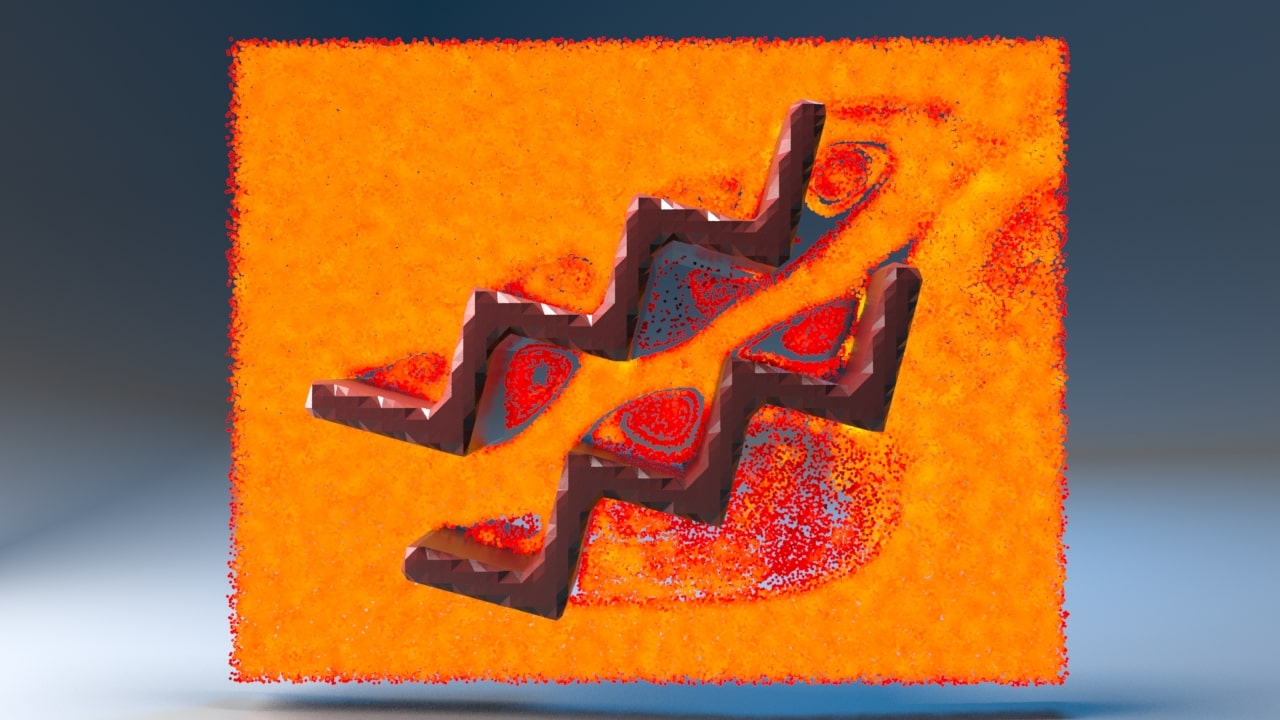}
    \end{minipage}    
    & 
    \begin{minipage}{.23\textwidth}
      \includegraphics[trim=7cm 1cm 7cm 1cm, clip, width=\textwidth]{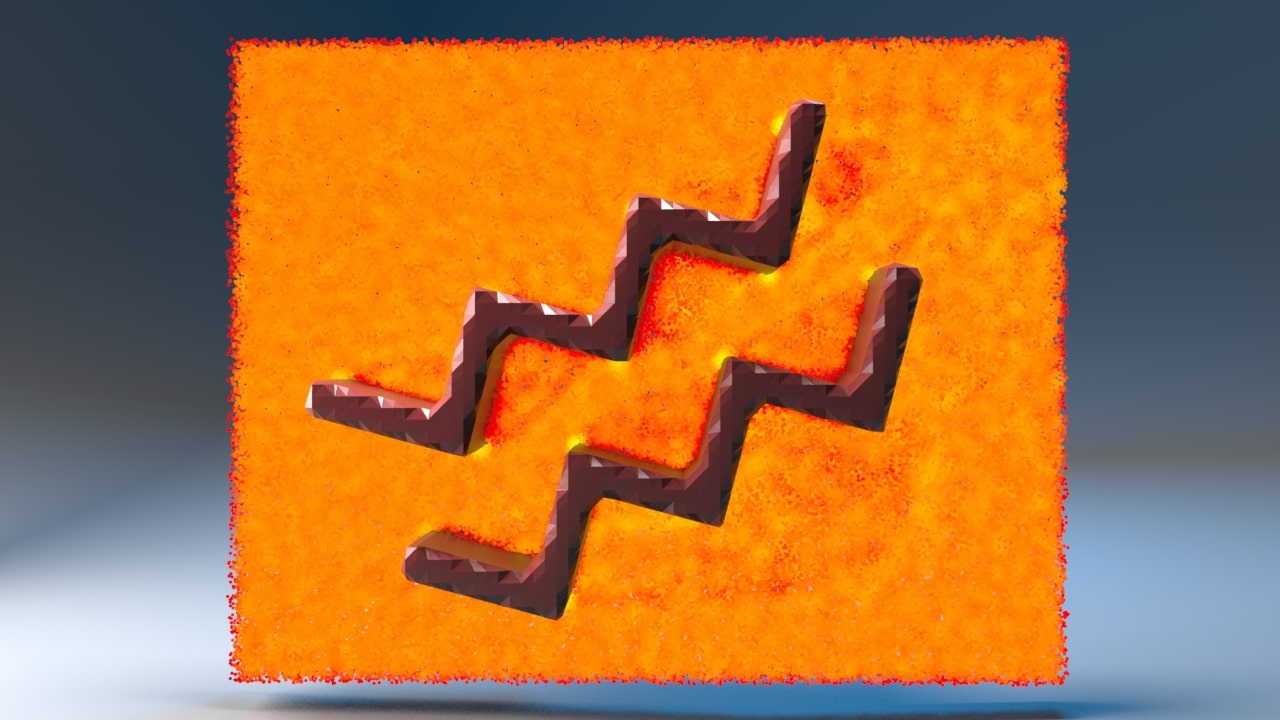}
    \end{minipage}    
    \\
    \hline
    \end{tabular}
	\caption{
	\textbf{Dynamic 3D Flow in A Wind Tunnel Past A Solid (middle frame):} 
	A slice-plane view of particles undergoing a time-dependent horizontal flow past a solid object in 3D.
	The color of the particles represents the particle densities: the density increases from red to yellow.
	Particles colliding into objects are halted in place for illustration. Regardless of the grid resolution, direct velocity interpolation creates spurious gaps, 
	while our Curl-Flow interpolation tightly follows the solid object, significantly reducing the gaps in the flow and maintaining better particle distribution.
	}
	\label{fig:solid3D}
\end{figure*}

\paragraph*{Dynamic Flows Past Smooth and Jagged Obstacles} 
At medium scales, we next compare Curl-Flow to direct velocity interpolation under separate 3D dynamic horizontal flows past a sphere and a jagged obstacle (Figure \ref{fig:solid3D}). Here we consider grids ranging from $40\times20\times20$ to $80\times60\times60$.
Only the particles within a narrow slice plane are shown to better visualize their density.
The particles are purely passive, but for illustration we seed them throughout the domain at a density representative of typical particle-in-cell schemes, i.e., 8 per cell \cite{Zhu2005}. As in the corresponding 2D examples, Curl-Flow retains more uniform particle sampling.

\begin{table*}
\centering  
\caption{Average computational time per timestep for didactic examples.}
\vspace{-3mm}
\begin{tabular}{c c c c c c c c}  
\hline\hline \\ [-2.5ex]
\thead{Examples}
& \thead{Interpolation \\ Method} 
& \thead{Pressure \\ Projection (s)} 
& \thead{Grid $\boldsymbol{\psi}$ \\ Construction (s)}
& \thead{Solid $\boldsymbol{\psi}$ \\ Construction (s)}
& \thead{Particle \\ Advection (s)}
& \thead{$\#$ Particles}
& \thead{$\#$ Solid Vertices \\ (Triangles)}
\\ [0.5ex]
\hline 
\multirow{2}{*}[-1.5mm]{\thead{ Flow Past A Sphere \\ (40 $\times$ 20 $\times$ 20) }} 
& \thead{ Direct Velocity \\ Interpolation } 
& \multirow{2}{*}[-1.5mm]{2.214$\times10^{-2}$}
& -- & -- & 3.962$\times10^{-1}$ & 117,482 & 
\multirowcell{2}[-1.5mm]{726 \\ (1448)} \\ 
& \thead{ Curl-Flow }
&& 9.040$\times10^{-3}$ & 5.887$\times10^{-3}$ & 4.443$\times10^{-1}$ & 117,483 & \\ [1.5ex]
\hline 
\multirow{2}{*}[-1.5mm]{\thead{ Flow Past A Sphere \\ (80 $\times$ 40 $\times$ 40) }} 
& \thead{ Direct Velocity \\ Interpolation } 
& \multirowcell{2}[-1.5mm]{4.268$\times10^{-1}$} & -- & -- & 3.010 & 949,272 &
\multirowcell{2}[-1.5mm]{2934 \\ (5864)} \\ 
& \thead{ Curl-Flow }
&& 1.812$\times10^{-1}$ & 3.888$\times10^{-2}$ & 3.708 & 949,277 & \\ [1.5ex]
\hline 

\multirow{2}{*}[-1.5mm]{\thead{ Flow Past A Jagged Obstacle \\ (40 $\times$ 30 $\times$ 30) }} 
& \thead{ Direct Velocity \\ Interpolation } 
& \multirowcell{2}[-1.5mm]{9.942$\times10^{-2}$} & -- & -- & 8.158$\times10^{-1}$ & 271,144 & 
\multirowcell{2}[-1.5mm]{3140 \\ (6272)} \\ 
& \thead{ Curl-Flow }
&& 4.262$\times10^{-2}$ & 2.907$\times10^{-2}$ & 1.050 & 271,148 & \\ [1.5ex]
\hline 
\multirow{2}{*}[-1.5mm]{\thead{ Flow Past A Jagged Obstacle \\ (80 $\times$ 60 $\times$ 60) }} 
& \thead{ Direct Velocity \\ Interpolation } 
& \multirowcell{2}[-1.5mm]{2.070} & -- & -- & 6.885 & 2,035,118 &
\multirowcell{2}[-1.5mm]{12454 \\ (24900)} \\ 
& \thead{ Curl-Flow }
&& 9.001$\times10^{-1}$ & 2.070$\times10^{-1}$ & 7.723 & 2,035,124 & \\ [1.5ex]
\hline 
\vspace{1mm}
\end{tabular}
\label{tab:didacticTiming}
\end{table*}
\

The average computational costs per timestep for our method applied to the dynamic flows above are given in Table \ref{tab:didacticTiming}. For reference, we also include timings for the underlying fluid simulator's pressure projection, as it is often a bottleneck in fluid solvers. For these tests, we solve \emph{both} pressure projection and gauge correction with the basic conjugate gradient method in the Eigen library \cite{Eigenweb}, with relative tolerance $1\times10^{-8}$. We used Eigen's LDLT direct solver for the solid surface problem \eqref{eq:solidKKT}, since we found it can be ill-conditioned. The observed total cost of reconstructing $\boldsymbol{\psi}$ (for the grid and solid surfaces) is noticeably less than for pressure projection (around $50\%-75\%$ as much). The cost of advecting particles with our interpolant is slightly more than for trilinear interpolation (around $10\%-30\%$ more).
\begin{figure*}[t]
  \setlength{\tabcolsep}{0em}
  \centering
  \begin{tabular}{| c | c | c | c |}
    \hline
     & \thead{ 3s } & \thead{ 6s } & \thead{ 9s } \\ \hline
    \thead{ Direct \\ Velocity \\ \ Interpolation \ \\ ($150^3)$ }
    &
    \begin{minipage}{.28\textwidth}
      \includegraphics[trim=350 50 300 0, clip,width=\textwidth]{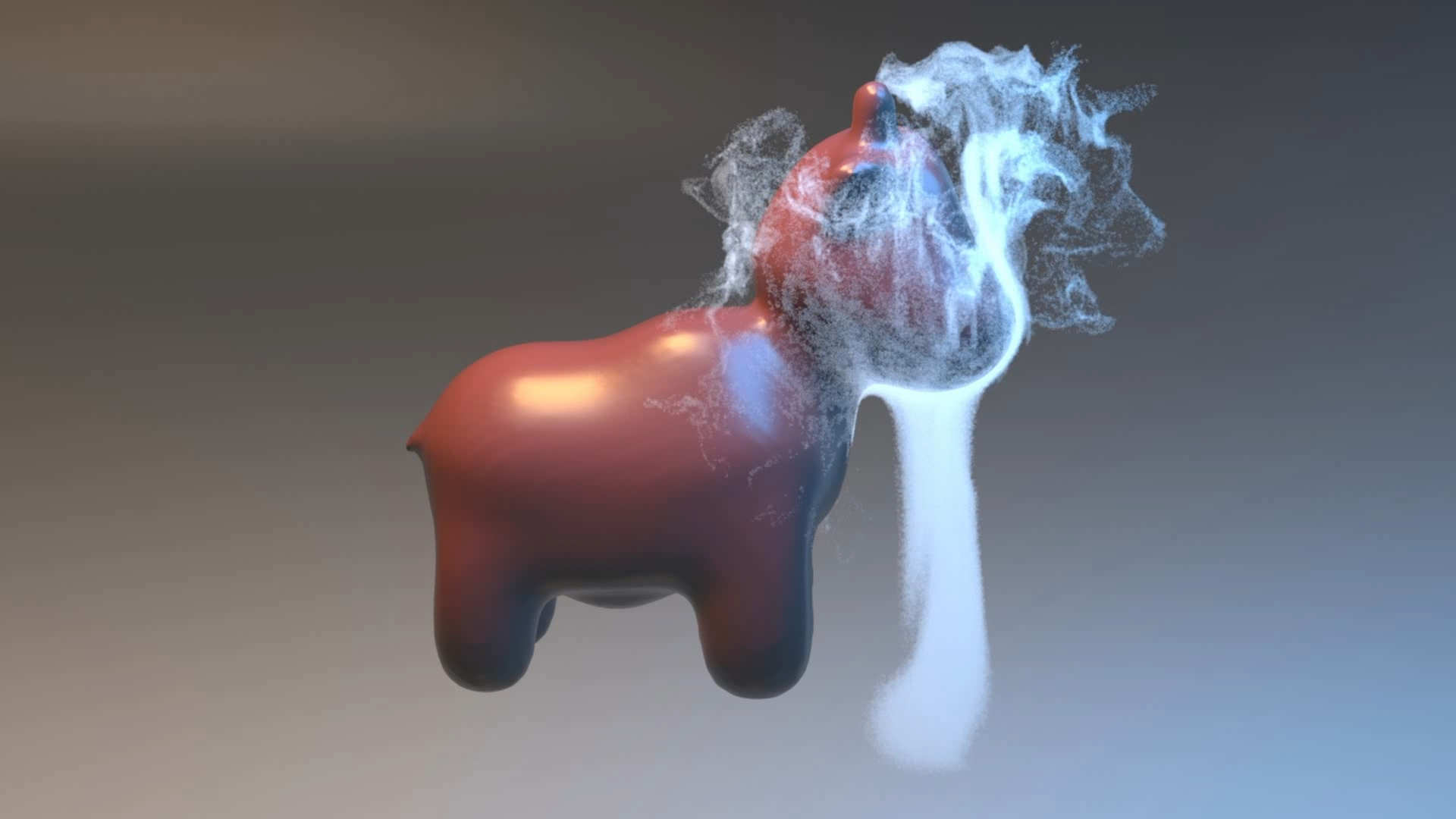}
    \end{minipage}
    &
    \begin{minipage}{.28\textwidth}
      \includegraphics[trim=350 50 300 0, clip,width=\textwidth]{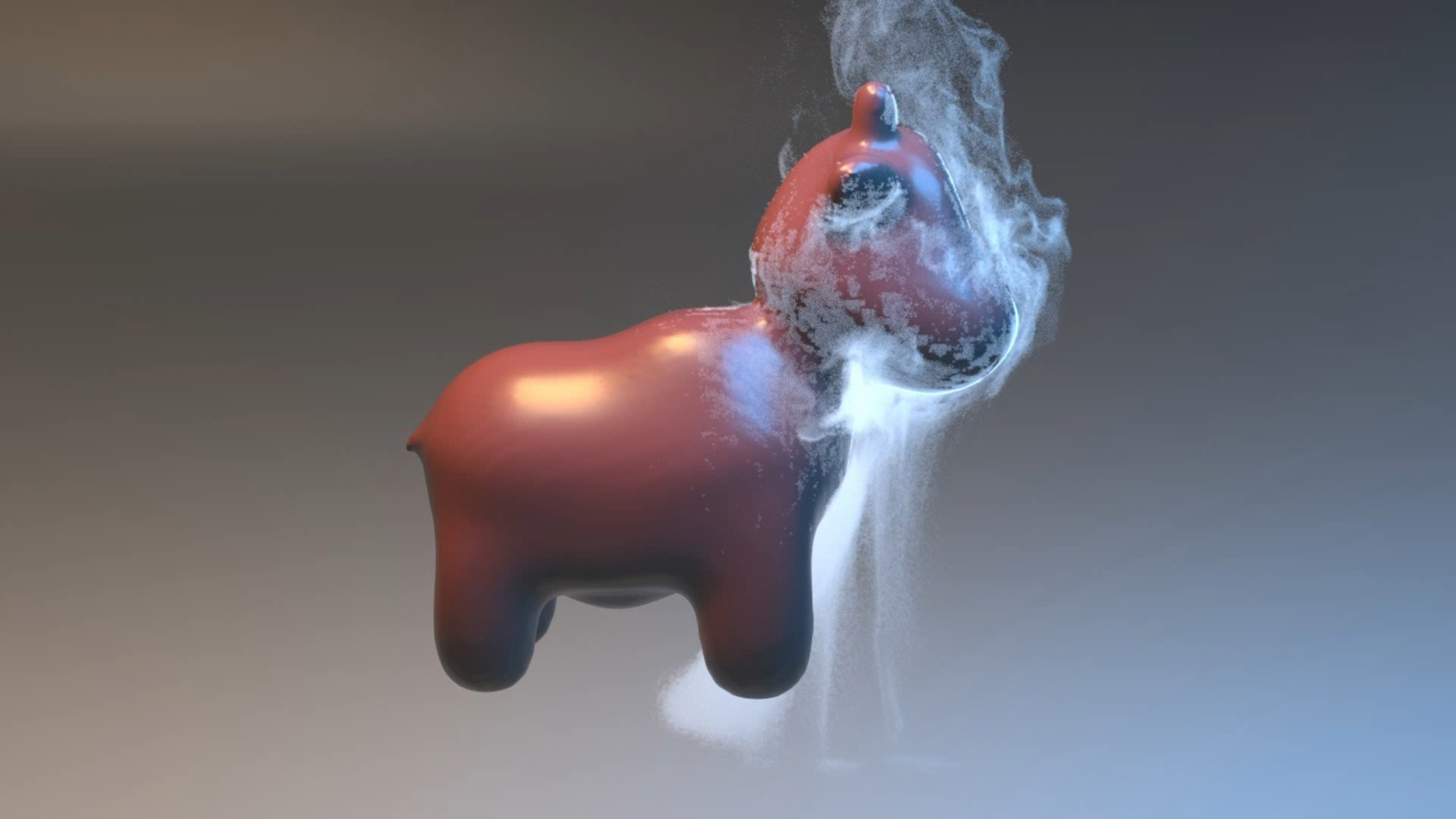}
    \end{minipage}
    &
    \begin{minipage}{.28\textwidth}
      \includegraphics[trim=350 50 300 0, clip,width=\textwidth]{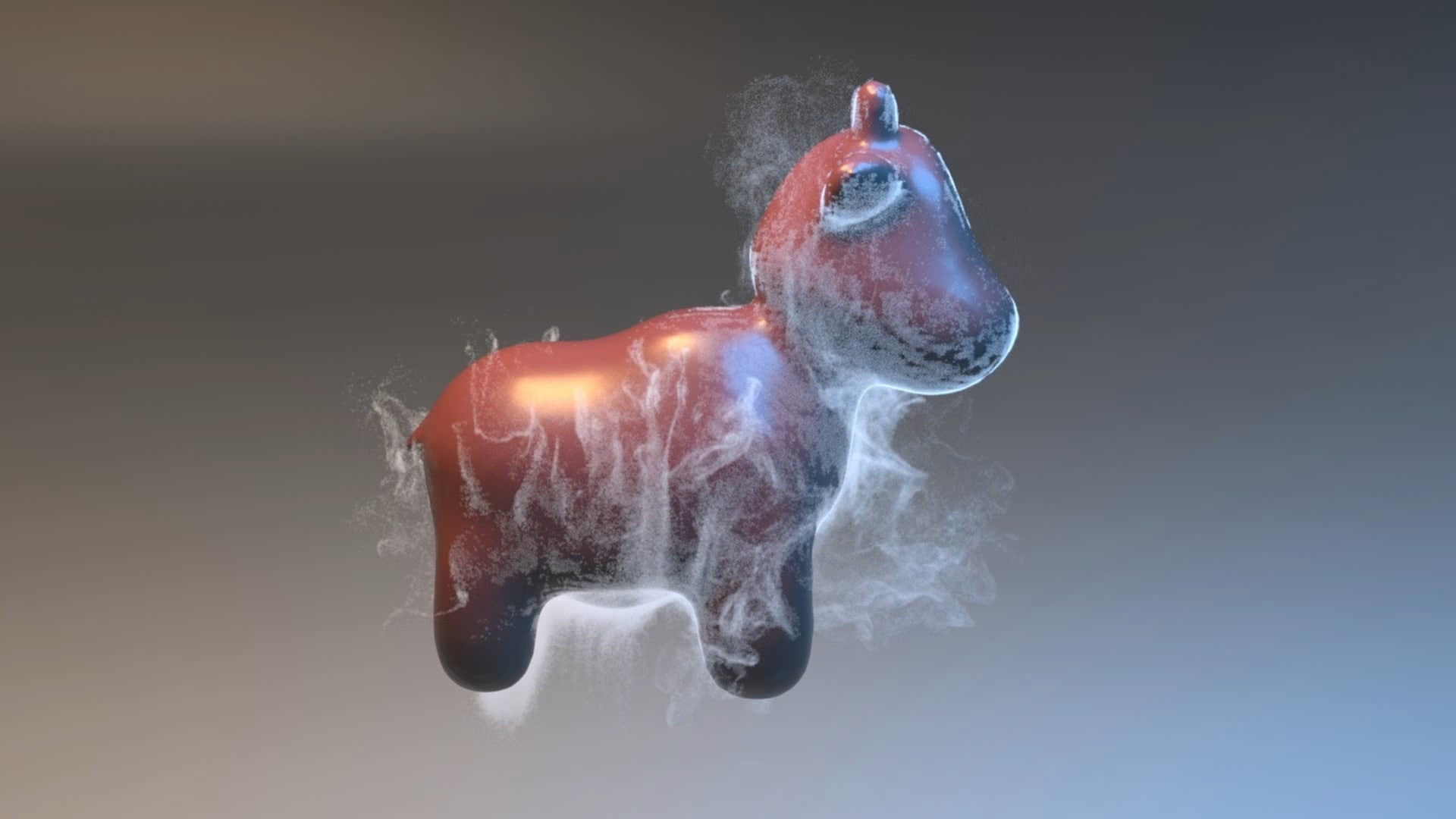}
    \end{minipage}
    \\
    \hline
    \thead{ Curl-Flow \\ ($150^3$) }
    &
    \begin{minipage}{.28\textwidth}
      \includegraphics[trim=300 50 300 0, clip,width=\textwidth]{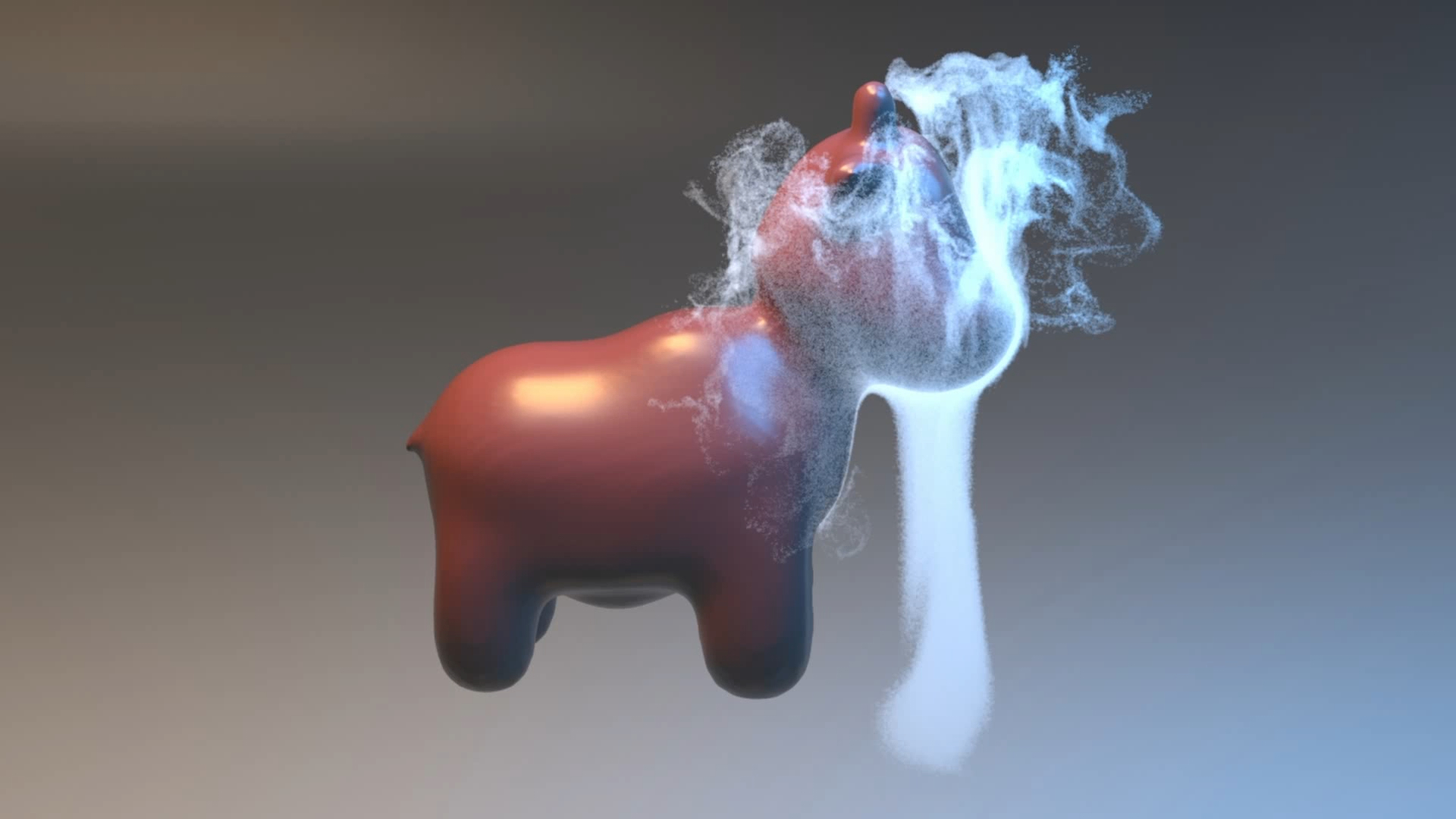}
    \end{minipage}
    &
    \begin{minipage}{.28\textwidth}
      \includegraphics[trim=300 50 300 0, clip,width=\textwidth]{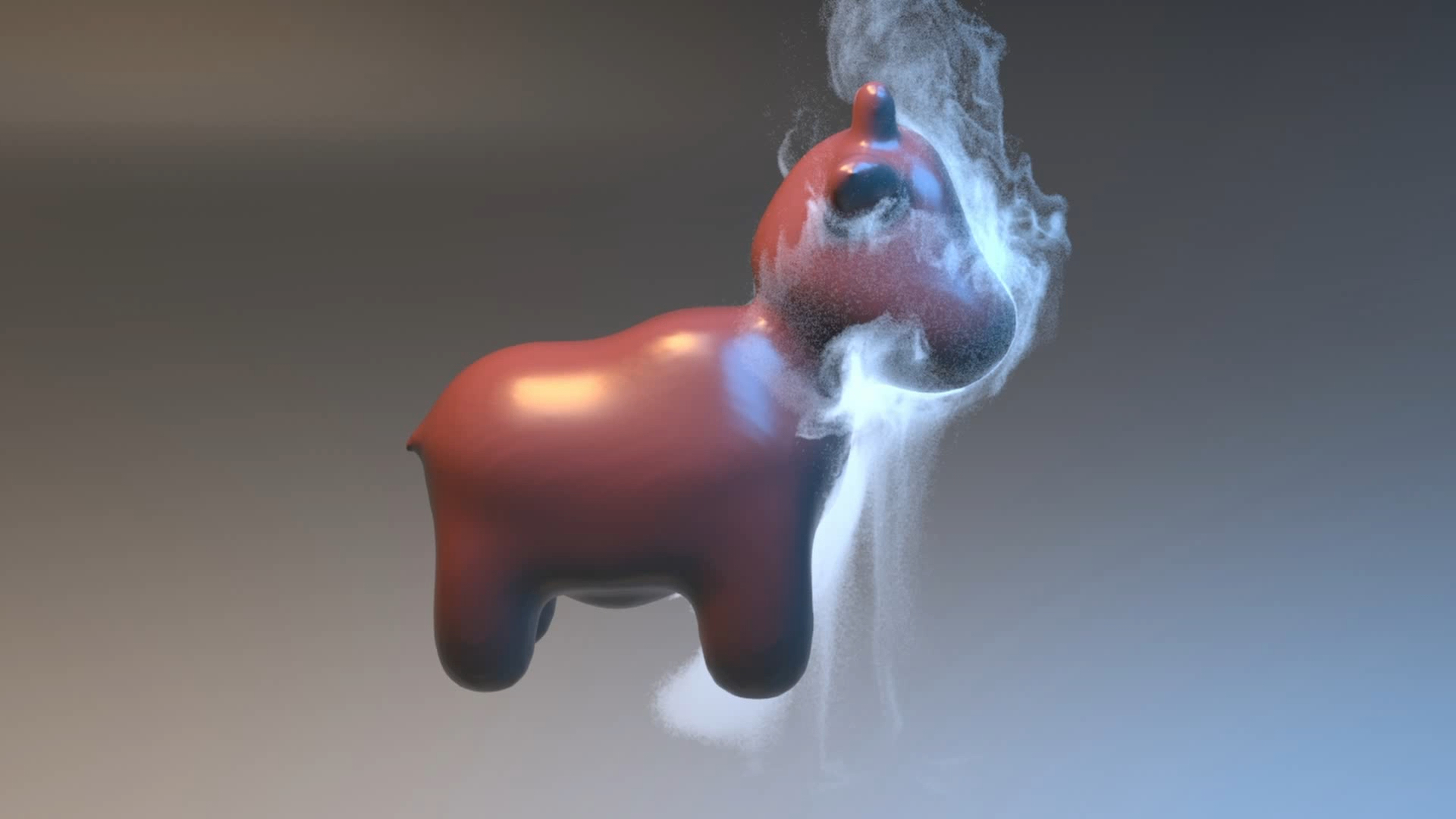}
    \end{minipage}
    &
    \begin{minipage}{.28\textwidth}
      \includegraphics[trim=300 50 300 0, clip,width=\textwidth]{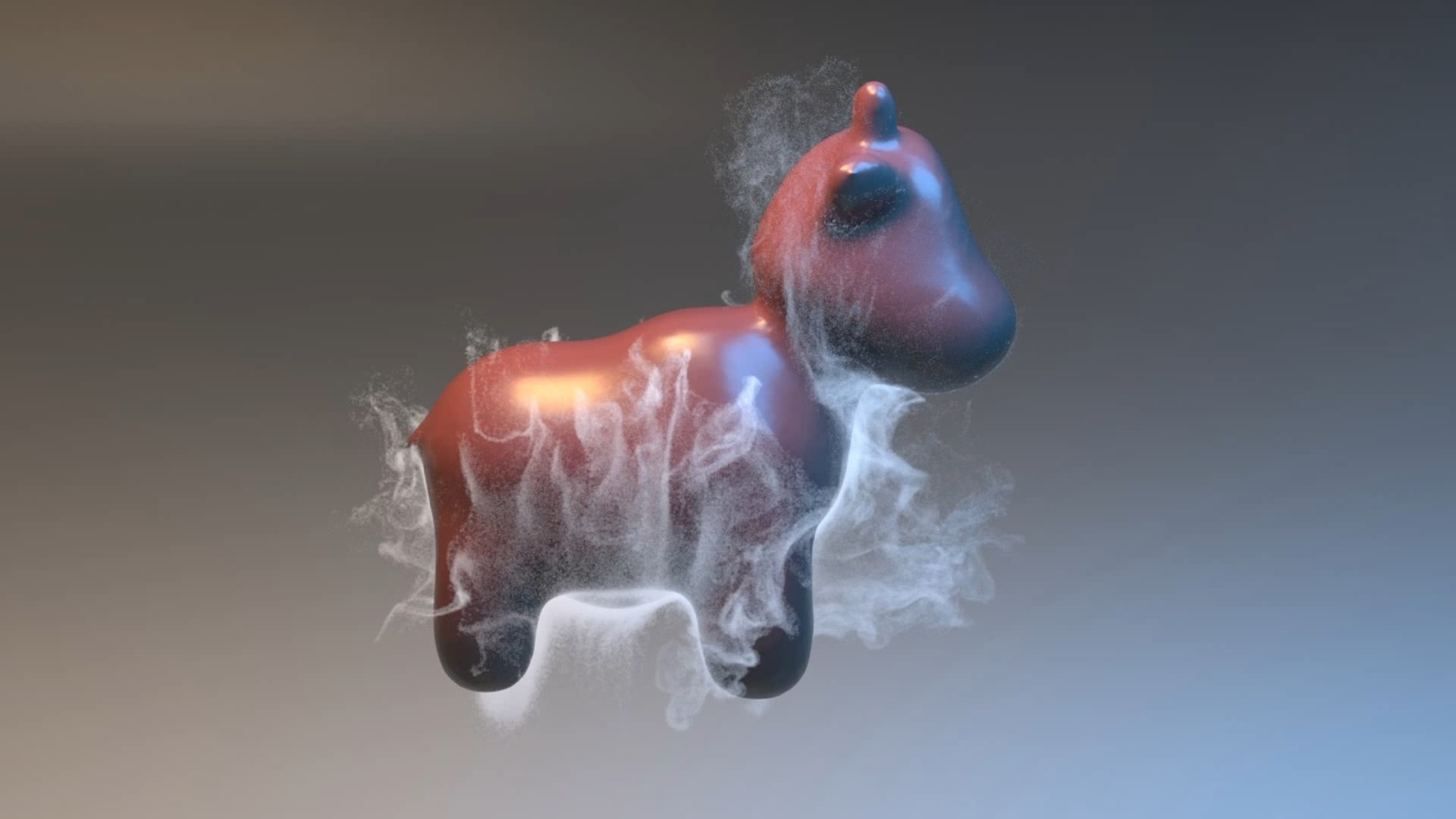}
    \end{minipage}
    \\ \hline
    \end{tabular}
	\caption{ \textbf{A Smoke Plume Simulation with A Solid Shaped like A Calf}.
	  In the direct velocity interpolation case, a large number of particles erroneously collide with the solid and are then frozen, especially around the ears and underside of the head; with Curl-Flow, more particles flow smoothly over and around the object's surface and since more smoke remains active it is denser and brighter (middle column, above the face, and see video).
	}
	\label{fig:calf}
\end{figure*}

\begin{figure*}[t]
  \setlength{\tabcolsep}{0em}
  \centering
  \begin{tabular}{| c | c  c  c |}
    \hline
    & \thead{4s} & \thead{8s} & \thead{10s} \\ \hline
    \thead{ Direct \\ Velocity \\ \ Interpolation \ \\ ($150^3$) }
    &
    \begin{minipage}{.28\textwidth}
      \includegraphics[trim=0 0 0 0, clip,width=\textwidth]{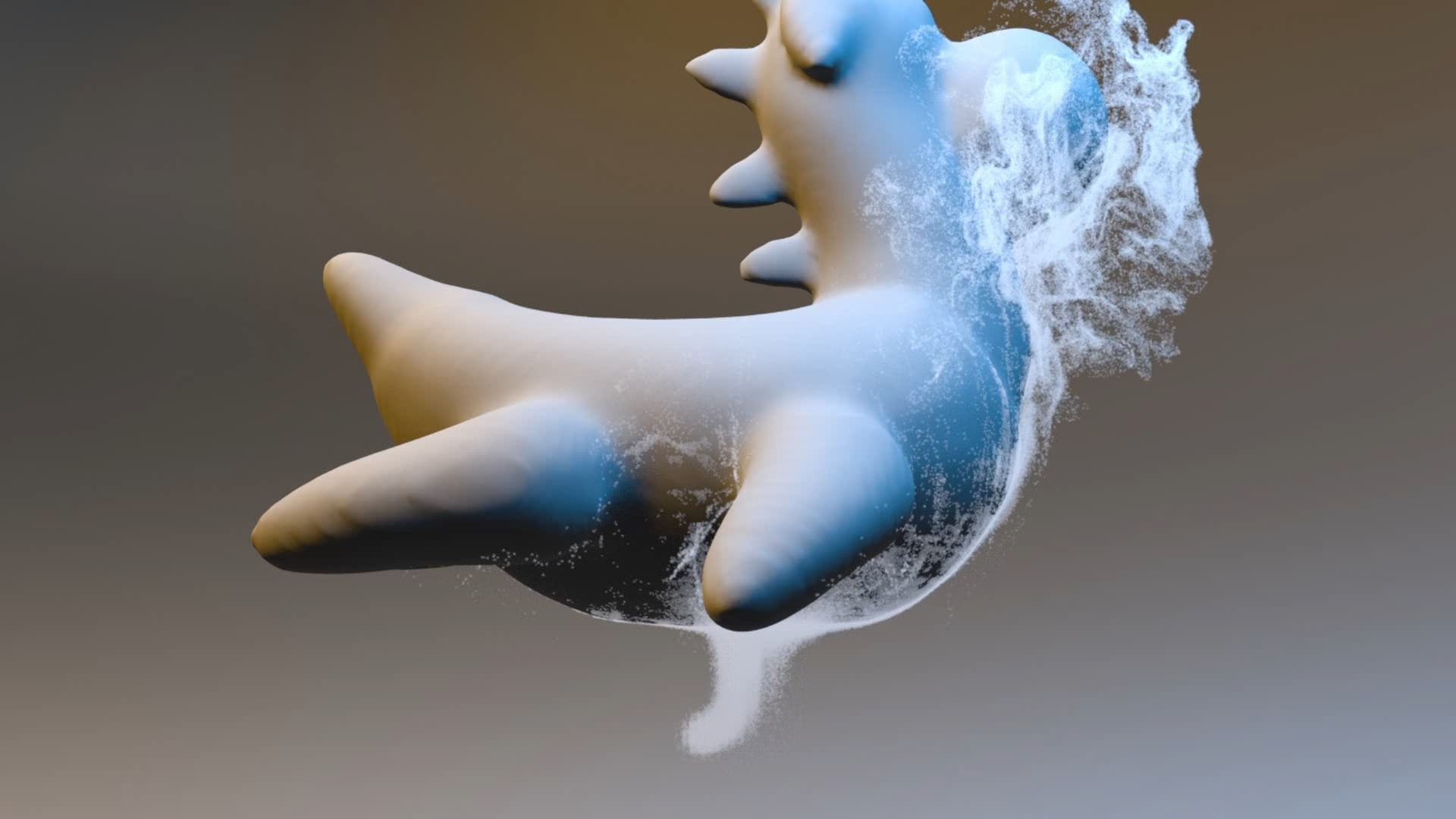}
    \end{minipage}
    &
    \begin{minipage}{.28\textwidth}
      \includegraphics[trim=0 0 0 0, clip,width=\textwidth]{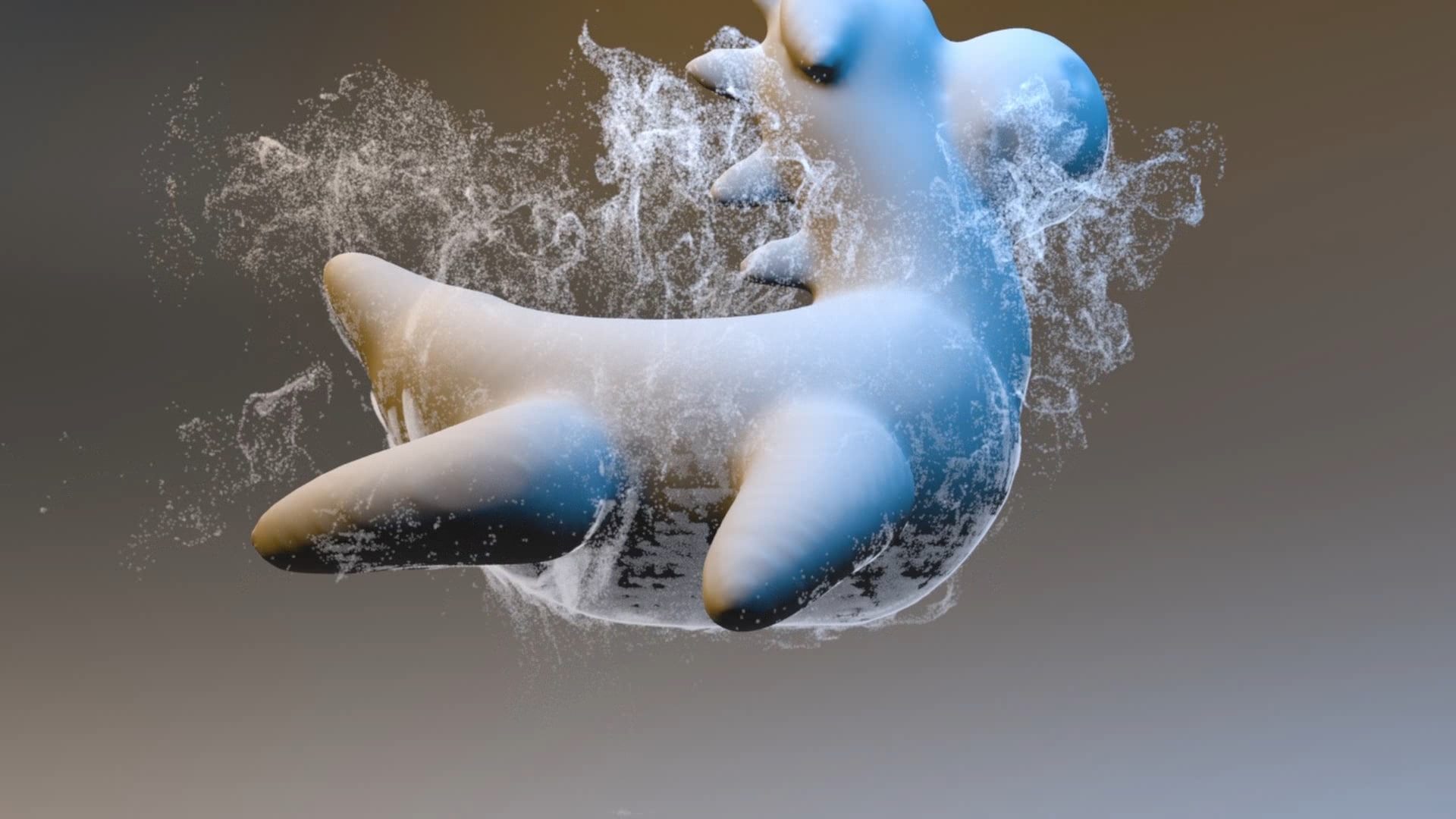}
    \end{minipage}
    &
    \begin{minipage}{.28\textwidth}
      \includegraphics[trim=0 0 0 0, clip,width=\textwidth]{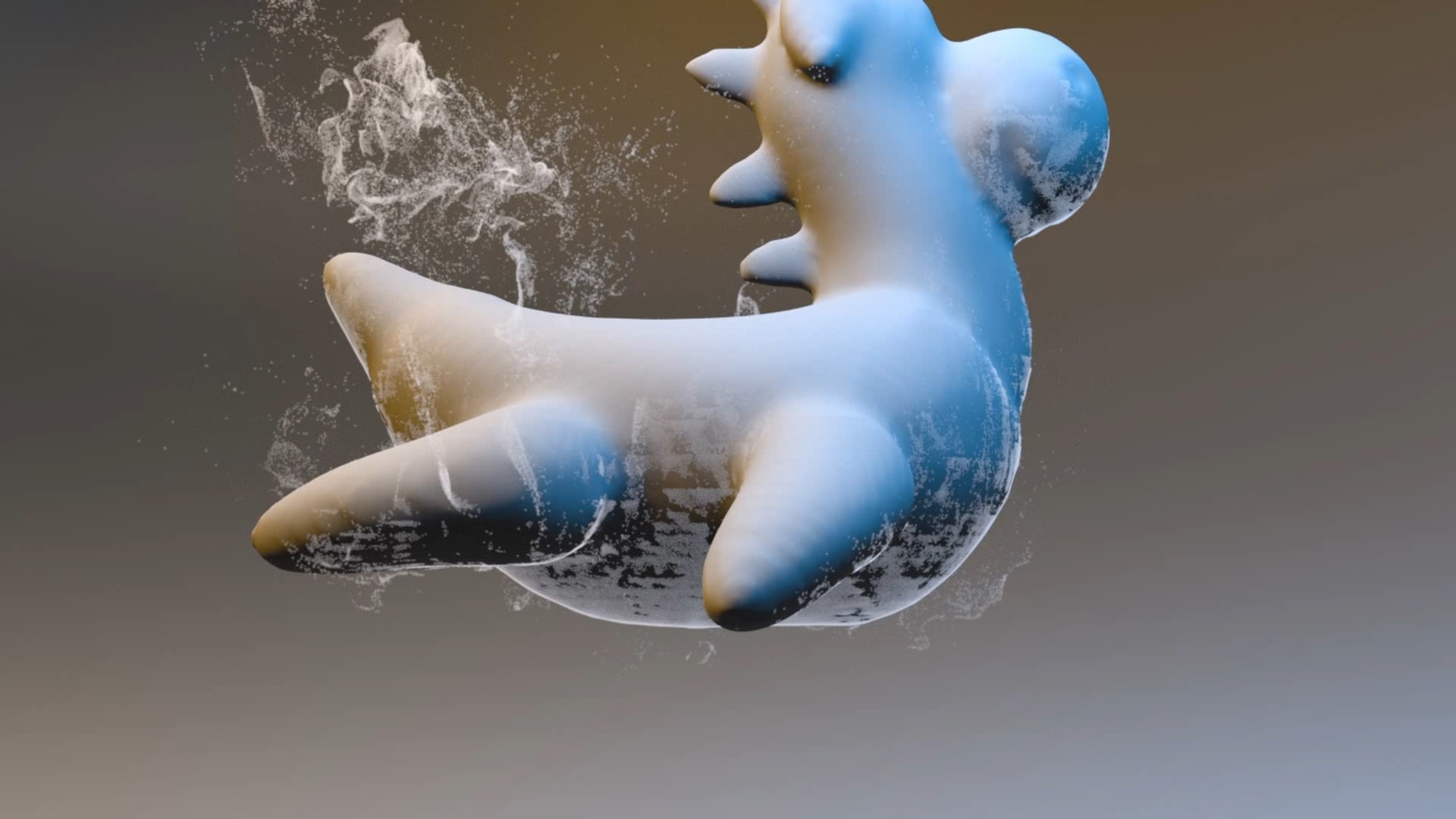}
    \end{minipage}
    \\
    \hline
    \thead{ Curl-Flow \\ ($150^3$) }
    &
    \begin{minipage}{.28\textwidth}
      \includegraphics[trim=0 0 0 0, clip,width=\textwidth]{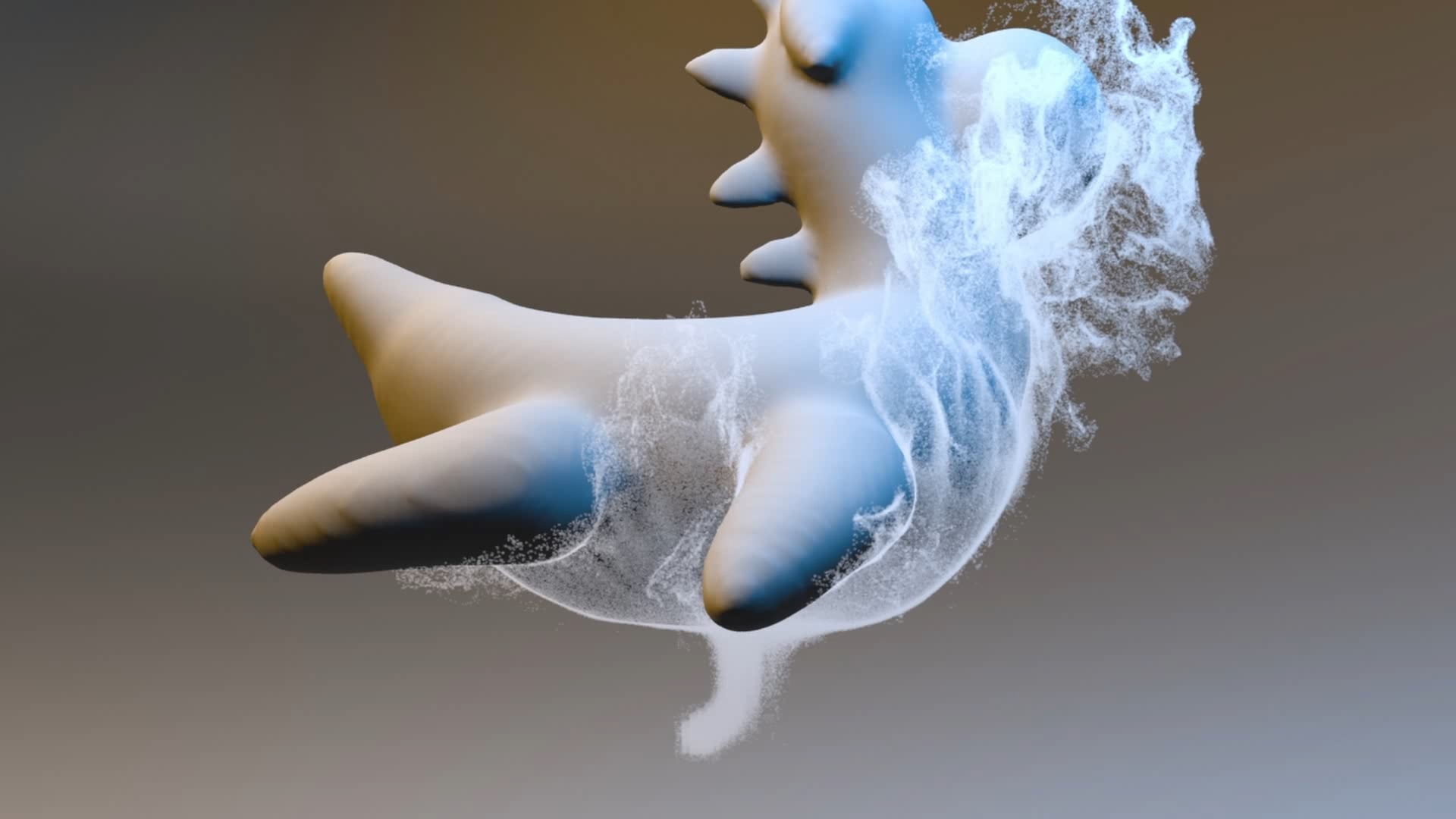}
    \end{minipage}
    &
    \begin{minipage}{.28\textwidth}
      \includegraphics[trim=0 0 0 0, clip,width=\textwidth]{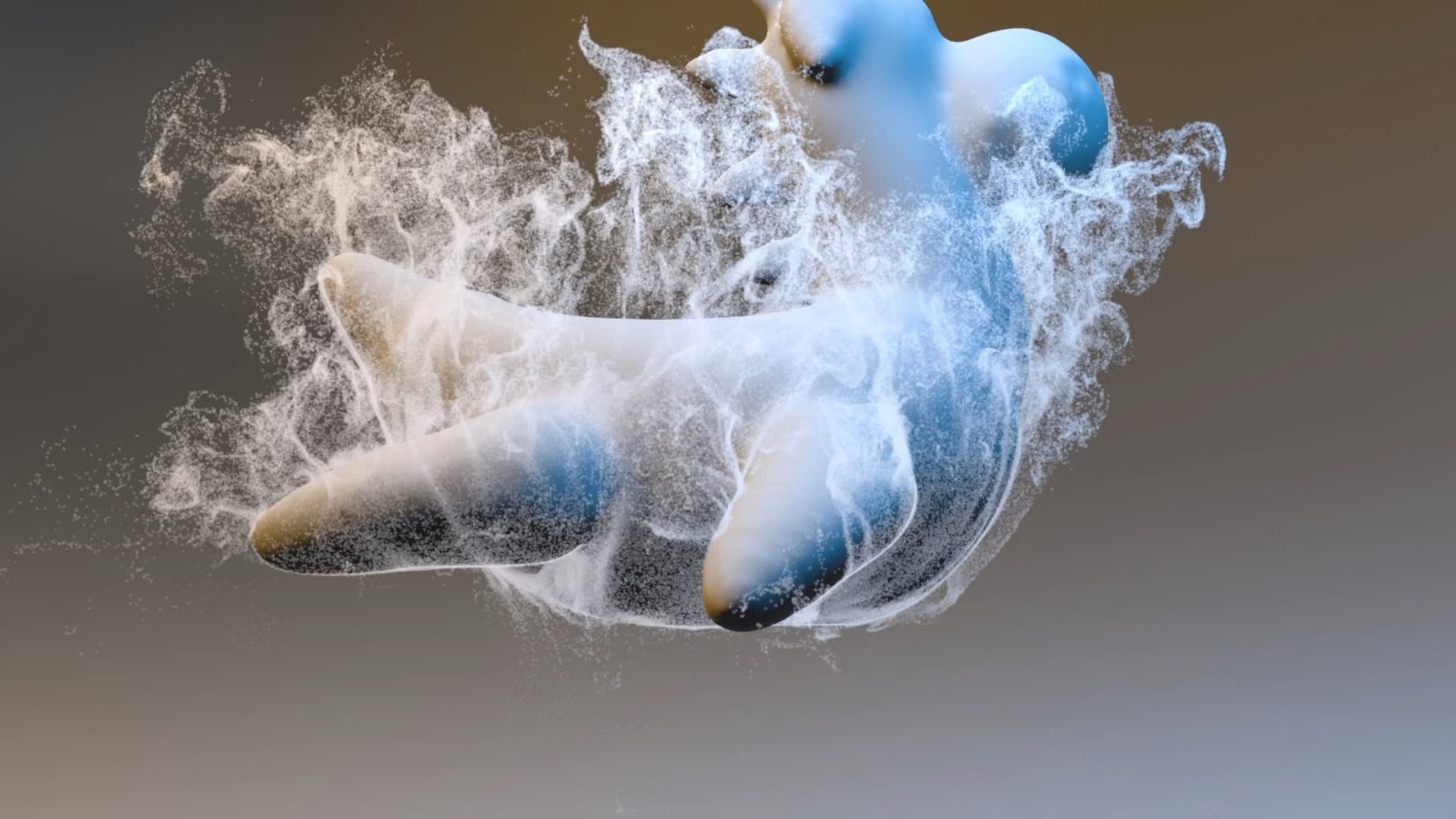}
    \end{minipage}
    &
    \begin{minipage}{.28\textwidth}
        \includegraphics[trim=0 0 0 0, clip,width=\textwidth]{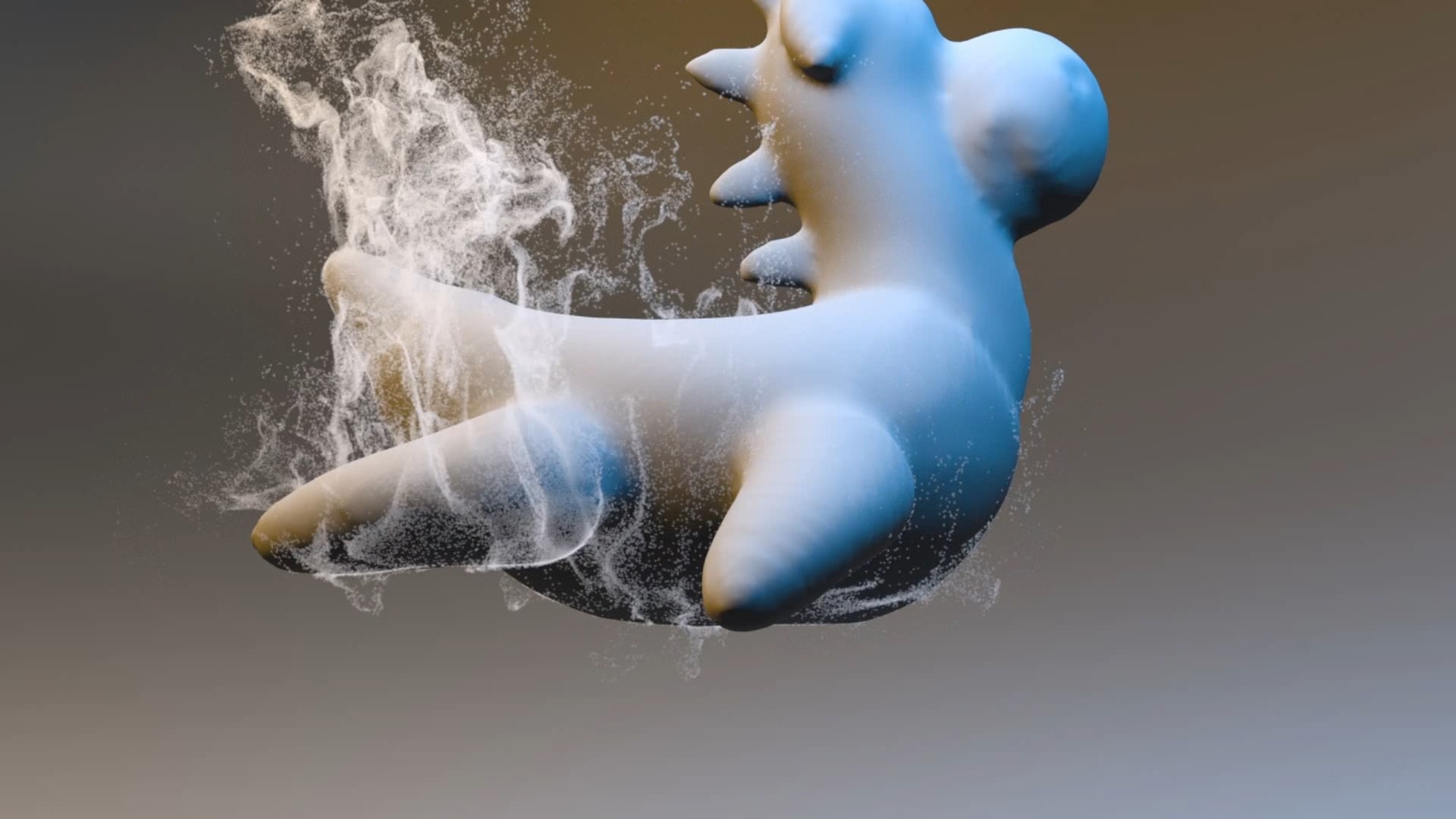}
    \end{minipage}
    \\ \hline
    \end{tabular}
	\caption{ \textbf{A Smoke Plume Simulation with A Rubber Toy}.
        The lingering outline of dense smoke from collided particles (frozen for emphasis) is conspicuous on the belly in the direct velocity interpolation case.  For Curl-Flow, the smoke instead flows around and past the object, so it remains visibly denser after going by (middle and right columns). 
	}
	\label{fig:dragon}
\vspace{1mm}
\end{figure*}

\subsection{Smoke Simulations and Performance}
\label{sec:smokeSimulation}
In the following two examples, we consider rising smoke plumes in $150^3$ grids with different obstacles: a stylized calf (Figure \ref{fig:calf}) and Houdini's rubber toy (Figure \ref{fig:dragon}).
For a simple age-based dissipation effect, we removed particles after 4 seconds of existence, but to emphasize collision errors we did not delete any particles that collided with the obstacle. We supplied up to 4M purely passive smoke particles for each example.

Our ramping method enforces a pointwise boundary-respecting velocity field under Curl-Flow, but particles advected with discrete timesteps will still collide with solid obstacles, especially with fast dynamic flows and/or complex solids.
For the smoke examples (our version and the baseline), 
we apply substepping to (only) particle advection to partially ameliorate this innate problem, constraining the timestep so that particles cannot move more than one grid cell per substep. Using more substeps would further reduce collisions, at the cost of increasing computational time. 
We additionally apply a simple "collision-aware" substepping upon first contact: 
if a particle hits an obstacle, we push it slightly back from the hit position (10\% of the distance the particle moved) and then perform a new partial substep from that position for the remaining time. (If it is still colliding afterwards, it is then frozen.)

To demonstrate Curl-Flow's practicality, these smoke plume tests were carried out by implementing our method into Houdini's smoke solver (\emph{Pyro}),
with various modifications (e.g., we used our own cut-cell pressure projection, solved with Eigen's CG).
Unlike the prior examples, we used a MacCormack scheme (Houdini's default) for velocity advection and we used our DST Poisson solver for fast gauge correction.
In both examples, Curl-Flow shows superior boundary-respecting behavior; particles flow over and around the shape, instead of into it and halting.  This effect can be observed by inspecting the density and brightness of smoke in the later images. Timings are provided in Table \ref{tab:smokeSimulationTiming}.

Curl-Flow incurs preprocessing costs at each step to reconstruct discrete $\boldsymbol{\psi}$ on the regular grid and on solid vertices, i.e., primarily a volumetric Poisson problem and a surface-based least squares problem.
Fortunately, our DST solver makes gauge correction efficient and the fact that the surface solve is lower dimensional likewise keeps its scaling reasonable. We investigate these costs in the $\boldsymbol{\psi}$ reconstruction scaling tests of Table \ref{tab:discreteFieldsTiming}. At $256^3$, $\boldsymbol{\psi}$ construction is two orders of magnitude faster than cut-cell pressure projection with Eigen's CG (Fourier methods do not support cut-cells). With major implementation effort, advanced multigrid and GPU solvers can sharply reduce the cost of pressure projection \cite{Shao:2022:Multigrid,raateland2022dcgrid}; however, even in the worst case, since both are Poisson problems, pressure projection and gauge correction would have around the same cost (e.g., see rows 3 and 4 of Table \ref{tab:AndoSilberman}).

The remaining additional overhead of our method arises during particle advection:
Curl-Flow has a larger stencil size for interpolation (quadratic vs.\ linear), 
and our ramping strategy requires additional operations, e.g., finding the closest point on the solid, evaluating a second interpolation, etc., unlike simple grid-based velocity interpolation.
In the calf and dragon examples, collided (frozen) particles also no longer undergo advection, which significantly benefits the direct approach.
In Figure \ref{fig:calf}, 1,508,617 (top) and 885 (bottom) particles are frozen and in Figure \ref{fig:dragon}, 2,136,520 (top) and 2,236 (bottom) particles are frozen at the last frame.
Advection costs are heavily dependent on the number of active particles and can become the bottleneck for massive particle counts. We anticipate that interpolation could be further optimized. Most importantly, since particle advection is done in a fully parallel manner, more compute cores immediately translate into speedups. Genuinely passive particles can also be traced in an entirely parallel post-process, after simulation completes.

\begin{table*}
\vspace{2mm}
\centering  
\caption{Average computational time per timestep for smoke simulations of Figure \ref{fig:calf} and Figure \ref{fig:dragon}. $~400\text{K}$ particles were supplied every second, and they fade away after 4 seconds.} 
\vspace{-7mm}
\begin{tabular}{c c c c c c c}  
\\ \hline\hline \\ [-2.5ex]
& \thead{Interpolation \\ Method} 
& \thead{Pressure \\ Projection (s)} 
& \thead{Grid $\boldsymbol{\psi}$ \\ Construction (s)}
& \thead{Solid $\boldsymbol{\psi}$ \\ Construction (s)}
& \thead{Particle \\ Advection (s)}
& \thead{$\#$ Solid Vertices  (Triangles)}
\\ [0.5ex]
\hline 
\multirow{2}{*}[-0.75mm]{\thead{ Stylized Calf ($150^3$) }}
& \thead{ Direct Velocity Interpolation }
& \multirow{2}{*}[-0.75mm]{3.898$\times10^{1}$}
& -- & -- & 1.213 & 
\multirow{2}{*}[-0.75mm]{16,450 (32,896)}  \\ 
&\thead{ Curl-Flow }
&& 3.807$\times10^{-1}$ & 3.571$\times10^{-1}$ & 2.749$\times10^{1}$ &\\ [0.5ex]
\hline 
\multirow{2}{*}[-0.75mm]{\thead{ Rubber Toy ($150^3$) }}
& \thead{ Direct Velocity Interpolation }
& \multirowcell{2}[-0.75mm]{ 3.847$\times10^{1}$ } & -- & -- & 4.049$\times10^{-1}$ 
& \multirow{2}{*}[-0.75mm]{21,050  (42,096)} \\
& \thead{ Curl-Flow }
&& 3.687$\times10^{-1}$ & 4.431$\times10^{-1}$ & 1.626$\times10^{1}$ &\\ 
\hline 

\end{tabular}
\label{tab:smokeSimulationTiming}
\end{table*}

\begin{table*}
\vspace{3mm}
\centering
\caption{Average computational time per timestep for discrete field construction of Curl-Flow at various scales.
}
\vspace{-7mm}
\begin{tabular}{c c c c c}  
\\ \hline\hline \\ [-2.5ex]
& \thead{Pressure \\ Projection (s)} 
& \thead{Grid $\boldsymbol{\psi}$ \\ Construction (s)}
& \thead{Solid $\boldsymbol{\psi}$ \\ Construction (s)}
& \thead{$\#$ Solid Vertices (Triangles)}
\\ [0.5ex]
\hline 
\thead{ Stylized Calf ($32^3$) }
& 4.875$\times10^{-2}$ & 6.540$\times10^{-3}$ & 4.026$\times10^{-3}$ & 726 (1,448) \\ 
\thead{ Stylized Calf ($64^3$) }
& 8.419$\times10^{-1}$ & 2.920$\times10^{-2}$ & 2.483$\times10^{-2}$ & 2,974 (5,944) \\ 
\thead{ Stylized Calf ($128^3$) }
& 1.984$\times10^{1}$ & 2.158$\times10^{-1}$ & 1.987$\times10^{-1}$ & 11,946 (23,888) \\ 
\thead{ Stylized Calf ($256^3$) }
& 3.685$\times10^{2}$ & 1.713 & 1.570 & 47,930 (95,856) \\ 
\hline 

\thead{ Rubber Toy ($32^3$) }
& 4.804$\times10^{-2}$ & 1.229$\times10^{-2}$ & 9.849$\times10^{-3}$ & 1,836 (3,664) \\ 
\thead{ Rubber Toy ($64^3$) }
& 8.998$\times10^{-1}$ & 2.954$\times10^{-2}$ & 3.257$\times10^{-2}$ & 7,508 (15,008) \\ 
\thead{ Rubber Toy ($128^3$) }
& 1.997$\times10^{1}$ & 2.111$\times10^{-1}$ & 2.509$\times10^{-1}$ & 30,600 (61,192) \\ 
\thead{ Rubber Toy ($256^3$) }
& 3.844$\times10^{2}$ & 1.705 & 2.166 & 123,284 (246,560) \\
\hline 
\end{tabular}
\label{tab:discreteFieldsTiming}
\end{table*}

\subsection{Vector Potential Reconstruction Comparisons }
\begin{figure}
    \centering	
    \captionsetup[subfigure]{labelformat=empty}
    \begin{subfigure}[t]{0.23\textwidth}
		\begin{overpic}[trim= 300 0 260 0, clip, width=\textwidth]{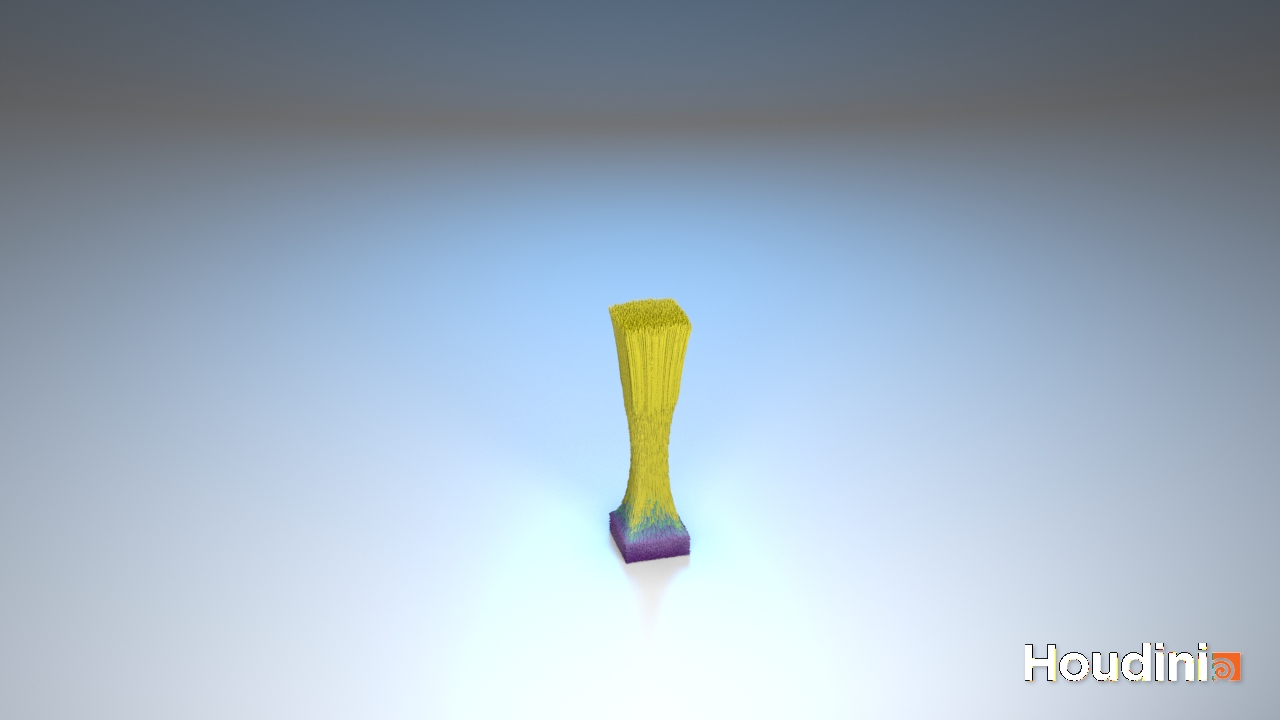}
		\end{overpic}
	\end{subfigure}~
	\begin{subfigure}[t]{0.23\textwidth}
		\begin{overpic}[trim= 300 0 260 0, clip, width=\textwidth]{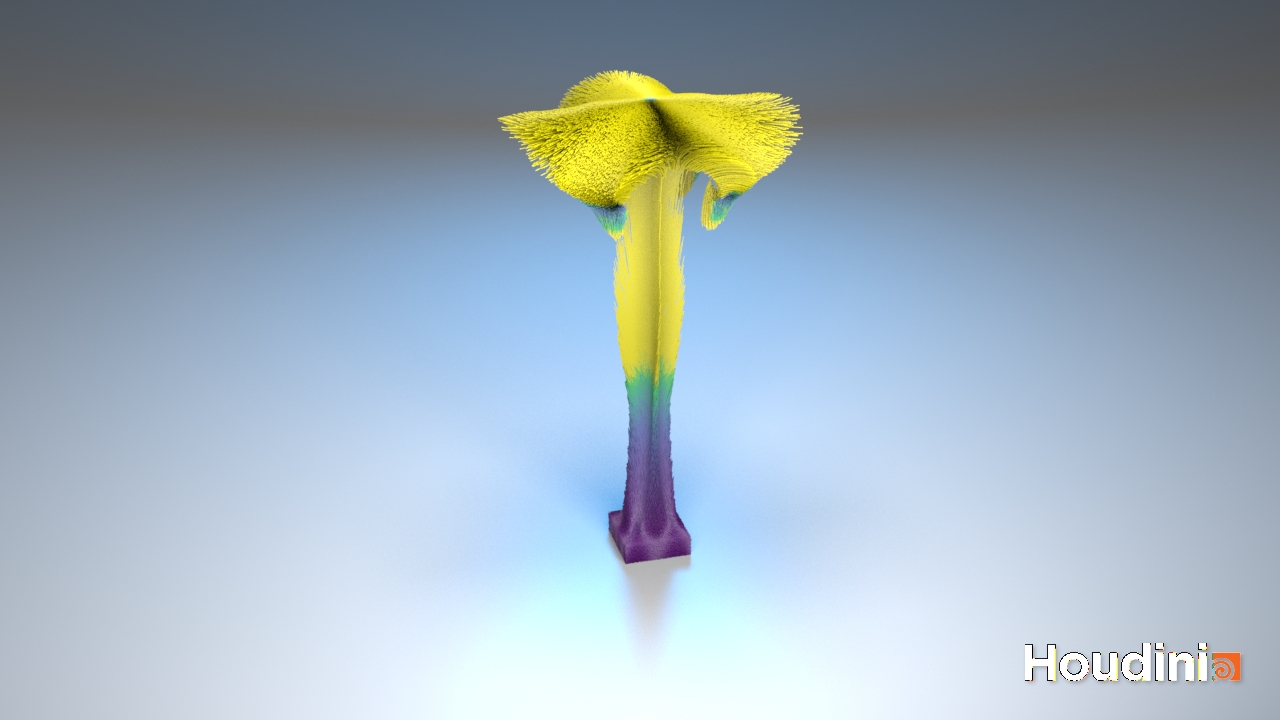}
		\end{overpic}
	\end{subfigure}~
	\\
	\begin{subfigure}[t]{0.23\textwidth}
		\begin{overpic}[trim= 300 0 260 0, clip, width=\textwidth]{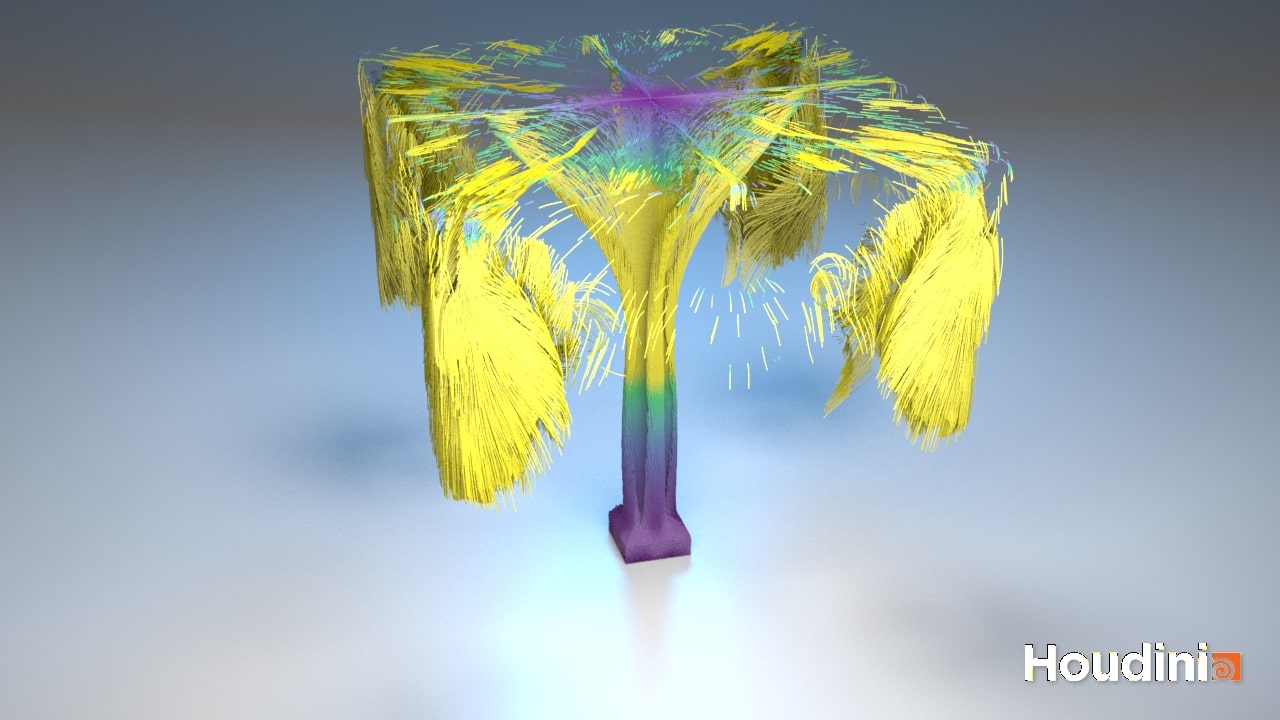}
		\end{overpic}
	\end{subfigure}~
	\begin{subfigure}[t]{0.23\textwidth}
		\begin{overpic}[trim= 300 0 260 0, clip, width=\textwidth]{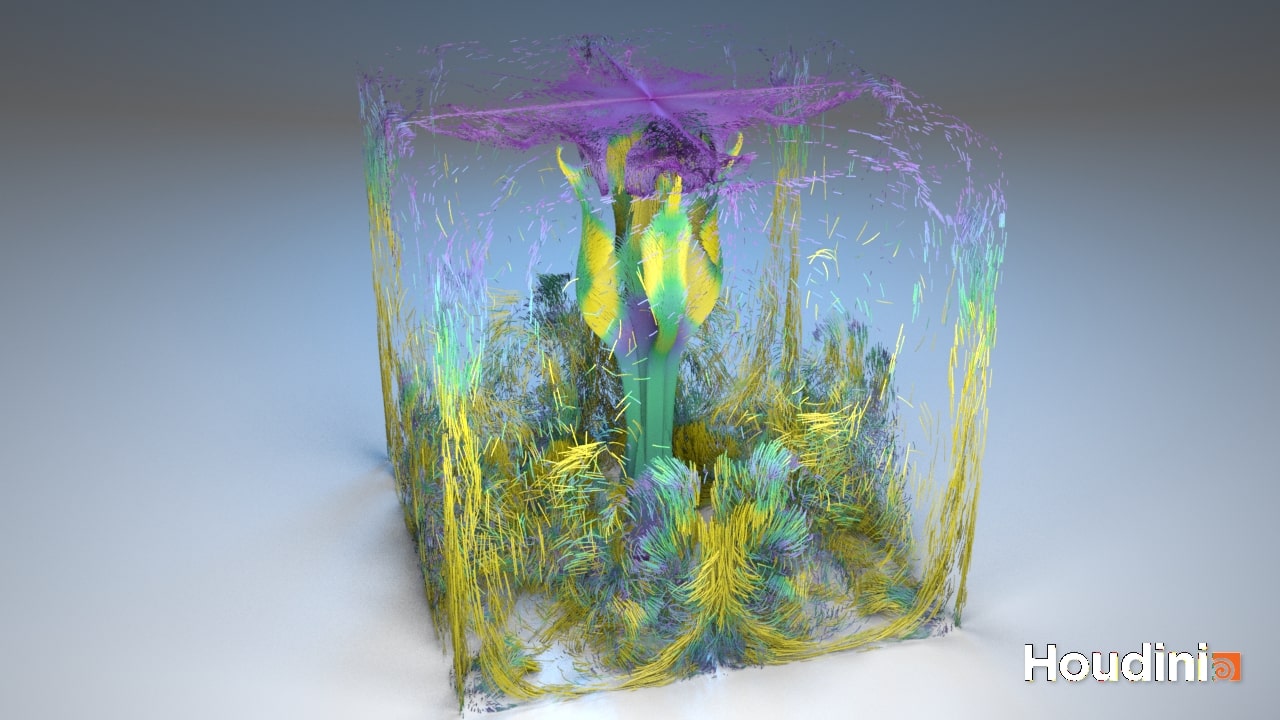}
		\end{overpic}
	\end{subfigure}
	\caption{ \textbf{A Rising Plume Using Curl-Flow} : 
	Five different $\boldsymbol{\psi}$ construction methods are plugged into Curl-Flow
	to produce this rising plume scene. 
	They all yield the same results shown above, with different computational costs (Table \ref{tab:AndoSilberman}).
	}
	\label{fig:AndoSilberman}
\end{figure}

\begin{table*}
 \centering 
  \caption{Average computational time per timestep for discrete $\boldsymbol{\psi}$ construction on the $150^3$ grid simulation of Figure \ref{fig:AndoSilberman} for various methods.}
  \vspace{-3mm}
 \begin{tabular}{c | c c c | c} 
 \hline\hline 
 & \thead{Pressure \\ Projection (s)} 
 & \thead{Initial $\boldsymbol{\psi}$ \\ Construction (s)}
 & \thead{Gauge \\ Correction (s)} 
 & \thead{Total $\boldsymbol{\psi}$ \\ Construction (s)}
 \\ [0.5ex]
 \hline 
 \thead{ Vector Poisson, Dimensionally-Coupled (Ando et al. [2015])} 
 & - & - & - & 152.0 \\
 \hline 
 \thead{Vector Poisson, Componentwise (e.g., Bao et al. [2017])} 
 & - & - & - & 81.87  \\ 
 \hline 
 \thead{Pressure Projection + Gauge Correction (CG) (Silberman et al.\ [2019])} 
 & 37.13 & 5.723 & 25.51 & 68.37 \\ 
 \hline 
 \thead{Ours: Pressure Projection + Gauge Correction (CG) + Parallel Sweeping}  
 & 37.13 & 0.05904 & 24.33 & 61.51 \\
  \thead{Ours: Pressure Projection + Gauge Correction (DST) + Parallel Sweeping}  
  & 37.13 & 0.05781 & 0.3304 & 37.52 \\
 \hline 
 \end{tabular}
 \label{tab:AndoSilberman}
 \vspace{1mm}
 \end{table*}
Several alternative, but more costly, methods to reconstruct edge-based vector potential fields from face-based vector fields have previously been proposed \cite{Sato2015,Ando2015,Bao2017,Silberman2019}. Since those methods mostly do not handle cut-cell obstacles, to fairly compare computational costs we consider a flow in an empty box domain. 
We adapted the methods of Ando et al.\ and Silberman et al.\ into our framework, replacing our $\boldsymbol{\psi}$ reconstruction step.
We consider the combined cost of pressure projection \emph{and} $\boldsymbol{\psi}$ reconstruction for each method in Table \ref{tab:AndoSilberman}, since Ando et al.\ achieve both simultaneously. For a fair comparison, the first four rows use Eigen's CG for all solves. Since our DST-based gauge correction can always be used whether obstacles are present or not, we present this result in its own row.
Despite different computational costs, the solutions are numerically consistent and the results visually indistinguishable (Figure \ref{fig:AndoSilberman}).

\paragraph*{Comparison to \citet{Ando2015}:}
Ando et al.\ solve a single, dimensionally coupled vector Poisson problem for $\boldsymbol{\psi}$ from a \emph{divergent} velocity field, avoiding a separate pressure projection.
We modified this approach to satisfy our domain boundary conditions (Section \ref{exteriorBoundary}).
As shown in Table \ref{tab:AndoSilberman}, its computational cost (topmost row) is expensive compared to our method (bottom two rows): about $2.5 - 4\times$ slower for the $150^3$ grid of Figure \ref{fig:AndoSilberman}. 
Unless boundary conditions couple them (as is often the case for free surfaces and some solid boundary treatments \cite{Ando2015}), the dimensionally-coupled vector Poisson equation can be split into three independent scalar Poisson equations for efficiency. Applying this special case optimization nearly halves the cost (Table \ref{tab:AndoSilberman}, second row), but it remains about ${1.3\times} - {2.2\times}$ the cost of our approach.

\paragraph*{Comparison to \citet{Silberman2019}:}
The reconstruction method of Silberman et al.\ from electromagnetics is most similar to ours: it recovers
$\boldsymbol{\psi}$ from an initially divergence-free input vector field, using a cell-by-cell construction of an initial vector potential, followed by Poisson-based gauge correction.
Silberman et al.\ assumed infinite boundary conditions to enable gauge correction by FFT.
We adapted this idea by matching our boundary conditions and linear solver (CG) -- see Table \ref{tab:AndoSilberman}, third row.
Compared to our sweeping approach, the initial cell-by-cell $\boldsymbol{\psi}$ construction of Silberman et al.\ is slower because
it is inherently serial and it builds all three components of $\boldsymbol{\psi}$, rather than the two in our sweeping scheme.
It also requires complicated case-by-case code: it constructs 12 edge values on a cell, and adjusts them for consistency depending on which neighbors have already been processed.
The cost of our parallel sweeping strategy is \emph{two orders of magnitude faster} than Silberman et al.\ (Table \ref{tab:AndoSilberman}, third column), and negligible compared to the Poisson solves.
Overall, the (adapted) method of \citet{Silberman2019} is $11\%$ slower than ours, or $82\%$ slower when we use DST. (We emphasize that they also do not handle our boundary conditions nor interior obstacles, so their FFT approach never applies to our setting.)

\paragraph*{Comparison to \citet{Bao2017}:}
Bao et al.\ use a vector potential in their immersed boundary method.
They assume divergence-free input and exploit the lack of (cut-cell) obstacles to use the componentwise vector Poisson form (three scalar Poisson problems). They assume periodic boundaries to solve with the FFT.
For a fair comparison, we used their (MATLAB) code and replaced only the discrete $\boldsymbol{\psi}$ construction step with our method, similarly using FFT for gauge correction.
Our method for $\boldsymbol{\psi}$ construction is ${\sim}2.5\times$ faster, as shown in Table ~\ref{tab:Bao}.
\begin{table}[t]
\centering 
\caption{Average computational time for constructing $\boldsymbol{\psi}$, 
within the code of \citet{Bao2017} for an immersed deforming membrane. For the "Curl-Flow" comparison, we replace only their $\boldsymbol{\psi}$ construction part.}
\vspace{-7mm}
\begin{tabular}{cc}  \\\toprule
& $\boldsymbol{\psi}$ construction time (s) \\\midrule
\citet{Bao2017} & 2.324 $\times 10^{-3}$ \\\midrule  
Curl-Flow      & 9.397 $\times 10^{-4}$ \\\bottomrule
\end{tabular}
\label{tab:Bao}
\end{table}

\begin{figure}
\vspace{3mm}
\centering	
    \begin{subfigure}{0.10\textwidth}
	    \begin{overpic}[trim= 9cm 5cm 28cm 6cm, clip, width=0.95\textwidth]{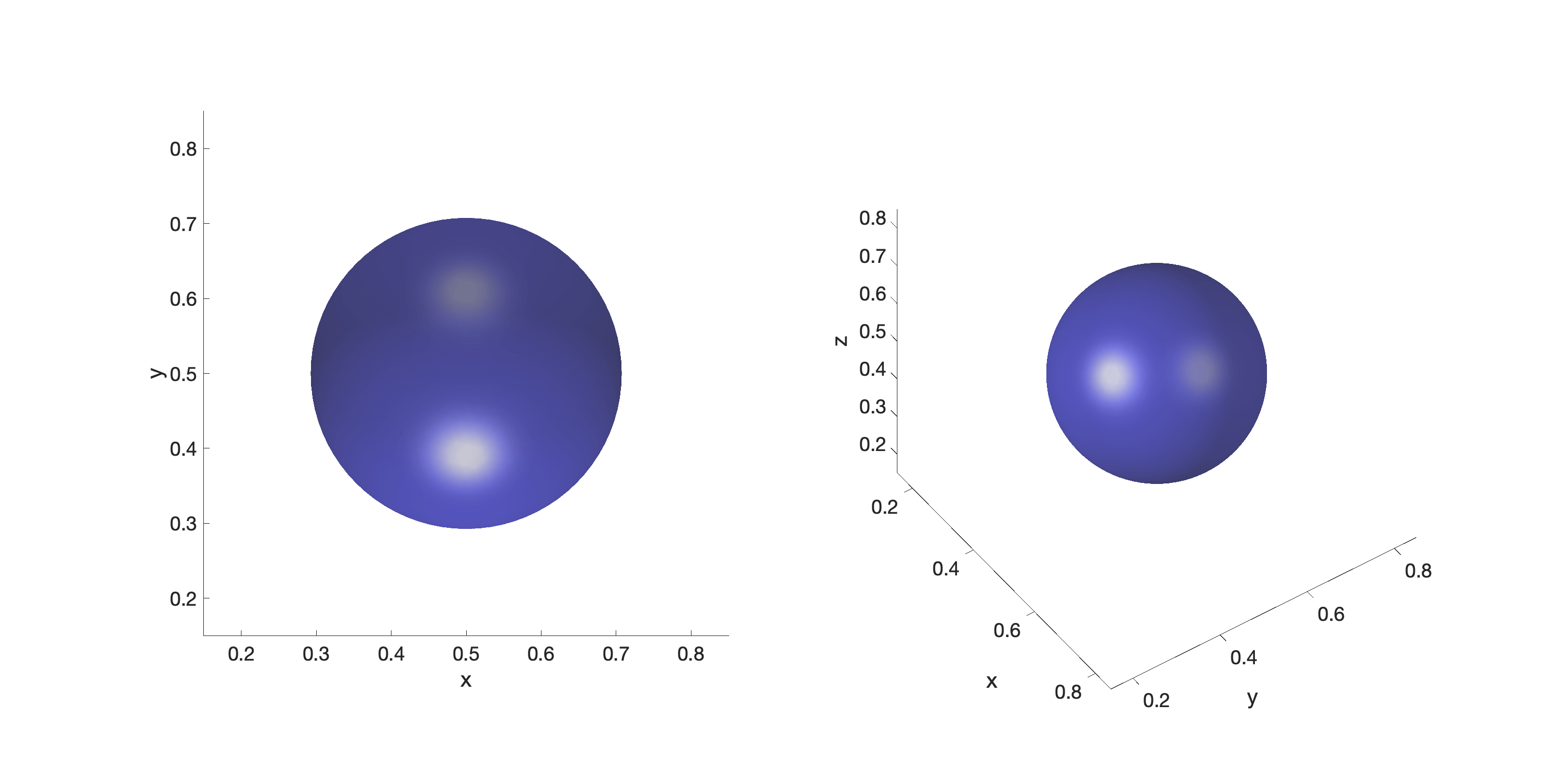}
		\put(22, -5) {\footnotesize Frame 0}
		\end{overpic}
	\end{subfigure}
	\begin{subfigure}{0.10\textwidth}
	    \begin{overpic}[trim= 9cm 5cm 28cm 6cm, clip, width=0.95\textwidth]{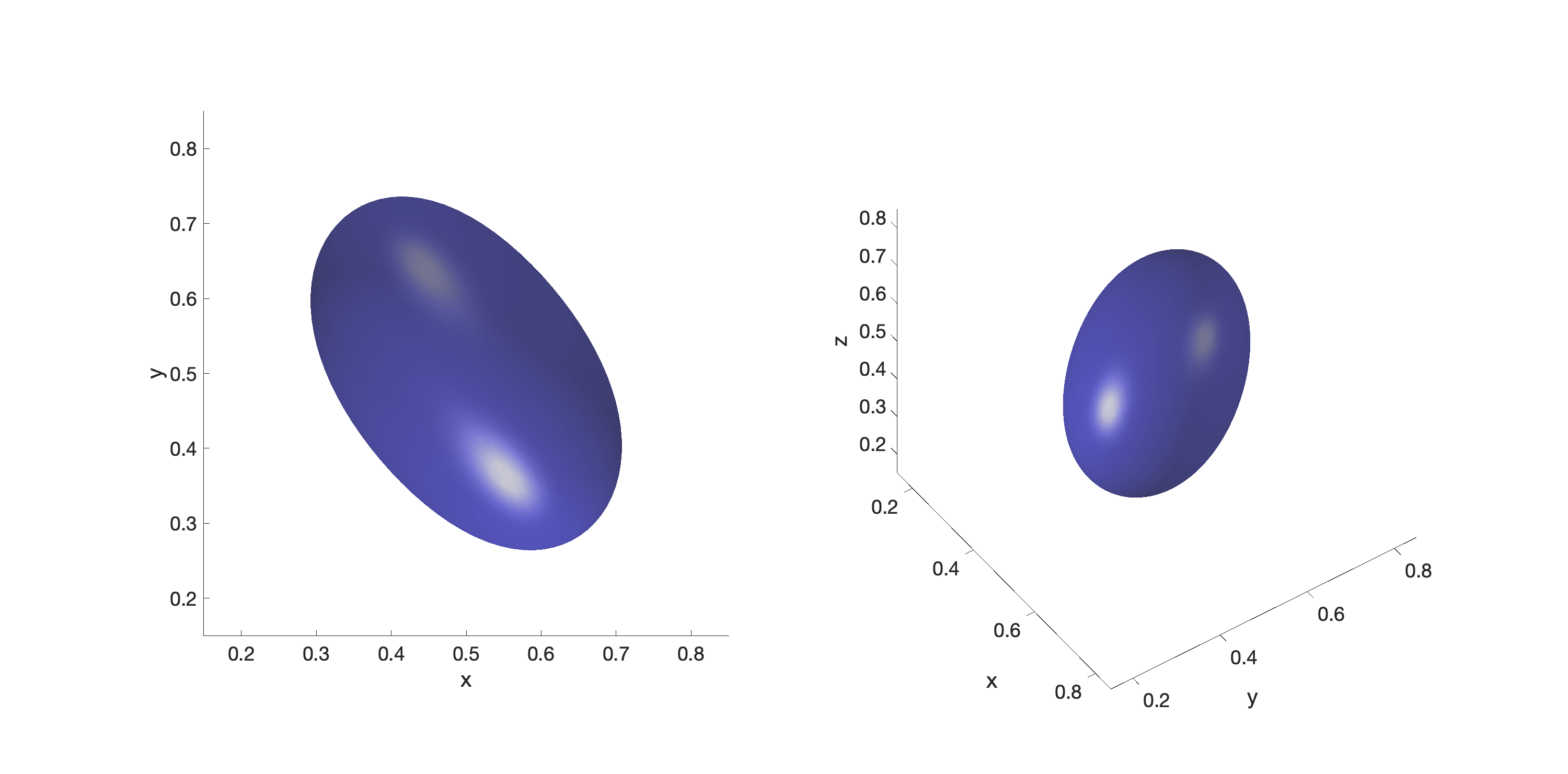}
		\put(22, -5) {\footnotesize Frame 15}
		\end{overpic}
	\end{subfigure}
	\begin{subfigure}{0.10\textwidth}
	    \begin{overpic}[trim= 9cm 5cm 28cm 6cm, clip, width=0.95\textwidth]{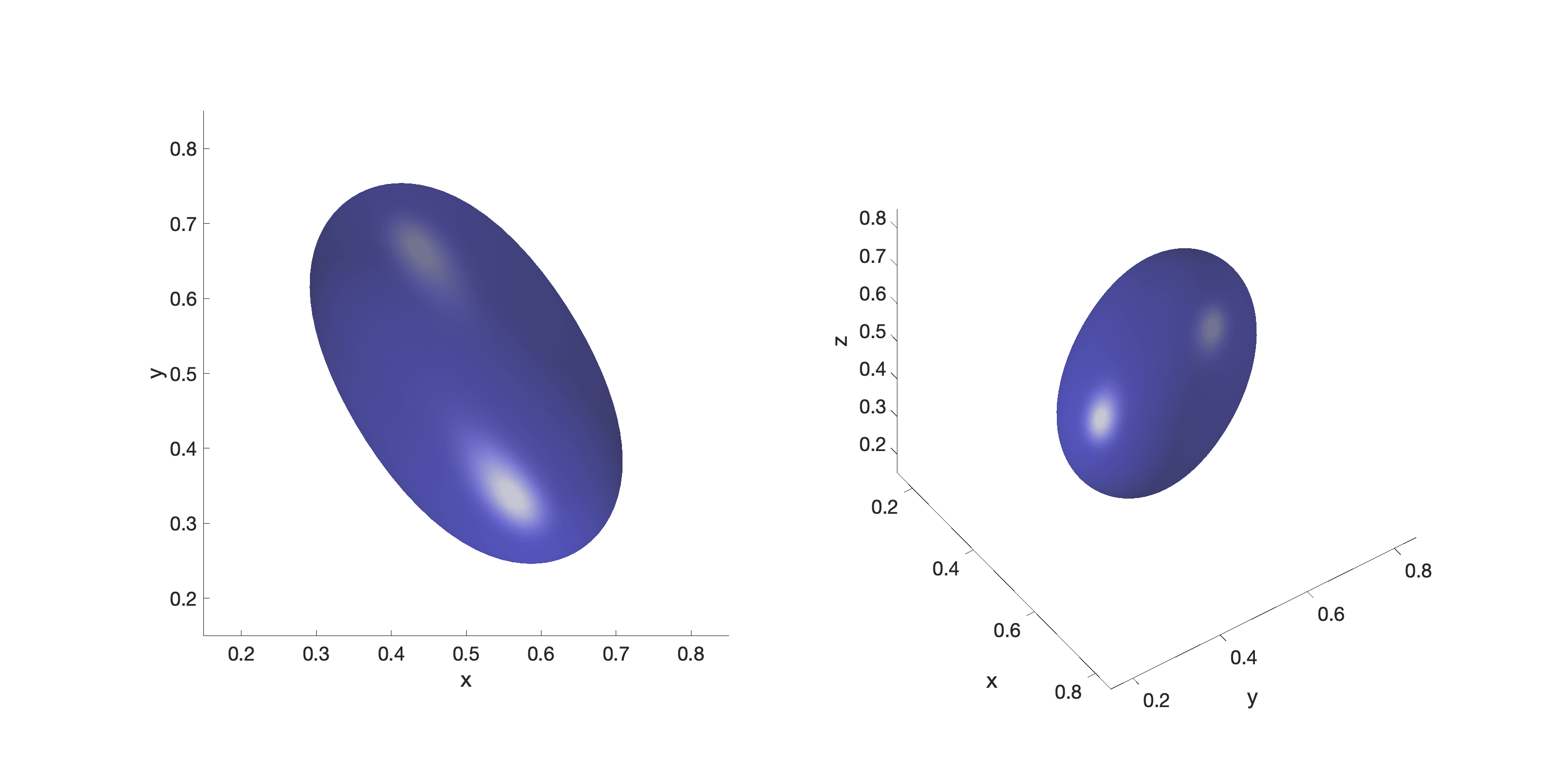}
		\put(22, -5) {\footnotesize Frame 30}
		\end{overpic}
	\end{subfigure}
	\begin{subfigure}{0.10\textwidth}
	    \begin{overpic}[trim= 9cm 5cm 28cm 6cm, clip, width=0.95\textwidth]{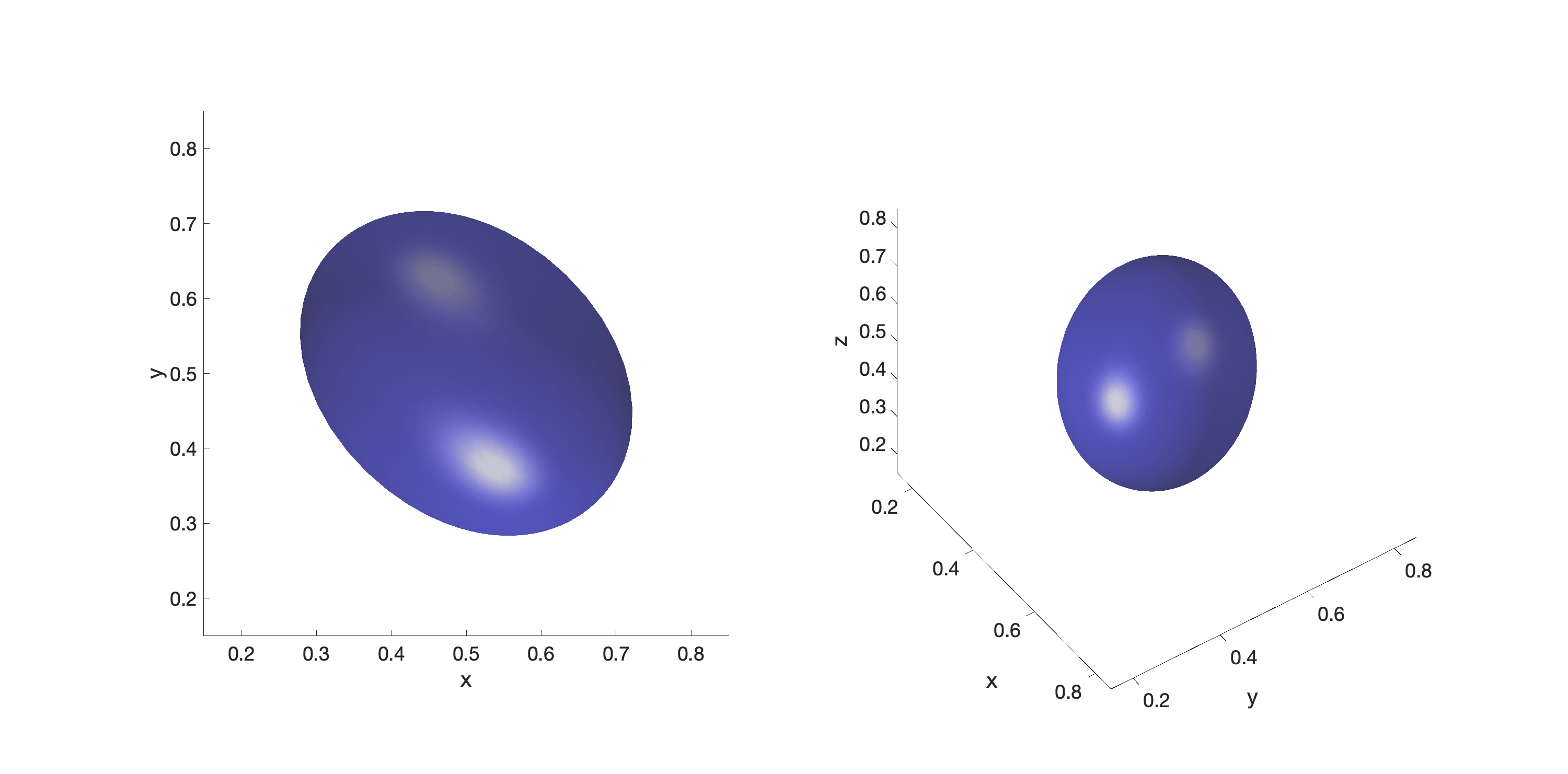}
		\put(22, -5) {\footnotesize Frame 60}
		\end{overpic}
	\end{subfigure}
	\caption{ \textbf{Deforming elastic membrane test from \protect\cite{Bao2017} :} 
	A spherical membrane immersed in fluid is deformed with an initial velocity, before
	surface tension returns it to a spherical shape. 
	We replace Bao's $\boldsymbol{\psi}$ reconstruction method with ours. Our methods produces identical results but faster (Table \ref{tab:Bao}).}
	\label{fig:Bao}
\end{figure}

Although our focus is fluid velocity interpolation, vector potential reconstruction has already been used in graphics (for fluid control \cite{Sato2015,Sato2021} and visualization \cite{Biswas2016}), as well as electromagnetics and beyond.
Our new approach can thus offer immediate speedups in all of these domains, e.g., based on Table \ref{tab:AndoSilberman}, perhaps doubling or quadrupling the speed of the most expensive step of the fluid guiding method of \citet{Sato2021}.

\subsection{Convergence of Curl-Flow Velocity Interpolation}
\begin{figure}
    \centering	
    \begin{subfigure}[t]{0.225\textwidth}
		\fbox{\begin{overpic}[width=0.99\textwidth]{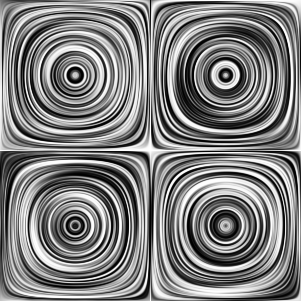}
		\end{overpic}}
		\subcaption{Analytical solution}
	\end{subfigure}
	\begin{subfigure}[t]{0.24\textwidth}
		\begin{overpic}[width=0.99\textwidth]{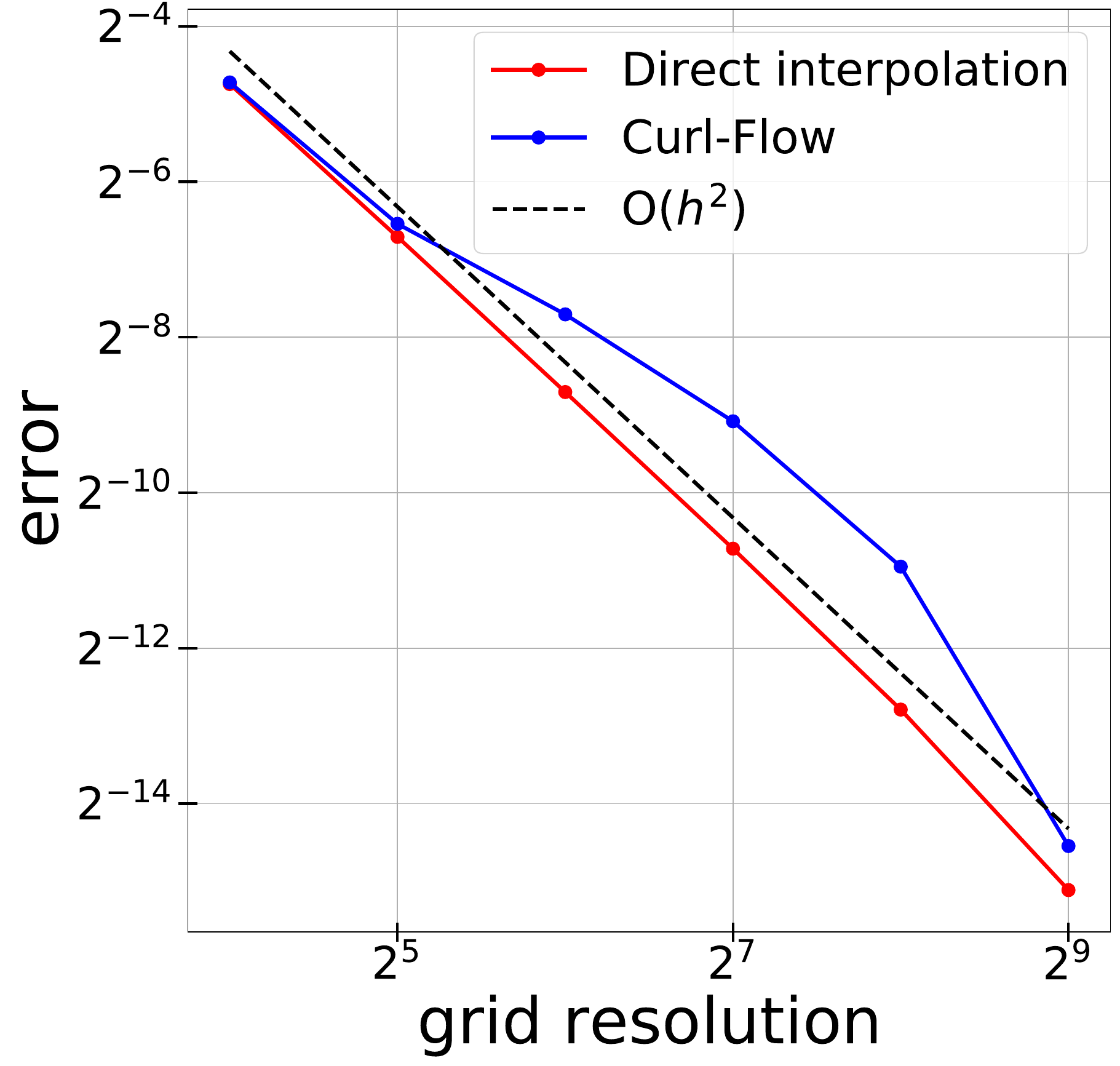}
		\end{overpic}
	\subcaption{$L_\infty$ norm}
	\end{subfigure}
	\caption{\textbf{Convergence of Velocity Interpolation:} 
        We sample discrete grid velocities from the analytically divergence-free field of (a), enforce discrete incompressibility,  interpolate using direct (bilinear) velocity interpolation and Curl-Flow interpolation, and measure the error.
        Both methods show approximately second-order convergence. 
        See error data in Table 1 in supplemental material.
    }
	\label{fig:velocityConvergence}
\end{figure}
Since Curl-Flow uses intermediate vector potentials rather than directly interpolating grid velocities,
one may ask if it is a valid interpolant at all: does the resulting continuous field converge? We study this question by evaluating how well the interpolated field agrees with an input analytically divergence-free field under refinement.
Since the method's input should be a discretely divergence-free velocity field defined on the staggered grid, we first sample grid velocities from an input analytical field and
apply pressure projection for discrete incompressibility, 
then apply the chosen interpolant (at a fine sampling of points) and evaluate their error with respect to the original analytical field. 
We use the field
\begin{equation*}
    \mathbf{u}(x, y) = ( \text{sin}(2 \pi x) \text{cos}(2 \pi y), -\text{cos}(2 \pi x) \text{sin}(2 \pi y)),
\end{equation*} 
shown in Figure \ref{fig:velocityConvergence}(a).
Since Curl-Flow's quadratic kernel will query some sample positions outside the domain, we linearly extrapolate $\psi$ values to such samples for the convergence test.
Direct (bilinear) velocity interpolation yields second-order convergence and Curl-Flow shows similar behavior (Figure \ref{fig:velocityConvergence}).

\section{Conclusions and Future Work}
Large time steps in particle tracing, poorly enforced boundaries, and inadequate pressure solver tolerances are known to cause density/volume drift; many post-compensation strategies have been developed \cite{Ando2012,Sato2018b,Takahashi2019,Kugelstadt2019}. Our work is the first in computer graphics to identify divergent velocity interpolation as an additional factor. Curl-Flow, tailored to plug into popular grid-based cut-cell fluid simulators, addresses this issue up front by constructing pointwise divergence-free velocity fields that also respect obstacles. It offers better long-term particle distributions and natural motions around obstacles and, when used for FLIP particles, yields more dynamic motion and/or reduces the need for post-compensation. We believe our work opens a new direction in the design of advection schemes for fluid animation and raises compelling questions for future study.

We did not consider moving obstacles, which are a useful element of many practical scenarios. Allowing for non-zero boundary fluxes during vector potential reconstruction seems feasible; however, ensuring exactly collision-safe trajectories near Eulerian cut-cell solids (i.e., the aim of the present work) is significantly complicated by the Lagrangian nature of moving obstacles. We have also only considered grid domains that are fully open, fully closed, or "wind-tunnel"-like. More general boundary conditions are an obvious next step, including free surface boundaries for liquid animation.

For our ramping $\boldsymbol{\psi}$ towards solids, we found that seeking a minimal perturbation to $\boldsymbol{\psi}$ gave visually natural results, but other criteria are possible (e.g., controlling $\nabla \times \boldsymbol{\psi}_t$). A drawback of boundary handling in Curl-Noise-type schemes is that, though always incompressible, they can yield discontinuities when the closest polygon facet changes. A challenging extension would be to develop a strictly continuous variant of Curl-Noise, perhaps by incorporating a smooth distance measure \cite{madan2021fast}.

While not a salient issue for simply tracing particle paths, interpolation of higher order than linear can yield overshooting. We sidestepped this possibility in our FLIP tests by not using the Curl-Flow field for P2G or G2P transfers: any new extrema will not feed back into the dynamic velocity field. Using Curl-Flow for G2P (or all the interpolations within a semi-Lagrangian update) may necessitate some form of limiter. Straightforward velocity clamping \cite{Selle2008} would sacrifice incompressibility, so falling back to linear Curl-Flow (with kinks) may be preferable.

Lastly, the ability to edit fluid simulations in an \emph{efficient} post-process remains highly desirable \cite{Pan2013,Sato2018} due to their cost and limited controllability. Our work, and that of Sato et al. \shortcite{Sato2015,Sato2021}, hint that vector potentials may unlock fast, divergence-free editing capabilities by combining discrete \emph{simulation} tools and continuous \emph{procedural} tools.


\bibliographystyle{ACM-Reference-Format}
\bibliography{bibliography2}


\begin{thebibliography}{78}


\ifx \showCODEN    \undefined \def \showCODEN     #1{\unskip}     \fi
\ifx \showDOI      \undefined \def \showDOI       #1{#1}\fi
\ifx \showISBNx    \undefined \def \showISBNx     #1{\unskip}     \fi
\ifx \showISBNxiii \undefined \def \showISBNxiii  #1{\unskip}     \fi
\ifx \showISSN     \undefined \def \showISSN      #1{\unskip}     \fi
\ifx \showLCCN     \undefined \def \showLCCN      #1{\unskip}     \fi
\ifx \shownote     \undefined \def \shownote      #1{#1}          \fi
\ifx \showarticletitle \undefined \def \showarticletitle #1{#1}   \fi
\ifx \showURL      \undefined \def \showURL       {\relax}        \fi
\providecommand\bibfield[2]{#2}
\providecommand\bibinfo[2]{#2}
\providecommand\natexlab[1]{#1}
\providecommand\showeprint[2][]{arXiv:#2}

\bibitem[\protect\citeauthoryear{Albanese and Rubinacci}{Albanese and Rubinacci}{1990}]%
        {Albanese1990}
\bibfield{author}{\bibinfo{person}{R. Albanese} {and} \bibinfo{person}{G. Rubinacci}.} \bibinfo{year}{1990}\natexlab{}.
\newblock \showarticletitle{{Magnetostatic field computations in terms of two-component vector potentials}}.
\newblock \bibinfo{journal}{\emph{Internat. J. Numer. Methods Engrg.}} \bibinfo{volume}{29}, \bibinfo{number}{3} (\bibinfo{date}{mar} \bibinfo{year}{1990}), \bibinfo{pages}{515--532}.
\newblock
\showISSN{0029-5981}
\urldef\tempurl%
\url{https://doi.org/10.1002/nme.1620290305}
\showDOI{\tempurl}


\bibitem[\protect\citeauthoryear{Ando, Thuerey, and Wojtan}{Ando et~al\mbox{.}}{2015}]%
        {Ando2015}
\bibfield{author}{\bibinfo{person}{Ryoichi Ando}, \bibinfo{person}{Nils Thuerey}, {and} \bibinfo{person}{Chris Wojtan}.} \bibinfo{year}{2015}\natexlab{}.
\newblock \showarticletitle{{A stream function solver for liquid simulations}}.
\newblock \bibinfo{journal}{\emph{ACM Transactions on Graphics}} \bibinfo{volume}{34}, \bibinfo{number}{4} (\bibinfo{date}{jul} \bibinfo{year}{2015}), \bibinfo{pages}{53:1--53:9}.
\newblock
\showISSN{07300301}


\bibitem[\protect\citeauthoryear{Ando, Thurey, and Tsuruno}{Ando et~al\mbox{.}}{2012}]%
        {Ando2012}
\bibfield{author}{\bibinfo{person}{Ryoichi Ando}, \bibinfo{person}{Nils Thurey}, {and} \bibinfo{person}{Reiji Tsuruno}.} \bibinfo{year}{2012}\natexlab{}.
\newblock \showarticletitle{Preserving fluid sheets with adaptively sampled anisotropic particles}.
\newblock \bibinfo{journal}{\emph{IEEE Transactions on Visualization and Computer Graphics}} \bibinfo{volume}{18}, \bibinfo{number}{8} (\bibinfo{year}{2012}), \bibinfo{pages}{1202--1214}.
\newblock


\bibitem[\protect\citeauthoryear{Angelidis and Neyret}{Angelidis and Neyret}{2005}]%
        {Angelidis2005}
\bibfield{author}{\bibinfo{person}{Alexis Angelidis} {and} \bibinfo{person}{Fabrice Neyret}.} \bibinfo{year}{2005}\natexlab{}.
\newblock \showarticletitle{Simulation of smoke based on vortex filament primitives}. In \bibinfo{booktitle}{\emph{Proceedings of the 2005 ACM SIGGRAPH/Eurographics symposium on Computer animation}}. ACM, \bibinfo{pages}{87--96}.
\newblock


\bibitem[\protect\citeauthoryear{Azevedo, Batty, and Oliveira}{Azevedo et~al\mbox{.}}{2016}]%
        {Azevedo2016}
\bibfield{author}{\bibinfo{person}{Vinicius~C Azevedo}, \bibinfo{person}{Christopher Batty}, {and} \bibinfo{person}{Manuel~M Oliveira}.} \bibinfo{year}{2016}\natexlab{}.
\newblock \showarticletitle{Preserving geometry and topology for fluid flows with thin obstacles and narrow gaps}.
\newblock \bibinfo{journal}{\emph{ACM Transactions on Graphics (TOG)}} \bibinfo{volume}{35}, \bibinfo{number}{4} (\bibinfo{year}{2016}), \bibinfo{pages}{97}.
\newblock


\bibitem[\protect\citeauthoryear{Balsara}{Balsara}{2001}]%
        {Balsara2001}
\bibfield{author}{\bibinfo{person}{Dinshaw~S Balsara}.} \bibinfo{year}{2001}\natexlab{}.
\newblock \showarticletitle{Divergence-free adaptive mesh refinement for magnetohydrodynamics}.
\newblock \bibinfo{journal}{\emph{J. Comput. Phys.}} \bibinfo{volume}{174}, \bibinfo{number}{2} (\bibinfo{year}{2001}), \bibinfo{pages}{614--648}.
\newblock


\bibitem[\protect\citeauthoryear{Balsara}{Balsara}{2004}]%
        {Balsara2004}
\bibfield{author}{\bibinfo{person}{Dinshaw~S Balsara}.} \bibinfo{year}{2004}\natexlab{}.
\newblock \showarticletitle{Second-order-accurate schemes for magnetohydrodynamics with divergence-free reconstruction}.
\newblock \bibinfo{journal}{\emph{The Astrophysical Journal Supplement Series}} \bibinfo{volume}{151}, \bibinfo{number}{1} (\bibinfo{year}{2004}), \bibinfo{pages}{149}.
\newblock


\bibitem[\protect\citeauthoryear{Balsara}{Balsara}{2009}]%
        {Balsara2009}
\bibfield{author}{\bibinfo{person}{Dinshaw~S Balsara}.} \bibinfo{year}{2009}\natexlab{}.
\newblock \showarticletitle{Divergence-free reconstruction of magnetic fields and WENO schemes for magnetohydrodynamics}.
\newblock \bibinfo{journal}{\emph{J. Comput. Phys.}} \bibinfo{volume}{228}, \bibinfo{number}{14} (\bibinfo{year}{2009}), \bibinfo{pages}{5040--5056}.
\newblock


\bibitem[\protect\citeauthoryear{Bao, Donev, Griffith, McQueen, and Peskin}{Bao et~al\mbox{.}}{2017}]%
        {Bao2017}
\bibfield{author}{\bibinfo{person}{Yuanxun Bao}, \bibinfo{person}{Aleksandar Donev}, \bibinfo{person}{Boyce~E Griffith}, \bibinfo{person}{David~M McQueen}, {and} \bibinfo{person}{Charles~S Peskin}.} \bibinfo{year}{2017}\natexlab{}.
\newblock \showarticletitle{An Immersed Boundary method with divergence-free velocity interpolation and force spreading}.
\newblock \bibinfo{journal}{\emph{Journal of computational physics}}  \bibinfo{volume}{347} (\bibinfo{year}{2017}), \bibinfo{pages}{183--206}.
\newblock


\bibitem[\protect\citeauthoryear{Biswas, Strelitz, Woodring, Chen, and Shen}{Biswas et~al\mbox{.}}{2016}]%
        {Biswas2016}
\bibfield{author}{\bibinfo{person}{Ayan Biswas}, \bibinfo{person}{Richard Strelitz}, \bibinfo{person}{Jonathan Woodring}, \bibinfo{person}{Chun-Ming Chen}, {and} \bibinfo{person}{Han-Wei Shen}.} \bibinfo{year}{2016}\natexlab{}.
\newblock \showarticletitle{A scalable streamline generation algorithm via flux-based isocontour extraction}. In \bibinfo{booktitle}{\emph{Proceedings of the 16th Eurographics Symposium on Parallel Graphics and Visualization}}. \bibinfo{pages}{69--78}.
\newblock


\bibitem[\protect\citeauthoryear{Bridson}{Bridson}{2015}]%
        {Bridson2015}
\bibfield{author}{\bibinfo{person}{R. Bridson}.} \bibinfo{year}{2015}\natexlab{}.
\newblock \bibinfo{booktitle}{\emph{Fluid Simulation for Computer Graphics, Second Edition}}.
\newblock \bibinfo{publisher}{Taylor \& Francis}.
\newblock
\showISBNx{9781482232837}
\showLCCN{2015452396}


\bibitem[\protect\citeauthoryear{Bridson, Houriham, and Nordenstam}{Bridson et~al\mbox{.}}{2007}]%
        {Bridson2007}
\bibfield{author}{\bibinfo{person}{Robert Bridson}, \bibinfo{person}{Jim Houriham}, {and} \bibinfo{person}{Marcus Nordenstam}.} \bibinfo{year}{2007}\natexlab{}.
\newblock \showarticletitle{Curl-Noise for Procedural Fluid Flow}.
\newblock \bibinfo{journal}{\emph{ACM Trans. Graph.}} \bibinfo{volume}{26}, \bibinfo{number}{3} (\bibinfo{date}{July} \bibinfo{year}{2007}), \bibinfo{pages}{46–es}.
\newblock
\showISSN{0730-0301}
\urldef\tempurl%
\url{https://doi.org/10.1145/1276377.1276435}
\showURL{%
\tempurl}


\bibitem[\protect\citeauthoryear{Brochu, Keeler, and Bridson}{Brochu et~al\mbox{.}}{2012}]%
        {Brochu2012}
\bibfield{author}{\bibinfo{person}{Tyson Brochu}, \bibinfo{person}{Todd Keeler}, {and} \bibinfo{person}{Robert Bridson}.} \bibinfo{year}{2012}\natexlab{}.
\newblock \showarticletitle{Linear-time smoke animation with vortex sheet meshes}. In \bibinfo{booktitle}{\emph{Proceedings of the ACM SIGGRAPH/Eurographics Symposium on Computer Animation}}. Eurographics Association, \bibinfo{pages}{87--95}.
\newblock


\bibitem[\protect\citeauthoryear{Bunge, Botsch, and Alexa}{Bunge et~al\mbox{.}}{2021}]%
        {bunge2021diamond}
\bibfield{author}{\bibinfo{person}{Astrid Bunge}, \bibinfo{person}{Mario Botsch}, {and} \bibinfo{person}{Marc Alexa}.} \bibinfo{year}{2021}\natexlab{}.
\newblock \showarticletitle{The Diamond Laplace for Polygonal and Polyhedral Meshes}. In \bibinfo{booktitle}{\emph{Computer Graphics Forum}}, Vol.~\bibinfo{volume}{40}. Wiley Online Library, \bibinfo{pages}{217--230}.
\newblock


\bibitem[\protect\citeauthoryear{Casquero, Zhang, Bona-Casas, Dalcin, and Gomez}{Casquero et~al\mbox{.}}{2018}]%
        {Casquero2018}
\bibfield{author}{\bibinfo{person}{Hugo Casquero}, \bibinfo{person}{Yongjie~Jessica Zhang}, \bibinfo{person}{Carles Bona-Casas}, \bibinfo{person}{Lisandro Dalcin}, {and} \bibinfo{person}{Hector Gomez}.} \bibinfo{year}{2018}\natexlab{}.
\newblock \showarticletitle{Non-body-fitted fluid--structure interaction: Divergence-conforming B-splines, fully-implicit dynamics, and variational formulation}.
\newblock \bibinfo{journal}{\emph{J. Comput. Phys.}}  \bibinfo{volume}{374} (\bibinfo{year}{2018}), \bibinfo{pages}{625--653}.
\newblock


\bibitem[\protect\citeauthoryear{Cockburn, Li, and Shu}{Cockburn et~al\mbox{.}}{2004}]%
        {Cockburn2004}
\bibfield{author}{\bibinfo{person}{Bernardo Cockburn}, \bibinfo{person}{Fengyan Li}, {and} \bibinfo{person}{Chi-Wang Shu}.} \bibinfo{year}{2004}\natexlab{}.
\newblock \showarticletitle{Locally divergence-free discontinuous Galerkin methods for the Maxwell equations}.
\newblock \bibinfo{journal}{\emph{J. Comput. Phys.}} \bibinfo{volume}{194}, \bibinfo{number}{2} (\bibinfo{year}{2004}), \bibinfo{pages}{588--610}.
\newblock


\bibitem[\protect\citeauthoryear{Cui, Langlois, Sen, and Kim}{Cui et~al\mbox{.}}{2021}]%
        {cui2021spiral}
\bibfield{author}{\bibinfo{person}{Qiaodong Cui}, \bibinfo{person}{Timothy Langlois}, \bibinfo{person}{Pradeep Sen}, {and} \bibinfo{person}{Theodore Kim}.} \bibinfo{year}{2021}\natexlab{}.
\newblock \showarticletitle{Spiral-spectral fluid simulation}.
\newblock \bibinfo{journal}{\emph{ACM Transactions on Graphics (TOG)}} \bibinfo{volume}{40}, \bibinfo{number}{6} (\bibinfo{year}{2021}), \bibinfo{pages}{1--16}.
\newblock


\bibitem[\protect\citeauthoryear{Da, Hahn, Batty, Wojtan, and Grinspun}{Da et~al\mbox{.}}{2016}]%
        {Da2016}
\bibfield{author}{\bibinfo{person}{Fang Da}, \bibinfo{person}{David Hahn}, \bibinfo{person}{Christopher Batty}, \bibinfo{person}{Chris Wojtan}, {and} \bibinfo{person}{Eitan Grinspun}.} \bibinfo{year}{2016}\natexlab{}.
\newblock \showarticletitle{Surface-only liquids}.
\newblock \bibinfo{journal}{\emph{ACM Transactions on Graphics (TOG)}} \bibinfo{volume}{35}, \bibinfo{number}{4} (\bibinfo{year}{2016}), \bibinfo{pages}{1--12}.
\newblock


\bibitem[\protect\citeauthoryear{De~Goes, Desbrun, Meyer, and DeRose}{De~Goes et~al\mbox{.}}{2016}]%
        {DeGoes2016}
\bibfield{author}{\bibinfo{person}{Fernando De~Goes}, \bibinfo{person}{Mathieu Desbrun}, \bibinfo{person}{Mark Meyer}, {and} \bibinfo{person}{Tony DeRose}.} \bibinfo{year}{2016}\natexlab{}.
\newblock \showarticletitle{Subdivision exterior calculus for geometry processing}.
\newblock \bibinfo{journal}{\emph{ACM Transactions on Graphics (TOG)}} \bibinfo{volume}{35}, \bibinfo{number}{4} (\bibinfo{year}{2016}), \bibinfo{pages}{1--11}.
\newblock


\bibitem[\protect\citeauthoryear{De~Witt, Lessig, and Fiume}{De~Witt et~al\mbox{.}}{2012}]%
        {DeWitt2012}
\bibfield{author}{\bibinfo{person}{Tyler De~Witt}, \bibinfo{person}{Christian Lessig}, {and} \bibinfo{person}{Eugene Fiume}.} \bibinfo{year}{2012}\natexlab{}.
\newblock \showarticletitle{Fluid simulation using Laplacian eigenfunctions}.
\newblock \bibinfo{journal}{\emph{ACM Transactions on Graphics (TOG)}} \bibinfo{volume}{31}, \bibinfo{number}{1} (\bibinfo{year}{2012}), \bibinfo{pages}{10}.
\newblock


\bibitem[\protect\citeauthoryear{DeWolf}{DeWolf}{2006}]%
        {DeWolf2006}
\bibfield{author}{\bibinfo{person}{Ivan DeWolf}.} \bibinfo{year}{2006}\natexlab{}.
\newblock \bibinfo{booktitle}{\emph{Divergence-free noise}}.
\newblock \bibinfo{type}{{T}echnical {R}eport}. \bibinfo{institution}{Martian Labs., 2005.}
\newblock


\bibitem[\protect\citeauthoryear{Eisenberger, L{\"a}hner, and Cremers}{Eisenberger et~al\mbox{.}}{2018}]%
        {Eisenberger2018}
\bibfield{author}{\bibinfo{person}{Marvin Eisenberger}, \bibinfo{person}{Zorah L{\"a}hner}, {and} \bibinfo{person}{Daniel Cremers}.} \bibinfo{year}{2018}\natexlab{}.
\newblock \showarticletitle{Divergence-Free Shape Interpolation and Correspondence}.
\newblock \bibinfo{journal}{\emph{arXiv preprint arXiv:1806.10417}} (\bibinfo{year}{2018}).
\newblock


\bibitem[\protect\citeauthoryear{Elcott, Tong, Kanso, Schr{\"o}der, and Desbrun}{Elcott et~al\mbox{.}}{2007}]%
        {Elcott2007}
\bibfield{author}{\bibinfo{person}{Sharif Elcott}, \bibinfo{person}{Yiying Tong}, \bibinfo{person}{Eva Kanso}, \bibinfo{person}{Peter Schr{\"o}der}, {and} \bibinfo{person}{Mathieu Desbrun}.} \bibinfo{year}{2007}\natexlab{}.
\newblock \showarticletitle{Stable, circulation-preserving, simplicial fluids}.
\newblock \bibinfo{journal}{\emph{ACM Transactions on Graphics (TOG)}} \bibinfo{volume}{26}, \bibinfo{number}{1} (\bibinfo{year}{2007}), \bibinfo{pages}{4}.
\newblock


\bibitem[\protect\citeauthoryear{Enright, Marschner, and Fedkiw}{Enright et~al\mbox{.}}{2002}]%
        {Enright2002}
\bibfield{author}{\bibinfo{person}{Douglas Enright}, \bibinfo{person}{Stephen Marschner}, {and} \bibinfo{person}{Ronald Fedkiw}.} \bibinfo{year}{2002}\natexlab{}.
\newblock \showarticletitle{Animation and rendering of complex water surfaces}.
\newblock \bibinfo{journal}{\emph{ACM Transactions on Graphics (TOG)}} \bibinfo{volume}{21}, \bibinfo{number}{3} (\bibinfo{year}{2002}), \bibinfo{pages}{736--744}.
\newblock


\bibitem[\protect\citeauthoryear{Evans and Hughes}{Evans and Hughes}{2013}]%
        {Evans2013}
\bibfield{author}{\bibinfo{person}{John~A Evans} {and} \bibinfo{person}{Thomas~JR Hughes}.} \bibinfo{year}{2013}\natexlab{}.
\newblock \showarticletitle{Isogeometric divergence-conforming B-splines for the unsteady Navier--Stokes equations}.
\newblock \bibinfo{journal}{\emph{J. Comput. Phys.}}  \bibinfo{volume}{241} (\bibinfo{year}{2013}), \bibinfo{pages}{141--167}.
\newblock


\bibitem[\protect\citeauthoryear{Fedkiw, Stam, and Jensen}{Fedkiw et~al\mbox{.}}{2001}]%
        {Fedkiw2001}
\bibfield{author}{\bibinfo{person}{Ronald Fedkiw}, \bibinfo{person}{Jos Stam}, {and} \bibinfo{person}{Henrik~Wann Jensen}.} \bibinfo{year}{2001}\natexlab{}.
\newblock \showarticletitle{Visual simulation of smoke}. In \bibinfo{booktitle}{\emph{Proceedings of the 28th annual conference on Computer graphics and interactive techniques}}. \bibinfo{pages}{15--22}.
\newblock


\bibitem[\protect\citeauthoryear{Fritsch and Carlson}{Fritsch and Carlson}{1980}]%
        {Fritsch1980}
\bibfield{author}{\bibinfo{person}{F.~N. Fritsch} {and} \bibinfo{person}{R.~E. Carlson}.} \bibinfo{year}{1980}\natexlab{}.
\newblock \showarticletitle{Monotone Piecewise Cubic Interpolation}.
\newblock \bibinfo{journal}{\emph{SIAM J. Numer. Anal.}} \bibinfo{volume}{17}, \bibinfo{number}{2} (\bibinfo{year}{1980}), \bibinfo{pages}{238--246}.
\newblock


\bibitem[\protect\citeauthoryear{Guennebaud, Jacob, et~al\mbox{.}}{Guennebaud et~al\mbox{.}}{2010}]%
        {Eigenweb}
\bibfield{author}{\bibinfo{person}{Ga\"{e}l Guennebaud}, \bibinfo{person}{Beno\^{i}t Jacob}, {et~al\mbox{.}}} \bibinfo{year}{2010}\natexlab{}.
\newblock \bibinfo{title}{Eigen v3}.
\newblock \bibinfo{howpublished}{http://eigen.tuxfamily.org}.
\newblock


\bibitem[\protect\citeauthoryear{Guzm{\'a}n and Neilan}{Guzm{\'a}n and Neilan}{2014}]%
        {Guzman2014}
\bibfield{author}{\bibinfo{person}{Johnny Guzm{\'a}n} {and} \bibinfo{person}{Michael Neilan}.} \bibinfo{year}{2014}\natexlab{}.
\newblock \showarticletitle{Conforming and divergence-free Stokes elements in three dimensions}.
\newblock \bibinfo{journal}{\emph{IMA J. Numer. Anal.}} \bibinfo{volume}{34}, \bibinfo{number}{4} (\bibinfo{year}{2014}), \bibinfo{pages}{1489--1508}.
\newblock


\bibitem[\protect\citeauthoryear{Houston, Bond, and Wiebe}{Houston et~al\mbox{.}}{2003}]%
        {Houston2003}
\bibfield{author}{\bibinfo{person}{Ben Houston}, \bibinfo{person}{Chris Bond}, {and} \bibinfo{person}{Mark Wiebe}.} \bibinfo{year}{2003}\natexlab{}.
\newblock \showarticletitle{A unified approach for modeling complex occlusions in fluid simulations}.
\newblock In \bibinfo{booktitle}{\emph{ACM SIGGRAPH 2003 Sketches}}. \bibinfo{pages}{1--1}.
\newblock


\bibitem[\protect\citeauthoryear{Huerta, Vidal, and Villon}{Huerta et~al\mbox{.}}{2004}]%
        {Huerta2004}
\bibfield{author}{\bibinfo{person}{Antonio Huerta}, \bibinfo{person}{Yolanda Vidal}, {and} \bibinfo{person}{Pierre Villon}.} \bibinfo{year}{2004}\natexlab{}.
\newblock \showarticletitle{Pseudo-divergence-free element free Galerkin method for incompressible fluid flow}.
\newblock \bibinfo{journal}{\emph{Computer Methods in Applied Mechanics and Engineering}} \bibinfo{volume}{193}, \bibinfo{number}{12} (\bibinfo{year}{2004}), \bibinfo{pages}{1119 -- 1136}.
\newblock
\showISSN{0045-7825}


\bibitem[\protect\citeauthoryear{Jenny, Pope, Muradoglu, and Caughey}{Jenny et~al\mbox{.}}{2001}]%
        {Jenny2001}
\bibfield{author}{\bibinfo{person}{P Jenny}, \bibinfo{person}{SB Pope}, \bibinfo{person}{M Muradoglu}, {and} \bibinfo{person}{DA Caughey}.} \bibinfo{year}{2001}\natexlab{}.
\newblock \showarticletitle{A hybrid algorithm for the joint PDF equation of turbulent reactive flows}.
\newblock \bibinfo{journal}{\emph{J. Comput. Phys.}} \bibinfo{volume}{166}, \bibinfo{number}{2} (\bibinfo{year}{2001}), \bibinfo{pages}{218--252}.
\newblock


\bibitem[\protect\citeauthoryear{Jiang, Schroeder, Selle, Teran, and Stomakhin}{Jiang et~al\mbox{.}}{2015}]%
        {Jiang2015}
\bibfield{author}{\bibinfo{person}{Chenfanfu Jiang}, \bibinfo{person}{Craig Schroeder}, \bibinfo{person}{Andrew Selle}, \bibinfo{person}{Joseph Teran}, {and} \bibinfo{person}{Alexey Stomakhin}.} \bibinfo{year}{2015}\natexlab{}.
\newblock \showarticletitle{The affine particle-in-cell method}.
\newblock \bibinfo{journal}{\emph{ACM Transactions on Graphics (TOG)}} \bibinfo{volume}{34}, \bibinfo{number}{4} (\bibinfo{year}{2015}), \bibinfo{pages}{51}.
\newblock


\bibitem[\protect\citeauthoryear{Jiang, Schroeder, Teran, Stomakhin, and Selle}{Jiang et~al\mbox{.}}{2016}]%
        {Jiang2016}
\bibfield{author}{\bibinfo{person}{Chenfanfu Jiang}, \bibinfo{person}{Craig Schroeder}, \bibinfo{person}{Joseph Teran}, \bibinfo{person}{Alexey Stomakhin}, {and} \bibinfo{person}{Andrew Selle}.} \bibinfo{year}{2016}\natexlab{}.
\newblock \showarticletitle{The material point method for simulating continuum materials}.
\newblock In \bibinfo{booktitle}{\emph{ACM SIGGRAPH 2016 Courses}}. \bibinfo{pages}{1--52}.
\newblock


\bibitem[\protect\citeauthoryear{Kim, Th\"{u}rey, James, and Gross}{Kim et~al\mbox{.}}{2008}]%
        {Kim2008}
\bibfield{author}{\bibinfo{person}{Theodore Kim}, \bibinfo{person}{Nils Th\"{u}rey}, \bibinfo{person}{Doug James}, {and} \bibinfo{person}{Markus Gross}.} \bibinfo{year}{2008}\natexlab{}.
\newblock \showarticletitle{Wavelet Turbulence for Fluid Simulation}.
\newblock \bibinfo{journal}{\emph{ACM Trans. Graph.}} \bibinfo{volume}{27}, \bibinfo{number}{3} (\bibinfo{date}{Aug.} \bibinfo{year}{2008}), \bibinfo{pages}{1–6}.
\newblock
\showISSN{0730-0301}
\urldef\tempurl%
\url{https://doi.org/10.1145/1360612.1360649}
\showDOI{\tempurl}


\bibitem[\protect\citeauthoryear{Kugelstadt, Longva, Thurey, and Bender}{Kugelstadt et~al\mbox{.}}{2019}]%
        {Kugelstadt2019}
\bibfield{author}{\bibinfo{person}{Tassilo Kugelstadt}, \bibinfo{person}{Andreas Longva}, \bibinfo{person}{Nils Thurey}, {and} \bibinfo{person}{Jan Bender}.} \bibinfo{year}{2019}\natexlab{}.
\newblock \showarticletitle{Implicit Density Projection for Volume Conserving Liquids}.
\newblock \bibinfo{journal}{\emph{IEEE Computer Architecture Letters}} \bibinfo{number}{01} (\bibinfo{year}{2019}), \bibinfo{pages}{1--1}.
\newblock


\bibitem[\protect\citeauthoryear{Larionov, Batty, and Bridson}{Larionov et~al\mbox{.}}{2017}]%
        {Larionov2017}
\bibfield{author}{\bibinfo{person}{Egor Larionov}, \bibinfo{person}{Christopher Batty}, {and} \bibinfo{person}{Robert Bridson}.} \bibinfo{year}{2017}\natexlab{}.
\newblock \showarticletitle{Variational stokes: a unified pressure-viscosity solver for accurate viscous liquids}.
\newblock \bibinfo{journal}{\emph{ACM Transactions on Graphics (TOG)}} \bibinfo{volume}{36}, \bibinfo{number}{4} (\bibinfo{year}{2017}), \bibinfo{pages}{101}.
\newblock


\bibitem[\protect\citeauthoryear{Lederer, Linke, Merdon, and Schoberl}{Lederer et~al\mbox{.}}{2017}]%
        {Lederer2017}
\bibfield{author}{\bibinfo{person}{Philip~L Lederer}, \bibinfo{person}{Alexander Linke}, \bibinfo{person}{Christian Merdon}, {and} \bibinfo{person}{Joachim Schoberl}.} \bibinfo{year}{2017}\natexlab{}.
\newblock \showarticletitle{Divergence-free reconstruction operators for pressure-robust Stokes discretizations with continuous pressure finite elements}.
\newblock \bibinfo{journal}{\emph{SIAM J. Numer. Anal.}} \bibinfo{volume}{55}, \bibinfo{number}{3} (\bibinfo{year}{2017}), \bibinfo{pages}{1291--1314}.
\newblock


\bibitem[\protect\citeauthoryear{Lehrenfeld and Sch{\"o}berl}{Lehrenfeld and Sch{\"o}berl}{2016}]%
        {Lehrenfeld2016}
\bibfield{author}{\bibinfo{person}{Christoph Lehrenfeld} {and} \bibinfo{person}{Joachim Sch{\"o}berl}.} \bibinfo{year}{2016}\natexlab{}.
\newblock \showarticletitle{High order exactly divergence-free hybrid discontinuous Galerkin methods for unsteady incompressible flows}.
\newblock \bibinfo{journal}{\emph{Computer Methods in Applied Mechanics and Engineering}}  \bibinfo{volume}{307} (\bibinfo{year}{2016}), \bibinfo{pages}{339--361}.
\newblock


\bibitem[\protect\citeauthoryear{Linke}{Linke}{2012}]%
        {Linke2012}
\bibfield{author}{\bibinfo{person}{Alexander Linke}.} \bibinfo{year}{2012}\natexlab{}.
\newblock \showarticletitle{A divergence-free velocity reconstruction for incompressible flows}.
\newblock \bibinfo{journal}{\emph{Comptes Rendus Mathematique}} \bibinfo{volume}{350}, \bibinfo{number}{17-18} (\bibinfo{year}{2012}), \bibinfo{pages}{837--840}.
\newblock


\bibitem[\protect\citeauthoryear{Lipnikov, Shashkov, and Svyatskiy}{Lipnikov et~al\mbox{.}}{2006}]%
        {Lipnikov2006}
\bibfield{author}{\bibinfo{person}{Konstantin Lipnikov}, \bibinfo{person}{Mikhail Shashkov}, {and} \bibinfo{person}{Daniil Svyatskiy}.} \bibinfo{year}{2006}\natexlab{}.
\newblock \showarticletitle{The mimetic finite difference discretization of diffusion problem on unstructured polyhedral meshes}.
\newblock \bibinfo{journal}{\emph{J. Comput. Phys.}} \bibinfo{volume}{211}, \bibinfo{number}{2} (\bibinfo{year}{2006}), \bibinfo{pages}{473--491}.
\newblock


\bibitem[\protect\citeauthoryear{Lowitzsch}{Lowitzsch}{2005}]%
        {Lowitzsch2005}
\bibfield{author}{\bibinfo{person}{Svenja Lowitzsch}.} \bibinfo{year}{2005}\natexlab{}.
\newblock \showarticletitle{Matrix-valued radial basis functions: stability estimates and applications}.
\newblock \bibinfo{journal}{\emph{Advances in Computational Mathematics}} \bibinfo{volume}{23}, \bibinfo{number}{3} (\bibinfo{year}{2005}), \bibinfo{pages}{299--315}.
\newblock
\showISBNx{1572-9044}
\urldef\tempurl%
\url{https://doi.org/10.1007/s10444-004-1786-8}
\showDOI{\tempurl}


\bibitem[\protect\citeauthoryear{Madan and Levin}{Madan and Levin}{2022}]%
        {madan2021fast}
\bibfield{author}{\bibinfo{person}{Abhishek Madan} {and} \bibinfo{person}{David~IW Levin}.} \bibinfo{year}{2022}\natexlab{}.
\newblock \showarticletitle{Fast Evaluation of Smooth Distance Constraints on Co-Dimensional Geometry}.
\newblock \bibinfo{journal}{\emph{ACM Transactions on Graphics (TOG)}}  \bibinfo{volume}{(to appear)} (\bibinfo{year}{2022}).
\newblock


\bibitem[\protect\citeauthoryear{Maljaars, Labeur, and M{\"o}ller}{Maljaars et~al\mbox{.}}{2018}]%
        {Maljaars2018}
\bibfield{author}{\bibinfo{person}{Jakob~M Maljaars}, \bibinfo{person}{Robert~Jan Labeur}, {and} \bibinfo{person}{Matthias M{\"o}ller}.} \bibinfo{year}{2018}\natexlab{}.
\newblock \showarticletitle{A hybridized discontinuous Galerkin framework for high-order particle--mesh operator splitting of the incompressible Navier--Stokes equations}.
\newblock \bibinfo{journal}{\emph{J. Comput. Phys.}}  \bibinfo{volume}{358} (\bibinfo{year}{2018}), \bibinfo{pages}{150--172}.
\newblock


\bibitem[\protect\citeauthoryear{Manges and Cendes}{Manges and Cendes}{1995}]%
        {Manges1995}
\bibfield{author}{\bibinfo{person}{J.B. Manges} {and} \bibinfo{person}{Z.J. Cendes}.} \bibinfo{year}{1995}\natexlab{}.
\newblock \showarticletitle{{A generalized tree-cotree gauge for magnetic field computation}}.
\newblock \bibinfo{journal}{\emph{IEEE Transactions on Magnetics}} \bibinfo{volume}{31}, \bibinfo{number}{3} (\bibinfo{date}{may} \bibinfo{year}{1995}), \bibinfo{pages}{1342--1347}.
\newblock
\showISSN{00189464}


\bibitem[\protect\citeauthoryear{McNally}{McNally}{2011}]%
        {McNally2011}
\bibfield{author}{\bibinfo{person}{Colin~P McNally}.} \bibinfo{year}{2011}\natexlab{}.
\newblock \showarticletitle{Divergence-free interpolation of vector fields from point values—exact div-B= 0 in numerical simulations}.
\newblock \bibinfo{journal}{\emph{Monthly Notices of the Royal Astronomical Society: Letters}} \bibinfo{volume}{413}, \bibinfo{number}{1} (\bibinfo{year}{2011}), \bibinfo{pages}{L76--L80}.
\newblock


\bibitem[\protect\citeauthoryear{Meyer and Jenny}{Meyer and Jenny}{2004}]%
        {Meyer2004}
\bibfield{author}{\bibinfo{person}{DW Meyer} {and} \bibinfo{person}{P Jenny}.} \bibinfo{year}{2004}\natexlab{}.
\newblock \showarticletitle{Conservative velocity interpolation for PDF methods}. In \bibinfo{booktitle}{\emph{PAMM: Proceedings in Applied Mathematics and Mechanics}}, Vol.~\bibinfo{volume}{4}. Wiley Online Library, \bibinfo{pages}{466--467}.
\newblock


\bibitem[\protect\citeauthoryear{Ng, Min, and Gibou}{Ng et~al\mbox{.}}{2009}]%
        {Ng2009}
\bibfield{author}{\bibinfo{person}{Yen~Ting Ng}, \bibinfo{person}{Chohong Min}, {and} \bibinfo{person}{Fr{\'e}d{\'e}ric Gibou}.} \bibinfo{year}{2009}\natexlab{}.
\newblock \showarticletitle{An Efficient Fluid-solid Coupling Algorithm for Single-phase Flows}.
\newblock \bibinfo{journal}{\emph{J. Comput. Phys.}} \bibinfo{volume}{228}, \bibinfo{number}{23} (\bibinfo{date}{Dec.} \bibinfo{year}{2009}), \bibinfo{pages}{8807--8829}.
\newblock
\showISSN{0021-9991}


\bibitem[\protect\citeauthoryear{Pan, Huang, Tong, Zheng, and Bao}{Pan et~al\mbox{.}}{2013}]%
        {Pan2013}
\bibfield{author}{\bibinfo{person}{Zherong Pan}, \bibinfo{person}{Jin Huang}, \bibinfo{person}{Yiying Tong}, \bibinfo{person}{Changxi Zheng}, {and} \bibinfo{person}{Hujun Bao}.} \bibinfo{year}{2013}\natexlab{}.
\newblock \showarticletitle{Interactive localized liquid motion editing}.
\newblock \bibinfo{journal}{\emph{ACM Transactions on Graphics (TOG)}} \bibinfo{volume}{32}, \bibinfo{number}{6} (\bibinfo{year}{2013}), \bibinfo{pages}{1--10}.
\newblock


\bibitem[\protect\citeauthoryear{Park and Kim}{Park and Kim}{2005}]%
        {Park2005}
\bibfield{author}{\bibinfo{person}{Sang~Il Park} {and} \bibinfo{person}{Myoung~Jun Kim}.} \bibinfo{year}{2005}\natexlab{}.
\newblock \showarticletitle{Vortex fluid for gaseous phenomena}. In \bibinfo{booktitle}{\emph{Proceedings of the 2005 ACM SIGGRAPH/Eurographics symposium on Computer animation}}. ACM, \bibinfo{pages}{261--270}.
\newblock


\bibitem[\protect\citeauthoryear{Peskin}{Peskin}{2002}]%
        {Peskin2002}
\bibfield{author}{\bibinfo{person}{Charles~S Peskin}.} \bibinfo{year}{2002}\natexlab{}.
\newblock \showarticletitle{The immersed boundary method}.
\newblock \bibinfo{journal}{\emph{Acta numerica}}  \bibinfo{volume}{11} (\bibinfo{year}{2002}), \bibinfo{pages}{479--517}.
\newblock


\bibitem[\protect\citeauthoryear{Pfaff, Thuerey, and Gross}{Pfaff et~al\mbox{.}}{2012}]%
        {Pfaff2012}
\bibfield{author}{\bibinfo{person}{Tobias Pfaff}, \bibinfo{person}{Nils Thuerey}, {and} \bibinfo{person}{Markus Gross}.} \bibinfo{year}{2012}\natexlab{}.
\newblock \showarticletitle{Lagrangian vortex sheets for animating fluids}.
\newblock \bibinfo{journal}{\emph{ACM Transactions on Graphics (TOG)}} \bibinfo{volume}{31}, \bibinfo{number}{4} (\bibinfo{year}{2012}), \bibinfo{pages}{112}.
\newblock


\bibitem[\protect\citeauthoryear{Poelke and Polthier}{Poelke and Polthier}{2016}]%
        {Poelke2016}
\bibfield{author}{\bibinfo{person}{Konstantin Poelke} {and} \bibinfo{person}{Konrad Polthier}.} \bibinfo{year}{2016}\natexlab{}.
\newblock \showarticletitle{Boundary-aware Hodge decompositions for piecewise constant vector fields}.
\newblock \bibinfo{journal}{\emph{Computer-Aided Design}}  \bibinfo{volume}{78} (\bibinfo{year}{2016}), \bibinfo{pages}{126--136}.
\newblock


\bibitem[\protect\citeauthoryear{Pusok, Kaus, and Popov}{Pusok et~al\mbox{.}}{2017}]%
        {Pusok2017}
\bibfield{author}{\bibinfo{person}{Adina~E Pusok}, \bibinfo{person}{Boris~JP Kaus}, {and} \bibinfo{person}{Anton~A Popov}.} \bibinfo{year}{2017}\natexlab{}.
\newblock \showarticletitle{On the quality of velocity interpolation schemes for marker-in-cell method and staggered grids}.
\newblock \bibinfo{journal}{\emph{Pure and Applied Geophysics}} \bibinfo{volume}{174}, \bibinfo{number}{3} (\bibinfo{year}{2017}), \bibinfo{pages}{1071--1089}.
\newblock


\bibitem[\protect\citeauthoryear{Raateland, Hadrich, Herrera, Banuti, Pa{\l}ubicki, Pirk, Hildebrandt, and Michels}{Raateland et~al\mbox{.}}{2022}]%
        {raateland2022dcgrid}
\bibfield{author}{\bibinfo{person}{Wouter Raateland}, \bibinfo{person}{Torsten Hadrich}, \bibinfo{person}{Jorge Alejandro~Amador Herrera}, \bibinfo{person}{Daniel~T Banuti}, \bibinfo{person}{Wojciech Pa{\l}ubicki}, \bibinfo{person}{S{\"o}ren Pirk}, \bibinfo{person}{Klaus Hildebrandt}, {and} \bibinfo{person}{Dominik~L Michels}.} \bibinfo{year}{2022}\natexlab{}.
\newblock \showarticletitle{DCGrid: An Adaptive Grid Structure for Memory-Constrained Fluid Simulation on the GPU}.
\newblock  (\bibinfo{year}{2022}).
\newblock


\bibitem[\protect\citeauthoryear{Rasmussen, Enright, Nguyen, Marino, Sumner, Geiger, Hoon, and Fedkiw}{Rasmussen et~al\mbox{.}}{2004}]%
        {Rasmussen2004}
\bibfield{author}{\bibinfo{person}{Nick Rasmussen}, \bibinfo{person}{Douglas Enright}, \bibinfo{person}{Duc Nguyen}, \bibinfo{person}{Sebastian Marino}, \bibinfo{person}{Nigel Sumner}, \bibinfo{person}{Willi Geiger}, \bibinfo{person}{Samir Hoon}, {and} \bibinfo{person}{Ronald Fedkiw}.} \bibinfo{year}{2004}\natexlab{}.
\newblock \showarticletitle{Directable photorealistic liquids}. In \bibinfo{booktitle}{\emph{Proceedings of the 2004 ACM SIGGRAPH/Eurographics symposium on Computer animation}}. \bibinfo{pages}{193--202}.
\newblock


\bibitem[\protect\citeauthoryear{Ravu, Rudman, Metcalfe, Lester, and Khakhar}{Ravu et~al\mbox{.}}{2016}]%
        {Ravu2016}
\bibfield{author}{\bibinfo{person}{Bharath Ravu}, \bibinfo{person}{Murray Rudman}, \bibinfo{person}{Guy Metcalfe}, \bibinfo{person}{Daniel Lester}, {and} \bibinfo{person}{Devang Khakhar}.} \bibinfo{year}{2016}\natexlab{}.
\newblock \showarticletitle{Creating analytically divergence-free velocity fields from grid-based data}.
\newblock \bibinfo{journal}{\emph{J. Comput. Phys.}}  \bibinfo{volume}{323} (\bibinfo{date}{07} \bibinfo{year}{2016}).
\newblock
\urldef\tempurl%
\url{https://doi.org/10.1016/j.jcp.2016.07.018}
\showDOI{\tempurl}


\bibitem[\protect\citeauthoryear{Rhebergen and Wells}{Rhebergen and Wells}{2018}]%
        {Rhebergen2018}
\bibfield{author}{\bibinfo{person}{Sander Rhebergen} {and} \bibinfo{person}{Garth~N Wells}.} \bibinfo{year}{2018}\natexlab{}.
\newblock \showarticletitle{A hybridizable discontinuous Galerkin method for the Navier--Stokes equations with pointwise divergence-free velocity field}.
\newblock \bibinfo{journal}{\emph{Journal of Scientific Computing}} \bibinfo{volume}{76}, \bibinfo{number}{3} (\bibinfo{year}{2018}), \bibinfo{pages}{1484--1501}.
\newblock


\bibitem[\protect\citeauthoryear{Sato, Dobashi, and Kim}{Sato et~al\mbox{.}}{2021}]%
        {Sato2021}
\bibfield{author}{\bibinfo{person}{Syuhei Sato}, \bibinfo{person}{Yoshinori Dobashi}, {and} \bibinfo{person}{Theodore Kim}.} \bibinfo{year}{2021}\natexlab{}.
\newblock \showarticletitle{Stream-Guided Smoke Simulation}.
\newblock \bibinfo{journal}{\emph{ACM Transactions on Graphics (TOG)}} \bibinfo{volume}{40}, \bibinfo{number}{4}, Article \bibinfo{articleno}{161} (\bibinfo{year}{2021}).
\newblock


\bibitem[\protect\citeauthoryear{Sato, Dobashi, and Nishita}{Sato et~al\mbox{.}}{2018a}]%
        {Sato2018}
\bibfield{author}{\bibinfo{person}{Syuhei Sato}, \bibinfo{person}{Yoshinori Dobashi}, {and} \bibinfo{person}{Tomoyuki Nishita}.} \bibinfo{year}{2018}\natexlab{a}.
\newblock \showarticletitle{Editing fluid animation using flow interpolation}.
\newblock \bibinfo{journal}{\emph{ACM Transactions on Graphics (TOG)}} \bibinfo{volume}{37}, \bibinfo{number}{5} (\bibinfo{year}{2018}), \bibinfo{pages}{1--12}.
\newblock


\bibitem[\protect\citeauthoryear{Sato, Dobashi, Yue, Iwasaki, and Nishita}{Sato et~al\mbox{.}}{2015}]%
        {Sato2015}
\bibfield{author}{\bibinfo{person}{Syuhei Sato}, \bibinfo{person}{Yoshinori Dobashi}, \bibinfo{person}{Yonghao Yue}, \bibinfo{person}{Kei Iwasaki}, {and} \bibinfo{person}{Tomoyuki Nishita}.} \bibinfo{year}{2015}\natexlab{}.
\newblock \showarticletitle{Incompressibility-preserving deformation for fluid flows using vector potentials}.
\newblock \bibinfo{journal}{\emph{The Visual Computer}} \bibinfo{volume}{31}, \bibinfo{number}{6} (\bibinfo{year}{2015}), \bibinfo{pages}{959--965}.
\newblock


\bibitem[\protect\citeauthoryear{Sato, Wojtan, Thuerey, Igarashi, and Ando}{Sato et~al\mbox{.}}{2018b}]%
        {Sato2018b}
\bibfield{author}{\bibinfo{person}{Takahiro Sato}, \bibinfo{person}{Christopher Wojtan}, \bibinfo{person}{Nils Thuerey}, \bibinfo{person}{Takeo Igarashi}, {and} \bibinfo{person}{Ryoichi Ando}.} \bibinfo{year}{2018}\natexlab{b}.
\newblock \showarticletitle{Extended narrow band FLIP for liquid simulations}. In \bibinfo{booktitle}{\emph{Computer Graphics Forum}}, Vol.~\bibinfo{volume}{37}. Wiley Online Library, \bibinfo{pages}{169--177}.
\newblock


\bibitem[\protect\citeauthoryear{Schechter and Bridson}{Schechter and Bridson}{2008}]%
        {Schechter2008}
\bibfield{author}{\bibinfo{person}{Hagit Schechter} {and} \bibinfo{person}{Robert Bridson}.} \bibinfo{year}{2008}\natexlab{}.
\newblock \showarticletitle{Evolving sub-grid turbulence for smoke animation}. In \bibinfo{booktitle}{\emph{Proceedings of the 2008 ACM SIGGRAPH/Eurographics symposium on Computer animation}}. Eurographics Association, \bibinfo{pages}{1--7}.
\newblock


\bibitem[\protect\citeauthoryear{Schroeder, Chowdhury, and Shinar}{Schroeder et~al\mbox{.}}{2022}]%
        {schroeder2022}
\bibfield{author}{\bibinfo{person}{Craig Schroeder}, \bibinfo{person}{Ritoban~Roy Chowdhury}, {and} \bibinfo{person}{Tamar Shinar}.} \bibinfo{year}{2022}\natexlab{}.
\newblock \showarticletitle{Local divergence-free polynomial interpolation on MAC grids}.
\newblock \bibinfo{journal}{\emph{J. Comput. Phys.}}  \bibinfo{volume}{468} (\bibinfo{year}{2022}), \bibinfo{pages}{111500}.
\newblock


\bibitem[\protect\citeauthoryear{Selle, Fedkiw, Kim, Liu, and Rossignac}{Selle et~al\mbox{.}}{2008}]%
        {Selle2008}
\bibfield{author}{\bibinfo{person}{Andrew Selle}, \bibinfo{person}{Ronald Fedkiw}, \bibinfo{person}{Byungmoon Kim}, \bibinfo{person}{Yingjie Liu}, {and} \bibinfo{person}{Jarek Rossignac}.} \bibinfo{year}{2008}\natexlab{}.
\newblock \showarticletitle{An unconditionally stable MacCormack method}.
\newblock \bibinfo{journal}{\emph{Journal of Scientific Computing}} \bibinfo{volume}{35}, \bibinfo{number}{2-3} (\bibinfo{year}{2008}), \bibinfo{pages}{350--371}.
\newblock


\bibitem[\protect\citeauthoryear{Shao, Huang, and Michels}{Shao et~al\mbox{.}}{2022}]%
        {Shao:2022:Multigrid}
\bibfield{author}{\bibinfo{person}{Han Shao}, \bibinfo{person}{Libo Huang}, {and} \bibinfo{person}{Dominik~L. Michels}.} \bibinfo{year}{2022}\natexlab{}.
\newblock \showarticletitle{A Fast Unsmoothed Aggregation Algebraic Multigrid Framework for the Large-Scale Simulation of Incompressible Flow}.
\newblock \bibinfo{journal}{\emph{ACM Transaction on Graphics}} \bibinfo{volume}{41}, \bibinfo{number}{4}, Article \bibinfo{articleno}{49} (\bibinfo{date}{07} \bibinfo{year}{2022}).
\newblock


\bibitem[\protect\citeauthoryear{Silberman, Adams, Faber, Etienne, and Ruchlin}{Silberman et~al\mbox{.}}{2019}]%
        {Silberman2019}
\bibfield{author}{\bibinfo{person}{Zachary~J. Silberman}, \bibinfo{person}{Thomas~R. Adams}, \bibinfo{person}{Joshua~A. Faber}, \bibinfo{person}{Zachariah~B. Etienne}, {and} \bibinfo{person}{Ian Ruchlin}.} \bibinfo{year}{2019}\natexlab{}.
\newblock \showarticletitle{Numerical generation of vector potentials from specified magnetic fields}.
\newblock \bibinfo{journal}{\emph{J. Comput. Phys.}}  \bibinfo{volume}{379} (\bibinfo{year}{2019}), \bibinfo{pages}{421 -- 437}.
\newblock
\showISSN{0021-9991}
\urldef\tempurl%
\url{https://doi.org/10.1016/j.jcp.2018.12.006}
\showDOI{\tempurl}


\bibitem[\protect\citeauthoryear{Stam}{Stam}{1999}]%
        {Stam1999}
\bibfield{author}{\bibinfo{person}{Jos Stam}.} \bibinfo{year}{1999}\natexlab{}.
\newblock \showarticletitle{Stable Fluids}. In \bibinfo{booktitle}{\emph{Proceedings of the 26th Annual Conference on Computer Graphics and Interactive Techniques}} \emph{(\bibinfo{series}{SIGGRAPH '99})}. \bibinfo{publisher}{ACM Press/Addison-Wesley Publishing Co.}, \bibinfo{address}{New York, NY, USA}, \bibinfo{pages}{121--128}.
\newblock
\showISBNx{0-201-48560-5}
\urldef\tempurl%
\url{https://doi.org/10.1145/311535.311548}
\showDOI{\tempurl}


\bibitem[\protect\citeauthoryear{Stam and Fiume}{Stam and Fiume}{1993}]%
        {Stam1993Turbulent}
\bibfield{author}{\bibinfo{person}{Jos Stam} {and} \bibinfo{person}{Eugene Fiume}.} \bibinfo{year}{1993}\natexlab{}.
\newblock \showarticletitle{Turbulent wind fields for gaseous phenomena}. In \bibinfo{booktitle}{\emph{Proceedings of the 20th annual conference on Computer graphics and interactive techniques}}. \bibinfo{pages}{369--376}.
\newblock


\bibitem[\protect\citeauthoryear{Steffen, Kirby, and Berzins}{Steffen et~al\mbox{.}}{2008}]%
        {Steffen2008}
\bibfield{author}{\bibinfo{person}{Michael Steffen}, \bibinfo{person}{Robert~M Kirby}, {and} \bibinfo{person}{Martin Berzins}.} \bibinfo{year}{2008}\natexlab{}.
\newblock \showarticletitle{Analysis and reduction of quadrature errors in the material point method (MPM)}.
\newblock \bibinfo{journal}{\emph{International journal for numerical methods in engineering}} \bibinfo{volume}{76}, \bibinfo{number}{6} (\bibinfo{year}{2008}), \bibinfo{pages}{922--948}.
\newblock


\bibitem[\protect\citeauthoryear{Takahashi and Lin}{Takahashi and Lin}{2019}]%
        {Takahashi2019}
\bibfield{author}{\bibinfo{person}{Tetsuya Takahashi} {and} \bibinfo{person}{Ming~C Lin}.} \bibinfo{year}{2019}\natexlab{}.
\newblock \showarticletitle{A geometrically consistent viscous fluid solver with two-way fluid-solid coupling}. In \bibinfo{booktitle}{\emph{Computer Graphics Forum}}, Vol.~\bibinfo{volume}{38}. Wiley Online Library, \bibinfo{pages}{49--58}.
\newblock


\bibitem[\protect\citeauthoryear{Tong, Lombeyda, Hirani, and Desbrun}{Tong et~al\mbox{.}}{2003}]%
        {Tong2003}
\bibfield{author}{\bibinfo{person}{Yiying Tong}, \bibinfo{person}{Santiago Lombeyda}, \bibinfo{person}{Anil~N Hirani}, {and} \bibinfo{person}{Mathieu Desbrun}.} \bibinfo{year}{2003}\natexlab{}.
\newblock \showarticletitle{Discrete multiscale vector field decomposition}.
\newblock \bibinfo{journal}{\emph{ACM Transaction on Graphics (TOG)}} \bibinfo{volume}{22}, \bibinfo{number}{3} (\bibinfo{year}{2003}), \bibinfo{pages}{445--452}.
\newblock


\bibitem[\protect\citeauthoryear{Van~Loan}{Van~Loan}{1992}]%
        {van1992computational}
\bibfield{author}{\bibinfo{person}{Charles Van~Loan}.} \bibinfo{year}{1992}\natexlab{}.
\newblock \bibinfo{booktitle}{\emph{Computational frameworks for the fast Fourier transform}}.
\newblock \bibinfo{publisher}{SIAM}.
\newblock


\bibitem[\protect\citeauthoryear{Von~Funck, Theisel, and Seidel}{Von~Funck et~al\mbox{.}}{2006}]%
        {VonFunck2006}
\bibfield{author}{\bibinfo{person}{Wolfram Von~Funck}, \bibinfo{person}{Holger Theisel}, {and} \bibinfo{person}{Hans-Peter Seidel}.} \bibinfo{year}{2006}\natexlab{}.
\newblock \showarticletitle{Vector field based shape deformations}. In \bibinfo{booktitle}{\emph{ACM Transactions on Graphics (TOG)}}, Vol.~\bibinfo{volume}{25}. ACM, \bibinfo{pages}{1118--1125}.
\newblock


\bibitem[\protect\citeauthoryear{Wang, Agrusta, and van Hunen}{Wang et~al\mbox{.}}{2015}]%
        {Wang2015}
\bibfield{author}{\bibinfo{person}{Hongliang Wang}, \bibinfo{person}{Roberto Agrusta}, {and} \bibinfo{person}{Jeroen van Hunen}.} \bibinfo{year}{2015}\natexlab{}.
\newblock \showarticletitle{Advantages of a conservative velocity interpolation (CVI) scheme for particle-in-cell methods with application in geodynamic modeling}.
\newblock \bibinfo{journal}{\emph{Geochemistry, Geophysics, Geosystems}} \bibinfo{volume}{16}, \bibinfo{number}{6} (\bibinfo{year}{2015}), \bibinfo{pages}{2015--2023}.
\newblock


\bibitem[\protect\citeauthoryear{Wang, Tong, Desbrun, and Schr{\"o}der}{Wang et~al\mbox{.}}{2006}]%
        {Wang2006}
\bibfield{author}{\bibinfo{person}{Ke Wang}, \bibinfo{person}{Yiying Tong}, \bibinfo{person}{Mathieu Desbrun}, {and} \bibinfo{person}{Peter Schr{\"o}der}.} \bibinfo{year}{2006}\natexlab{}.
\newblock \showarticletitle{Edge subdivision schemes and the construction of smooth vector fields}.
\newblock \bibinfo{journal}{\emph{ACM Transactions on Graphics (TOG)}} \bibinfo{volume}{25}, \bibinfo{number}{3} (\bibinfo{year}{2006}), \bibinfo{pages}{1041--1048}.
\newblock


\bibitem[\protect\citeauthoryear{Zhao, Desbrun, Wei, and Tong}{Zhao et~al\mbox{.}}{2019}]%
        {Zhao2019}
\bibfield{author}{\bibinfo{person}{Rundong Zhao}, \bibinfo{person}{Mathieu Desbrun}, \bibinfo{person}{Guo-Wei Wei}, {and} \bibinfo{person}{Yiying Tong}.} \bibinfo{year}{2019}\natexlab{}.
\newblock \showarticletitle{3D Hodge decompositions of edge-and face-based vector fields}.
\newblock \bibinfo{journal}{\emph{ACM Transactions on Graphics (TOG)}} \bibinfo{volume}{38}, \bibinfo{number}{6} (\bibinfo{year}{2019}), \bibinfo{pages}{1--13}.
\newblock


\bibitem[\protect\citeauthoryear{Zhu and Bridson}{Zhu and Bridson}{2005}]%
        {Zhu2005}
\bibfield{author}{\bibinfo{person}{Yongning Zhu} {and} \bibinfo{person}{Robert Bridson}.} \bibinfo{year}{2005}\natexlab{}.
\newblock \showarticletitle{Animating sand as a fluid}.
\newblock \bibinfo{journal}{\emph{ACM Transactions on Graphics (TOG)}} \bibinfo{volume}{24}, \bibinfo{number}{3} (\bibinfo{year}{2005}), \bibinfo{pages}{965--972}.
\newblock


\end{thebibliography}

\appendix

\end{document}